%% file: lectures6.tex
\documentclass[aps,nofootinbib,preprintnumbers,eqsecnum]{revtex4-1}

\input HLdefs

\usepackage{tikz}

\newcommand{\vNa}{von Neumann algebra}

\newcommand{\fri}{{\mathfrak{i}}}

\newcommand{\fc}{{\mathfrak{c}}}

\newcommand{\wt}{\widetilde}

\newtheorem{theorem}{Theorem}[section]
\newtheorem{Prop}{Proposition}[section]
\newtheorem{lem}{Lemma}[section]
\newtheorem{defn}{Definition}[section]

\usepackage{graphicx}
\usepackage{subcaption}

\usepackage{float}
\usepackage{ragged2e}

\newcommand{\calh}{\mathcal{H}}
\newcommand{\cala}{\mathcal{A}}

\newcommand{\calhobsR}{\mathcal{H}_{\text{obs,R}}}

\begin{document}

\title{Lectures on entanglement, von Neumann algebras, and emergence of spacetime}

\preprint{MIT-CTP/5938}

\author{Hong Liu}
\affiliation{MIT Center for Theoretical Physics---a Leinweber Institute,\\ 
Massachusetts
Institute of Technology, \\
77 Massachusetts Ave.,  Cambridge, MA 02139 }

\begin{abstract}

 \noindent 
 
We review recent developments in the use of von Neumann algebras to analyze the entanglement structure of quantum gravity and the emergence of spacetime in the semi-classical limit. Von Neumann algebras provide a natural framework for describing quantum subsystems when standard tensor factorizations are unavailable, capturing both kinematic and dynamical aspects of entanglement. 

The first part of the review introduces the fundamentals of von Neumann algebras, including their classification, and explains how they can be applied to characterize entanglement. Topics covered include modular and half-sided modular flows and their role in the emergence of time, as well as the crossed-product construction of von Neumann algebras. 

The second part turns to applications in quantum gravity, including an algebraic formulation of AdS/CFT in the large-$N$ limit, the emergence of bulk spacetime structure through subregion-subalgebra duality, and an operator-algebraic perspective on gravitational entropy. It also discusses simple operator-algebraic models of quantum gravity, which provide concrete settings in which to explore these ideas. In addition, several original conceptual contributions are presented, including a diagnostic of firewalls and an algebraic formulation of entanglement islands. The review concludes with some speculative remarks on the mathematical structures underlying quantum gravity.



\bigskip

{\it Expanded from lectures at TASI 2023 (Boulder, June 2023), the New York University Workshop on Operator Algebras (August 2023), the Asian Winter School (Kyoto, December 2023), and the China-India-UK School in Mathematical Physics at ICMS (Edinburgh, June 2025).}

\end{abstract}

\today

\maketitle

\tableofcontents

\section{Introduction and Motivations}

Entanglement characterizes intrinsic quantum correlations, which have no classical counterparts. 
While entanglement was already recognized in the early days of quantum mechanics by the EPR thought experiment~\cite{EinPod35} and Schr\"odinger~\cite{Sch35}, its importance in quantum many-body systems has only been understood since the 1990's.
It now plays key roles in quantum information and quantum computations, condensed matter physics, and also in understanding quantum gravity. In the context of the AdS/CFT duality~\cite{Mal97,GubKle98,Wit98}, the Ryu-Takayanagi proposal~\cite{RyuTak06,RyuTak06-2,HubRan07,Wal11,EngWal14,LewMal13,FauLew13} strongly indicates that in a quantum gravitational system the spacetime structure is tied to the entanglement structure of the system. 
While much has been learned through subsequent developments, a precise and complete understanding of the relationship between entanglement and spacetime structure remains elusive.




This lack of a precise relation reflects a deeper issue: we currently do not have a fully developed language for describing the quantum gravitational regime,\footnote{Throughout these lectures, we set $\hbar = 1$. The quantum gravity regime is characterized by a finite Newton constant $G_N$, with the semiclassical limit corresponding to $G_N \to 0$.} where spacetime itself (e.g., the metric) is expected to undergo large fluctuations.

Geometric concepts fundamental to classical gravity and quantum field theory in curved spacetime---such as spacetime geometry, causal structure, notions of time, and local regions---are sharply defined only in the $G_N \to 0$ limit. These notions typically serve as starting points in the search for a quantum theory of gravity---for example, in canonical quantization or gravitational path integral approaches---where the challenge is to understand how geometry becomes ``fuzzy'' as one moves away from the semiclassical regime.

Here, however, we adopt the opposite perspective: suppose we are given a complete theory of quantum gravity at finite $G_N$, in which conventional geometric structures are not defined a priori. Then how do such structures emerge in the $G_N \to 0$ limit? What are the underlying physical and mathematical mechanisms responsible for the emergence of spacetime geometry, causal structure, and locality? In particular, what role does entanglement play in this emergence? Hopefully, answers to these questions will help guide us toward a formulation of the quantum gravitational regime.

This latter perspective is, in fact, the natural one in the context of the AdS/CFT duality, where quantum gravity in an asymptotically anti-de Sitter~(AdS) spacetime is described by a conformal field theory~(CFT) living on its boundary.
The boundary CFT contains a parameter $N$ that characterizes the number of field degrees of freedom, and this parameter maps to a positive power of $1/G_N$ on the gravity side. Consequently, the quantum gravity regime with a finite $G_N$ corresponds to a CFT with a finite $N$, and the semi-classical $G_N \to 0$ limit corresponds to the $N \to \infty$ limit. 
In this context, the questions posed in the previous paragraph become: How do the bulk spacetime and the associated geometric concepts~(such as locality and causal structure)  emerge from these CFT degrees of freedom in the $N \to \infty$ limit?

It turns out that in the $N \to \infty$ limit, the entanglement structures of a boundary CFT undergo profound changes, and these emergent entanglement features have been argued to give rise to bulk spacetime structures~\cite{LeuLiu21a,LeuLiu21b,LeuLiu22}.

The emergent entanglement structures in the $N \to \infty$ limit cannot be captured by the standard formulation of entanglement, which defines subsystems through Hilbert space factorization.
In the standard formulation, we say that a subsystem $A$ is entangled with its complement $\oA$ if the state $\ket{\Psi}$ of the combined system $L = A \cup \oA$,  
\be \label{entaSa}
\ket{\Psi}=\sum_{n} c_n \ket{\psi_{n}}_A  \otimes \ket{\chi_{n}}_{\overline A} 
\ee
cannot be factorized into a simple product state of $A$ and $\overline{A}$. 
The tensor factorization,
\be\label{0fact}
\mathcal{H}=\mathcal{H}_{A}\otimes\mathcal{H}_{\overline A},
\ee
where $\sH$, $\sH_A$, and $\sH_\oA$ are the Hilbert spaces associated with $L$, $A$ and $\overline{A}$ respectively, is necessary for writing down~\eqref{entaSa}. Without~\eqref{0fact}, entanglement cannot be defined.

\begin{figure}[H]
        \centering
		\includegraphics[width=5cm]{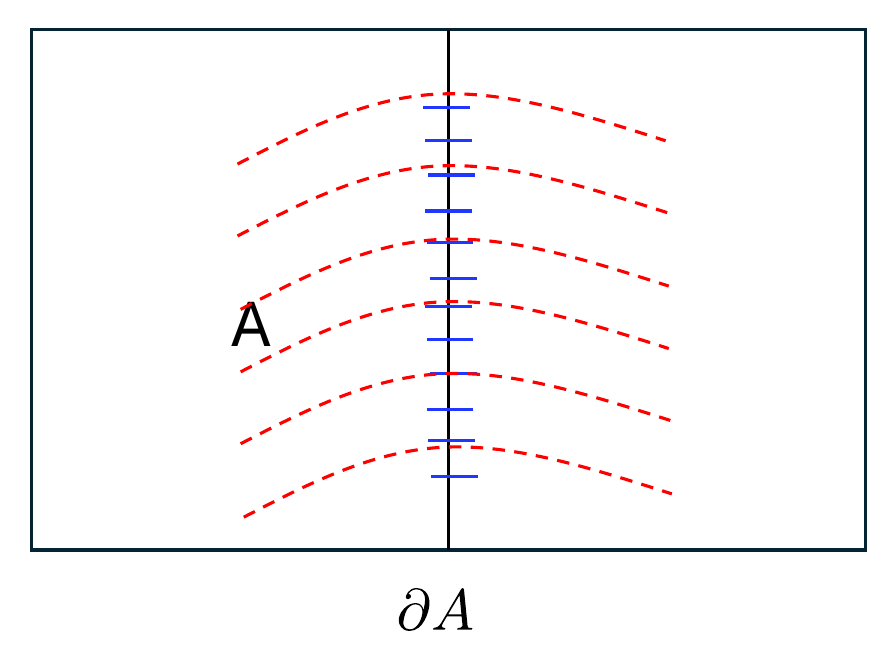}
                \caption[  ]
    {\small Cartoon of entanglement structure of a CFT for a subregion $A$. There are two types of entanglement: short-range entanglement near the boundary $\p A$ (represented by short blue lines)  due to local interactions there and long-range entanglement (represented by long red dashed lines). We assume that the CFT is put on a lattice, so that the former is regularized. In the $N \to \infty$ limit, both types of entanglement become infinite as they are proportional to the number of field theoretical degrees of freedom, leading to new entanglement structures. In particular, as we will discuss in detail in this review, the divergence of the long-range entanglement is responsible for the emergence of local bulk physics and bulk causal structure
        in the $G_N \to 0$ limit.}
\label{fig:newSt}
\end{figure}

However, as we will discuss in detail in Sec.~\ref{sec:inf}, the factorization of $\mathcal{H}$ can break down when there is infinite entanglement between $A$ and $\overline{A}$, making it impossible to define separate Hilbert spaces  $\mathcal{H}_A$,$\mathcal{H}_{\overline{A}}$ for $A$ and $\overline{A}$.
As illustrated in Fig.~\ref{fig:newSt}, in the $N \to \infty$ limit, new forms of infinite entanglement emerge in the boundary CFT, giving rise to novel structural properties.

In situations involving infinite entanglement, the conventional approach has been to introduce regularizations to render entanglement measures (such as R\'enyi  and entanglement entropies) finite.  However, such regularizations can obscure entanglement structures that are intrinsically tied to the underlying physical system. To characterize entanglement in a more innate and universal manner, a generalized notion of subsystems is needed---one that goes beyond the traditional framework of Hilbert space factorization.
Fortunately, this is possible by drawing on another early development of quantum mechanics. It involves a shift in perspective: employing operator subalgebras to define and analyze subsystems.

In the late 1920s and early 1930s,  von Neumann, while formulating the mathematical foundations of quantum mechanics, embarked on the task of classifying operator subalgebras for general quantum systems~\cite{Neu30,Neu36}. With Murray, he established the fundamental theory for what is now known as von Neumann algebras, including a classification scheme for them~\cite{MurNeu36,MurNeu37}. Over the years, von Neumann algebras have undergone extensive development and found numerous applications in both mathematics and physics. What has gone largely unappreciated for many years is that subsystems can, in fact, be defined in terms of von Neumann algebras, and that the structure of these algebras reflects the underlying entanglement structure of the system. 
In particular, in these lectures we will elaborate on the following points: 

\ben 

\item The classification of von Neumann algebras can be understood as a broad classification of entanglement types, i.e., 
 \be
\begin{bmatrix} 
\text{classification of} \cr
\text{von Neumann algebras}
\end{bmatrix} 
\quad \lra \quad
\begin{bmatrix} 
\text{broad classification of} \cr
\text{types of entanglement} 
\end{bmatrix}  \ .
\label{keyS}
 \ee
 
 \item  Beyond this, more sophisticated tools from von Neumann algebra theory provide access to significantly finer entanglement features than those captured by the broad classification~\eqref{keyS}.  

\een 



\begin{figure}[H]
        \centering
			\includegraphics[width=8cm]{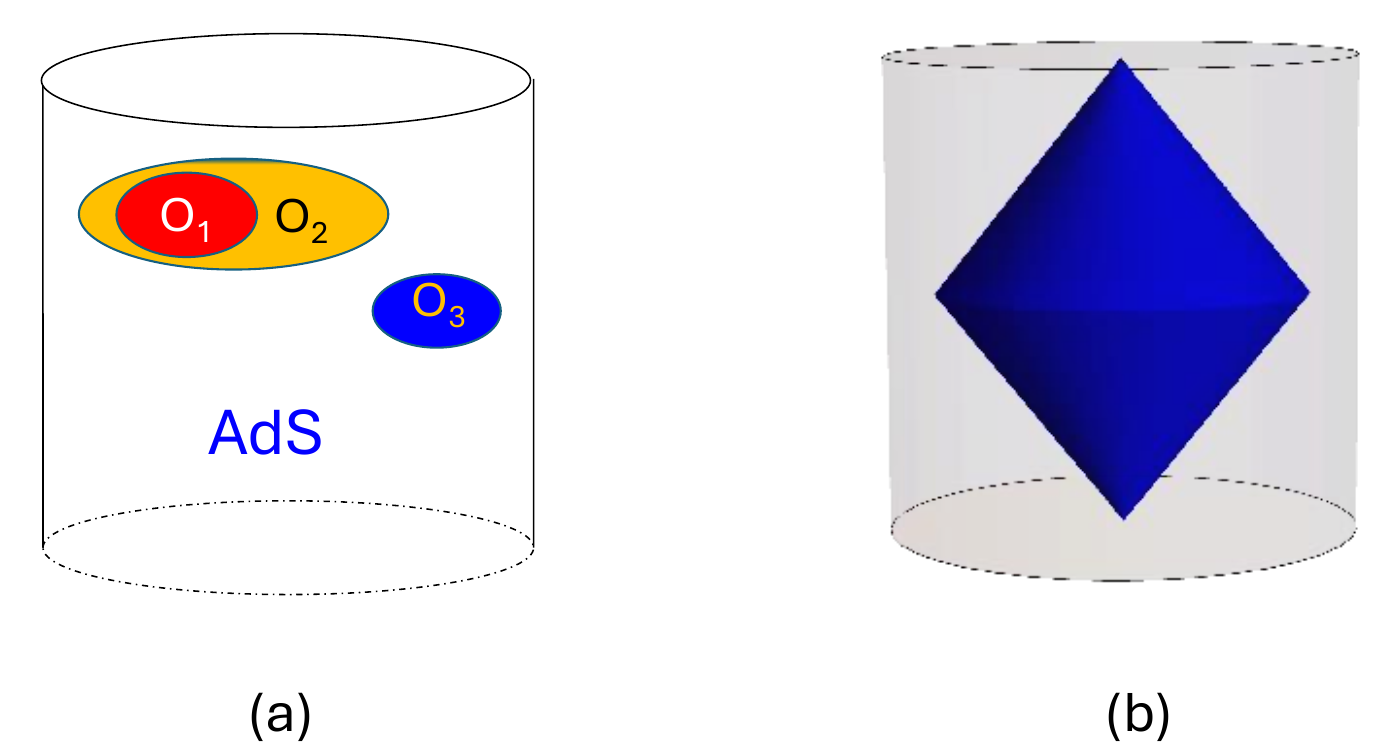}
                \caption[  ]
        {\small (a) Cartoon of subregion-subalgebra duality. A bulk subregion $O_1$ is dual to a boundary subalgebra $\sA_1$, which captures all bulk physical operations in $O_1$. Similarly bulk subregions $O_2$ and $O_3$
correspond to boundary algebras $\sA_2$ and $\sA_3$, respectively. Bulk geometric and causal relations among these regions translate into algebraic relations among the corresponding algebras. For example, $O_1 \subset O_2$ becomes $\sA_1 \subset \sA_2$. That $O_1$ and $O_3$ are spacelike separated is captured by $[\sA_1 , \sA_3] =0$, i.e., the bulk causal structure in one higher dimension is captured by the emergent boundary commutant structure in the large $N$ limit. That these regions can be sharply defined in the $G_N \to 0$ limit requires a specific bulk entanglement structure. This structure is captured by the type III$_1$ nature of the $\sA_i$'s, which arises from the infinite long-range entanglement described in Fig.~\ref{fig:newSt}.
    (b) A causal diamond in the global AdS that does not touch the boundary is dual to an emergent subalgebra in the boundary system that does not have a geometric description.} 
\label{fig:subregion}
\end{figure}

We will utilize \vNa s to describe the emergent entanglement structures of the large $N$ limit of a boundary CFT and explain how such structures can be used to understand the emergence of bulk spacetime. 
A key concept is called the subregion-subalgebra duality~\cite{LeuLiu22}, which equates an arbitrary causally complete bulk spacetime subregion with a specific type (type III$_1$) of emergent von Neumann subalgebra of the boundary CFT.
See Fig.~\ref{fig:subregion}(a) for illustrations.
The boundary subalgebra dual to a bulk subregion is emergent in the following sense:
(i) it can only be precisely defined in the $N \to \infty$ limit; and (ii) its entanglement structure undergoes qualitative changes, such as a transition from a type I von Neumann algebra to type III$_1$ in the large-$N$ limit.

The subregion-subalgebra duality opens new avenues for understanding the emergence of bulk locality, causal structure, and, more generally, spacetime itself~\cite{LeuLiu21b,LeuLiu22}. 
It also provides novel approaches for defining bulk causal and event horizons in the stringy regime~\cite{GesLiu24,HerKud25} and fresh insights into spacetime connectivity~\cite{EngLiu23}.

The subregion-subalgebra duality reformulates entanglement wedge reconstruction~\cite{Van09,CzeKar12,CzeKar12b,Wal12,HeaHub14,AlmDon14,JafSuh14,PasYos15,JafLew15,DonHar16,Har16,FauLew17,CotHay17} and causal wedge reconstruction~\cite{BanDou98,Ben99,BalKraLaw98,HamKab05,HamKab06,KabLif11,Hee12,HeeMar12,PapRaj12,Mor14,Hub14,EngPen21a,Wit23,EngLiu25} in algebraic terms, extending them to arbitrary bulk subregions such as those shown in Fig.~\ref{fig:subregion}(b). It also provides new insights~\cite{LeuLiu22,LeuLiu24} into features of entanglement wedge reconstruction, including its quantum error-correcting properties~\cite{AlmDon14} and the existence of islands~\cite{AlmMah19a}.


It may sound surprising that a subalgebra of a certain type can serve as a dual description of a bulk spacetime region. Such relationships, however, have deep roots in modern mathematics. The Gelfand duality, for instance, asserts that the algebra of continuous functions on a topological space can be employed as a description of the space itself. This duality has been a starting point for defining noncommutative geometries using noncommutative algebras (see e.g.~\cite{Con95}).
The subregion-subalgebra duality shares a similar spirit and generalizes such relations. It posits that an abstract algebra in the boundary CFT (possibly without a boundary geometric interpretation) describes all the physical operations of a bulk gravity theory in a local spacetime region in the $G_N \to 0$ limit. In other words, the bulk spacetime region may be considered as a geometrization of certain abstract algebra from the boundary theory. 

Tools from the theory of von Neumann algebras have also opened up many other exciting explorations of AdS/CFT and more broadly quantum gravity, including the construction of quantum error correcting codes for embedding bulk physics into the boundary system~\cite{Har16,KanKol18,Fau20,KanKol21,GesKan21,FauLi22}; understanding observables' algebras~\cite{PenWit23,Kol23,Lin22,Gao24,PenWit24,Ban25j} in low-dimensional quantum gravity theories such as Jackiw-Teitelboim gravity~\cite{Tei83,Jackiw,AlmPol14};
constructing new models of quantum gravity and diffeomorphism invariant observables~\cite{Wit21b,ChaLon22,Wit23b,KolLiu24,AliJef23,KliLei23,KliLei23a,AliChe24a,AliChe24,AliKli24,Gom22,Gom23,Gom23a,Gom23c,Fio23,MerTap25}; understanding the physical origin of generalized gravitational entropies, both for asymptotically AdS spacetimes and for spacetimes with other asymptotic structures~\cite{ChaPen22,JenSor23,KudLeu23,ChaLon22,KudLeu24,FewJan24,vanVer24,DeVEcc24b,DeVEcc24,DeVEcc24a}; 
describing closed universes~\cite{CheJun25,Liu25,KudWit25}, and giving boundary interpretation of bulk volume~\cite{LeuLiu25}. 
See also~\cite{BahBel22,NogBan21,BanDor22,BanMor23,BurDas23,KriMoh23,BanVos23,JenRaj24,Bah25,PenTab25,Sia25,GenJia25,deBBah25,DiGDor25}.

The roles played by  algebraic structures of the boundary theory in describing emergent bulk spacetime structures in the $G_N \to 0$ limit, as well as generalized gravitational entropies, strongly suggest that operator algebras should play crucial roles at finite $G_N$ as well, where conventional geometric notions are no longer applicable.
They offer hints about the mathematical structure of quantum gravity. We provide some perspectives on these aspects in the conclusion section.

Operator algebras have long been used to characterize relativistic quantum field theory, an approach known as algebraic QFT (see e.g.~\cite{Haa92,Ara99,Bor00,Yng04,HolWal08,HolWal14,Yng14,Wit18,Wit21a} for textbooks and reviews), to treat statistical systems (see e.g.~\cite{BraRobV2,Sew02}), and to extract quantum information (see e.g.~\cite{OhyPet}). More recently, they have also played important roles specifically in algebraic formulations and proofs of the Average Null Energy Condition (ANEC)~\cite{CasTes17,FauLei16} and the Quantum Null Energy Condition (QNEC)~\cite{BalFau17,CeyFau18,FauLi18,HolLon25}.
See also~\cite{Ges23b,FurLas23,OusFur23,GesSan24} for discussions of quantum chaotic hierarchies within the algebraic approach.

The use of von Neumann algebras to understand the AdS/CFT duality was pioneered by Rehren and collaborators in a series of early papers~\cite{Reh99a,Reh00,DueReh02a,DueReh02b}. Further pioneering contributions include the work of~\cite{PapRaj13b}, which connected black hole physics with Tomita-Takesaki modular theory, and of~\cite{Har16,KanKol18,Fau20,KanKol21,GesKan21}, which employed von Neumann algebras to construct toy models of holography as quantum error-correcting codes, and~\cite{Jef18}.




\subsection{Plan of the article}


This article contains two main parts: 

\ben 

\item {\bf Sec.~\ref{sec:ivna}---\ref{sec:crossed}: Entanglement and von Neumann algebras}

This part introduces the relevant background on von Neumann algebras and elucidates how they can be used to capture entanglement.

Sec.~\ref{sec:ivna} introduces the basics of von Neumann algebras, including their classification. Secs.~\ref{sec:iandii} and \ref{sec:III} then elaborate on how these algebras can be used to characterize entanglement. A central takeaway is that standard quantum information techniques are only applicable to type I von Neumann algebras, while many physically relevant situations require type II or type III algebras.
The tools for analyzing the entanglement structure of type III algebras are discussed in Sec.~\ref{sec:III}. There we introduce modular theory, a powerful framework for understanding entanglement in type III von Neumann algebras. We explain how it gives rise to an emergent notion of time and provides a way to further classify the entanglement structures of type III algebras. Among the various type III algebras, type III$_1$ is of particular importance, as local operator algebras in relativistic QFT are of this type. We also discuss half-sided modular flows, a powerful structure that leads to another notion of emergent time.
Finally, Sec.~\ref{sec:crossed} introduces the crossed product construction for type III$_1$ algebras, which will play an important role in later discussions of quantum gravity models and in elucidating generalized entropies.

\item {\bf Sec.~\ref{sec:ads/cft}---\ref{sec:diff}: Applications to AdS/CFT and quantum gravity} 

This part discusses recent applications of operator algebras to the emergence of spacetime structure and gravitational entropies. 

Sec.~\ref{sec:ads/cft} develops the formulation of the AdS/CFT duality in the large-$N$ limit using operator algebras. In Sec.~\ref{sec:sub}, we introduce the notion of subregion-subalgebra duality, and reformulate entanglement wedge and causal wedge reconstruction in algebraic terms.
Sec.~\ref{sec:Eme} is devoted to understanding how various geometric features of bulk spacetime---such as local regions, causal structure, notions of time, and spacetime connectivity---emerge from the boundary theory.

Sec.~\ref{sec:diff} presents simple models of quantum gravity constructed from operator algebras. The first model, involving static observers, is used to clarify the physical origin of generalized entropies of black holes, static patch of de Sitter spacetime, and arbitrary bulk regions. The section then turns to a model of dynamical observers in de Sitter spacetime, which illustrates the emergence of cosmological horizons, and concludes with a discussion of the operator algebras of JT gravity.

Finally, Sec.~\ref{sec:diss} offers some speculations on the mathematical structure of quantum gravity.

\een










\subsection{Conventions and Notations}


\noindent $\bullet$ The large-$N$ limit (equivalently, the $G_N \to 0$ limit) generally refers to the perturbative expansion in $1/N$, and does not necessarily mean only the leading-order term. By contrast, finite $N$ (or finite $G_N$) refers to treating $N$ (or $G_N$) non-perturbatively.

\noindent $\bullet$ $\sH$ denotes a Hilbert space, and $\sB(\sH)$ the set of bounded operators acting on $\sH$. In the context of field theory, we assume that any operator under consideration has been suitably regularized and is bounded. $\bid$ denotes the identity operator. 

\noindent $\bullet$ $\tilde J^\pm(Y)$ denotes the causal future/past of a region $Y$ in the conformal completion of the bulk spacetime. A prime on a region $Y$ denotes its causal complement (with the bulk or boundary context understood), while a double prime denotes its causal completion.

\noindent $\bullet$ Suppose $R$ is a subregion of a Cauchy slice. Its complement on the slice is denoted by $\bar{R}$, and its domain of dependence by $\hat{R}$. The same notation applies to both bulk and boundary subregions, with the context understood.

\noindent $\bullet$ $\subset$ always means proper subset. 



\section{Introduction to von Neumann algebras}  \label{sec:ivna}

In this section we discuss basic elements of \vNa s and their classifications.  In Sec.~\ref{sec:iandii} and Sec.~\ref{sec:III} we discuss how they can be used to characterize entanglement structure  of a quantum system. These sections are not meant as a comprehensive review of the theory of von Neumann algebras, which is a vast subject.\footnote{See~\cite{BraRobV1,TakV1,TakV2,TakV3} for textbooks.  A recent very nice review for physicists is~\cite{Sor23a}.}  It is our hope to convey the basic physical ideas with as little formal mathematical language as possible. In doing so, we may inevitably sacrifice some mathematical rigor. For example, we will often say ``it can be shown'' without giving a proof, and instead use heuristic arguments to convey the physical and mathematical intuitions behind a statement. Some of the proofs are simple, which can be filled in by readers without using significant extra mathematical machinery. In such cases we will often use the phrase ``it can be {\it readily} shown'' to encourage readers to give a try.


\subsection{Systems with an infinite amount of entanglement} \label{sec:inf}

In this subsection, we motivate the operator algebraic approach to entanglement by discussing some simple examples where the standard quantum informational techniques prove inadequate.



In the standard treatment of entanglement of a quantum system (e.g. bi-partite entanglement), we
divide the system into two subsystems $R \cup L$ and 
 assume that the full Hilbert space $\sH$ can be factorized as 
 a tensor product of those of $R$ and $L$, i.e. 
\be \label{fact} 
\sH = \sH_R \otimes \sH_L \ .
\ee
 Suppose the system is in a state $\ket{\Psi}$
 from which we can construct the reduced density operator for $R$ 
\be \label{reDe}
\rho_{R} \equiv {\rm Tr}_{L}    \le( |\Psi\rangle\langle\Psi| \ri) \ .   
\ee
$\rho_R$ is the key object from which entanglement information of the system can be extracted through various entanglement measures.   
Examples include the von Neumann entropy and R\'enyi entropies of $\rho_R$, 
\bega \label{EE}
S_{R}=-{\rm Tr}_R \rho_{R}\log\rho_{R} , \qquad S_R^{(n)} = - {1 \ov n-1} \log {\rm Tr}_R \rho_R^n , \quad n=2,3, \cdots 
\end{gather}
and the relative entropy 
between two such reduced density operators $\rho_R$ and $\sig_R$,
\be\label{REE}
S (\rho_R||\sig_R) =  {\rm Tr}_R  \rho_R (\log \rho_R - \log \sig_R)  \ .
\ee 

\begin{figure}[H]
\centering
	 \includegraphics[width=15cm]{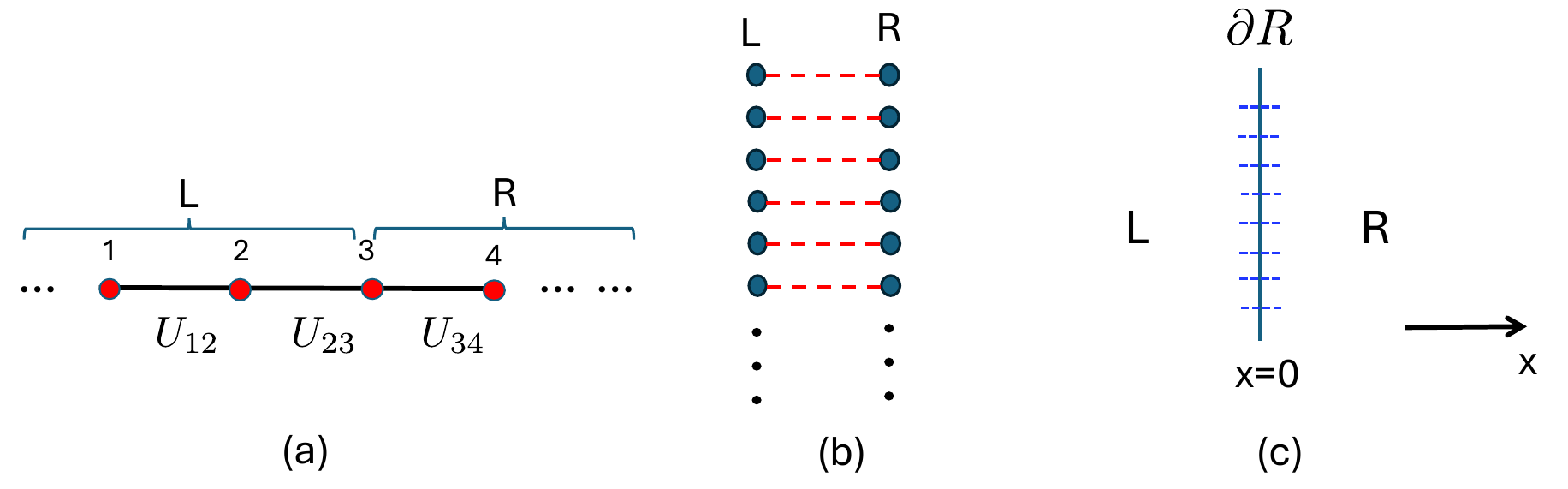}
\caption[  ]
{\small (a) A one-dimensional lattice gauge theory: dynamical variables $U_{ij} \in G$ 
are assigned to each oriented link $ij$ joining lattice points $i$ and $j$, and gauge transformations are 
$U_{ij} \to V_i U_{ij} V_j^\da$ with $V_i, V_j \in G$, where $G$ is the gauge group. 
(b) Two copies of a system of $N$ spins~(blue dots) in an entangled state. Red dashed lines represent entanglement between pairs of spins. 
 (c) A spatial slice of a relativistic quantum field theory is separated into $R$ and $L$ regions by surface $x=0$. Dashed blue lines represent short-range entanglement near $x=0$. }
\label{fig:latt}
\end{figure}

There are, however, situations where the factorization of Hilbert space~\eqref{fact} does not exist, in which case~\eqref{reDe}--\eqref{REE} as well as many other quantum informational measures cannot be immediately defined. 
 Here are some examples:

\ben 


\item[Ex. 1:] \label{ex:0} Consider the  lattice gauge system of Fig.~\ref{fig:latt}(a).
Variables $U_{ij}$ are not gauge invariant. That physical states must be gauge invariant means that the Hilbert space $\sH$ can cannot be factorized into $\sH_R \otimes \sH_L$ for $R$ and $L$ regions.

\item[Ex. 2:] \label{ex:1}
Consider two copies of a system of $N$ spins, as indicated in Fig.~\ref{fig:latt}(b). 
Suppose each pair of spins of $R$ and $L$ systems is in an entanglement state 
\be \label{ijn}
\ket{\phi_\th} = \cos \th \ket{00} + \sin \th \ket{11},
\ee 
with the state for the full system given by 
\be \label{une} 
\ket{\Phi_\th} = \ket{\phi_\th}_1 \otimes \ket{\phi_\th}_2  \otimes \cdots \ket{\phi_\th}_N  \ .
\ee
For $\th ={\pi \ov 4}$ the two spins in a pair are maximally entangled, so are the $R$ and $L$ systems.\footnote{As a result, $\ket{\Phi_{\pi \ov 4}}$ is the thermofield double (TFD) state of the $R$ and $L$ systems at an infinite temperature.}
For a general $\th \in (0, \pi/4)$, the two systems are entangled, but not maximally. 

Now consider the ``thermodynamic'' limit $N \to \infty$. Since physical processes involve only finite energy, we restrict our attention to finite-energy excitations around the state~\eqref{une}, which can be completed into a Hilbert space $\sH_{\Phi_\th}$. 
For a generic Hamiltonian\footnote{Consider for example, $H_R = H_L = f \sum_{i=1}^N Z_i$ where $Z_i$ denotes Pauli matrix $\sig_z$ for $i$-th spin, where $H_R, H_L$ are respectively the Hamiltonians of $R$ and $L$ systems.}, any finite-energy excitation around the state~\eqref{une} can flip only a finite number of spins. As a consequence:

(i) All states in $\sH_{\Phi_\th}$ possess an infinite amount of entanglement. Unentangling the 
$R$ and $L$ subsystems would require flipping an infinite number of spins---thus costing infinite energy.

(ii) The Hilbert space $\sH_{\Phi_\th}$ cannot be factorized into separate Hilbert spaces for the 
$R$ and $L$ systems. Such a factorization would require the existence of product states with no entanglement between 
$R$ and $L$, which are absent.

(iii) The states $\ket{\Phi_{\theta}}$ corresponding to different values of  $\theta$
differ by an infinite amount of entanglement, and are therefore expected to belong to distinct Hilbert spaces.

In the $N \to \infty$ limit, the $2^{2N}$-dimensional Hilbert space $\sH_N$ of the original system of $N$ spin pairs is no longer relevant for the physics of interest, as it includes states with infinite energy. In fact, the resulting vector space has uncountable dimension and is not a separable Hilbert space.

\item[Ex. 3:] 
Consider dividing a time slice of a {\it relativistic} QFT (RQFT), say a scalar field theory in a $d$-dimensional Minkowski spacetime,  into $R$ and $L$ halves (see Fig.~\ref{fig:latt}(c)).
Across the surface $x=0$ separating $R$ and $L$ regions, there are an infinite number of local degrees of freedom coupled together, which results 
in an infinite amount of entanglement between $R$ and $L$ in {\it any finite energy state}\footnote{Here, the energy corresponds to the Hamiltonian associated with Minkowski time, measured relative to the vacuum.}. 
For example, when the short-distance cutoff $\ep$ approaches zero (the continuum limit), the entanglement entropy of the $R$-region in a general state contains a universal divergent term given by
\begin{equation}
S_{R}= b \frac{\text{Area}(\p R)}{\epsilon^{d-2}}+\cdots  \ .
\label{eq:161}
\end{equation}
Similar to the entangled-spins example, the Hilbert space of the QFT cannot be factorized into those of $R$ and $L$, as product states in such a factorized space would possess infinite energies.





\een 


Without factorization of $\sH$, the reduced density operator~\eqref{reDe} for a subsystem cannot be defined, and the standard approach of extracting entanglement correlations of a system cannot be used.

We stress that the physics underlying the non-factorization of $\sH$ in the lattice gauge system of Ex. 1 is fundamentally different from that of Ex. 2 and 3. The former arises from gauge constraints. The latter is due to infinite entanglement from infinite number of degrees of freedom.  This latter situation, which will be the main focus of these lectures, occurs generically in statistical systems in the thermodynamic limit and in QFTs.\footnote{Using the argument given earlier for the entangled spin example, it is expected that in generic situations, infinite entanglement prevents factorization of Hilbert space---states separated by infinite entanglement are expected to be separated by infinite energy barriers, and thus do not coexist in the same Hilbert space. Nevertheless, (fine-tuned) counterexamples exist, see Appendix~\ref{app:counter}.  Below in Sec.~\ref{sec:vNCl}, we will discuss a precise criterion for when factorization exists.} 
 As we will see in later discussions, the treatments of these two situations differ significantly: the lattice gauge system requires only modest modifications to the standard procedure, while Ex. 2 and 3 demand fundamental changes in perspective and new theoretical tools.

The usual strategy for dealing with systems with an infinite amount of entanglement, such as those of Ex. 2 and Ex. 3,  is to introduce a regulator $\ep$ such that the amount of entanglement becomes finite, and as a result  $\sH^\ep = \sH^\ep_R \otimes \sH^\ep_L$, where $\sH^\ep$ denotes the Hilbert space in the presence of regulator $\ep$.
 In the entangled spin examples, this can be done by first calculating quantum informational quantities for a finite $N$~(whose Hilbert space clearly has a factorization structure),  and then taking $N \to \infty$ at the end of the calculation. In a QFT, the system can be put on a lattice, with the short-distance cutoff $\ep$ in~\eqref{eq:161} being the lattice spacing. 
While the entanglement entropies or other quantum informational measures become divergent when the regulator is removed, sometimes finite values that are independent of regularization procedure can be extracted. Powerful techniques, such as the replica trick, have been developed to calculate various regularized entanglement measures and to extract from them regularization-independent quantities. 
Many important results and insights have been obtained this way.

However, introducing a regulator risks obscuring physics inherent to the continuum or thermodynamic limit. As we will see in later sections, this is indeed the case. Divergences often serve as reminders that the appropriate tools or language have not yet been employed. To intrinsically characterize entanglement in such situations, it is necessary to generalize the definition of a subsystem beyond Hilbert space factorization. Below we will  demonstrate that von Neumann algebras provide both the language and the powerful tools needed to capture the structure of entanglement in these contexts.

\subsection{What is a von Neumann algebra}

\subsubsection{Von Neumann algebras as subsystems}

Consider a quantum system with Hilbert space $\sH$, with $\sB (\sH)$ denoting the set of all bounded operators. 
For technical simplicity, we will always restrict our discussion to bounded operators, whose actions on the Hilbert space are continuous. 

Suppose we have only partial access to the system, i.e. we have access to only a subset of operators of the system, the collection of which will be denoted as $\sM \subset \sB (\sH)$. $\sM$ contains all manipulations we can perform to either extract information from or to influence the system. For example, given a state $\ket{\Psi} \in \sH$, we can obtain expectation values $\vev{\Psi|A|\Psi}$  and  perform operations $B \ket{\Psi}$, with $A, B \in \sM$.    

We will assume that $\sM$ is closed  
under operator products, additions, and hermitian conjugate (i.e. if $A \in \sM$, $A^\da \in \sM$),  which means that it is a {\bf $*$-subalgebra} of $\sB(\sH)$. 
In addition to being closed, it is also useful to have a notion of completeness, i.e. the limit of a convergent sequence of operations in $\sM$ should lie in $\sM$. 

There are multiple ways to define the convergence of a sequence of operators. Bounded operators on $\sH$ have a well-defined norm $||\cdot||$, which satisfies 
\be \label{csno}
||O^\da O|| = ||O||^2  \ .
\ee
We say a sequence of operators $\{A_n\}  \equiv \{A_1, A_2, \cdots\}$
is {\bf norm convergent} if there exists an operator $A$ such that $\lim_{n \to \infty} ||A_n-A|| = 0$. $\sM$ is complete with respect to norm convergence if $\{A_n \} \in \sM$ implies $A \in \sM$, and such an algebra is called a {\bf $C^*$ algebra}.

A different notion of completeness can be introduced by using instead matrix elements. 
We say a sequence of operators $\{A_n\} = \{A_1, A_2, \cdots\}$ is {\bf weakly convergent} if there exists an operator $A$ such that $\lim_{n \to \infty} \vev{\xi|A_n |\eta} = \vev{\xi|A|\eta}$ for 
all $\ket{\xi}, \ket{\eta} \in \sH$. If $\sM$ is complete under weak convergence, i.e. a weakly convergent sequence $\{A_n\} \in \sM$ implies $A \in \sM$,  it is called a {\bf \vNa}. 
Since norm convergence implies weak convergence, a von Neumann algebra is a $C^*$ algebra, but the converse is not true. 
 
$C^*$ algebras can in fact be defined abstractly without a Hilbert space as long as a norm satisfying~\eqref{csno} exists. 
But the definition of  \vNa\  requires a Hilbert space.

Since matrix elements are, in principle, physically observable, requiring weak convergence is often more appropriate in most physical situations. For this reason, we will primarily focus on von Neumann algebras.

Now, instead of using tensor factorization of the Hilbert space to define a subsystem, we define a subsystem using a von Neumann algebra $\sM$, i.e.,
\be\label{subvon}
\text{subsystem} \quad = \quad \text{von Neumann algebra} \ .
\ee
$\sM$ represents the collection of all possible physical operations that can be performed on a subsystem, closed under Hermitian conjugation, operator products, and weak convergence, providing an operational and more physical definition of a subsystem.
We define the {\bf complement} of a subsystem---specified by a von Neumann algebra $\sM$---in terms of its commutant $\sM'$, which is the set of operators that commute with $\sM$, i.e.,
\be
\sM' = \{B \in \sB (\sH),  BA = AB , \forall A \in \sM\} \ .
\ee
We will see that the definition~\eqref{subvon} includes the factorization definition~\eqref{fact} as a sub-case, but is also applicable to non-factorized situations.

For the rest of this section, we discuss basic properties of von Neumann algebras and their classification, establishing the necessary background for the examination of their entanglement properties in Sec.~\ref{sec:iandii} and Sec.~\ref{sec:III}.

\subsubsection{Basic properties}

Despite seemingly small differences in the definitions of von Neumann and $C^*$ algebras, their properties differ significantly. 
 A fundamental statement about \vNa\ is von Neumann's double commutant theorem which says that a \vNa\ satisfies\footnote{For a general $*$-algebra $\sM$, it follows immediately from the definition of commutant that $\sM \subseteq \sM''$.} 
 \be \label{dcth}
\sM = \sM''  ,
\ee
where $\sM''$ means taking the commutant twice.  Equation~\eqref{dcth} is a highly remarkable and powerful result, as it converts the completeness under weak convergence, which is a topological statement, into an algebraic statement. Operationally, checking whether $\sM$ coincides with its double commutant is in general  simpler than checking the completeness condition. 
 
It can be readily shown that for a $*$-subalgebra $\sA$ 
\be
\sA' = \sA''', 
\ee
so the commutant of a $*$-subalgebra is always a \vNa. From $\sA$ we can also construct a \vNa\ by taking its double commutant 
\be
\sM = \sA'' \ .
\ee
It can be readily shown that $\sM$ is the smallest \vNa\ containing $\sA$, and $\sM' = \sA'$.

Consider two \vNa s, $\sM_1, \sM_2$. Denote the smallest von Neumann algebra containing $\sM_1$ and $\sM_2$ as $\sM_1 \vee \sM_2$ and the 
largest von Neumann algebra contained in $\sM_1$ and in $\sM_2$ as $\sM_1 \wedge \sM_2$. We then have 
\be 
\sM_1 \vee \sM_2 = (\sM_1 \cup \sM_2)'' , \quad \sM_1 \wedge \sM_2 = \sM_1 \cap \sM_2  \ .
\ee
There is no need to have double commutant on the right hand side of the second equation as it can be readily seen that the intersection of two von Neumann algebras is a von Neumann algebra. 
Also note that 
\be 
(\sM_1 \vee \sM_2)' = \sM_1' \wedge \sM_2' = \sM_1' \cap \sM_2' \ .
\ee

The center of $\sZ (\sM)$ of a von Neumann algebra $\sM$ is defined as the collection of elements of $\sM$ that commute with $\sM$, i.e., 
\be 
\sZ (\sM) = \sM \cap \sM'  \ .
\ee
 A von Neumann algebra is called {\bf primary} or {\bf a factor} if it has a trivial center, i.e. $\sZ (\sM) = \CC \bid$, or equivalently 
$\sM \vee \sM' = \sB (\sH)$. A general von Neumann algebra $\sM$ can be written as direct sums (integrals) of factors, so below we will mostly concentrate on factors.

\subsubsection{Weights and states}\label{sec:states}

In quantum mechanics, a key characterization of an operator is its expectation values in states of interest. 
For $C^*$ and \vNa s, the definition of a state does not require introducing vectors or density operators in a Hilbert space and can be 
formulated abstractly. 

A {\bf linear functional} $\om$ on a von Neumann algebra $\sM$ is a continuous map: $\om: \sM \to \CC$ satisfying
\be \label{iwn}
\om (aA + b B) = a \om (A) + b \om (B) , \quad \om (A^\da) = (\om (A))^*, \quad a, b \in \CC, \; A, B \in \sM \ .
\ee
A {\bf weight} $\om$ on $\sM$ is a linear functional that satisfies in addition to~\eqref{iwn}, 
\be \label{pou}
\om (A^\da A) \geq 0, \quad \forall A \in \sM \ .
\ee
Equations~\eqref{iwn}--\eqref{pou} imply that
a weight $\om$ satisfies the Cauchy-Schwarz inequality 
\be \label{cahy}
|\om (A^\da B) |^2 \leq \om (A^\da A) \om (B^\da B)  \ .
\ee

A weight $\om$ on $\sM$ is called a {\bf trace}  if it further 
satisfies 
\be \label{exch}
\om (AB) = \om (BA), \quad \forall A, B \in \sM \ .
\ee
The trace of an operator does not have to be finite. 
A {\bf positive linear functional} is a weight with $\om(\bid)$ finite.
A {\bf state} is a weight with 
\be\label{pnor}
\om (\bid) = 1 \ .
\ee
That is, a state is properly normalized. 
A {\bf tracial state} is a trace with $\om (\bid) = 1$. We can interpret $\om (A)$ as the expectation value of $A$ in $\om$.  

A {\bf normal} state is defined as a state $\om$ satisfying the condition\footnote{$\sup$ denotes the supremum.} 
\be \label{norde}
\sup \om (A_i) = \om (\sup A_i) 
\ee
where $\{A_i\}$ is a bounded sequence of {\it positive} operators in $\sM$. There exists a partial ordering among positive operators, i.e. $A_{i_1} > A_{i_2}$ if $A_{i_1} - A_{i_2}$ is a positive operator. Equation~\eqref{norde} means that the partial orderings of positive operators are preserved by the expectation values in the state $\om$. 
It can be shown that a state is normal if and only if there exists a density operator $\rho$ on $\sH$ with 
\be \label{norS}
\om (A) = \Tr (\rho A), \quad A \in \sM \ .
\ee
From now on we will only consider normal states. 
 
A state $\om$ is {\bf faithful}, if $\om(A^\da A ) = 0$ implies that $A=0$.  
 
 For two states $\om_1, \om_2$, and some $ \lam \in (0,1)$, 
 \be \label{mixes}
 \om = \lam \om_1 + (1-\lam) \om_2 
 \ee
 is again a state. If $\om$ can be written in the form~\eqref{mixes}, we say $\om$ {\bf dominates} $\om_1$ and $\om_2$. 
 An $\om$ which cannot be written as~\eqref{mixes} is called a {\bf pure} state.


\subsubsection{$C^*$ and von Neumann algebras generated by a single operator} \label{sec:sigO}

To get some intuition for the norm and weak completeness used to define respectively a $C^*$ algebra and a von Neumann algebra, here we consider the simplest example of such algebras. 
Consider a bounded  Hermitian operator $\sO$. We denote the $C^*$ and \vNa s generated by $\sO$ (i.e. the algebras built from $\sO$ alone) respectively by $\sA (\sO)$ and $\sM (\sO)$.  
Since $\sO$ commutes with itself, these are Abelian algebras. 
Both $\sA (\sO)$ and $\sM (\sO)$ should include polynomials of $\sO$, i.e., operators of the form $\sum_{n=0} c_n \sO^n$ for $c_n \in \CC$, but can contain additional elements 
arising from the norm and weak completeness.

Denote the spectrum of $\sO$ as $\sig (\sO)$, which is a compact subset of $\RR$.\footnote{Recall that the spectrum $\sig (A)$ for an operator $A \in \sB (\sH)$ is the set of complex numbers $\ka$  for which $A - \ka \bid$ is not invertible. For a bounded Hermitian operator $\sO$, $\sig (\sO)$ is a closed compact subset of $\RR$.}
It can be shown\footnote{See e.g. Theorem A.8 in Appendix A of~\cite{Sor23a}.} that $\sA(\sO)$ is isomorphic to the algebra of {\it continuous} functions on $\sig (\sO)$.
That is, for any continuous function
$f (\ka): \ka \in \sig(\sO) \to \CC$, 
there exists some operator $f(\sO) \in \sA (\sO)$. 
In contrast, it can be shown\footnote{See e.g. Theorem A.11 in Appendix A of~\cite{Sor23a}.} that $\sM (\sO)$ is isomorphic to the algebra of {\it bounded} functions\footnote{More precisely, bounded measurable functions up to equivalence on sets of measure zero.} on $\sig (\sO)$.
Since continuous functions on a compact set are bounded, the bounded functions include continuous functions as a subset. Thus,  $\sA (\sO)$ is a subset of $\sM (\sO)$.

In particular,  the von Neumann algebra $\sM (\sO)$ includes all the projection operators in the spectral decomposition of $\sO$, while $\sA (\sO)$ in general does {\it not} include them. Such projection operators 
correspond to indicator functions on $\sig (\sO)$, which are in general discontinuous. 
For example, consider the projection operator $P_I$ that projects to a 
subset $I$  of $\sig (\sO)$, the corresponding indicator function for $I$ is 
\be\label{1hn}
f_I (\ka) = \bca 
1 & \ka \in I \cr
0 & \ka \not \in I
\eca \ ,
\ee
and $P_I = f_I (\sO)$. 

\subsubsection{Projections}

The discussion of Sec.~\ref{sec:sigO} implies that a \vNa\ can be considered as being spanned by its 
projection operators (or simply projections). More explicitly, for a Hermitian operator $A \in \sM$, $\sM$ includes the von Neumann algebra generated by $A$, and thus contains all the spectral projections of $A$. That is, $A$ can be expanded in terms of projections of $\sM$.  A (non-Hermitian) operator $B \in \sM$ can be written as $B = B_R + i B_I$, where $B_R = \ha (B + B^\da)$ and $B_I = {1 \ov 2i} (B-B^\da)$ are Hermitian and are in $\sM$ (as $B^\da \in \sM$ by definition). Thus $B$ can also be expanded in terms of projections of $\sM$.


Projections are familiar objects in quantum mechanics. Recall that there is a one-to-one correspondence between projections and closed subspaces of $\sH$:  
a projection $P$ projects to the subspace $P \sH \subseteq \sH$, and conversely, for any closed subspace of $\sH$ we can associate a projection. If $P \in \sM$, we say the subspace $P \sH$ belongs to $\sM$. Heuristically, this is a subspace that observers who control $\sM$  have access to. 
The correspondence between projections and closed subspaces induces a partial ordering on the projections. We say\footnote{Alternatively it can be defined as $
P \leq Q  \Leftrightarrow  Q-P \; \text{is a positive operator}$.
}
\be \label{yrw}
P \leq Q \quad \text{if} \quad PQ = P ,
\ee
which means that $P \sH \subseteq Q \sH$. Clearly, the largest projection is the identity $\bid$. For two projections $P$ and $Q$, 
$P \land Q$ denotes the projection onto $P \sH \cap Q\sH$, and $P \lor Q$ is the projection for the closure of 
the span of $P \sH \cup Q \sH$. 
The orthocomplementation $\bot$ is defined as $P^\bot = \bid - P$.
 If $P, Q \in \sM$, then $P\land Q$, $P \lor Q$, $P_\perp$ are also in $\sM$. That is, projections in $\sM$ form a complete 
lattice with respect to the operations $\land, \lor$.\footnote{A lattice is a partially ordered set in which every pair of elements has a least upper bound and a greatest lower bound. A complete lattice extends this property to every subset, not just pairs of elements.}

Given the fundamental role of projections, the properties of a von Neumann algebra---and indeed its classification---can be characterized in terms of the properties of its projections.
Two characterizations of projections will play important roles: minimal and finite projections.

The notion of minimality is straightforward to define: a projection 
$P$ is said to be {\bf minimal} {\it in $\sM$} if it is nonzero and $\sM$ contains no projection $Q < P$.

The definition of finiteness is more subtle. Consider, for example, the projection 
 \be \label{trex} 
 P = \ket{\psi} \bra{\psi} \otimes \bid_2
 \ee
on a Hilbert space $\sH = \sH_1 \otimes \sH_2$, with $\ket{\psi} \in \sH_1$ and $\bid_2$ the identity on $\sH_2$. Suppose $\sH_2$ is infinite dimensional.  While $P$ may be an infinite-rank projection in the full Hilbert space $\sH$,
it is a finite projection of rank one when considered within the subspace $\sH_1$---that is, within the subalgebra $\sB (\sH_1) \otimes \bid_2$. This example illustrates that the notion of finiteness depends not solely on its rank but on the algebraic context in which the projection is considered. We therefore require a notion of finiteness that reflects the intrinsic properties of a projection within an algebra itself.

For this purpose, we first introduce the notion of equivalence: two projections $P, Q \in \sM$ are called {\bf equivalent} {\it with respect to $\sM$}, denoted as $P \sim Q$, if there is an operator $V \in \sM$ such that 
\be\label{eqpro}
P = V^\da V, \quad Q = VV^\da \ .
\ee
This means that there exists a partial isometry $V$ that 
maps the subspace $\sH_1 = P \sH$ isometrically onto the subspace $\sH_2 = Q \sH$. 
A nonzero projection $P$ is called {\bf finite} if it is not equivalent with a smaller projection.
 Conversely,
$P$ is called {\bf infinite} if there exists a projector $Q \in \sM$ such that $Q \sim P$ and
$Q<P$.


Under this definition,~\eqref{trex} is a finite projection in the algebra $\sM = \sB (\sH_1) \otimes \bid_2$.

For a von Neumann factor $\sM$, the identity operator $\bid$ is the maximal projection.
Depending on the nature of $\sM$, this projection can be either finite or infinite. If 
$\bid$ is finite, then every projection in $\sM$ must also be finite; in this case, 
$\sM$ is called a {\bf finite von Neumann factor}. Conversely, if the identity is infinite, we say that 
$\sM$ is an {\bf infinite von Neumann factor}. It can be shown that in such cases, any two infinite projections are equivalent, and in particular, every infinite projection is equivalent to the identity $\bid$.

\subsection{Classifications of von Neumann factors}\label{sec:vNCl}

We now turn to the classification of von Neumann factors, based on the properties of their projections.


The dimension of a projection $P$ 
is usually defined as $\Tr P$, where $\Tr$ is the  trace of $\sH$. The value of  $\Tr P$ gives the dimension 
of the subspace $P \sH$. For example, $P = \ket{\chi} \bra{\chi}$ has dimension $1$, while $P= \sum_{i=1}^n \ket{\chi_i} \bra{\chi_i}$ with $\vev{\chi_i|\chi_j} = \de_{ij}$ has dimension $n$. The dimension as defined is given by an 
integer, which can be infinite. But this dimension is in general inadequate as it  does not reflect intrinsic properties 
of a projection in an algebra. In the example~\eqref{trex}, $\Tr P = {\rm dim} \, \sH_2$, which is infinite. However, 
from the perspective of $\sH_1$---or equivalently the algebra  $\sB (\sH_1) \otimes \bid_2$---it is a rank-one projection.
Therefore, to characterize the dimension of a projection $P$ in a von Neumann algebra, we need a definition that reflects intrinsically its nature in the algebra.

The equivalence relation~\eqref{eqpro} together with the partial ordering~\eqref{yrw} of projections implies that we can introduce  a dimension function $d(P)$ in $\sM$ such that it is positive for any nonzero finite projection and satisfies 
\be\label{eug}
d (P) < d (Q) \quad \text{for} \quad P < Q , \quad \text{and} \quad d (P) = d (Q) \quad \text{for} \quad P \sim Q \ .
\ee
For an infinite projection $P$, then there exists a $Q \in \sM$ such that  $d (Q) < d(P)$ and $d(P) = d (Q)$ at the same time, which means $d(P)$ and $d(Q)$ have to be infinite, 
 \be \label{inftr}
 d (P) = \infty, \quad P: \; \text{infinite projection} \ .
 \ee

Such a dimension function can be used to demonstrate the existence of a {\it normal and faithful} trace $\tr (\cdot)$
on the algebra $\sM$, 
with
\be
 \tr P = d(P)  , \quad \forall P \in \sM, 
\ee 
and this trace is unique up to an overall normalization (as $d(P)$ is only defined up to a scaling).
The trace $\tr (\cdot)$ on $\sM$ generally differs from the standard Hilbert space trace $\Tr$ on $\sH$, and, as we will see below, can be understood as a kind of ``renormalized'' trace.

In the example of~\eqref{trex} on $\sM = \sB (\sH_1) \otimes \bid_2$, since $P$ is a rank-1 projection on $\sH_1$, we can define $d (P) =1$ and the corresponding trace $\tr$ can be defined as
\be \label{simtr}
\tr (\cdot) = {\rm Tr}_{\sH_1} (\cdot) 
\ee
 by simply stripping off the factor $\bid_2$ in~\eqref{trex}, as all operators in $\sM$ has such a trivial factor.  



We are now ready to discuss the classification of a von Neumann factor $\sM$:
\ben 

\item {\bf Type I:} $\sM$ contains nonzero minimal projection(s). 

Using the scaling freedom of $\tr$ (or $d(P)$), we can normalize $\tr$ such that $d (P_0) = \tr P_0 =1$ for a minimal projection $P_0$. In this case, it can be further shown that all the projections can be built from superpositions of minimal projections, and 
$d(P)$ is given by a positive integer. 

Depending on the value of $n = d(\bid)$, a type I factor can be further classified as
\bea \label{I10}
& \text{I$_n$:} & \quad \text{$d (P)= \{0,1,...,n\}$, where $n$ is a natural number. } \\
&\text{I$_\infty$:} & \quad  \text{$d (P) = \{0,1 ,\cdots, n, \cdots ,\infty\}$.}
\label{I11}
\eea
$\sB (\sH)$ for a separable Hilbert space $\sH$ is trivially type I: it is type $I_n$ if $\sH$ is $n$-dimensional, and is type $I_\infty$ for an infinite-dimensional $\sH$.  In this case, $\tr$ is just the standard trace $\Tr_{\sH}$ of $\sH$. 
 
Since the trace $\tr$ normalized to unity for minimal projections and all dimensions in~\eqref{I10}--\eqref{I11} are integers, we expect that $\tr$ should have the usual interpretation of ``counting'' states.  
Indeed, it can be shown that for a type I factor $\sM$, there exists a factorization of the Hilbert space 
$\sH = \sH_R\otimes  \sH_L$ such that 
\be \label{tyia} 
\sM = \sB (\sH_R) \otimes \bid_L, \quad \tr = {\rm Tr}_{\sH_R} , \quad \sM' = \bid_R \otimes \sB (\sH_L)  \ .
\ee 
Conversely, if $\sM$ can be written in the form~\eqref{tyia}, it is clearly type I.

Thus, Hilbert space factorization exists if and only if the von Neumann algebra associated with a subsystem is a type I factor. This provides a precise mathematical criterion for determining when Hilbert space factorization is well-defined. Consequently, the conventional notion of a subsystem defined via Hilbert space factorization is a special case of the more general algebraic formulation given in~\eqref{subvon}.



\item {\bf Type II:}  $\sM$ contains finite projections, but does not contain any minimal projection.

In the absence of minimal projections, equation~\eqref{eug} implies that the dimension function $d(P)$
can take arbitrarily small values and spans a continuous range beginning at zero. This allows us to assign a {\it real} number as the dimension of the Hilbert subspace $P \sH$, in contrast to the discrete dimensions familiar from finite-dimensional settings.

Depending on whether $d(\bid)$ is bounded or not, we can subdivide type II factors into: 

\ben 

\item II$_1$: $d (\bid)$ is bounded. We can  normalize the trace $\tr$ such that 
\be\label{norII}
d(\bid) = \tr \bid =1
\ee
 and thus 
$d (P) \in (0,1]$. 

\item  II$_\infty$: $d (\bid)$ is unbounded. In this case, $d (P)\in (0, \infty]$. 
\een

\item {\bf Type III}: All nonzero projections are infinite. 

In this case, equation~\eqref{inftr} implies $d (P) = \infty, \forall P \in \sM$. Therefore, a meaningful dimension function and trace $\tr$ do not exist. New conceptual and mathematical tools are required to describe and classify such algebras---these will be discussed later in Sec.~\ref{sec:III}.

\een 


Type II and III algebras may appear unintuitive, but we will see they arise naturally in many physical contexts, including the entangled spin example and quantum field theory discussed in Sec.~\ref{sec:inf}.

The definitions of type II and III algebras have some immediate implications. For example, it can be shown that the absence of a minimal projection for type II and III factors implies that such an algebra does not admit a pure state (recall the definition below~\eqref{mixes}). 
This can be intuitively understood in terms of lack of factorization of Hilbert space in such situations.
When there is factorization of Hilbert space, as in~\eqref{tyia}, a pure state on $\sM$ exists, corresponding to a product state in $\sH_R \otimes \sH_L$ (i.e., an unentangled state). Non-existence of a pure state on $\sM$ can thus be interpreted as non-existence of a product state. (Recall that in the examples of entangled spins and quantum field theory of Sec.~\ref{sec:inf}
all states in the Hilbert space are infinitely entangled).
 
 For type III algebras, that all  projections are infinite implies for any projection $P \in \sM$ there exists an isometry $W \in \sM$ such that 
\be \label{e1hn}
W^\da W =\bid, \qquad W W^\da = P \ .
\ee
This has some interesting consequences. For example, given any state $\om$ on $\sM$, we can use the isometry operation $W$ to construct a new state 
$\om_W$ which is an eigenstate of $P$,\footnote{We say a state $\vp$ on $\sM$ is an eigenstate of a projection $P$ if 
$\vp (P) =1$.}
\be 
\om_W (A) \equiv \om (W^\da A W) \quad \to \quad \om_W (P) =1 \ .
\ee
The ``local''\footnote{
Here ``local'' simply refers to that it belongs to $\sM$.} isometry operation $W$ does not affect the commutant of $\sM$, with $\om_W (A') = \om (A')$ for $A' \in \sM'$.

\subsection{``Emergent'' Hilbert space and von Neumann algebra: GNS construction} \label{sec:GNS}


 The definition of von Neumann algebras requires a Hilbert space as its completeness is defined in terms of matrix elements. In contrast, $C^*$ algebras can be defined without a Hilbert space as only a norm is needed. In this subsection we discuss a  construction where a Hilbert space and thus von Neumann algebras can emerge from a $C^*$ algebra $\sA$ and a state $\om$ on $\sA$. This construction will play important roles in later discussions.
 
 The basic result is that given a $C^*$ algebra $\sA$ and a state $\om$ on $\sA$, it is possible to construct a Hilbert space $\sH_\om$ and a representation $\pi_\om (\sA)$ of $\sA$  acting on $\sH_\om$ (i.e., an operator $A \in \sA$ is represented by an operator  $\pi_\om (A) \in \sB (\sH_\om)$). Heuristically, $\sH_\om$ can be viewed as being obtained by ``acting'' elements of $\sA$ on $\om$. Given the algebra $\pi_\om (\sA) \subseteq \sB (\sH_\om)$, there is  a natural von Neumann algebra associated with it by taking 
the double commutant in $\sB (\sH_\om)$~(or equivalently weak closure in $\sH_\om$), 
\be \label{vnaA} 
\sM = \pi_\om (\sA)'' \ .
\ee
That is, from $\sA$ and $\om$, there emerges a Hilbert space $\sH_\om$ and a von Neumann algebra $\sM$. 

Below we first give the abstract construction of $\sH_\om$ and $\pi_\om$, which is called the Gelfand-Naimark-Segal (GNS) construction, and then consider a simple example. 

For $A, B \in \sA$, we can define an inner product as 
\be
\vev{A|B} \equiv \om(A^\da B) ,  \quad \vev{A|A} \geq 0 , \quad \vev{B|A} = \om (B^\da A) = (\om(A^\da B))^* = \vev{A|B}^*  , 
\ee
where the last two equations follow respectively from the positivity of $\om$ and~\eqref{iwn}. 

This does not yet define a Hilbert space as there may be elements $X \in \sA$ with $\om (X^\da X) =0$. 
Denote $\sJ \subset \sA$ to be set of all such elements. Note that 
if $X \in \sJ$, then 
\be \label{xhg}
 A X \in \sJ, \qquad \forall A \in \sA \ .
\ee
To see~\eqref{xhg}, we have from~\eqref{cahy} 
\be 
0 \leq \om((AX)^\da (AX)) = \om(X^\da A^\da A X) \leq \le(\om (X^\da X) \om ((A^\da AX)^\da A^\da AX)\ri)^\ha = 0 \ .
\ee 
Similarly,  $\om (X B) = 0$ and $\om (A X) =0$, for any $A,B \in \sA$. 
It then follows that 
$\vev{A|B} = \vev{A + X_1|B + X_2}$ with 
$X_{1,2} \in \sJ$. Accordingly, we can define an equivalence relation 
\be 
A \sim A + X , \qquad A \in \sA, \quad \forall X \in \sJ   
\ee
and 
\be 
\vev{A|A} = 0 , \quad {\rm iff} \quad A\sim 0  \ .
\ee
Denote the equivalence class of $A$ as $[A]$. The set of $[A]$, which is the quotient $\sA/\sJ$, 
has a positive norm for any nonzero element. 

The completion of $\sA/\sJ$ then defines a Hilbert space $\sH_\om$.  
Denote $\ket{C}$ as the vector associated with $[C]$. The representation $\pi_\om (A)$ of $A \in \sA$ on $\sH_\om$ can be defined as 
\be 
\pi_\om (A)\ket{C} = \ket{AC}
, \quad \forall C \in \sA  
\ee
 which satisfies by definition 
 \be 
 \pi_\om (A) \pi_\om (B) = \pi_\om (AB) \ .   
\ee
The vector $\ket{\bid}$ for the equivalence class $[\bid]$ of the identity is special, and we introduce a special notation for it 
\be 
\ket{\Om} \equiv \ket{\bid} \ .
\ee
It then follows that 
\bega
\pi_\om (A) \ket{\Om} = \ket{A}, \quad A \in \sA , \\
\om (A) = \vev{\Om|\pi_\om (A)|\Om}  \ .
\label{exoO}
\end{gather} 
Thus $\om$ is represented by the state vector $\ket{\Om}$ in $\sH_\om$. 

$\ket{\Om}$ has the property that the set $\{\pi_\om (A) \ket{\Om} , A \in \sA\}$ is dense in $\sH_\om$ as by definition 
$\sH_\om$ is the completion of $\{[A], A \in \sA\}$. A vector with such property is called  {\bf cyclic} with respect to $\pi_\om (\sA)$, and a representation of $\sA$ with a cyclic vector is called a {\bf cyclic representation}.  
Thus the trio $(\sH_\om, \pi_\om (\sA), \ket{\Om})$ provides a cyclic representation for $\sA$.

It can further be shown that the representation $(\sH_\om, \pi_\om (\sA), \ket{\Om})$ is {\it unique} up to unitary equivalence. 
Properties of $\om$ affect the nature of the representation $(\sH_\om, \pi_\om (\sA), \ket{\Om})$, and it can be readily shown that:
\begin{Prop} \label{wpro0}
The representation $(\sH_\om, \pi_\om (\sA), \ket{\Om})$  is irreducible iff $\om$ is pure, i.e.,
\be \label{wpro1}
 \sB (\sH_\om) = (\pi_\om (\sA))'' \ .
 \ee
\end{Prop} 

 \begin{Prop} \label{wpro2}

$\ket{\Om}$ is separating with respect to $\pi_\om (\sA)$ if and only if $\om$ is faithful.
\end{Prop}

We define a vector $\ket{\Om}$ to be {\bf separating} with respect to an algebra, if $A \ket{\Om} \neq 0$ for any nonzero element $A$ of the algebra. When $\om$ is faithful, the set $\sJ$ is trivial, i.e. $\om (A^\da A) \neq 0$ for any nonzero $A \in \sA$, and the separating property of $\ket{\Om}$ then follows from~\eqref{exoO}.

To get some intuition on the GNS construction, it is instructive to consider a simple example.  
Suppose $\sA$ is the matrix algebra $M_n$ of $n \times n$ matrices, and take $\om$ to be 
\be 
\om (A) \equiv \Tr (\rho A), \quad A \in \sA = M_n 
\ee
where $\rho$ is a positive-definite Hermitian $n \times n$ matrix with unit trace and $\Tr$ is the matrix trace.  Clearly, $\om$ satisfies the conditions for a state given in Sec.~\ref{sec:states}. Suppose $\rho$ has rank $m$ with $1 \leq m \leq n$. 
From our definitions around~\eqref{mixes}, when $m = 1$, $\om$ corresponds to a pure state, while $\om$ is a 
faithful state for $m = n$, i.e., for an invertible $\rho$.

The constructions of the GNS Hilbert space $\sH_\om$ and the representation $\pi_\om (\sA)$ are straightforward and are left as an exercise. Here we merely state the results (up to unitary equivalence). We find $\sH_\om = \sH_n \otimes \sH_m$ where $\sH_m$ is the Hilbert space of $m$ complex vectors, and $\ket{\Om}$ and $\ket{A}$ corresponding to $[A]$ can be written as 
\be
\ket{\Om} = \sum_{a=1}^m \rho_a^\ha \ket{a}_R \ket{a}_L , \quad \ket{A} =  \sum_{i=1}^n \sum_{a=1}^m A_{ia} \rho_a^\ha \ket{i}_R \ket{a}_L   \ .
\ee
In the above equations we have labelled basis vectors of $\sH_n$ with a subscript $R$, while those of $\sH_m$ 
with a subscript $L$. 
$\{\rho_a, \, a=1, \cdots, m\}$ are the nonzero eigenvalues of $\rho$,  $\{\ket{i}, i=1, \cdots n\}$ is the basis where $\rho$ is diagonalized, and $A_{ij}$ are the matrix elements of $A \in M_n$ in that basis. The representation of $A$ in $\sH_\om$ is 
\be \label{repA} 
\pi_\om (A) = A \otimes \bid_m   \ .
\ee
From~\eqref{repA}, the action of $\pi_\om (\sA)$ on $\sH_\om$ is irreducible only for $m=1$. This is consistent with Proposition~\ref{wpro0} and equation~\eqref{wpro1}.  For $m=n$, $\ket{\Om}$ is both cyclic and separating with respect to $\pi_\om (\sA)$, consistent with Proposition~\ref{wpro2}. 

This simple example also illustrates another important application of the GNS construction. Here 
$\sA$ can also be viewed as a von Neumann algebra acting on the Hilbert space $\sH_n$ spanned by $n$ complex vectors, and $\rho$ is a density operator on $\sH_n$. For $m > 1$, $\rho$ is not pure, and $\ket{\Om}$ provides a canonical purification of $\rho$ (the GNS Hilbert space $\sH_\om$ is larger than the Hilbert space $\sH_n$ we start with). 

More generally, consider a \vNa\ algebra $\sA$ 
and a normal state $\om$ on $\sA$. 
 Using the GNS construction we can construct a GNS Hilbert space 
 $\sH_\om$ in which $\om$ is represented by a vector $\ket{\Om} \in \sH_\om$. Thus $\ket{\Om}$ provides a purification of $\om$. When $\om$ is faithful, $\ket{\Om}$ is  cyclic and separating with respect to $\pi_\om (\sA)$.  

\subsection{Emergent von Neumann algebras of the entangled spin example} \label{sec:entsp}

Consider the entangled spin example of Sec.~\ref{sec:inf} in the $N \to \infty$ limit. There we mentioned that for a generic Hamiltonian, the space of finite-energy excitations around the state~\eqref{une} forms a Hilbert space $\sH_{\Phi_\th}$. With the GNS construction, we can now give an explicit description of this Hilbert space. 

For this purpose, consider the algebra $\sA$ of finite-energy operations, which consists of operators of the form 
\bega\label{phop}
O= \al_1  \otimes \al_2 \otimes \cdots \otimes \al_n \otimes \cdots , \\ 
\text{all but finitely many of the $\al_i$ are  equal to $\bid_2 \otimes \bid_2$}, \nonumber 
\end{gather}
where $\al_i$ is an operator acting on $i$-th spin pair and $\bid_2$ is the identity operator acting on a single spin.
 The requirement that only a finite number of $\al_i$ can be nontrivial (not equal to the identity) comes from that finite energy processes can only flip a finite number of spins from $\ket{\Phi_\th}$. In the $N \to \infty$ limit, 
$O$ has a well-defined norm descended from the finite $N$ system. 
Upon completion in terms of norm convergence, $\sA$ is then a $C^*$-algebra.
$\ket{\Phi_\th}$ defines a state $\om_\th (\sA)$ on $\sA$ as 
\be 
\om_\th (O) = \vev{\Phi_\th |O|\Phi_\th} , \quad O \in \sA \ .
\ee
Using the GNS construction of Sec.~\ref{sec:GNS}, a Hilbert space $\sH_{\Phi_\th}$ can be obtained from $\sA$ and $\om_\th$. Heuristically, $\sH_{\Phi_\th}$ can be understood as obtained from acting $\sA$ on $\ket{\Phi_\th}$, i.e., the completion of the set of states obtained from $\ket{\Phi_\th}$ by flipping a finite number of spins. 
In this case,~\eqref{vnaA} is simply $\sB (\sH_{\Phi_\th})$, the set of bounded operators on $\sH_{\Phi_\th}$. 

Suppose we have access only to the $R$-system. The subalgebra $\sM_R$ of $\sB (\sH_{\Phi_\th})$ consisting of operators acting on the right spin system, completed under weak convergence on $\sH_{\Phi_\th}$, is then a von Neumann algebra. We can similarly define a \vNa\ $\sM_L$ consisting of operators of $\sB (\sH_{\Phi_\th})$ acting on the left spin system. Since actions on the left and right spins commute, we should have 
\be 
\sM_L = \sM_R'  \ .
\ee
In Sec.~\ref{sec:II} and~\ref{sec:ensR} we will discuss entanglement between $R$ and $L$ systems using the structure of $\sM_R$ for different $\th$.

The entangled spin example  is a bit special as there is a complete symmetry between the $R$ and $L$ spin systems. We can also obtain $\sH_{\Phi_\th}$  using the algebra of the $R$-system alone. Instead of~\eqref{phop}, consider  the algebra $\sA_R$ consisting of finite-energy operations on the $R$-systems, i.e. operators of the form 
\be\label{phop00}
A= a_1  \otimes a_2 \otimes \cdots \otimes a_n \otimes \cdots , \quad \text{all but finitely many of the $a_i$ are  equal to $\bid_2$},
\ee
where $a_i$ is an operator acting on the $i$-th spin in the $R$-system.  With the norm descended from the large $N$ limit, $\sA_R$ is again a $C^*$-algebra, and $\ket{\Phi_\th}$ defines a state $\om_\th$ on $\sA_R$ as 
\be 
\om_\th (A) = \vev{\Phi_\th |A|\Phi_\th} , \quad A \in \sA_R \ .
\ee
Note that $\om_\th$ is faithful state of $\sA_R$, but not $\sA$. Now using the GNS construction, we can build a GNS Hilbert space from $\sA_R$ and $\om_\th$, and it can be readily shown that the Hilbert space coincides with $\sH_{\Phi_\th}$.\footnote{To have some intuition on this, consider a single spin pair. Acting on $\ket{\phi_\th}$ using operators of the right spin system can generate all the states from acting on $\ket{\phi_\th}$ using operators of the left spin.}  The von Neumann algebra $\sM_R$ defined earlier can be thought as the \vNa\ corresponding to $\sA_R$
obtained through~\eqref{vnaA}.

\section{Von Neumann algebras and entanglement: type I and II} \label{sec:iandii}


After giving a brief introduction to von Neumann algebras, we now proceed to discuss how they can be used to capture entanglement. As discussed earlier, a von Neumann algebra $\sM$ defines a subsystem, and vice versa. 
 We will show that properties of $\sM$ reflect the entanglement structure of the subsystem. 
 In particular, we will see type II and III algebras provide new perspectives and tools for dealing with systems with infinite amount of entanglement. 
 
 A key difference between a type I or type II algebra and a type III one is that there exists a trace $\tr$ associated with the algebra, which has important implications. In this section we discuss how to capture entanglement for type I and II algebras. Type III algebras are discussed in the next section.   

\subsection{Density operators for type I and II algebras} 


Suppose the system is in a state $\ket{\Psi}$. 
When a \vNa\  $\sM$ has a trace, it is possible to define a density operator $\rho_\sM$ associated with $\sM$ for $\ket{\Psi}$. 
More explicitly, define $\rho_\sM \in \sM$ as the operator satisfying the equation
\be \label{rhps}
\tr (A \rho_\sM) =  
\vev{\Psi|A |\Psi} 
, \quad \forall  A \in \sM  \ .
\ee
Given the positive properties of $\tr$, equation~\eqref{rhps} has a unique solution\footnote{It is important that $\rho_\sM$ belongs to $\sM$.}, and $\rho_\sM$ is normalized and positive, thus is indeed a density operator.\footnote{The definition~\eqref{rhps} applies to a general state $\om$ on $\sM$ (which from~\eqref{norS} corresponds to a general density operator $\rho$ on $\sH$), with the replacement of  $\vev{\Psi|A |\Psi}$ by $\om (A)$. Properties~\eqref{iwn}--\eqref{pou} and~\eqref{pnor} of $\om$ implies that the resulting $\rho_\sM$ is normalized and positive.}

We stress that the definition~\eqref{rhps} of $\rho_\sM$  does {\it not} require factorization structure~\eqref{fact}. 
We can interpret $\rho_\sM$ as the reduced density operator of $\ket{\Psi}$ with respect to the algebra $\sM$, and use it to calculate entropies  and other quantum informational measures. 
For example, the entanglement entropy $S_\sM$ is defined as 
\be \label{Mee} 
S_\sM  
\equiv -\tr \rho_\sM \log \rho_\sM \ .
\ee

The interpretation of $\rho_\sM$ as a reduced density operator may seem puzzling at first sight, as the definition of~\eqref{rhps} 
is very different from~\eqref{reDe}. In particular, since we have only access to $\sM$, in what sense does $\rho_\sM$ capture the entanglement of the system? We will now answer this question based on the classification of $\sM$, which also shows how different types of \vNa s capture different entanglement structures. 

\subsection{Type I algebras} \label{sec:type I}

\subsubsection{Type I factor}


For $\sM$ being a type I factor, as mentioned earlier, there exists a factorization of the Hilbert space 
$\sH = \sH_R\otimes  \sH_L$ such that 
\be \label{0tyia} 
\sM = \sB (\sH_R) \otimes \bid_L, \quad \tr = {\rm Tr}_{\sH_R},  \quad \sM' = \bid_R \otimes \sB (\sH_L) \ .
\ee 
If $\sH_R$ is an $n$-dimensional Hilbert space with $n$ finite, then $\sM$ is type I$_n$, otherwise it is I$_\infty$. 

Given a state $\ket{\Psi}$ on $\sH$, it can be readily seen that the definition~\eqref{rhps} in fact coincides with~\eqref{reDe}: 
using~\eqref{reDe} the RHS of~\eqref{rhps} is simply 
\be \label{rhps1}
\vev{\Psi|A|\Psi} 
= {\rm Tr}_{\sH_R} (\rho_R A), \quad \text{for all } A \in \sM  \ .
\ee
Comparing~\eqref{rhps} with~\eqref{0tyia}--\eqref{rhps1}, we conclude that $\rho_\sM = \rho_R = 
\Tr_{\sH_L} \ket{\Psi}\bra{\Psi}$, 
showing that~\eqref{rhps} gives an equivalent description of the reduced density operator and hence captures exactly the same quantum entanglement information.


{\bf Remarks:} 

\ben 
\item We stress that the two approaches to obtaining the reduced density operator are conceptually very different. In~\eqref{reDe} we need the knowledge of the global state $\ket{\Psi}$ for the whole system,
including the part in $\sH_L$ which we cannot access. 
It is striking that using~\eqref{rhps} we can deduce $\rho_R$ and hence all the entanglement information 
encoded in it using only data concerning $\sM$ (expectation values of operators in $\sM$), without any knowledge of the complement. That is, quantum information of $\rho_R$ is fully encoded in the structure of $\sM$.

\item  That the trace $\tr$ can be associated with a Hilbert space implies that it can be used to ``count'' states. Thus the corresponding entropy can be given a statistical interpretation. 

\een

\subsubsection{General type I} \label{sec:InonF}

Now suppose $\sM$ is type I, but with a nontrivial center. It can be shown that there exists a decomposition 
\bega \label{deco1}
\sH = \oplus_\al \sH_\al, \quad \sH_\al = \le(\sH_{R_\al} \otimes \sH_{L_\al} \ri), \\
\sM = \oplus_\al \le( \sB (\sH_{R_\al} )\otimes \bid_{L_\al} \ri), \quad \sM' =  \oplus_\al \le( \bid_{R_\al}   \otimes \sB (\sH_{L_\al} ) \ri) \ .
\end{gather} 
A trace on $\sM$ can be defined as a sum of the trace in~\eqref{0tyia} for each factor in the decomposition\footnote{such that minimal projections again have dimension $1$},
\be \label{htr1}
\tr A =\sum_\al {\rm tr}_\al A_\al \equiv  \sum_\al {\rm Tr}_{\sH_{R_\al}}  A_\al, \quad A = \sum_\al (A_\al \otimes \bid_{L_\al}), \; A_\al \in \sB (\sH_{R_\al} ) \ .
\ee

Now consider a state $\rho$ (which can be pure or mixed) on $\sH$, which from~\eqref{deco1} can be decomposed as 
\bega
\rho = \oplus_\al p_\al \rho_\al  , \quad {\rm Tr}_\al \rho_\al =1, \quad \sum_\al p_\al =1, \quad p_\al \in [0,1]
\end{gather} 
where $\rho_\al$ is a density operator on $\sH_\al$ and $ {\rm Tr}_\al$ denotes the trace of $\sH_\al$.
Equations~\eqref{htr1} and~\eqref{rhps} (with expectation value replaced by that in $\rho$) then give 
\be 
\rho_\sM = \sum_\al p_\al (\rho_{R\al}  \otimes \bid_{L_\al}), \quad \rho_{R\al}  = {\rm Tr}_{\sH_{L_\al}}\rho_\al  \in \sB (\sH_{R_\al} )  \ .
\ee
From~\eqref{Mee}, the entanglement entropy of the subsystem $\sM$ in the state $\rho$ is then given by 
\be 
S_\sM  \equiv -\tr \rho_\sM \log \rho_\sM = - \sum_\al p_\al \log p_\al + \sum_\al p_\al S_\al, \quad 
S_\al = -{\rm tr}_\al \rho_{R\al} \log \rho_{R\al}  \ .
\ee
The entropy consists of two parts, one part from the ``statistical'' entropy of $\{p_\al\}$, and the other part from weighted sum of $S_\al$.

In this case the non-factorization is rather mild, and the algebraic formulation~\eqref{rhps} of the density operator gives a unified treatment 
of both the factorized and non-factorized case. An example of the non-factorized type I case is the lattice gauge theory of Fig.~\ref{fig:latt}(a). The non-factorization due to gauge constraints can be understood in the algebraic language in terms of that the corresponding algebra is not a factor (see e.g.~\cite{CasHue13}).

\subsection{Type II algebras} \label{sec:II}

Now suppose $\sM$ is a type II factor. To illustrate its entanglement structure, we consider an explicit example of 
type II$_1$ algebra, which is provided by the entangled spin system of Sec.~\ref{sec:inf} with $\th = {\pi \ov 4}$. As discussed there, in the $N \to \infty$ limit, the physical Hilbert space $\sH_{\Phi_\th}$ of finite energy excitations cannot be factorized, due to  infinite entanglement. 
In Sec.~\ref{sec:entsp} we discussed the construction of $\sH_{\Phi_\th}$ and introduced the von Neumann algebra $\sM_R$ for operations on $\sH_{\Phi_\th}$ in the $R$-system. 
Below we show that for $\th = {\pi \ov 4}$, $\sM = \sM_R$ is type II$_1$, and use this example to illustrate physical interpretations of the reduced density operator $\rho_\sM$ and entanglement entropy $S_\sM$
introduced in~\eqref{rhps}--\eqref{Mee}. 


\subsubsection{Trace for $\th = {\pi \ov 4}$}



We first show that for $\th = {\pi \ov 4}$, when $R$ and $L$ systems are maximally entangled, 
it is possible to define a trace on $\sM = \sM_R$ as follows, 
\be \label{ttr} 
\tr A \equiv \vev{\Phi_{\th = {\pi \ov 4}} |A|\Phi_{\th = {\pi \ov 4}}}, \quad A \in \sM  \ .
\ee
To see that~\eqref{ttr} satisfies all the properties of a trace, note: 

\ben 

\item Suppose $a$ is an operator acting on the right spin of a 2-spin system 
\be \label{yn1}
\vev{\phi_{\pi \ov 4} |a|\phi_{\pi \ov 4}}  = \ha {\rm Tr}_2 a 
\ee
where $\ket{\phi_\th}$ was defined in~\eqref{ijn} and $\Tr_2$ is the standard matrix trace on $2 \times 2$ matrices. 

\item From~\eqref{une} and~\eqref{phop00} we have 
\be \label{yn2}
\vev{\Phi_{\pi \ov 4} | A|\Phi_{\pi \ov 4}} = 
{1 \ov 2^k} 
 {\rm Tr}_2 a_{i_1} \cdots {\rm Tr}_2 a_{i_k}
\ee
where $a_{i_\al}, \al=1, \cdots, k$ denote those $a$'s in $A$ that are not equal to the identity.

\item Since the expectation value of an operator in $\Phi_{\pi \ov 4}$ reduces to products of traces of $2\times 2$ matrices, we then have
\be 
\vev{\Phi_{\pi \ov 4} | AB |\Phi_{\pi \ov 4}} = \vev{\Phi_{\pi \ov 4} |B A|\Phi_{\pi \ov 4}}
\ee
which establishes~\eqref{ttr} as a trace.  

\een

\subsubsection{Type II$_1$ for $\th = {\pi \ov 4}$}

Now consider projections in $\sM$. The maximal projection is the identity operator 
\be 
\bid \equiv \bid_2 \otimes \bid_2 \cdots ,
\ee
and other examples include
\be \label{pex}
P_1 = \bid_2 \otimes P_\uparrow \otimes  \bid_2 \otimes \bid_2\otimes \cdots  , \quad 
P_2 = P_\uparrow   \otimes \bid_2  \otimes P_\uparrow \otimes \bid_2 \otimes \bid_2 \otimes\cdots  ,  \quad \cdots, \quad P_\uparrow \equiv \bma 1 & 0 \cr 0 & 0 \ema \ .
\ee
We can similarly write down more projections by changing any finite number of $\bid_2$ to a two-dimensional projection. 

Clearly there is no minimal projection, as taking any projection $P$, we can always construct a projection $\tilde P$ smaller than $P$ by changing one of $\bid_2$ in $P$ to $P_\uparrow$. Thus $\sM$ should be type II. 
$\sM$ is in fact type II$_1$, as the identity has a finite dimension 
\be \label{trni}
d (\bid) = \tr \bid = 1 
\ee
which is already normalized. For those in~\eqref{pex}, 
\be 
d (P_1) = \tr P_1 = \ha , \quad d (P_2) = {1 \ov 4} , \quad \cdots 
\ee
By considering superpositions of orthogonal projections like those in~\eqref{pex} and the limits of such superpositions 
we can construct $P$ with $d(P)$ given by any real number between $0$ and~$1$. 

The existence of trace~\eqref{ttr} comes from the right hand side of~\eqref{yn1} being a trace, which only happens for $\th = {\pi \ov 4}$. For $\th\neq {\pi \ov 4}$, a sensible trace cannot be defined, and we will see in Sec.~\ref{sec:ensR} the corresponding $\sM$ is type III.

For a vector $\ket{\Psi} \in \sH_{\Phi_{\pi /4}}$, we can use the trace~\eqref{ttr} and~\eqref{rhps}--\eqref{Mee} to define a reduced density operator and the entanglement entropy. Consider, for example, 
\be \label{eunj}
\ket{\Psi} = \eta_{1} \otimes \eta_{2} \otimes \cdots \otimes \eta_{n} \otimes \cdots, \quad
\eta_i = \ket{\phi_{\pi \ov 4}} \;\; \text{except for $i =i_1, \cdots, i_k$}  ,  \quad k > 0
\ee
for which 
\be \label{rerm}
\rho_\sM (\Psi) = 2 \rho_1 \otimes 2 \rho_2 \otimes \cdots \otimes 2 \rho_{n} \otimes \cdots, 
\ee
where $\rho_i$ is the reduced density operator corresponding to $\eta_i$ in the two-spin system. 
The factors $2$ come from the $\ha$ factor on the right hand side of~\eqref{yn1}. 
Except for $i =i_1, \cdots, i_k$, $2 \rho_i =  \bid_2$. Note that since expectation values of $\ket{\Phi_{\pi \ov 4}}$ define
the trace on $\sM$, i.e. it is a tracial state, we have  
\be\label{tracia}
 \rho_\sM  (\Phi_{\pi \ov 4}) = \bid \ .
 \ee

From~\eqref{rerm}, we can then have the following entanglement entropy 
\be \label{enre}
S_\sM (\Psi) = \sum_{s=1}^k S_2 (\rho_{i_s})  - k \log 2 < 0
\ee
for $\sM$, where $S_2 (\rho_{i_s})$ is the von Neumann entropy for $\rho_{i_s}$ in the two-spin system. 
Equation~\eqref{enre} can also be written as 
\be \label{enre1}
S_\sM (\Psi) = \sum_{i=1}^\infty \le(S_2 (\rho_i) - S_2 (\bid_2/2) \ri)
= \lim_{N \to \infty} \le(S_R^{(N)} (\Psi) - S_R^{(N)} (\Phi_{\pi \ov 4}) \ri) 
\ee 
where $S_R^{(N)} (\Psi)$ denotes the entanglement entropy for $R$ if we truncate the system to $N$ pair of spins. 
Equation~\eqref{enre1} says that $S_\sM (\Psi)$ can be understood as the difference between the entanglement entropy of $R$ in $\ket{\Psi}$ with that of the maximally entangled state $\ket{\Phi_{\pi \ov 4}}$, although neither $S_R^{(N)} (\Psi)$ nor $S_R^{(N)} (\Phi_{\pi \ov 4}) $ has a sensible $N \to \infty$ limit.  
Using~\eqref{tracia} we also find 
\be \label{Reen}
S_\sM (\Psi)=
- S_\sM (\Psi||\Phi_{\pi \ov 4}) = -  \tr \le[\rho_\sM (\Psi) (\log \rho_\sM (\Psi) - \log \rho_\sM (\Phi_{\pi \ov 4})) \ri], 
\ee
i.e. the von Neumann entropy for $\rho_\sM (\Psi)$ is equal to the negative of the relative entropy with respect to 
$\ket{\Phi_{\pi \ov 4}}$.  

We can thus interpret $\rho_\sM (\Psi)$  as a ``renormalized'' density operator that captures the entanglement structure on top of the infinite entanglement background provided by $\ket{\Phi_{\pi \ov 4}}$, and $S_\sM (\Psi)$ as a ``renormalized'' entropy relative to the maximally entangled state $\ket{\Phi_{\pi \ov 4}}$.  
 The normalization of $\tr$ specified in~\eqref{ttr} and~\eqref{trni} give 
$S_\sM =0$ for the maximally entangled state $\ket{\Phi_{\pi \ov 4}}$. 
We should caution that equation~\eqref{enre1} has to do with the simple form of state~\eqref{eunj}. For a general vector $\ket{\Psi}$, we may not be able to write $S_\sM (\Psi)$ as such a difference, but~\eqref{Reen} always holds.


Similar interpretations can be given to $\rho_\sM$ and $S_\sM (\Psi)$ for a type II$_\infty$ algebra except in this case there does not exist a preferred way to normalize the trace, and the entropy is only defined up to a state-independent constant. 

In contrast to the type I case, for type II factors  the trace $\tr$ does not count states in some Hilbert space. 
This explains that why in~\eqref{enre} and~\eqref{Reen}, the entropy is negative even though $\rho$ is positive definite and normalized to have trace $1$. In type I algebras, an entropy may be given a statistical interpretation as counting states, while in type II it measures a certain ``distance'' from the maximally entangled state.

Here we only discussed type II factors. Type II algebras that are not factors can be expressed as direct sums (or direct integrals) of factors and the discussion is similar to that of Sec.~\ref{sec:InonF}.  

\subsection{Summary}

From the above discussion of type I and type II algebras, we see that the type of a von Neumann algebra determines the general entanglement structure of all states in the physical Hilbert space. For a type I factor, the Hilbert space can be tensor factorized, meaning that any state can be expressed as a superposition of product states. In particular, pure states on the algebra (as defined below~\eqref{mixes}) correspond to unentangled product states in the Hilbert space.

For a type II factor $\sM$, the Hilbert space does not admit a tensor factorization, meaning that  any state in the Hilbert space is necessarily entangled between the subsystems described by $\sM$ and its commutant $\sM'$; no product state exists.
Correspondingly, $\sM$ does not allow a pure state, as can be seen as follows.

A state $\om$ for type II algebra can be represented by a density operator $\rho_\sM \in \sM$, which can be decomposed in terms of projections in $\sM$. However,  since a type II algebra does not have a minimal projection,  $\rho_\sM$
can always be written as a {convex combination}, $\lam \rho_1 + (1-\lam) \rho_2$, with $\lam \in (0,1)$ and $\rho_1, \rho_2$ density operators in $\sM$. As a result, the corresponding state $\omega$ is always expressible as a nontrivial convex combination of other states: $\omega = \lambda \omega_1 + (1 - \lambda) \omega_2$, 
for some states $\omega_1, \omega_2 \) on \( \mathcal{M}$. This implies that no state on \( \mathcal{M} \) is pure, in contrast to the situation in type I algebras.

While each state in the Hilbert space has its own specific entanglement correlations, the structure of the algebra determines the general pattern of entanglement for all states.

\section{Von Neumann algebras and entanglement: type III} \label{sec:III}


For type III algebras, there is no trace. Therefore,~\eqref{rhps} cannot be applied, and 
it is not possible to define a density operator $\rho_\sM$ (even in the renormalized sense) associated with $\sM$ for a state. 
In this section we discuss how to capture their entanglement structure. 
A key idea is that entanglement can manifest physically through a ``local'' {\it temperature}. 
Heuristically, when a subsystem $A$ is highly entangled with its complement $\bar A$, ignorance of $\bar A$ leads to thermal behavior for observables in $A$.
 A precise mathematical formulation of this intuitive picture is given by Tomita-Takesaki theory, 
 which we discuss first.   
We will see that type III algebras can be further divided into distinct subtypes. In the entangled spin system, the operator algebras $\sM_R$ and $\sM_L$ defined in Sec.~\ref{sec:entsp} correspond to different type III subtypes for each $\theta \in (0, \pi/4)$~\cite{Pow67,AraWoo68}. The algebra associated with a subregion in a quantum field theory (example 3 in Sec.~\ref{sec:inf}) belongs to yet another subtype of type III~\cite{Ara64,Lon82,Fre84}. These distinctions reflect different underlying entanglement structures.

\subsection{Emergent times from entanglement: modular flows} \label{sec:modular}

To motivate the description of entanglement structure for type III algebras, we first return to the type I case~\eqref{fact} and reformulate the characterization of entanglement there using a slightly different language.
Instead of studying the von Neumann and R\'enyi entropies~\eqref{EE} of the reduced density operator $\rho_R \in \sB (\sH_R)$, we consider 
\be \label{entHam}
\rho_R ={\rm Tr}_{L}    \le( |\Psi\rangle\langle\Psi| \ri) \equiv e^{- K_R}, \quad K_R \equiv - \log \rho_R  \ .
\ee
The information contained in $S_R$ and $S_R^{(n)}$ is fully captured by the spectrum of $K_R$ (known as the entanglement spectrum), which  provides a much more detailed characterization of the entanglement structure than the entanglement entropy or any finite set of Rényi entropies alone.\footnote{Exploring the entanglement spectrum of condensed matter systems has been very fruitful and is a large field of research (e.g., the first paper~\cite{LiHal08} investigating it has been cited 1907 times to this day).}
$K_R \geq 0$ itself is referred to as the entanglement Hamiltonian.

The spectrum of $K_R$ can also be probed by flow generated by $K_R$, 
\be \label{mdieo}
A(s) = e^{i K_R s} A e^{- i K_R s} \in \sB (\sH_R) , \quad A \in \sB (\sH_R) \ .
\ee
The flow acts within the system---specifically, the flow of an operator acting on $\sH_R$
remains in the $R$-system. For an observer in the 
$R$-system using the flow parameter 
$s$ as a notion of ``time,'' the system appears to be in a thermal state~\eqref{entHam} with inverse temperature $\b = 1$. That is, correlation functions of flowed operators should satisfy the Kubo-Martin-Schwinger~(KMS) relations, which are characteristic of thermal equilibrium.

Similarly, we can introduce $K_L = - \log \rho_L, \; \rho_L = {\rm Tr}_{R}    \le( |\Psi\rangle\langle\Psi| \ri)$ and study the flow generated by it for operators in $\sH_L$. 
In fact, we can treat the $R$ and $L$ systems together by introducing\footnote{Below $K_R$ should be understood as $K_R \otimes \bid_L$, where for notational simplicity we always suppress the identity part. Similarly with $K_L$.} 
\be \label{splK}
\De_\Psi  \equiv \rho_R \otimes \rho_{L}^{-1} , 
\quad
- \log \De_\Psi = K_R - K_{L}  ,
\ee
that acts jointly on both subsystems, with the flow~\eqref{mdieo} and its counterpart in the $L$ subsystem written as 
\bega
\label{umos}
\sig_s (A)  \equiv 
\De_\Psi^{-i s}A \De_\Psi^{is} 
\in \sB (\sH_R) , \quad \forall A \in \sB (\sH_R) ,\;\; s \in \RR , \\
\sig_s (A')  \equiv  
\De_\Psi^{-i s} A' \De_\Psi^{is} 
 \in \sB (\sH_L) , \quad \forall  A' \in \sB (\sH_L) \ .
 \label{umos1}
 \end{gather} 
 $\De_\Psi$ is called the {\bf modular operator}, the action $\sig_s$ {\bf modular flow}, and $s$ the {\bf modular time}.  
 
 The fact that in a pure state $S_R = S_L$ and $S_R^{(n)} = S_L^{(n)}$---a consequence of that $\rho_R$ and $\rho_L$ have the same set of eigenvalues---also has a nice reformulation in terms of $\De_\Psi$. Specifically, $\De_\Psi$ leaves $\ket{\Psi}$ invariant, 
\be \label{Dinv}
\De_\Psi \ket{\Psi} = \De_\Psi^{-1} \ket{\Psi} =  \ket{\Psi}, \quad \text{i.e.} \quad (K_R -K_L) \ket{\Psi} =0  \ .
\ee
This implies that the modular flow generated by $\De_\Psi$ is a symmetry of the state $\ket{\Psi}$. 


It should be noted that the definitions of $K_R$ and $K_L$, and the modular operator $\De_\Psi$ given in~\eqref{splK} are well-defined only when the reduced density matrices $\rho_R$ and $\rho_L$ 
are full-rank~(i.e., invertible). This implies that the Hilbert spaces $\sH_R$ and $\sH_L$  have the same dimension (which may be infinite), and in turn ensures the existence of a correspondence between states and operators in $\sH_R$ and $\sH_L$. In particular, it becomes possible to define a ``swap'' operator $J_\Psi$ that exchange the two subsystems. Suppose $\ket{\Psi}$ has the Schmidt decomposition $\ket{\Psi} = \sum_n \sqrt{\lam_n}  \ket{n}_R \otimes   \ket{ n}_L$. Then, in the basis $\{\ket{n}\}$, 
for a state $\ket{\phi} = \sum_{m,n} \phi_{mn} \ket{m}_R \otimes  \ket{ n}_L  \in \sH$, 
we define $J_\Psi$ by
\be 
J_\Psi \ket{\phi} =  \sum_{m,n} \phi^*_{mn} \ket{n}_R \otimes   \ket{ m}_L ,
\ee
where $J_\Psi$ is {\it anti-unitary} and satisfies 
\be\label{iunl}
J^2_\Psi = 1 \ .
\ee
It can be readily checked acting with $J_\Psi$ on an operator in $\sM = \sB (\sH_R) \otimes \bid_L$ yields an operator in  $\sM = \bid_R \otimes \sB (\sH_L)$.  $J_\Psi$ is known as the {\bf modular conjugation operator}.

The spectrum of $\De_\Psi$,  the flows it generates~\eqref{umos}--\eqref{umos1}, and the existence of the modular conjugation operator $J_\Psi$, together provide a powerful framework for characterizing the entanglement structure of the system.
Importantly, it is this modular structure---rather than the reduced density matrices $\rho_R$ and $\rho_L$ themselves and the associated entanglement entropies---that admits a generalization to the type III case, where $\rho_R$ and $\rho_L$ are no longer defined.

To describe the generalization to type III situations, we need to rephrase the condition that $\rho_R$ and $\rho_{L}$ 
are full-rank using language that does not rely on the existence of these density matrices. The relevant concepts are cyclicity and separating properties, which were introduced in Sec.~\ref{sec:GNS}.
Recall that a state $\ket{\Psi}$ is 
is said to be {\bf cyclic} with respect to an algebra $\sM$ if the set  $\{A \ket{\Psi}, A \in \sM\}$ is dense in the Hilbert space $\sH$.
It can be readily checked that the condition that  $\rho_R$ and $\rho_{L}$ are full-rank is equivalent to the statement that  $\ket{\Psi}$ is cyclic with respect to both $\sM = \sB (\sH_R) \otimes \bid_L$  and its commutant $\sM' = \bid_R \otimes \sB (\sH_L)$. The condition can then be rephrased as requiring that $\ket{\Psi}$ be cyclic with respect to both $\sM$ and $\sM'$.
Moreover, it is a simple exercise to show that a state $\ket{\Psi}$ is cyclic with respect to $\sM'$, if and only if it is separating with respect to $\sM$.\footnote{Recall that a state $\ket{\Psi}$  is {\bf separating} with respect to $\sM$ if $A \ket{\Psi}  \neq 0$ for any nonzero $A \in \sM$.}  Therefore, the condition can be expressed entirely in terms of $\sM$: \emph{$\ket{\Psi}$ is cyclic and separating with respect to $\sM$.} This provides another manifestation of the idea that the entanglement structure of a bipartite system is fully encoded in the properties of the algebra associated with one side.


{\bf Tomita-Takesaki} theory says that for a von Neumann algebra $\sM$ with a cyclic and separating
vector $\ket{\Psi}$, then: 

\ben 
\item There exists a {\it positive} modular operator $\De_\Psi$ leaving $\ket{\Psi}$ invariant 
\be 
 \De_\Psi \ket{\Psi} = \ket{\Psi} ,
\ee
and $K_\Psi \equiv -\log \De_\Psi$  generates unitary automorphisms (modular flows) for $\sM, \sM'$,
\bega \label{umos01}
\sig_s (A)  \equiv e^{i K_\Psi  s} A e^{- i K_\Psi  s} = 
\De_\Psi^{-i s}A \De_\Psi^{is} 
\in \sM, \quad \forall A \in \sM, \\
\sig_s (A')  \equiv e^{i K_\Psi  s} A' e^{- i K_\Psi  s} = 
\De_\Psi^{-i s}A' \De_\Psi^{is} 
\in \sM', \quad \forall A' \in \sM' \ .
\label{umos11}
\end{gather}

\item There exists an anti-unitary modular conjugation operator $J_\Psi$, with the properties 
\bega \label{j01}
J_\Psi \ket{\Psi} = \ket{\Psi},  \quad J_\Psi=J_\Psi^{-1}=J_\Psi^\dagger, 
\quad  J_\Psi \De_\Psi J_\Psi = \De_\Psi^{-1}, \\  
J_\Psi \sM  J_\Psi = \sM'  , \quad 
J_\Psi \sM' J_\Psi = \sM  
 \ .
 \label{j02}
\end{gather} 

\item The vector 
\be 
\De^{-is}_\Psi A \ket{\Psi} , \quad A \in \sM 
\ee
can be analytically continued into the strip ${\rm Im}\, s \in (0, \ha)$, and 
\be 
\De^{-i(t + {i \ov 2})}_\Psi A \ket{\Psi} = \De_{\Psi}^{-it} J_\Psi A^\da \ket{\Psi}, \quad t \in \RR  \ .
\ee

\item 
Correlation functions of modular flowed operators 
\be \label{cowq}
f_{AB} (s) \equiv  \vev{\Psi|\sig_s (A) B |\Psi} , \quad A,B \in \sM  ,
\ee
can be analytically continued into the strip ${\rm Im}\, s \in (-1,0)$ in the complex $s$-plane, and satisfy the KMS relations
\be  \label{kmsm}
f_{AB} (s) = f_{BA} (-s-i) \ .
\ee


\een

Some remarks: 

\ben[(a)]

\item Physically, Tomita-Takesaki theory says that, given a von Neumann algebra $\sM$ and 
a cyclic and separating vector $\ket{\Psi}$, there is an ``emergent'' modular time evolution that operates within $\sM$ or $\sM'$
and leaves $\ket{\Psi}$ itself invariant. In particular, with respect to the modular time, observers who have access to only $\sM$ or $\sM'$ feel at a finite temperature with inverse temperature $\b =1$. 

\item While for type I factors the cyclic and separating condition is rather stringent---requiring $\sH_R$ and $\sH_L$ to have the same dimension---in type III situations of physical interest, such as quantum field theory and statistical systems, we will see that this condition is often naturally satisfied.

\item Tomita-Takesaki theory is usually established (see e.g.~\cite{BraRobV1})  by first defining an anti-linear operator $S_\Psi$, known as the Tomita operator, with the properties:
\bega \label{DefS}
S_\Psi A \ket{\Psi}=A^\dagger\ket{\Psi} ,  \quad 
S_\Psi^\da A' \ket{\Psi}=A'^\dagger\ket{\Psi} , \quad   \forall A\in \mathcal{M} , \forall A' \in \mathcal{M}'  , \\
\label{defS1}
S_\Psi^2 = 1 , \quad S_\Psi \ket{\Psi} = \ket{\Psi} \ .
\end{gather} 
One then introduces $J_\Psi$ and $\De_\Psi$ respectively as the unitary and self-adjoint part of the polar decomposition of $S_\Psi$,
\be \label{defS0}
S_\Psi = J_\Psi \De_\Psi^\ha \ .
\ee
The statements 1-4 above on $\De_\Psi$ and $J_\Psi$, and the KMS relations can then be proved  from~\eqref{DefS}.

\item It can be shown that if there exists an operator that generates automorphisms~\eqref{umos01}--\eqref{umos11} of the algebras, and satisfies the KMS condition~\eqref{kmsm}, then it must be the modular operator for $\ket{\Psi}$~\cite{TakV2} (see~\cite{Sor23b} for a recent discussion). This statement can be used to find the modular operator.

\item  For $\sM$ being a type II factor, we can define density operators $\rho_\sM, \rho_{\sM'}$ for $\ket{\Psi}$, and $\De_\Psi, K_\Psi $ can still be defined using~\eqref{splK}.  If $\ket{\Psi}$ is a tracial state, $\rho_\sM = \rho_{\sM'} = \bid$, then it follows that $\De_\Psi =\bid$. 

\item For a type III factor, $\De_\Psi$ and $K_\Psi$ cannot be split as in~\eqref{splK}. 


\een

To conclude this discussion, we mention a useful lemma~\cite{Fau25} that reflects an ``ergodic'' property of modular flows.
\begin{lem}\label{le:erg}
{\it Consider two von Neumann algebras $\sY \subset \sX$ where $\ket{\Psi}$ is jointly cyclic and separating for both algebras. Let $\sig_s$ denote the modular flow associated with $\sX$. Then $\sX$ can be generated by acting with $\sig_s$ on $\sY$.}
\end{lem}

\subsection{Classifications of type III factors}



The existence of the modular operator and modular flows provide powerful tools for characterizing type III algebras. 
In particular, properties of modular flows can be used to further classify type III factors. 

Firstly, modular flows can be used distinguished type III from type I and II factors. It can be shown that: 
\bega
\text{\it Modular flows $\sig_s (\sM)$ for all $s \in \RR$ are inner automorphisms of $\sM$} \cr
\text{\it if and only $\sM$ is type I or type II.}
\label{inner}
\end{gather} 
We say $\sig_s$ is an {\bf inner automorphism} if there exists a unitary operator $U_s \in \sM$ such that 
\be 
\sig_s (A) \equiv \De_\Psi^{-is} A \De_\Psi^{is} = U_s^\da A U_s , \quad A \in \sM  
\ .
\ee
For type I and type II factors, the modular operators can be split as in~\eqref{splK} and thus  modular flows 
are  inner automorphisms with $U_s = e^{-i K_R s} \in \sM$ for all $s \in \RR$. 
The statement~\eqref{inner} means that for a type III algebra, there should exist some $s$ such that $\sig_s (\sM)$ are not inner automorphisms. 

The modular operator $\De_\Psi$ and the corresponding flows $\sig_s^{\Psi}$ depend on 
the specific cyclic and separating vector $\ket{\Psi}$. For a different cyclic and separating vector $\ket{\Om}$, we have the corresponding $\De_\Om$ and $\sig_s^{\Om}$.  It is natural to wonder whether there are relations between these modular flows. Indeed, there are. 
It can be shown that they are related by {\it inner} automorphisms of $\sM$. 
More explicitly, there exists a one-parameter unitaries $u_{\Psi \Om} (s) \in \sM$ that relate the actions of $\sig^\Om_s$ and $\sig^\Psi_s$, 
\bega \label{nkd}
\sig_s^\Psi (A) = u_{\Psi \Om} (s)  \sig_s^\Om (A) u_{\Psi \Om}^\da (s) 
 , \quad \forall s\in \RR, \; \forall A \in \sM , \\
 \text{i.e.} \quad \De_\Psi^{-is} A \De_\Psi^{is}  = u_{\Psi \Om} (s)  (\De_\Om^{-i s} A \De_\Om^{i s} ) u_{\Psi \Om}^\da (s)  \ .
 \label{nkd1}
\end{gather} 
From~\eqref{nkd} it follows immediately that $u_{\Psi \Om} (t)$ should satisfy the chain rule
\bega
 u_{\Psi \Om} (t) u_{\Om \Phi} (t) = u_{\Psi \Phi} (t)  \ .
 \label{ss11}
  \end{gather}
Similarly, there exists a one-parameter family of unitaries $u'_{\Psi \Om} (s) \in \sM'$ relating modular flows of $\sM'$ generated by $\De_\Om$ and $\De_\Psi$. 

Now denote $T (\sM)$ as the collection of $\sig_t^\Psi$ that are inner automorphisms 
of $\sM$, i.e. 
\be \label{tmi}
T(\sM) = \{t \in \RR : \sig^\Psi_t \text{ is an inner automorphism of }\sM \} \ .
\ee
From~\eqref{nkd}, $T (\sM)$ is independent of the choice of $\Psi$, and thus is intrinsic to 
$\sM$. We call such an object an invariant of $\sM$, and $T(\sM)$ is one of Connes' two
fundamental invariants. From~\eqref{inner}, for type I and type II factors, $T (\sM) = \RR$, while for a type III factor $T (\sM)$ is a proper subset of $\RR$ (in fact it can be shown it is a measure zero subset).


The other invariant is defined as 
\be 
S(\sM) = \cap_\Psi \sig (\De_\Psi) \subset \RR_+  \equiv  [0, \infty) 
\ee
where $\sig (\De_\Psi)$ is the spectrum of $\De_\Psi$ and {intersection $\cap_\Psi$ is taken over all cyclic and separating vectors $\ket{\Psi}$.}  $S (\sM)$ is state-independent and reflects intrinsic properties of the algebra itself. 

For $\sM$ being type I or II,  $S(\sM) = \{1\}$, as the algebra contains a tracial state for which $\De_\Psi = \bid$.\footnote{For type I$_\infty$ and II$_\infty$, the tracial state is not a normalizable state, but can be approached by a sequence of normalizable states.} 
Connes proved that $S(\sM)$ is a closed multiplicaltive subgroup of $\RR_+$, and the only other possibilities are:
\bea
&& \text{Type III$_0$: \quad $S(\sM)=  \{0,1\}$}   \\
\label{iiil}
&&\text{Type III$_\lam$: \quad $S(\sM) = \{0\} \cup \{\lam^n | n \in \ZZ\}$,} \quad \lam \in (0,1) \\
&&\text{Type III$_1$:  \quad $S(\sM) = \RR_+$} \ .
\label{iii1} 
\eea

The classification of type III factors provides a classification of the entanglement structure for systems with type III factors, and thus leads to new handles for dealing with such systems.


To conclude this subsection, we mention some further properties of $u_{\Psi \Om} (s)$ introduced in~\eqref{nkd}. 
From~\eqref{nkd1}, it can be readily shown that $u_{\Psi \Om} (t)$ satisfies the  identity 
\bega \label{cocy}
u_{\Psi \Om} (t_1 + t_2) =  u_{\Psi \Om} (t_1) \le( \De_\Om^{-i t_1} u_{\Psi \Om} (t_2)  \De_\Om^{i t_1} \ri) 
= u_{\Psi \Om} (t_1) \sig_{t_1}^\Om ( u_{\Psi \Om} (t_2) ) \ .
\end{gather}
Denoting 
 \be 
 \sig^{\Psi \Om}_t (A) \equiv  \sig_t^\Psi (A)   u_{\Psi \Om} (t)  = u_{\Psi \Om} (t)  \sig_t^\Om (A),
 \ee
 we then have from~\eqref{cocy}
 \be   \label{coy1}
    \sig^{\Psi \Om}_{t +s} (A) =  \sig^{\Psi \Om}_{t} ( \sig^{\Psi \Om}_{s} (A)) \ .
    \ee
Conversely, it can be shown that  for any $u(t)$ satisfying the equation 
\be
u (t_1 + t_2) = u (t_1) \sig_{t_1}^\Om ( u (t_2) ), \quad \text{for all } t, s \in \RR
\ee
then there exists a unique cyclic and separating vector $\ket{\Psi}$ for $\sM$ with 
$u_{\Psi \Om} (t) = u (t)$.

\subsection{The entangled spin example revisited} \label{sec:ensR}

As an illustration, in this subsection we discuss $S(\sM)$ and $T (\sM)$ for the algebra $\sM_R$ of the entangled spin example in Sec.~\ref{sec:entsp}. We will see that each different 
$\th \in (0, \pi/4)$, $\sM_\th  \equiv \sM_R (\th)$ belongs to a different type III$_\lam$ algebra with $\lam = \tan^2 \th$. 

We will first work out the spectrum for the modular operator of $\sM_\th$ at a finite $N$, and then take the $N \to \infty$ limit. 
For a finite $N$, the operator algebra for the $R$-system is type I, and 
the reduced density operators for the $R$ and $L$ systems are given by 
\bega \label{rr11}
\rho_R (\Phi_\th) = \rho_r (\phi_\th) \otimes \rho_r (\phi_\th) \otimes \cdots  \otimes \rho_r (\phi_\th), \quad
\rho_r (\phi_\th) = \bma \cos^2 \th & 0 \cr 0 & \sin^2 \th \ema , \\
\rho_L (\Phi_\th) = \rho_l (\phi_\th) \otimes \rho_l (\phi_\th) \otimes \cdots  \otimes \rho_l (\phi_\th), \quad
\rho_l (\phi_\th) = \bma \cos^2 \th & 0 \cr 0 & \sin^2 \th \ema , 
\label{ll11}
\end{gather} 
where $\rho_r (\phi_\th)$ and $\rho_l (\phi_\th)$ are respectively the reduced density operators of the right and left spins of a spin pair in the state $\ket{\phi_\th}$. For $\th \in (0,{\pi \ov 4})$, both $\rho_R$ and $\rho_L$ are invertible, thus
$\ket{\Phi_\th}$ is cyclic and separating with respect to the operator algebra of the $R$-system. 
The modular operator is then given by 
\be \label{dd11}
\De_{\Phi_{\th}} = \rho_R (\Phi_\th) \otimes \rho_L^{-1} (\Phi_\th) = \de_\th \otimes \de_\th \otimes \cdots  \otimes \de_\th ,
\quad
\de_\th \equiv \rho_r (\phi_\th) \otimes \rho_l^{-1} (\phi_\th)  \ .
\ee
$\de_\th$ has eigenvalues $(1, \lam, \lam^{-1})$ with $\lam = \tan^2 \th \in (0,1)$. 
The eigenvalues of $\De_{\Phi_{\th}}$ are then given by 
\be \label{ebi}
\otimes_{i=1}^N (1, \lam, \lam^{-1}) = \prod_{i=1}^N \lam^{\al_i}, \quad \al_i = 0,1,-1 \ .
\ee 
Now considering the $N \to \infty$ limit, we find that the spectrum of $\De_{\Phi_\th}$ is 
\be \label{speDe}
\sig (\De_{\Phi_\th}) = \{0\} \cup \{\lam^n | n \in \ZZ\} \ .
\ee
Equation~\eqref{speDe} does not yet give $S(\sM_\th)$ which is given by the intersection of the spectrum for all cyclic and separating vectors. It can, however, be shown that $S (\sM_\th)$ in fact coincides with~\eqref{speDe}. Comparing with~\eqref{iiil}, we thus find that $\sM_\th$ is type III$_\lam$. That is, different $\th$'s give rise to different entanglement structures. 

From~\eqref{ebi}, $\De_{\Phi_\th}^{it} = \bid$ for $t$
satisfying $\lam^{it} = 1$ for any $N$, and accordingly this should also be the case in the $N \to \infty$ limit. 
Such $t$'s are then contained 
 in $T (\sM_\th)$, i.e.
\be
\le\{{2 \pi n \ov \log \lam}: n \in \ZZ \ri\} \subseteq T(\sM_\th)\ .
\ee
It can be shown that these in fact are the only inner automorphisms, i.e. 
\be
T (\sM_\th) = \le\{{2 \pi n \ov \log \lam}: n \in \ZZ \ri\} , \quad \lam = \tan^2 \th  \ .
\ee

Recall that for $\th = {\pi \ov 4}$, the algebra is type II$_1$, rather than type III. For $\th =0$, the two spins in a pair are unentangled. Therefore, the Hilbert space can still be factorized in the $N \to \infty$ limit, and $\sM_R$ remains a type I algebra. 
III$_0$ and III$_1$ algebras can also be obtained by slightly generalizing the example. Instead of having each pair of spins in the same entangled state $\ket{\phi_\th}$, we can take the $i$-th pair to be in  $\ket{\phi_{\th_i}}$, with an $i$-dependent $\th_i$~\cite{AraWoo68}.  Equations~\eqref{rr11}--\eqref{dd11} still apply with replacing $\th$ by $\th_i$, and~\eqref{ebi} becomes 
\be 
\label{ebi1}
\otimes_{i=1}^N (1, \lam, \lam^{-1}) = \prod_{i=1}^N \lam_i^{\al_i}, \quad \al_i = 0,1,-1 , \quad \lam_i = \tan^2 \th_i \ .
\ee 
In the $N \to \infty$ limit, all different types, including III$_0$ and III$_1$, can arise, depending on the asymptotic behavior of the infinite sequence $\{\lambda_i, i=1,2, \cdots , \infty\}$.\footnote{Araki and Woods~\cite{AraWoo68}  showed that if the sequence $\lam_1, \lam_2, \cdots$ converges to some $\lam \in (0,1)$, then this construction gives a Type III$_\lam$ algebra. If the sequence converges to $0$, one gets a type I$_\infty$ algebra, if the convergence is fast enough. If it is not fast enough, one gets type III$_0$.} In fact, for a generic choice of $\{\lambda_i\}$, the resulting algebra is of type III$_1$.

It is also possible to obtain III$_1$ algebra by considering another generalization. In each pair of entangled spins, instead of spin-$\ha$, suppose we consider two spin-$1$ system in a general entangled state. For a finite $N$, $\rho_R$ and $\rho_L$ still have the form~\eqref{rr11}--\eqref{ll11}, but with $\rho_r$ and $\rho_l$ replaced a $3 \times 3$ matrix
\be 
\rho_r = {1 \ov 1+\lam + \tilde \lam} {\rm diag} (1, \lam , \tilde \lam) = \rho_l, \quad \lam, \tilde \lam > 0 \ .
\ee
In the $N \to \infty$ limit, the eigenvalues of 
 $\De_{\Psi}$  consist of the numbers $\lam^n \tilde \lam^m$, $n, m \in \ZZ$. If  $\lam$ and $\tilde \lam$ are generic,  every non-negative real number can be approximated arbitrarily well by $\lam^n \tilde \lam^m$. 
Thus $\sig (\De_\Psi) =[0, +\infty) =  \RR_+$. It can be further shown that $S (\sM)$ in fact coincides with $\sig (\De_\Psi)$, which gives  type III$_1$. 

\subsection{Local algebras in a RQFT} \label{sec:qft}

\subsubsection{Local algebraic structure of a relativistic QFT} \label{sec:qftS}

We now consider Ex. 3 of Sec.~\ref{sec:inf}, the entanglement for the half space $x > 0$ in the vacuum state $\ket{\Om}$
of a RQFT. For definiteness, we will now restrict to $d=2$ on $\RR^{1,1}$ (with coordinates $x^\mu = (t, x)$). 

\begin{figure}[H]
\begin{centering}
	\includegraphics[width=8cm]{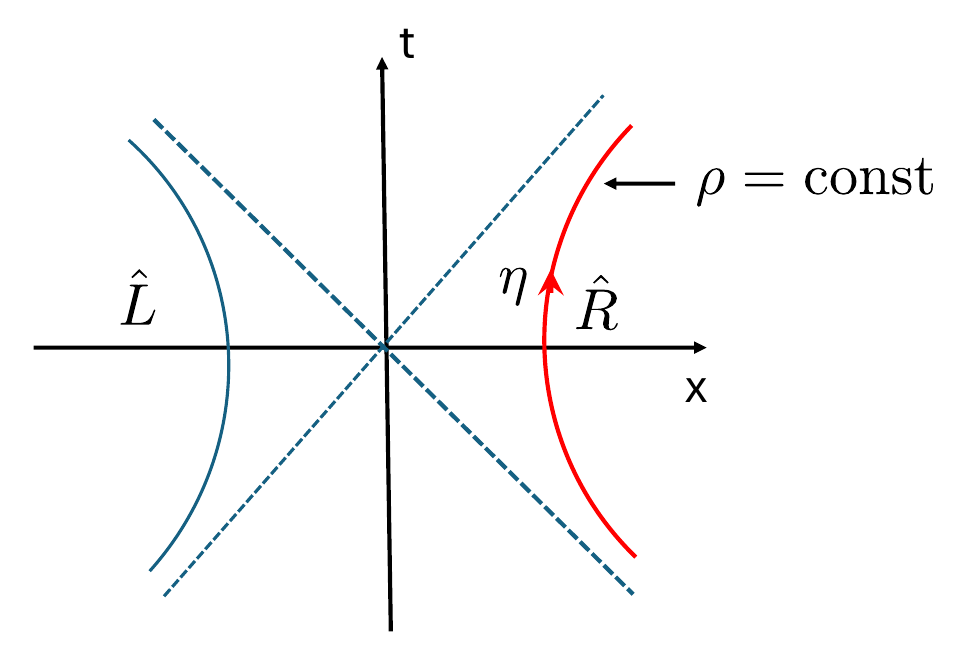}
\par\end{centering}
\caption{\small Rindler regions of $(1+1)$-dimensional Minkowski spacetime, whose metric can be written as  
$ds^2 = - dt^2 +d x^2 = - \rho^2 d \eta^2 + d \rho^2$. A Rindler observer's worldline is 
a line of constant $\rho$,  corresponding to a hyperbola 
parameterized by Rindler time $\eta$, which is the worldline of a Rindler observer. 
The proper time of a Rindler observer is thus given by $d \tau = \rho d \eta$. 
A boost generates a translation in $\eta$, and maps the right Rindler region $\hat R$  to itself.
}
\label{fig:rind}
\end{figure}

As discussed in Sec.~\ref{sec:inf}, the Hilbert space for the vacuum sector cannot be factorized into those for the $R$ and $L$ half spaces due to infinite entanglement between them in the continuum.  We can characterize the entanglement between $R$ and $L$ regions in  $\ket{\Om}$ by examining the algebra $\sM_R$ of operators localized in the $R$-region. 
 In a RQFT, the Heisenberg evolution is causal, which means that $\sM_R $ is equivalent to the algebra of operators localized in the domain of dependence $\hat R$ of the $R$-region, which is the right Rindler region (see Fig.~\ref{fig:rind}). From causality we also have $\sM_R' = \sM_L = \sM_{\hat L}$, where $\sM_L$ is the algebra of operators localized in the $L$-region.

The Reeh-Schlieder theorem states that in a relativistic quantum field theory, states obtained by applying operators localized in an open region to the vacuum state $\ket{\Om}$ form a dense set in the Hilbert space. This implies that $\ket{\Om}$ is cyclic respect to both $\sM_R$ and $\sM_L$,  and thus it is cyclic and separating with respect to $\sM_R$.

In this case, the modular operator $\De_\Om$ for $\sM_R$  can be shown to be proportional to the boost operator $K$~\cite{BisWic76}, 
\be \label{booS}
K_\Om \equiv  - \log \De_\Om 
= 2 \pi K \ .  
\ee
A modular flow then corresponds to a boost. 
To justify equation~\eqref{booS}, one needs to show: (i) Flows generated by $K$ are automorphisms of $\sM_R$. (ii) 
Correlation functions, such as 
\be 
\vev{\Om|e^{i K\eta} A (x^\mu) e^{- i K\eta}  B (y^\mu)|\Om}, \quad x^\mu, y^\mu \in \hat R ,
\ee
where $A$ and $B$ are some local operators, satisfy the KMS relation with $\b =2 \pi$ with respect to $\eta$. (i) can be readily checked as a boost takes a point of $\hat R$ to another point within the region. 
(ii) means that in the vacuum state $\ket{\Om}$ observers who use $\eta$ as their time ``feel'' a nonzero temperature $T ={1 \ov 2 \pi}$. From Fig.~\ref{fig:rind}, such observers should follow trajectories with $\rho={\rm const}$, i.e. the so-called Rindler observers. (ii) is equivalent to Unruh's observation~\cite{Unr76} that in $\ket{\Om}$, a Rindler observer experiences a finite temperature $T_\rho = {1 \ov 2 \pi \rho}$. 

The boost operator $K$ has a continuous spectrum $(-\infty, +\infty)$, which implies $\sig (\De_\Om) = \RR_+$. 
It can be further shown that $S (\sM_R)$ coincides with $\sig (\De_\Om)$, and then from~\eqref{iii1},
$\sM_R$ is type III$_1$. 
The modular conjugation operator is also simple, given by $J_\Om = \sC \sR \sT$ where $\sC$ is charge conjugation, $\sR$ is the reflection is the reflection $x \to - x$, and $\sT$ is time reversal. It can be readily checked that $J_\Om$ satisfies~\eqref{j01}--\eqref{j02}.  

The above discussion applies to half spaces in higher dimensions, with dimensions transverse to the $t-x$ plane going trivially along for the ride.

Now consider a general open region $O$ on a Cauchy slice, and the algebra 
 $\sM_O$ of operators localized in $O$.  From the Reeh-Schlieder theorem,  $ \ket{\Om}$ is cyclic and separating with respect to $\sM_O$. It is no longer possible to construct $\De_\Om$ and $J_\Om$ explicitly, since they depend on the specific theory and specific shape of $O$. Nevertheless, it is possible to deduce that $\sM_\sO$ should be type III$_1$ with the assumption that the theory has a scale invariant UV fixed point~\cite{Fre84}. A rough argument is as follows.  Consider the region very close to the boundary $\Sig_O$ of $O$, where locally $\Sig_O$ can be approximated by a half-plane. Assuming at such distances the physics can be locally described by that of the Rindler region discussed above, we can then conclude that  when acting on operators very close to $\Sig_O$ 
\be \label{joe1}
-\log \De_\Om = 2 \pi K,\quad \text{very close to $\Sig_O$}
\ee
where $K$ is the boost operator that leaves $\Sig_O$ invariant, and thus $\De_\Om$ should again have the spectrum $\RR_+$.

Type III$_1$ structure of a local operator algebra $\sM_O$ in a RQFT can be attributed to the local Rindler structure close to $\Sig_O$, which in turn arises from the relativistic causal structure of the theory. 
Accordingly,  we can say that the causal structure of a RQFT requires the type III$_1$ structure. Indeed, for a non-relativistic QFT, algebras of local operators do not have to be type III$_1$. 


\subsubsection{Split property} 

\begin{figure}[H]
\begin{centering}
	\includegraphics[width=2.0in]{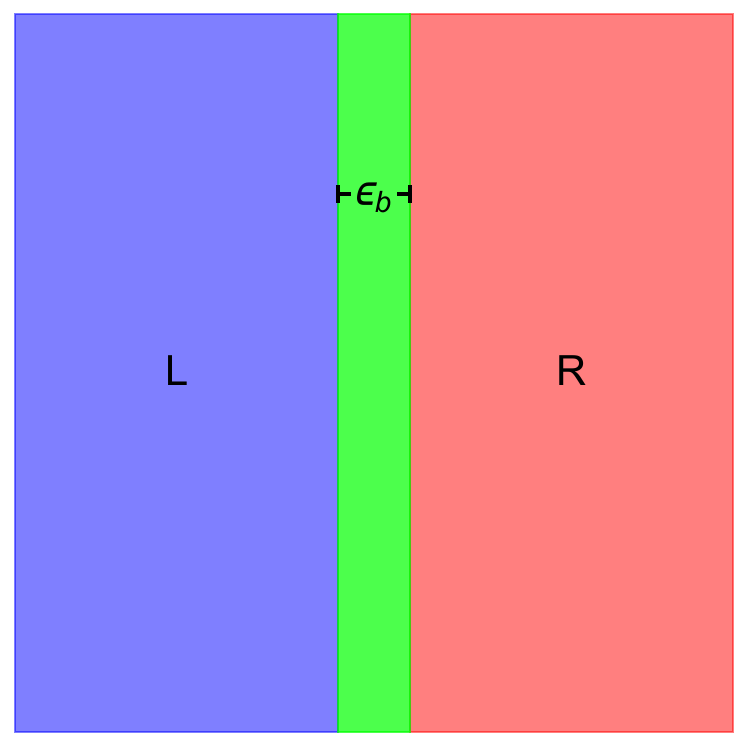} 
	\par\end{centering}
\caption{\small Slightly separated $R$ and $L$ regions on a spatial slice. We denote the green region as $I_\ep$. 
$R_\ep = R \cup I_\ep$ denotes the union of $R$ and the green region while $L_\ep = L \cup I_\ep$ denotes the union of $L$ and the green region. 
}
\label{fig:split}
\end{figure}

Heuristically, the non-factorization of the Hilbert space in a RQFT with respect to an open region and the associated type III structure 
can be attributed to infinite entanglement among short-distance degrees of freedom near the boundary of a region. It is possible to make this connection precise using the {\bf split property}~\cite{Buc74,DopLon84,BucWic86,BucFre87} (see also~\cite{Few16} for a recent discussion).
Consider the setup in Fig.~\ref{fig:split},
where we separate the two regions $R$ and $L$ by an infinitesimal distance $\ep_b$.
The split property says that there exists a tensor product decomposition of the Hilbert space $\sH = \sH_1 \otimes \sH_{2}$, 
 with  
\be\label{spli}
\sM_R \subset \sB (\sH_1) \otimes \bid_{\sH_2} \subset \sM_L' = \sM_{R_\ep}, \quad  \sM_L \subset \bid_{\sH_1} \otimes \sB (\sH_{2})  
\subset \sM_R' = \sM_{L_\ep},
\ee
where $\sM_{R_\ep}$ denotes the algebra in the region $R_\ep \equiv R \cup I_\ep$ (and similarly for $\sM_{L_\ep}$). 
$\sB (\sH_1) \otimes \bid_{\sH_2}$ is a type I algebra sandwiched between two type III$_1$ algebras $\sM_R$, $\sM_{R_\ep}$.
Equation~\eqref{spli} means that once we separate $R$ and $L$ regions by an infinitesimal distance, {\it they can be disentangled!} That is,  in $\sH$ 
there are unentangled products states with respect to subsystems described by $\sH_1$ and $\sH_2$, and $\sM_R$ and $\sM_L$ act respectively only on $\sH_1$  and $\sH_{2}$.

More generally, for any two open regions $O_1 \subset O_2$ whose boundaries do not touch (i.e., the closure of $O_1$ is contained in the interior of $O_2$), the split property says there exists a type I factor $\sN$ such that  
\be 
\sM_{O_1} \subset \sN \subset \sM_{O_2} \ .
\ee
 The split property can be shown~\cite{BucFre87} to follow from the {\bf nuclear property}~\cite{BucWic86}, which is a condition on the high energy behavior and is believed to be satisfied by ``reasonable'' RQFTs. 

If we additionally assume a RQFT to have the split property, it can be shown~\cite{BucFre87}  that 
$\sM_O$ for an open region $O$ is not only type III$_1$, but also hyperfinite. 
A von Neumann algebra $\sM$ is hyperfinite if it can be written as the weak closure of an increasing sequence of finite-dimensional matrix algebras. Hyperfinite type III$_1$ has been shown to be unique up to isomorphism~\cite{Haa87}. That is, local algebras in a RQFT are isomorphic to one another, and thus different theories are distinguished only by the relations among local algebras.

The combination of split property and type III structure of local algebras also implies a surprising property called {\bf strong local preparability}~\cite{BucDop86} (here our discussion follows that of~\cite{Yng14}). 
Consider a general state $\om$ acting on $\sM_R, \sM_L$ of Fig.~\ref{fig:split}, we expect in general 
\be 
\om (A B) \neq \om (A) \om (B), \quad A \in \sM_R, B \in \sM_L 
\ee
which reflects correlations between operators in $R$ and $L$. 
We may ask whether it is possible to use local operations $W$ in $\sM_R$ to construct  a new state $\om_W$ from $\om$ such that: (i) there is no correlation between operators in $\sM_R$ and $\sM_L$
i.e., $\om_W (AB) = \om_W (A) \om_W (B)$; (ii) on $\sM_R$ it coincides with some desired state $\phi$, but  on $\sM_L$ it remains unchanged from $\om$, i.e., $\om_W (A) = \phi (A)$,  $\om_W (B) = \om (B)$. These are very strong requirements, and it is hard to imagine that they could be satisfied.
It turns out these requirements can be achieved if $W$ lies in a slightly larger region, in $R_\ep$. 
More explicitly, for any states $\om$ and $\phi$ we can choose an {\it isometry} $W \in \sM_{R_\ep}$ (which depends on $\phi$ but not $\om$) such that 
\be 
\om_W (A B) = \phi (A) \om (B), \quad \om_W(\cdot) \equiv  \om (W^\da \cdot W) , \quad W^\da W = \bid ,
\quad A \in \sM_R, B \in \sM_L  \ .
\ee 

{\bf Proof:} From the split property~\eqref{spli}, $\sM_R$ is contained in $\sB (\sH_1)$. It can be shown there exists a vector $\ket{\xi} \in \sH_1$ such that $\phi (A) = \vev{\xi|A|\xi}$, which will be justified later in Sec.~\ref{sec:cone}. Consider the projection $P_\xi = \ket{\xi}\bra{\xi} \otimes \bid_{\sH_2} \subset \sM_{R_\ep}$. 
Since $\sM_{R_\ep}$ is type III, any projection is equivalent to the identity. 
There then exists an isometry $W \in \sM_{R_\ep}$ with $WW^\da = P_\xi, W^\da W =\bid$, 
and
\be 
\om_W = \om (W^\da B W) = \om (B), \quad B \in \sM_L \subseteq \sM_{R_\ep}'  \ .
\ee
We also have for $A \in \sM_R$
\be 
P_\xi A P_\xi = \phi (A) P_\xi  \quad  \to \quad W W^\da A W W^\da = \phi (A) W W^\da 
\quad  \to \quad W^\da A W = \phi (A) \bid \ .
\ee
We thus find 
\be 
\om (W^\da AB W) = \om (W^\da  A W  B ) = \phi (A) \om (B)  \ .
\ee


Another similar, but distinct property of type III$_1$ algebras is the so-called Connes-Stormer transitive property~\cite{ConSto78}: 

{\it Theorem (Connes-Stormer)} If $\sM$ is a type III$_1$ factor, then for every $\ep > 0$ and 
any state $\phi, \om$, there exists a {\it unitary} $W \in \sM$ such that 
\be
||\phi - \om_W|| < \ep \ .
\ee
In other words: Every state can be prepared {\it locally}, with arbitrary precision, from any other state. This can also be viewed as a statement of ergodicity. The state space is called {\bf homogeneous}.






\subsubsection{Collection of von Neumann algebras in a RQFT} \label{sec:ceRQ}

The von Neumann algebra $\sM (O)$ associated with an open region $O$ encodes all the physics associated with the region including the local spacetime geometry and its entanglement structure. The collection $\{\sM (O)\}$ for all choices of open regions $O$ in a RQFT should then 
provide a complete description of the theory. 
We conclude this subsection with a brief discussion of relations among $\sM (O)$'s. 

By definition, we should have 
\bea 
\label{he1}
	&\sM (O_1) \subseteq \sM(O_2), \quad {\rm for} \quad O_1 \subseteq O_2 , \\
	\label{Hev}
	&\sM (O) = \sM (\hat O)  , \\
	\label{Hev1}
	&\sM(O') \subseteq (\sM (O))'  \ .
\eea
Equation~\eqref{he1} is a statement of locality, i.e., if a region $O_1$ is contained in $O_2$ then all operations that can be performed in $O_1$ should be a subset of those in $O_2$. 
Equation~\eqref{Hev} is a consequence of the causal nature of the equations of motion, i.e., they evolve operators within the lightcone of the initial points. We can thus express an operator in the domain of dependence $\hat O$ of $O$ in terms of those in $O$ via equations of motion. It implies that, to describe $\{\sM (O)\}$, it is enough to restrict to a single Cauchy slice, which is also known as the  time-slice axiom.
Equation~\eqref{Hev1} is also statement of causality: operators in the causal complement $O'$ of $O$ (i.e., the set of spacetime points spacelike separated from $O$), should commute with those in $O$. 


When the equality sign in~\eqref{Hev1} is satisfied we have 
\be \label{nal}
\sM(O') = \sM (O)' ,
\ee
which is called Haag's duality. It implies a stronger form of locality, as it says that all operators that commute with those in 
$\sM (O)$ lie in $O'$. Haag's duality is not always true, but for topologically trivial regions, the duality is expected to hold for a general RQFT in the vacuum sector.

Now consider $O_1, O_2$ on a Cauchy slice. From~\eqref{he1}, we have 
\be
\label{Hev2}
	\sM (O_1) \lor \sM (O_2)\equiv \le(\sM (O_1) \cup \sM (O_2)\ri)'' \subseteq \sM (O_1 \cup O_2),
\ee
which follows since the left hand side (LHS)  is the smallest von Neumann algebra containing both $\sM (O_1)$ and $\sM (O_2)$, while from~\eqref{he1}, both are contained in $\sM (O_1 \cup O_2)$. In fact, for topologically trivial regions, 
it is expected that the algebras of $\{\sM (O)\}$ are additive, i.e., 
\be \label{nal1}
\sM (O_1 \cup O_2) = \sM (O_1) \lor \sM (O_2)  \ .
\ee
Equation~\eqref{nal1} is a stronger statement of locality as it says that there are no additional operators in a joint subregion that cannot separately be expressed in terms of~(limits of) sums of products of operators in the individual subregions making up the union. For topologically nontrivial regions, there can be nonlocal operators such as non-contractible Wilson lines that cannot be generated by local operators, which can spoil the additivity property. See~\cite{CasHue19,CasHue20} for a recent discussion.

 If both~\eqref{nal} and~\eqref{nal1} are satisfied, by taking the commutant of~\eqref{nal1} on both sides,
 we also have  the intersection property
\be \label{nal2}
\sM(O_1 \cap O_2) = \sM(O_1)\cap \sM(O_2) ,
\ee
which follows from 
\bega
(\sM (O_1 \cup O_2))' = \sM ((O_1 \cup O_2)') = \sM (O_1' \cap O_2') , \\
(\sM (O_1) \lor \sM (O_2))' =\sM (O_1)' \cap \sM (O_2)'  =  \sM (O_1') \cap \sM (O_2')  \ .
\end{gather}

When the Haag duality~\eqref{nal} and the additivity~\eqref{nal1} are satisfied for all regions (not just topologically trivial regions), we refer to such theories as `complete' theories (see e.g., a discussion in~\cite{Wit23}).

\subsection{Emergent times from subalgebras} \label{sec:halfs}

In this subsection, we show that additional notions of ``emergent'' time beyond the modular flow can arise from subalgebras of a von Neumann algebra. In particular, we will introduce a unique feature of type III$_1$ algebras called half-sided modular inclusion/translation~\cite{Bor92,Bor93,Wie93a,Wie93b,Bor96,Bor98,Bor00,Ara05}.

Suppose $\sM$ is a  von Neumann algebra and $\ket{\Om}$ is a cyclic and separating vector for $\sM$, with the associated modular and modular conjugation operators denoted as $\De_\sM \equiv e^{-K_\sM}$ and $J_\sM$. 
Now suppose there exists a von Neumann subalgebra $\sN \subset \sM$, and 
$\ket{\Om}$ is cyclic for $\sN$ (it is automatically separating for $\sN$ given $\sN$ is a subalgebra of $\sM$). 
We will denote the modular and modular conjugation operators of $\sN$ with respect to $\ket{\Om}$ as $\De_\sN = e^{-K_\sN}$ and $J_\sN$. The existence of such a subalgebra brings new structures. 

Firstly, there exists a positive operator that annihilates $\ket{\Om}$. Since $\sN \subset \sM$, from the definition~\eqref{DefS} of the Tomita operator, $S_\sM$ provides an extension of $S_\sN$.\footnote{$S_\sM$ and $S_\sN$ act respectively on $\sM \ket{\Om}$ and $\sN \ket{\Om}$, and $\sN \ket{\Om} \subseteq \sM \ket{\Om}$.}
There is a mathematical theorem saying that if an (unbounded operator) $X_1$ is an extension of $X_0$, then $X_0^\da X_0 \geq X_1^\da X_1$ (see e.g. Sec. II.1 of~\cite{Bor00}). Given that  $\De_\sM = S_\sM^\da S_\sM$, we thus have 
\be 
\De_\sN \geq \De_\sM  \ .
\ee
Therefore (the ${1 \ov 2 \pi}$ factor in the definition of $G$ is for later convenience),
\bega \label{motr0} 
 G \equiv  {1 \ov 2 \pi} (K_\sM - K_\sN)  \geq 0, \quad G \ket{\Om} = 0 
\ .
\end{gather} 
Given that $G$ is bounded from below, it can be interpreted as a ``Hamiltonian'' and the flow $e^{i G s}$  generated by it can be interpreted as a ``time'' flow. 
 In particular, the flow leaves $\ket{\Om}$ invariant. 
We thus see that existence of a subalgebra with the same cyclic vector leads to an ``emergent'' time $s$ and the vector $\ket{\Om}$ is time translation invariant. 

Another interesting object is~\cite{Lon84,Lon87} 
\be \label{defV}
V = J_\sM J_\sN , 
\ee
which leads to\footnote{Note that $V \sM V^\da$ and $V J_\sN V^\da$ cannot be defined.}
\bega 
\hat \sN_1 \equiv V^\da \sM V = J_\sN \sM' J_\sN \subset \sN  \\
\hat \sN_2 \equiv V^\da \sN V = J_\sN J_\sM \sN J_\sM  J_\sN \subset \hat \sN_1  , \\
\hat \sN_3 \equiv V^\da \hat \sN_1 V = J_\sN J_\sM \hat \sN_1 J_\sM  J_\sN \subset \hat \sN_2 
\end{gather} 
and so on. By acting repeatedly $V^\da \cdot V$ on $\sM$ we get a sequence $\hat \sN_{2n-1}, n=1,2,\cdots$, while 
acting repeatedly $V^\da \cdot V$ on $\sN$ we get another sequence $\hat \sN_{2n}, n=1,2,\cdots$, with the combined sequence nested
\be \label{etg}
\hat \sN_a \subset \hat \sN_b, \quad a < b \in \NN \ .
\ee

A third object is~\cite{Bor99} 
\be \label{defD}
D (t) \equiv \De_\sM^{- it} \De_\sN^{it} \ .
\ee
For small $t$, we have 
\be \label{eynw}
D (t) = 1 + 2 \pi i G t + O(t^2) \ .
\ee
Using $D(t)$ we can also generate new subalgebras of $\sM$
\bega\label{ehq}
\sN_t \equiv D(t) \sN (D(t))^\da = \De_\sM^{- it}  \sN \De_{\sM}^{it} 
\subseteq \sM, 
\end{gather} 
for which $\ket{\Om}$ is  cyclic and separating. It can shown that $D(t)$ the following nice  properties:
\ben 

\item It is unitary and strongly continuous in $t$, with $D(0) =\bid$. 

\item $D (t) \ket{\Om} = \ket{\Om}$ for all $t \in \RR$.

\item It can be analytically continued to the strip $S(0, \ha) = \{z \in \CC; {\rm Im} z \in (0, \ha)\}$ as holomorphic 
functions with values in $\sB (\sH)$. Furthermore, in the strip, 
\be 
||D (z)|| \leq 1 \ .
\ee

\item At the upper boundary $E (t) \equiv D(t+\ha i)$ is unitary and strongly continuous in $t$. \label{item:4}

\item It satisfies the relation 
\be 
D (s+t) = \De_\sM^{-it} D(s) \De_\sM^{it} D (t)   \ .
\ee

\item If we introduce the anti-unitary operator 
\be 
Q (t) \equiv E(t)^\da J_\sM D (t) , 
\ee
then $Q(t) = (Q (t))^\da$ and  is independent of $t$.

\item Defining a unitary operator 
\be
P(t) \equiv  D(t) E(0)^\da
\ee
we have 
\be\label{kia22}
P (t) \sM P^\da  (t) \subset \sM \ .
\ee

\een
{Most of the above statements  follow straightforwardly from the definition~\eqref{defD}.} In particular, for item~\ref{item:4}, it can be shown that 
\be  \label{kia}
E (t) =  J_\sM D(t) J_\sN  \ .
\ee
It then immediately follows that $Q(t) = J_\sN$. Also note from~\eqref{defV} and~\eqref{kia} 
\be \label{kia1}
V = E (0) = D\le({i \ov 2}\ri)\ .
\ee

It turns out the converse is also true~\cite{Bor99}: for any function $D(t)$ satisfying all the above properties,  there exists a unique von Neumann subalgebra $\sN \subset \sM$ for which  $\ket{\Om}$ is a cyclic vector, and $D(t) = \De_\sM^{-it} \De_\sN^{it}$.

From~\eqref{eynw}, for an infinitesimal $t$, $D(t)$ is related to the positive operator $G$ defined in~\eqref{motr0}, but for finite $t$, there is in general no simple relation between them. 

Surprisingly, $D(t)$ turns out be {\it generated by $G$}, if $\sN$ satisfies the following additional condition: 
\be \label{ghb}
\sN_t \equiv \De^{-it}_\sM \sN \De^{it}_\sM = D(t) \sN (D(t))^\da\subset \sN , \quad t \leq 0 \ .
\ee
This condition says that $\sN$ is preserved under modular flows generated by $\De_\sM$ for half of the $t$-axis, and is called half-sided modular inclusion. 
It can then be shown that~\cite{Bor00,Wie93a,Wie93b,Bor96,Bor98,Ara05}:
\ben

\item $K_\sM$ and $K_\sN$ obey the following algebra
 \bega  \label{mtr1} 
[K_\sM , K_{\sN}] = - 2 \pi i (K_\sM - K_\sN)  
= - i 4 \pi^2 G , 
 \\
J_\sM e^{i Gs}  J_\sM = e^{- i Gs}  , \quad
J_\sN e^{i Gs}  J_\sN = e^{- i Gs}  \ .
 \end{gather}

\item The flow generated by $G$ takes $\sM$ to itself for half of the $s$-axis, 
\be \label{halT}
e^{i G s} \sM e^{-i G s} \subset \sM , \quad s < 0,  
\ee
and in particular, 
\be\label{halT1}
 \sN = e^{-i G} \sM e^{iG}  \ .
\ee

\item $\sM$ must be type III$_1$.

\een

Fom~\eqref{mtr1}
\bega  \label{mtr2} 
[K_\sM, G] 
 = 2 \pi i G , \quad  \Delta^{-it}_\sM G \Delta^{it}_\sM =  e^{-2\pi t} G   \ .
 \end{gather} 
Using the Baker-Campbell-Hausdorff formula in~\eqref{defD}, we find that  
\be \label{kia2}
D(t)  = e^{ i G \lam_- (t)}  , \quad E(t) = D\le(t + {i \ov 2}\ri) = e^{i G \lam_+ (t)} \quad \lam_\pm (t) = 
 1\pm e^{-2 \pi t}  ,
\ee
and thus $D(t)$ is generated by $G$. Note that $\lam_- (t) \in (-\infty, 1)$ for $t \in (-\infty, +\infty)$, and $\lam_- (t) < 0$ for $t < 0$, while $\lam_+ (t) \in (1, +\infty)$. In particular, 
\be \label{kia3} 
D(+\infty) = e^{iG} = E (+\infty), \quad V = E(0) = e^{2 i G} 
\ee
where in the second equation we have used~\eqref{kia1}. 

From~\eqref{halT1} and~\eqref{kia2} 
\be 
\sN_t = \De_\sM^{-it} \sN \De_\sM^{it} = \De_\sM^{-it}  e^{-i G} \sM e^{iG}   \De_\sM^{it} 
= e^{-i G e^{-2 \pi t}} \sM e^{iG e^{-2 \pi t}} 
\ee
which in turn implies 
\be 
\sN_{\infty} = \sM , \quad \sN_{t_1} \subset \sN_{t_2}  \subset \sM, \quad  t_1 < t_2 \in \RR ,
\ee
where in the second equation we have used~\eqref{halT}. That is, the set $\{\sN_t, t \in \RR\}$ can be obtained by acting ``time translation'' generated from $G$ on $\sM$, and has a nested structure. The statement~\eqref{halT} is referred to as half-sided modular translation.

To summarize,  the half-sided modular inclusion structure~\eqref{ghb}: (i) relates both $D(t)$ of~\eqref{defD} and $V$ of~\eqref{defV} to $G$; (ii) leads to the half-sided translation structure~\eqref{halT}, which in turn extends the discrete nested sequence~\eqref{etg} into a continuous nested family labelled by $t \in \RR$ (i.e. not only makes it continuous, but also extends it to the other direction).

It can be shown that the half-sided modular translation structure, if exists, is unique. 

{\bf Theorem~\cite{Bor93}}  Suppose we have (i) nested von Neumann algebras $\sN_a, a \in \RR, \; \sN_a \subset \sN_b , a < b$
with common cyclic and separating vector $\ket{\Om}$; (ii) a one-parameter unitary group $T(a)$ with a positive generator
and $T(a) \ket{\Om} = \ket{\Om}$; (iii) $T(a)$ translates the algebras
\be 
\sN_a = T(a) \sN_0 T(-a) \ .
\ee
Then $T(a)$ is unique.

We can also consider half-sided modular inclusion for the positive half $t$-axis, 
\be \label{ghb1}
\sN_t \equiv \De^{-it}_\sM \sN \De^{it}_\sM  \subset \sM , \quad t \geq 0 \ .
\ee
Then we have (with certain sign changes from the previous case)
\bega  \label{motr3} 
[K_\sM , K_{\sN}]  
= i 4 \pi^2 G , \quad
[K_\sM, G] = - 2 \pi i G, 
\quad \Delta^{-it}_\sM G  \Delta^{it}_\sM =   G e^{2\pi t}  , \\
D(t) = e^{ i G \eta_+(t)}  , \quad E(t) =  e^{ i G \eta_-(t)} , \quad \eta_\pm (t) = 
-1 \pm  e^{2 \pi t}   , \quad V = E (0) = e^{-2 i G} \ .
\end{gather} 

\begin{figure}[H]
\begin{centering}
	\includegraphics[width=2.0in]{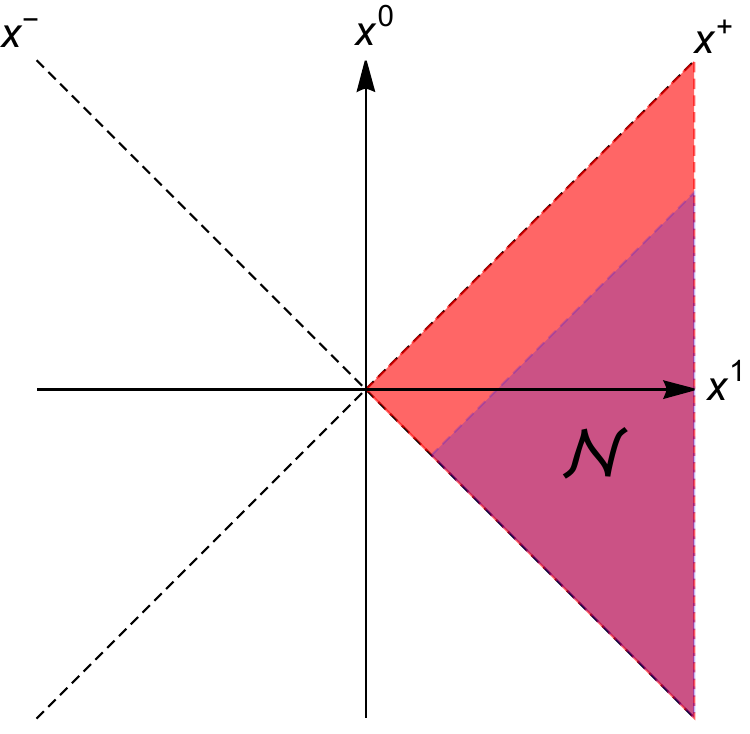} \qquad \qquad  \includegraphics[width=2.0in]{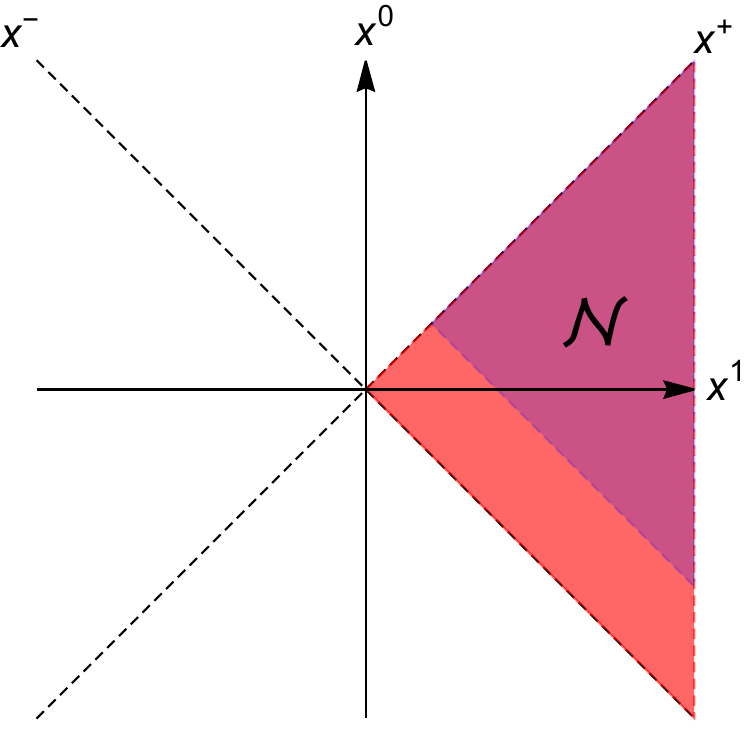}
\par\end{centering}
\caption{Left: The algebra of the subregion $\sN$ that leads to the half-sided modular inclusion structure for $x^-$ translation. Right: The algebra of the subregion $\sN$ that leads to the half-sided modular inclusion structure for $x^+$ translation. 
}
\label{fig:shiftedWedge}
\end{figure}

As an illustration, consider the example of $(1+1)$-dimensional QFT in the vacuum state of Sec.~\ref{sec:qft}. 
We take $\sM$ to be the operator algebra in the Rindler $R$-region, and $\sN$  
to be the operator algebra in the region $\{x^+ > 0, x^- < - 1\}$ (see {Fig~\ref{fig:shiftedWedge} Left}), where $x^\pm = x^0 \pm x^1$ are light-cone coordinates. From the identification~\eqref{booS}, we thus have 
\be 
\sN_t \equiv  e^{i K_\sM t}  \sN  e^{-i K_\sM t}= \text{operator algebra in region } \{x^+ > 0, x^- < -  e^{- 2 \pi t}\}   
\ee
with $\sN_t \subset  \sN$  for $t < 0$. We thus have the half-sided modular inclusion structure~\eqref{ghb}.

In this case the modular operator of $\sN$ can be found explicitly and existence of the positive generator $G$ can be directly verified. 
More explicitly, flows generated by the modular operator of $\sN$ correspond to boosts which which leave the point $a^\mu = (a^+, a^-) = (0, -1)$ invariant.  
Denote the translation operator of the QFT as $P^\mu = (P^0 = H, P^1 = P)$ and in the light-come coordinates as $P^\pm =\ha (P^0 \pm P^1)$. Then  $K_\sN$ is given by 
\be
e^{i t K_\sN} = e^{-i a^\mu P_\mu} e^{i t K_\sM} e^{i a^\mu P_\mu} = e^{-i  P^+} e^{i t K_\sM} e^{i  P^+}  \ .
\ee
From the commutations between $P^\pm$ and the boost operator $K$  
\bega 
[K, P^\pm] = \pm i P^\pm , 
\end{gather}
we thus find 
\be 
K_\sN = K_\sM - 2 \pi  P^+ \ .
\ee
From~\eqref{motr0} we conclude that the corresponding $G$ is given by 
\be 
G =  P^+  \ .
\ee
Thus in this case, $G$ simply generates a translation in $x^-$.  All the statements~\eqref{mtr1}--\eqref{halT1} can be readily verified.

By taking $\sN$ to be the operator algebra associated with the region in Fig.~\ref{fig:shiftedWedge} (b), there is a half-sided modular inclusion structure~\eqref{ghb1} with $t \geq 0$, and the corresponding modular translation operator is given by $G =  P^-$.

\subsection{Relative modular flows and relative entropy}

Tomita-Takesaki theory can be generalized to give relative quantum information between two states $\ket{\Psi}$ and $\ket{\Om}$
that are cyclic and separating with respect to a \vNa\ $\sM$. 
As a generalization of~\eqref{DefS} we introduce the Tomita operator $S_{\Psi \Om}$ between two states 
\be \label{defRE}
S_{\Psi \Om} A \ket{\Om} = A^\da \ket{\Psi} , \qquad S_{\Psi \Om}^\da  A' \ket{\Om}  = A'^\da \ket{\Psi}  ,\quad \forall A \in \sM, \, \forall A' \in \sM' \ . 
\ee
By definition $S_{\Psi \Om}$ is again anti-linear, and satisfies 
\be \label{cio} 
S_{\Psi\Om} S_{\Om\Psi} = \bid   \ .
\ee
Consider the polar decomposition 
\be
S_{\Psi\Om}  = J_{\Psi \Om} \De_{\Psi \Om}^\ha , \quad \De_{\Psi \Om} = S^\da_{\Psi \Om} S_{\Psi \Om} \ .
\ee
where $J_{\Psi \Om}$ (relative conjugation) is an anti-unitary operator and $\De_{\Psi \Om}$ (relative modular operator) is a positive operator. Equation~\eqref{cio} implies 
\be 
J_{\Psi \Om} \De_{\Psi \Om}^\ha J_{\Om \Psi } = \De^{-\ha}_{\Om \Psi} \ .
\ee
Writing the left hand side of the above equation as $J_{\Psi \Om} J_{\Om \Psi}  J_{\Om \Psi}^\da \De_{\Psi \Om}^\ha J_{\Om \Psi }$, from 
 the uniqueness of the polar decomposition of $\De^{-\ha}_{\Om \Psi}$ we find 
\be \label{usRe}
J_{\Psi \Om}  J_{ \Om \Psi} = \bid  \quad \to \quad J_{\Psi \Om} = J_{\Om \Psi}^\da , \quad \text{and} \quad J_{\Psi \Om} \De_{\Psi \Om} J^\da_{\Psi \Om} = \De^{-1}_{\Om \Psi} \ .
\ee
 There is an analogue of the KMS relation 
\be \label{hwn}
\vev{\Psi|A B |\Psi} = \vev{\Om|B \De_{\Psi \Om} A|\Om}, \quad A, B \in \sM  ,
\ee
which follows from 
\be 
{\rm LHS} =\vev{A^\da \Psi|J_{\Psi\Om} \De^\ha_{\Psi\Om} B^\da |\Om}   =  \vev{J_{\Psi\Om} \De^\ha_{\Psi\Om} A \Om| J_{\Psi\Om} \De^\ha_{\Psi\Om}B^\da |\Om} = {\rm RHS } \ .
\ee
Equation~\eqref{hwn}  means that using the relative modular operator we can convert correlation functions in $\ket{\Psi}$ to those $\ket{\Om}$ and vice versa, which can be very useful. 

We can also consider flows generated by the relative modular operator $\De_{\Psi \Phi}$, {which can be shown to satisfy}
\bega \label{ueb}
\De_{\Psi \Phi}^{-is} A \De_{\Psi \Phi}^{is}  = \De_\Psi^{-is} A \De_{\Psi}^{is} \in \sM , \quad A \in \sM, \; s \in \RR \\
\De_{\Psi \Phi}^{-is} A' \De_{\Psi \Phi}^{is}  = \De_\Phi^{-is} A' \De_{\Phi}^{is} \in \sM' , \quad A' \in \sM' ,  \; s \in \RR\ .
\end{gather}

Introduce 
\be \label{deRe}
u_{\Om \Psi} (s) \equiv \De_{\Om \Phi}^{-is} \De^{is}_{\Psi \Phi} \ .
\ee
It can be shown that the definition does not depend on the choice $\ket{\Phi}$.
Acting $u_{\Om \Psi} (s) \cdot u_{\Om \Psi} (s)^\da$ on both sides of~\eqref{ueb}, we find 
\be
\De_{\Om \Phi}^{-is} A \De_{\Om \Phi}^{is}  = \De_\Om^{-is} A \De_{\Om}^{is} =
u_{\Om \Psi} (s) \sig_s^{\Psi} (A) u_{\Om \Psi} (s)^\da \ .
\ee 
That is, $u_{\Om \Psi} (s) $ relates the modular flows generated by $\De_\Om$ and $\De_\Psi$. A nontrivial statement is that 
\be 
u_{\Om \Psi} (s)  \in \sM
\ee
and is independent of $\Phi$. This provides a justification of the claim made around~\eqref{nkd} that 
modular flows of $\De_\Om$ and $\De_\Psi$ are related by  inner automorphisms, and~\eqref{deRe} gives an explicit construction of the inner automorphisms using relative modular flows. 
It can be readily checked that the properties~\eqref{ss11} and~\eqref{cocy}--\eqref{coy1} indeed follow from the definition~\eqref{deRe}.  
Similarly, 
\be \label{deRe0}
u'_{\Om \Psi } (t) = \De^{-it}_{\Phi \Om} \De^{it}_{\Phi \Psi }  \in \sM'  
\ee
relate modular flows of $\sM'$ generated by $\De_\Om$ and $\De_\Psi$.  By choosing $\Phi$ to be either $\Om$ or $\Psi$ we can also write $u_{\Psi \Om}$ and $u_{\Psi \Om}'$ in various ways  
\bega \label{deRe1}  
u_{\Psi \Om} (t) = \De^{-it}_{\Psi \Om}  \De^{it}_{\Om }  =  \De^{-it}_{\Psi}  \De^{it}_{\Om\Psi },
  \\
 u'_{\Psi \Om} (t) =  \De^{it}_{\Om \Psi} \De^{-it}_{\Om}  =  \De^{it}_{\Psi} \De^{-it}_{\Psi \Om}  \ .
 \label{deRe2}  
 \end{gather}
 
Introduce 
 \be 
 u_{\Psi \Om} (t)  = e^{i h_{\Psi \Om} t}, \quad  u'_{\Psi \Om} (t)  = e^{i h'_{\Psi \Om} t}, \quad K_{\Psi \Om} = - \log \De_{\Psi \Om}  \ .
 \ee
 Taking derivatives on on $t$ and then setting $t$ to zero in~\eqref{deRe1}--\eqref{deRe2}, we find that 
 \be \label{cocy1}
 h_{\Psi \Om} = K_{\Psi \Om} - K_{\Om} = K_\Psi - K_{\Om \Psi} , \quad
 h_{\Psi \Om}' = -K_{\Om \Psi} + K_{\Om} = - K_\Psi +  K_{\Psi \Om} \ .
 \ee
 

For a type III algebra, while no entropy can be associated with $\sM$ in a state $\ket{\Om}$, a relative entropy between two states $\ket{\Psi}$
and $\ket{\Om}$ can be defined using the relative modular operator
\be \label{rela1}
S_\sM (\Psi\|\Om) \equiv  -\vev{\Psi|\log \De_{ \Om\Psi} |\Psi}  \ .
\ee
The above definition reduces to the standard relative entropy in situations where a trace $\tr$ is defined. More explicitly, 
suppose  $\rho_\Om, \rho_\Om'$ and $\rho_\Psi, \rho_\Psi'$ are respectively density operators associated with $\sM, \sM'$ in $\ket{\Om}$ 
and  $\ket{\Psi}$, and let 
\be 
\De_{ \Om\Psi} = \rho_\Om \rho'^{-1}_\Psi  \ .
\ee
We then find that\footnote{That $\tr (\rho'_\Psi \log \rho'_\Psi) =  \tr (\rho_\Psi \log \rho_\Psi) $ follows from $\vev{\Psi|\log \De_\Psi|\Psi} =0$.}
\be 
S_\sM (\Psi\|\Om) = \tr (\rho'_\Psi \log \rho'_\Psi) - \tr (\rho_\Psi \log \rho_\Om) 
= \tr (\rho_\Psi \log \rho_\Psi) - \tr (\rho_\Psi \log \rho_\Om)  \ .
\ee
The relative entropy~\eqref{rela1} is non-negative as can be seen from
\be 
 -\vev{\Psi|\log \De_{ \Om\Psi} |\Psi}  \geq  \vev{\Psi|1 - \De_{ \Om\Psi} |\Psi} =
 -\vev{\Psi|\Psi} + \vev{\Om|\Om} 
 = 0 
 \ee
 where we have used that $\log x \leq x-1$ for a positive $x$ 
 and~\eqref{hwn}. 


\subsection{A canonical purification: the natural cone} \label{sec:cone}

Consider a \vNa\ algebra $\sM$ acting on a Hilbert space $\sH$, and a normal state $\om$ on $\sM$. 
From~\eqref{norS}, $\om$ can be associated with a density operator $\rho$ on $\sH$. In general, there can be an infinite number of choices. As an illustration, consider a simple type I example, with $\sM = \sB (\sH_1)\otimes \bid_{\sH_2}$. 
A state $\om$ on $\sM$ can be associated with a reduced density operator $\rho_\om \in \sB (\sH_1)$. 
On $\sH = \sH_1 \otimes \sH_2$, there is in general an infinite number of $\rho$'s~(which can be either pure or mixed) that reduce to $\rho_\om$ when tracing out $\sH_2$. 
It is then natural to ask: does there exist a canonical choice of $\rho$? Can $\rho$ be a pure state, i.e., a vector in $\sH$?

It turns out these questions can be answered explicitly, if an additional condition is imposed: $\sH$ contains a cyclic and separating vector with respect to $\sM$. In such a case, $\sM$ is said to be in the {\bf standard form}, and there is a canonical {\it vector} $\ket{\xi_\om} \in \sH$ such that 
\be 
\om (A) = \vev{\xi_\om|A|\xi_\om} , \quad \forall A \in \sM \ .
\ee
An algebra $\sM_O$ associated with an open region $O$ in a RQFT (in the vacuum sector) is in the standard form as the vacuum state is a cyclic and separating vector. From the discussion of Sec.~\ref{sec:GNS}, $\sM = \pi_\om (\sA)''$  constructed from a {\it faithful} state $\om$ on a $C^*$ algebra $\sA$ is in the standard form on the GNS Hilbert space $\sH_\om$. 

We will now describe $\ket{\xi_\om}$ explicitly. Suppose there exists a cyclic and separating vector $\ket{\Om}$ for $\sM$. Denote $\sP_\Om$ as the closure of the set of vectors 
\be \label{uwg0}
 \{A j_\Om (A)  \ket{\Om}, \;\; A \in \sM\}  , \quad j_\Om (A) \equiv J_\Om A J_\Om ,
\ee
which is called the natural cone of $\ket{\Om}$. An equivalent definition of $\sP_\Om$ is the closure of the set 
\be 
\{\De_\Om^{1 \ov 4} A^\da A \ket{\Om}, \;\;  A \in \sM\}  \ .
\label{uwg}
\ee
It can be shown that every normal state $\om$ on $\sM$ has a unique representative $\ket{\xi_\om}$ in $\sP_\Om$. 
We stress that this ``canonical'' purification $\ket{\xi_\om}$  depends on $\ket{\Om}$; a different cyclic and separating state can 
lead to a different purification.

From~\eqref{uwg0}--\eqref{uwg} it can be readily checked that for a $\ket{\Psi} \in \sP_\Om$
\bega 
\De_\Om^{it} \ket{\Psi} \in \sP_\Om, \forall t \in \RR,  \quad
J_\Om  \ket{\Psi}  = \ket{\Psi}, \quad B j_\Om (B) \ket{\Psi} \in \sP_\Om, \; B \in \sM, \\
\text{If it  is separating for $\sM$ then it is also cyclic for $\sM$ and vice versa.}
\end{gather}  

Vectors in the natural cone have many nice properties, which we list here (see~\cite{BraRobV1} for proofs) 

\ben


\item For any $ \ket{\Phi}, \ket{\Psi} \in \sP_\Om$, 
\be
\vev{\Phi|\Psi} = \vev{\Psi|\Phi} \geq 0 \ . 
\ee

\item If $\ket{\Phi} \in \sH$ satisfies $J_\Om \ket{\Phi} = \ket{\Phi}$, then it can be uniquely decomposed as
\be 
\ket{\Phi} = \ket{\Phi}_+ - \ket{\Phi}_- , \qquad \ket{\Phi}_+ , \ket{\Phi}_- \in \sP_\Om, \qquad \vev{\Phi_+|\Phi_-} = 0\ . 
\ee

\item Let $\ket{\Phi_{1,2}} \in \sP_\Om$ be the  vector representatives of the states $\vp_{1,2}$, then 
\be 
||\, \ket{\Phi_1} - \ket{\Phi_2}||^2 \leq ||\vp_1 - \vp_2||^2 \leq || \, \ket{\Phi_1} - \ket{\Phi_2} || \, || \, \ket{\Phi_1} + \ket{\Phi_2} || \ .
\ee

\item Suppose $\ket{\Psi} \in \sP_\Om$ and is cyclic and separating for $\sM$, then
\be \label{jj1}
J_{\Psi \Om} = J_{\Om}  = J_{\Om \Psi} = J_{\Psi} \equiv J , \quad \sP_\Om = \sP_\Psi \ .
\ee
From~\eqref{usRe} and~\eqref{defRE} we then have 
\bega 
J \De_{\Psi \Om} J = \De_{\Om \Psi}^{-1} , \quad J \ket{\Psi }= \ket{\Psi } , \quad J \ket{\Om} = \ket{\Om} , \\
\De^\ha_{\Psi \Om} \ket{\Om} = \ket{\Psi } , \qquad \De^\ha_{\Om \Psi} \ket{\Psi } = \ket{\Om}  \ .
\end{gather} 

\item Suppose $\ket{\Phi}$ is a cyclic and separating vector not in the natural cone $\sP_\Om$. We can write it as 
\be 
\ket{\Phi} = u' \ket{\Phi_c}, \quad u' = J_{\Phi\Om} J_\Om \in \sM', \quad \ket{\Phi_c} \in \sP_\Om \ .
\ee

\item Every automorphism of $\sM$ can be written in terms of a unitary action,
\be 
\al (A) = U_\al A U^\da_\al, \quad A \in \sM ,
\ee
and the unitary $U_\al$ may be chosen so that 
\be 
 U_\al  \sP_\Om = \sP_\Om, \qquad J  U_\al  =  U_\al  J  \ .
\ee

\een


 \section{Crossed product by modular group} \label{sec:crossed}

Consider a \vNa\ $\sM$ acting on Hilbert space $\sH$. In many physical contexts, there are ``symmetries''  acting on the algebra. This means that there exists a group $G$, and a continuous unitary action $\al$ of $G$ on $\sM$, 
\be
\al_g (A) = U_g A U_g^\da , \quad A \in \sM , \; g \in G , \; U_g \in \sB(\sH), \text{unitary}  \ .
\ee
For simplicity, we will assume that $G$ is locally compact. 
Given such a triple $(\sM, G, \al)$, it is possible to construct a new \vNa, called the crossed product of $\sM$ by the action $\al$ of $G$, denoted as $\wh \sM = \sM \otimes_\al G$. $\wh \sM$ acts on the extended Hilbert space $\wh \sH = \sH \otimes L^2 (G)$. Reviews on crossed product can be found in~\cite{BraRobV1,Dae78,Wit21b}.

For example, take $G = \RR$, with generator $K$, and the action on $\sM$  
\be \label{mdeQ}
\al_t (A) = e^{i K t} A e^{- i Kt} , \quad A \in \sM , \; t \in \RR \ .
\ee
The crossed product $\wh \sM = \sM \otimes_\al \RR$ acts on $\wh \sH = \sH \otimes L^2 (\RR)$. 

For our later applications, we will be mainly interested in the case $K = - \log \De_\Psi$, where $\De_\Psi$ is the modular operator of $\sM$ for a cyclic and separating vector $\ket{\Psi}$, and~\eqref{mdeQ} then corresponds to the action of the modular group. In this case, it turns out that, for a  type III algebra $\sM$, $\wh \sM$ is type II~\cite{Tak73}. 
This means that a trace can be defined for 
$\wh \sM$, allowing for the introduction of density operators and entanglement entropies. Moreover, 
$\wh \sM$ depends only on the algebra 
$\sM$ itself, not on the choice of the state $\ket{\Psi}$. Thus, the type III algebra 
$\sM$ can be studied through its type II counterpart 
$\wh \sM$. 

In Sec.~\ref{sec:diff}, we will see that the crossed product with the modular group arises naturally in various gravitational contexts, and can be used to shed light on the black hole and de Sitter entropies.

Below we will discuss: (i) how to construct $\wh \sM$;  (ii) show that it is type II by demonstrating the existence of a trace; (iii) use the trace to define the density operator for the subsystem $\wh \sM$ in a class of states in $\wh \sH$; (iv) use the density operator to calculate the entanglement entropy of $\wh \sM$ in the class of states. We will see that the entanglement entropy of $\wh \sM$ reduces to the relative entropy of the original type III algebra $\sM$.

\subsection{Construction of $\wh \sM$}\label{sec:crossed1}

To describe $\wh \sM$, we add a one-dimensional quantum mechanical system to the original system, 
with the new Hilbert space given by $\sH \otimes L^2 (\RR)$. We denote $\hat q$ and $\hat p$ respectively as the position and momentum operators acting on $L^2 (\RR)$ (with eigenvalues $q$ and $p$ and eigenbasis $\ket{q}, \ket{p}$), with 
\be \label{ccp}
[\hat q , \hat p] = i \ .
\ee

The crossed product $\wh \sM$ can be defined as the subalgebra of $\sM \otimes \sB(L^2 (\RR))$ consisting of elements that are invariant under the action of unitaries generated by
\be
C \equiv K + \hat q \ .
\ee
That is, $a \in \wh \sM$ iff $[a, C] =0$. Mathematically, this condition mixes nontrivially $\sM$ and $ \sB(L^2 (\RR))$, generating new structure. Physically,  since $C$ generates diagonal translations of $p$ and time $t$ of~\eqref{mdeQ}, the invariance condition can be interpreted as ``gauging'' such translations. In our later examples of Sec.~\ref{sec:diff}, $\hat q$ can be the $O(1)$ part of an asymptotic Hamiltonian or the Hamiltonian of an external observer, in which case $p$ is the time of such an observer, and invariance under the actions generated by $C$ becomes time diffeomorphism invariance.

Clearly $\hat q$ commutes with $C$, thus operators generated by $e^{- i \hat q s}, s \in \RR$ should belong to $\wh \sM$.\footnote{$\hat q$ is not bounded, so we will use its Weyl form $e^{- i \hat q s}$.} $\sM$ does not commute with $C$, but it can be readily checked that the following dressed form of $\sM$ commutes with $C$, 
\be 
[e^{i K \hat p} A e^{- i K \hat p} , C] =0, \quad A \in \sM \ .
\ee
Thus $\wh \sM$ can be described as 
\be \label{n00} 
\wh \sM = \{e^{i K \hat p} A e^{- i K \hat p}, e^{- i \hat q s}| A \in \sM, s \in \RR\}'' \ .
\ee
A general element of $\wh \sM$ can be written as 
\be 
\hat A = \int ds \, A (\hat p; s) e^{- i \hat q s}, \quad A (\hat p; s) \equiv e^{i K \hat p} A (s) e^{- i K \hat p}
\ee
where $A(s) \in \sM$ is an operator-valued function on $\RR$. 
Recall $K$ generates automorphisms on $\sM$, so $A (\hat p; s) \in \sM$; it is a ``function'' of $\hat p$ which is now an operator. We may say $A (\hat p; s) $ lives in a quantum ``spacetime.'' 
Now consider the commutant $\wh \sM'$ of $\wh \sM$. $A' \in \sM'$ commutes with $A (\hat p) \in \sM$, and it trivially commutes with $\hat q$. Also by definition $C$ commutes with $\wh \sM$, thus $\wh \sM'$ can be written as 
\be \label{n01}
\wh \sM' = \{A', e^{i C s}|A' \in \sM', s \in \RR\}'' \ .
\ee
Note that $\wh \sM'$ does not commute with $C$. 

We can also consider $\wh \sM$ in other forms (in ``different frames'') by performing a unitary rotation $U$, i.e. 
\be 
\wh \sM \to U \wh \sM U^\da, \quad C \to U C U^\da , \quad \wh \sM' \to U \wh \sM' U^\da \ .
\ee
A convenient choice is $U = e^{- i K \hat p}$, which gives 
\bega \label{n1}
\wh \sM = \{A, e^{i (K - \hat q) s}|A \in \sM, s \in \RR\}'', \quad C = \hat q , \\
\wh \sM' = \{e^{- i K \hat p} A' e^{ i K \hat p}, e^{i \hat q s}|A' \in \sM', s \in \RR\}'' \ .
\label{n2}
\end{gather}
A general element then has the form 
\be\label{etp}
\hat A = \int ds \, A(s) e^{i s (K -\hat q)} , \quad A (s) \in \sM \ .
\ee
We will often use this form below.

\subsection{$\wh \sM$ is type II for modular group}\label{sec:IIcro}

So far the discussion applies to any generator $K$ that generates automorphisms of $\sM$. 
We now consider $K = - \log \De_\Psi$, and show that the corresponding $\wh \sM$ is type II.  For this purpose, consider the following state 
\be \label{haSa}
\ket{\hat \Psi} = \ket{\Psi} \otimes \ket{p=0} ,
\ee
where $\ket{p=0}$ is the zero eigenvalue state for $\hat p$.\footnote{Strictly speaking, $\ket{\hat \Psi}$ is only plane wave normalizable and thus does not lie in $\sH \otimes L^2 (\RR)$. Thus  $\ket{\hat \Psi}$ is a weight, not a state on $\wh \sM$. 
But our conclusion is not affected by this. We use the normalization $\vev{q|p=0} = {1 \ov \sqrt{2 \pi}}$.} $\ket{\hat \Psi}$ is cyclic and separating with respect to $\wh \sM$. We can find the corresponding modular operator $\hat \De$ by ``solving'' the KMS relation 
\be \label{neso}
\vev{\hat \Psi|\hat A \hat B|\hat \Psi} = 
\vev{\hat \Psi|\hat B \hat \De \hat A|\hat \Psi}, \quad \hat A , \hat B \in \wh \sM \ .
\ee
Using the KMS relation for $\ket{\Psi}$, it can be shown that the above equation is solved by (see Appendix~\ref{app:mod} for a derivation) 
\be \label{neso1}
\hat \De = \De_\Psi \ . 
\ee
This result does not depend on whether we use~\eqref{n00} or~\eqref{n1}. Below we will use~\eqref{n1}.  

Now here is an elementary but key observation 
\bega 
\hat \De = e^{-K} = e^{- (K-\hat q)} e^{-\hat q}  = \rho \rho'^{-1}, \quad \rho = e^{-(K-\hat q)} , \quad \rho' = e^{\hat q} \\
\hat \De^{-is} = e^{i (K - \hat q) s} e^{i \hat q s} 
\label{n3}
\end{gather} 
Now comparing with~\eqref{n1}--\eqref{n2}, equation~\eqref{n3} is a product of unitary elements of $\wh \sM$ and $\wh \sM'$. In other words, acting on $\wh \sM$ and $\wh \sM'$, $\hat \De^{-is}$ corresponds to an inner automorphism. 
Thus we conclude that $\wh \sM$ cannot be type III. Since $\hat \sH$ cannot be factorized with respect $\wh \sM$ (which follows from that $\sH$ cannot be factorized with respect to $\sM$), then it must be type~II. 
While $\hat \De$ does not depend on whether we use~\eqref{n00} or~\eqref{n1}. $\rho$ and $\rho'$ do. 
In going from~\eqref{n00} to~\eqref{n1} we need to take 
\be 
\rho \to U \rho U^\da , \quad \rho' \to U \rho' U^\da
\ee
Indeed for~\eqref{n00} we have instead
\be 
\hat \De = e^{-K} = e^{\hat q} e^{- (K + \hat q)}  \quad \to \quad  \rho = e^{\hat q} , \quad \rho' = e^{K + \hat q} 
= U^\da e^{\hat q}  U 
\ee

Since the algebra is type II, it should allow a trace, which can be defined as\footnote{The definition can be understood as follows. If the algebra allows a trace $\tr$, we expect $\vev{\hat \Psi|\hat A |\hat \Psi} = \tr (\hat A \rho)$. To obtain the trace, we thus multiply a factor $\rho^{-1}$.} 
\be \label{dtra}
\tr \hat A \equiv \vev{\hat \Psi|\hat A \rho^{-1}|\hat \Psi} = \vev{\hat \Psi|\hat A \rho'^{-1}|\hat \Psi} ,
\ee
where in the second equality we have used that $\hat \Psi$ is invariant under $\hat \De$. For an element of the form~\eqref{etp}, we can write the above expression more explicitly as 
\be 
\tr \hat A = \int_{-\infty}^\infty dq \, e^{-q} \vev{\Psi|\hat A |\Psi} , 
\quad \hat A  \equiv \int ds \, A (s) e^{i (K -q) s}  \ .
\ee
It can readily checked that $\tr$ indeed satisfies the cyclic permutation property, 
\bega
\tr (\hat A \hat B) = \vev{\hat \Psi|\hat A \hat B \rho^{-1}|\hat \Psi} = \vev{\hat B \rho^{-1} \hat \De \hat A} = 
\vev{\hat B \rho'^{-1} \hat A} = \vev{\hat B\hat A \rho'^{-1} }  = \tr (\hat B \hat A)  ,
\end{gather} 
where we have used~\eqref{neso} in the second equality and suppressed $\hat \Psi$ in subsequent expectation values for notation simplicity. 
We can equivalently write~\eqref{dtra} as 
\be 
\tr \hat A = \vev{\hat \Psi_{\rm T} |\hat A|\hat \Psi_{\rm T}}
\ee
where $\ket{\hat \Psi_{\rm T}}$ can be written as 
\be \label{tyitr}
\ket{ \hat \Psi_{\rm T} }= \ket{\Psi} \otimes \int dq \, e^{- {q \ov 2}} \ket{q} \ .
 \ee
Since 
\be 
\tr \bid =  \int_{-\infty}^\infty dq \, e^{-q} \vev{\Psi|\bid|\Psi}  = \int_{-\infty}^\infty dq \, e^{-q} = \infty,
\ee
the algebra is type II$_\infty$.



 Interestingly, we can obtain a type II$_1$ algebra if we restrict the spectrum of $\hat q$ to be bounded below, e.g. imposing $q \geq 0$. This can be achieved by introducing a projector $\Pi = \th (\hat q)$ 
on $L^2 (\RR)$, where $\th (x)$ is the step function that is $1$ for $x \geq 0$ and $0$ for $x< 0$.
An operator $\hat A \in \wh \sM$ then becomes $\Pi \hat A \Pi$ acting on $\sH \otimes L^2(\RR_+)$ (where $\RR_+$ is the half-line $q> 0$), and the algebra becomes $\wh \sM_+ = \Pi \wh \sM \Pi$. The identity operator in the algebra $\wh \sM_+$ is now given by $\Pi = \th (\hat q)$, and its trace is given by 
 \be\label{tyii1}
{\rm Tr}_{\wh \sM_+} \bid ={\rm Tr}_{\wh \sM} \Pi =  \int_{-\infty}^\infty dq \, e^{-q} \th (q) =  \int_{0}^\infty dq \, e^{-q}   = 1 \ .
\ee
As discussed earlier in Sec.~\ref{sec:vNCl}, a type II algebra with a normalizable trace is type II$_1$, and thus $\wh \sM_+$ is type II$_1$.

To conclude this subsection, we note that while in the construction above, the cyclic and separating state $\ket{\Psi}$
played a central role, the crossed product, in fact, does not depend on the choice of $\ket{\Psi}$, and can be viewed as intrinsically defined for the algebra $\sM$ itself. Consider $\wh \sM_\Phi$ obtained from a different cyclic and separating state $\ket{\Phi}$. It can be shown to be unitarily equivalent to $\wh \sM_\Psi$ obtained from $\ket{\Psi}$, 
\be \label{crow1}
\wh \sM_\Phi = u'_{\Phi \Psi} (\hat p) \wh \sM_\Psi u'^\da_{\Phi \Psi} (\hat p) 
\ee
where $u'_{\Phi \Psi} (\hat p)$ is the unitary~\eqref{deRe2} relating modular flows of $\sM'$ in state $\ket{\Phi}$ and $\ket{\Psi}$
with the flow parameter taken to be $\hat p$, i.e.  
\be 
u'_{\Phi \Psi} (\hat p) =  \De^{i \hat p}_{ \Psi \Phi} \De^{-i\hat p}_{\Psi}  =  \De^{i\hat p}_{\Phi} \De^{-i \hat p}_{\Phi \Psi}  \in \sM' \  .
 \ee

To see~\eqref{crow1}, consider $\hat A = A e^{i t (K_\Psi - \hat q )} \in \wh \sM_\Psi$, 
\ie \label{excy}
& u'_{\Phi \Psi} (\hat p) A e^{i (K_\Psi -\hat q) t} u'^\da_{\Phi \Psi} (\hat p) 
 = A u'_{\Phi \Psi} (\hat p) e^{i K_\Psi t} u'^\da_{\Phi \Psi} (\hat p + t)  e^{- i \hat q t}  \cr
& =A\De^{i \hat p}_{ \Psi \Phi} \De^{-i\hat p}_{\Psi}  e^{i K_\Psi t} \De^{i(\hat p +t)}_{\Psi}  \De^{-i (\hat p+t)}_{ \Psi \Phi} e^{- i \hat q t} 
= A \De^{-i t}_{ \Psi \Phi} e^{- i \hat q t}  = A\De_\Phi^{-it} \De_\Phi^{it}  \De^{-i t}_{ \Psi \Phi} e^{- i \hat q t}
 \cr
& = A e^{i (K_\Phi - \hat q) t} u_{\Phi \Psi} (t) \in \wh \sM_\Phi 
\fe
where in the last line we have used~\eqref{deRe1} and that $u_{\Phi \Psi} (t) \in \sM$.

\subsection{Density operator for $\wh \sM$ in a general semi-classical state}

Now consider states in $\sH \otimes L^2 (\RR)$ of the form 
\be 
\ket{\hat \Phi} = \ket{\Phi}  \otimes \ket{g} \equiv \ket{\Phi, g}, \quad 
\ket{\Phi} \in \sH, \quad  \ket{g} \in L^2 (\RR)  , 
\ee
where $\ket{g}$ has coordinate space wave function $g(q)$ which is taken to be nonzero everywhere and normalized as 
\be \label{gnirZ}
\int dq \, |g (q)|^2 = 1 \ .
\ee


The density operator $\rho_{\hat \Phi} (\wh \sM) $ associated with $\wh \sM$ in the state $\ket{\hat \Phi}$ can be found from
\be \label{twg}
\tr (\rho_{\hat \Phi} \hat A ) = \vev{\hat \Phi|  \hat A |\hat \Phi} , \quad \forall \hat A \in \wh \sM ,
\ee
where the trace $\tr$ is defined by~\eqref{dtra}. 
It is convenient to use the form~\eqref{n1}, with $\hat A = A e^{i s (K - \hat q )}$ and $\ket{\hat \Phi} \to e^{- i K \hat p} \ket{\hat \Phi}$ and~\eqref{twg} becomes 
\be 
 \vev{\hat \Psi|  \rho_{\hat \Phi} A e^{i s (K -\hat q )} e^{-\hat q} | \hat \Psi} 
=  \vev{\hat \Phi| e^{i K \hat p}  A e^{i s (K - \hat q )}  e^{-i K \hat p} |\hat \Phi} 
 \ .
\label{deneq1}
\ee

We can further write the RHS of~\eqref{deneq1} as 
\bega \label{rhs1}
RHS =   \vev{\Phi,g| e^{i K \hat p}  A e^{i s (K - \hat q )}  e^{-i K \hat p}  |\Phi,g} 
=  \vev{\Psi,g| \De_{\Phi \Psi} e^{i K \hat p}  A e^{i s (K - \hat q )}  e^{-i K \hat p}   |\Psi,g} \\
= \int dq \,  \vev{\Psi,g|  \De_{\Phi \Psi} e^{i K \hat p} A e^{i s (K - \hat q )}  |q}\vev{q| e^{-i K \hat p}  |\Psi,g} \\
= \int dq \, \vev{\Psi |  \De_{\Phi \Psi} g^* (q - K)  A  e^{i s (K-q) } g (q-K)   |\Psi}
 \label{rhs2}
\end{gather} 
where we have used~\eqref{hwn} in the second equality of~\eqref{rhs1} and used 
\be\label{ienn}
\vev{q|e^{- i K \hat p} |g} = g(q-K) 
\ee
in obtaining~\eqref{rhs2}. The LHS of~\eqref{deneq1} can be written as 
\bega   \label{lhs1}
LHS = {1 \ov 2 \pi} \int dq dq' \, e^{-q} \vev{\Psi,q'| \rho_{\hat \Phi} A e^{is (K-q)}  |\Psi, q} 
\end{gather} 

Comparing~\eqref{rhs2} with~\eqref{lhs1}, we conclude that 
\be \label{des1}
\rho_{\hat \Phi} =2 \pi g (\hat q - K) e^{ \hat q} \De_{\Phi \Psi}  g^* (\hat q - K) , 
\ee
which is the unique expression that is Hermitian.\footnote{To see that with~\eqref{des1},~\eqref{lhs1} is equal to~\eqref{rhs2}, use~\eqref{hwn} twice in the resulting~\eqref{lhs1}.} To see that~\eqref{des1} is affiliated with $\sM$, note that 
\be
\le(\De_{\Phi \Psi} e^{\hat q}\ri)^{is} = \De^{is}_{\Phi \Psi} \De_\Psi^{-is}  e^{i (\hat q - K) s} 
= u_{\Phi \Psi} (s) e^{i (\hat q - K) s} \in \wh \sM  \ .
\ee

To obtain $\rho_{\hat \Phi}$ for the representation~\eqref{n00}, we conjugate~\eqref{des1} by $U^\da$, 
\be 
 \rho_{\hat \Phi} =2 \pi e^{i K \hat p} g (\hat q - K) e^{ \hat q} \De_{\Phi \Psi}  g^* (\hat q - K) e^{-i K \hat p} \ .
\ee

\subsection{Entanglement entropy for $\wt \sM$ in a general semi-classical state} 

Now consider the entanglement entropy for $\wh \sM$ in the state $\ket{\hat \Phi} = \ket{\Phi, g}$, which is defined as the von Neumann entropy of $\rho_{\hat \Phi}$, 
\be 
S_{\wh \sM} = - \tr \le(\rho_{\hat \Phi} \log \rho_{\hat \Phi}\ri)
= - \vev{\hat \Phi | e^{i K \hat p}  \log \rho_{\hat \Phi} e^{-i K \hat p} |\hat \Phi} \ .
\ee
Note that the entropy does not depend on which representation we use. 
Below we will assume that $\ket{\Phi}$ is also cyclic and separating with respect to $\sM$, and that $g$ is slowly varying, i.e., $g' (q) \propto \ep$ where $\ep$ is a small parameter. We will work to zeroth order 
$\ep$. 

We then find that 
\be 
- \log \rho_{\hat \Phi}  = -\hat q - \log |g(\hat q - K)|^2 + K_{\Phi \Psi} + O(\ep)
\ee
where the $O(\ep)$ contribution comes from commuting $g (\hat q -K)$ with $K_{\Phi \Psi} = - \log \De_{\Phi \Psi}$. 
Now using~\eqref{cocy1} 
\be 
K_{\Phi \Psi} = K_\Phi + K - K_{\Psi \Phi}, 
\ee
we have 
\bega
-   \log \rho_{\hat \Phi}   = -(\hat q -K)  - \log |g(\hat q - K)|^2 + K_\Phi - K_{\Psi \Phi} + O(\ep) \\
- e^{i K \hat p}  \log \rho_{\hat \Phi} e^{-i K \hat p}  = -\hat q - \log |g (\hat q)|^2 + 
e^{i K \hat p}  (K_\Phi - K_{\Psi \Phi} ) e^{-i K \hat p} 
\end{gather} 

Recall that 
\be 
\vev{q|e^{- i K \hat p}|\Phi, g}  = g (q - K) \ket{\Phi} 
\ee
Using~\eqref{ienn}, we then find 
thus 
\bega 
\vev{\Phi, g|e^{i K \hat p}  (K_\Phi - K_{\Psi \Phi} ) e^{-i K \hat p} |\Phi, g}
= \int dq \, \vev{\Phi|g^* (q-K)  (K_\Phi - K_{\Psi \Phi} ) g (q - K) |\Phi} \cr
= - S_\sM (\Phi||\Psi) + O(\ep)  \ .
\end{gather} 
In the last step we have expanded $g (q-K)$ around $g(q)$ (note the normalization~\eqref{gnirZ}), and used $K_\Phi \ket{\Phi} =0$ and~\eqref{rela1}. 

We then have 
\bega \label{finE}
S_{\wh \sM} = - S_\sM (\Phi||\Psi) - \bar q + S_o  , \\
 \bar q = \int dq \, q |g(q)|^2, \quad 
S_o = -\int dq \, |g(q)|^2 \log |g(q)|^2  \ .
\end{gather} 
Thus,  up to some ($\ket{g}$-dependent) constants, the entanglement entropy for the type II$_\infty$ algebra $\wh \sM$ in the state $\ket{\hat \Phi}$ is given by the negative of the relative entropy between $\ket{\Phi}$ and $\ket{\Psi}$. 
This result can in fact be expected, as $\wh \sM$ depends only  on the $\sM$, and the only entropy that can be defined for $\sM$ is its relative entropy.

\section{AdS/CFT duality in the large $N$ limit: algebraic formulation} \label{sec:ads/cft}

 Having elucidated the connections between entanglement and von Neumann algebras, and having introduced the relevant background on von Neumann algebras, we now turn to their applications in the context of AdS/CFT duality and quantum gravity.
In this section, we examine the AdS/CFT correspondence in the large-$N$ limit from an operator-algebraic perspective~\cite{LeuLiu21b,LeuLiu22,FauLi22}. Some of the earliest results along these lines were obtained by Rehren and collaborators~\cite{Reh99a,Reh00,DueReh02a,DueReh02b} shortly after the discovery of the duality. Notably, it was argued that the duality requires the boundary CFT to violate the additivity condition~\eqref{nal1}~\cite{Reh99a}, and that it should be formulated as a theory of generalized fields that fails to satisfy the time-slice axiom~\eqref{Hev}~\cite{DueReh02b}.



 \subsection{General description}  \label{sec:genG}


In the AdS/CFT duality, a bulk quantum gravity system in AdS$_{d+1}$ is equivalent to a CFT$_d$ on $\RR\times S_{d-1}$~(which is the conformal boundary of AdS$_{d+1}$).  A prototypical example is the duality between 
the $\sN =4$ Super-Yang-Mills~(SYM) theory with gauge group $SU(N)$ on $\RR \times S_3$ and 
the IIB superstring in AdS$_5 \times S_5$. Here are some entries of the dictionary between the two theories:
\bea \label{du1}
\sH_{\rm bulk} \quad & = & \quad \sH_{\rm CFT} \\
G_N \to 0 \quad & \lra & \quad N \to \infty  \\
\label{du3}
\apr \to 0 \quad  & \lra & \quad \lam \to \infty  \; \text{($\lam$: t' Hooft coupling)} \\
\label{du4}
\text{elementary field} \quad  & \lra & \quad \text{single-trace operator} \\
\text{classical bulk geometry $\phi_c$} \quad  & \lra & \quad \ket{\Psi} \; \text{(semi-classical state)} 
\label{du5}
\eea 
Since the bulk and boundary Hilbert spaces are identified, a reference to a state implies both the bulk and boundary.  Below we simply denote $\sH_{\rm bulk} = \sH_{\rm CFT}$ as $\sH$. 
The correspondence~\eqref{du4} can be written explicitly in terms of the extrapolation dictionary,   
\be \label{exT}
\sO (x) =  \lim_{r \to \infty} r^{-\De} \phi (X) ,  \quad X = (r, x)
\ee
where $\phi (X)$ denotes a bulk elementary field, with $r$ the bulk radial coordinate and $x$ denoting a boundary point. For the $\sN = 4$ SYM theory, $\sO (x)$ is a boundary operator of the form 
\be
\sO (x) = \Tr (\cdots), 
\ee
and $\De$ is the conformal dimension of $\sO$.  
For general holographic systems where such a description is not known explicitly, we will use~\eqref{exT} as the  {\bf definition of  single-trace operators}, i.e., operators in the boundary system corresponding to the boundary limit of bulk elementary fields. 
In~\eqref{du5}, we use $\phi$ to collectively denote all bulk fields---a notation we will continue to use below---with $\phi_c$ denoting their configurations in a classical solution.


As mentioned in the Introduction, our goal is to understand how bulk geometric concepts emerge in the semi-classical $G_N \to 0$ limit. We will thus be mainly concerned with the $N \to \infty$ limit of the boundary theory, where we will see that the structures of the Hilbert space and operator algebras undergo significant changes. Unless mentioned explicitly, we will also restrict to the $\lam \to \infty$ limit, where the bulk theory is described by the Einstein gravity coupled to matter fields. In some theories, such as theories from M2-branes or M5-branes, $N$ is the only parameter, for which we only need to take the $N \to \infty$ to get the Einstein gravity regime.




We refer to a state $\ket{\Psi} $ that has a bulk description in terms of a classical geometry as a semi-classical state. 
By definition, such a state has a well-defined large $N$ limit.  
That is,  
$\ket{\Psi}$ can be regarded as the $N \to \infty$ limit of a sequence of states $\{\ket{\Psi_N}\}$, one for each $SU(N)$ theory.
An example of a semi-classical state is the CFT vacuum $\ket{\Om}$ which corresponds to the empty AdS spacetime. 
Another example is the thermofield double state $\ket{\Psi_{\beta}}$ which at a sufficiently high temperature is dual to the eternal black hole~\cite{Mal01}.  A single-sided black hole formed from gravitational collapse should also correspond to a semi-classical state, even though finding an explicit boundary description of this state is in general difficult. 

We say that an operator $A$ has a well-defined large $N$ limit in a semi-classical state $\ket{\Psi}$ if it can be regarded as the $N \to \infty$ limit\footnote{In light of the recent discussion of~\cite{SchWit22,Liu25}, the $N \to \infty$ limit may involve averages over $N$.} of a sequence of operators $\{A_N\}$ in the finite-$N$ theory, with corresponding states $\{\ket{\Psi_N}\}$, such that the expectation value converges to a finite result
\be
\lim_{N \to \infty} \bra{\Psi_N} A_N \ket{\Psi_N} < \infty \ .
\ee
The collection $\sA_{\Psi}$ of such operators may depend on the specific semi-classical state $\ket{\Psi}$, which we will discuss more explicitly below. From the extrapolation dictionary~\eqref{exT}, a universal set that exists for all semi-classical states is  
\be \label{SingOp}
\sS \equiv 
\text{$*$-algebra generated by single-trace operators} \subseteq \sA_\Psi \ .
\ee
$\sA_\Psi$ may coincide with $\sS$, but may in principle contain additional operators whose existence depends on the specific semi-classical state $\ket{\Psi}$. Below, we will denote the restriction of $\sS$ to a boundary subregion $O$ as $\sS_O$.


We will further assume that $\sA_{\Psi}$ is closed under products of operators, in which case $\sA_{\Psi}$ is a $*$-algebra. Operators in $\sA_{\Psi}$ should in principle also carry a norm inherited from the finite-$N$ theories. Completing it with respect to this norm, we then obtain a $C^*$-algebra.
Expectation values in $\ket{\Psi}$ define a state $\om_\Psi$ on $\sA_\Psi$. 
We can then build a GNS Hilbert space $\sH_{\Psi}^{({\rm GNS})}$ using the action of $\om_\Psi$ on $\sA_\Psi$ , which can be viewed as the space of ``small'' excitations around $\ket{\Psi}$, with the representation of $\sA_\Psi$ on $\sH_{\Psi}^{({\rm GNS})}$ given by $\pi_\Psi (\sA_\Psi)$.

If the bulk geometry dual to $\ket{\Psi}$ is smooth and $\ket{\Psi}$ is pure, we expect $\om_\Psi$ to be a pure state with respect to the algebra $\sA_\Psi$. From Proposition~\ref{wpro0}, this implies
\be \label{pureS}
\sB(\sH_{\Psi}^{({\rm GNS})}) = \bigl(\pi_\Psi(\sA_\Psi)\bigr)'' \ .
\ee
When the corresponding bulk geometry contains a singularity, however,~\eqref{pureS} is not guaranteed. If the geometry contains a complete smooth Cauchy slice (e.g. with spacelike singularities only in the past and/or future), then~\eqref{pureS} should still hold. Unless otherwise noted, we will assume~\eqref{pureS} below.\footnote{In Sec.~\ref{sec:firewall}, we discuss a diagnostic of ``firewalls'' at horizons, which is associated with the failure of~\eqref{pureS}.}
We will denote~(where $O$ is a boundary region)
\be \label{Sejne}
\sY = (\pi_\Psi (\sS))'' , \quad \sY_O =  (\pi_\Psi (\sS_O))'' \ .
\ee

In the case that $\sS$ is a proper subalgebra of $\sA_\Psi$, $\om_\Psi$ should be a mixed state for $\sS$, which can be used to diagnose existence of operators in $\sA_\Psi$ outside $\sS$. More explicitly, we can build a GNS Hilbert space $\sH_{\Psi, \sS}^{({\rm GNS})}$ using the action of $\om_\Psi$ on $\sS$. If $\om_\Psi$ is not pure, the completion of the representation of $\sS$ on $\sH_{\Psi, \sS}^{({\rm GNS})}$ is a strict subset of $\sB (\sH_{\Psi, \sS}^{({\rm GNS})})$. Note also that for $\sS \subset \sA_\Psi$,  $\sH_{\Psi, \sS}^{({\rm GNS})}$ is different from $\sH_{\Psi}^{({\rm GNS})}$.


On the gravity side, we take all bulk fields to be gravitationally coupled, with the ${1 \ov 16 \pi G_N}$ factor
in front of the full bulk action. Let
\be \label{1sper}
\phi = \phi_c +\ka  \de \phi , \quad \ka = \sqrt{8 \pi G_N},
\ee
and expand the gravity action $S[\phi]$ in $\de \phi$ to get, schematically,
\bega 
\label{freac}
S [\phi] = S [\phi_c] + S_2 [\de \phi] + S_{\rm int} [\de \phi] , 
  \\
S_{\rm int}= 
 \ka S_3 + \ka^2 S_4 + \cdots + \ka^{n-2} S_n + \cdots, 
 \label{0Inact}
\end{gather} 
where $S_2$ is the quadratic action of $\de \phi$ with no dependence on $ \ka$, and $S_n$ contains terms with $n$ factors of $\de \phi$. At the leading order in the $\ka \to 0$ limit, only the quadratic action $S_2$ survives, resulting in a free quantum field theory of $\de \phi$ in the curved spacetime described by $\phi_c$. $S_2$ can be quantized using the standard formalism. 
 With the appropriate selection of a ``vacuum'' state $\ket{0}_{\phi_c}$,
a Fock space $\sH_{\Psi}^{\rm (Fock)}$ can be constructed by applying $\de \phi$'s to $\ket{0}_{\phi_c}$.

 For the duality to hold, we must have 
\be \label{fockid} 
\sH_\Psi^{\rm (Fock)} = \sH_{\Psi}^{({\rm GNS})},
\ee
which means that the GNS Hilbert space $ \sH_{\Psi}^{({\rm GNS})}$ must have the structure of a Fock space, i.e. 
the representation of $\sA_{\Psi}$ on $ \sH_{\Psi}^{({\rm GNS})}$ should be described by a Gaussian theory. 
For products of single-trace operators in the vacuum or the thermofield double state, this is the standard story of the large $N$ factorization. The duality~\eqref{fockid} implies that this should be true also for any semi-classical states and for those operators that lie outside $\sS$.\footnote{In fact, we could use the large $N$ factorization as the definition of 
a semi-classical state purely from the boundary perspective. For example, that is the definition used in~\cite{LeuLiu22}.}
It is also natural to identify the vacua on the two sides, i.e. 
\be \label{vacd}
\ket{0}_{\phi_c} = \ket{\bid}_{\Psi},
\ee
where $\ket{\bid}_{\Psi}$ denotes the state corresponding to the identity operator in $\sH_{\Psi}^{({\rm GNS})}$.

Each semi-classical state $\ket{\Psi}$ gives rise to a GNS Hilbert space $\sH_{\Psi}^{({\rm GNS})}$ (or equivalently, a Fock space $\sH_\Psi^{(\rm Fock)}$). Distinct classical geometries typically differ in energy by an amount of order  $O(1/G_N)$, whereas states within  $\sH_{\Psi}^{({\rm GNS})}$ differ in energy from $\ket{\Psi}$ only by $O(G_N^0)$ by construction. Consequently, the GNS Hilbert spaces $\sH_{\Psi_{1}}$ and 
$\sH_{\Psi_{2}}$, associated with two such classical geometries, cannot overlap at any finite order in $G_N$-perturbation theory. 
Moreover, even if two semi-classical states have the same energy, they may still belong to different GNS Hilbert spaces if they are separated by an infinite energy barrier in the configuration space. 

\begin{figure}[H]
\begin{center}
\includegraphics[width=10cm]{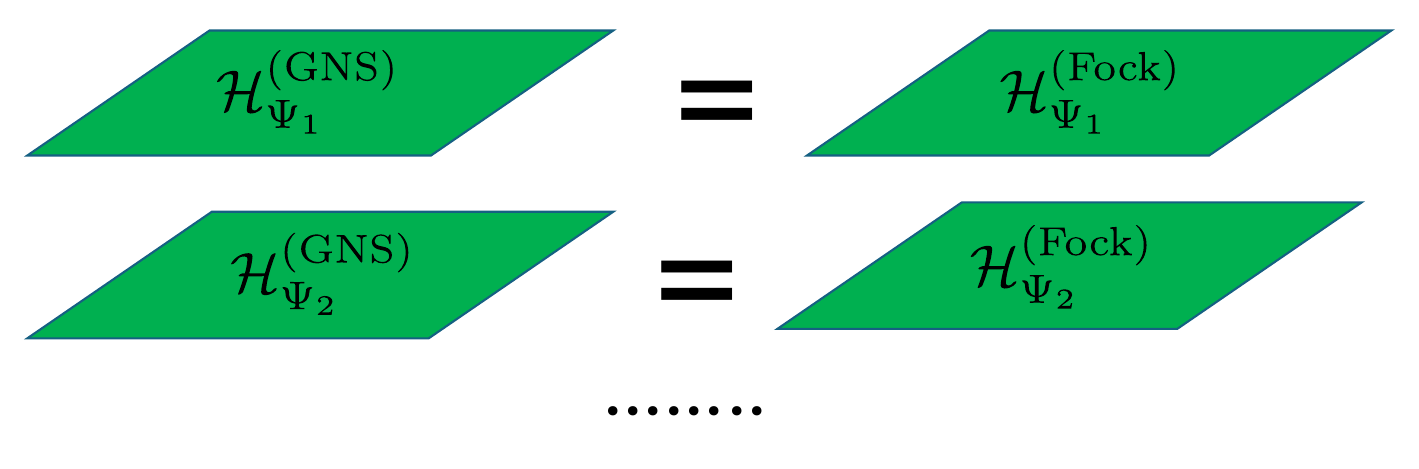}
\caption[]{
\small In the large $N$ limit, the space of states separates into disjoint sectors around semi-classical states.}
\label{fig:sectors}
\end{center}
\end{figure}

Therefore,  in the large $N$ limit, the system no longer has a single Hilbert space. Instead, the space of states separates into 
{\it disjoint sectors} around different semi-classical states.  See Fig.~\ref{fig:sectors}.
This is rather similar to the entangled spin example in the thermodynamic limit  discussed in Sec.~\ref{sec:inf} and~\ref{sec:entsp}. For different values of $\th$, the states $\ket{\Phi_\th}$ lead to disjoint sectors, each with its own GNS Hilbert space and distinct operator algebraic structures.

\begin{figure}[H]
\begin{center}
\includegraphics[width=2.5cm]{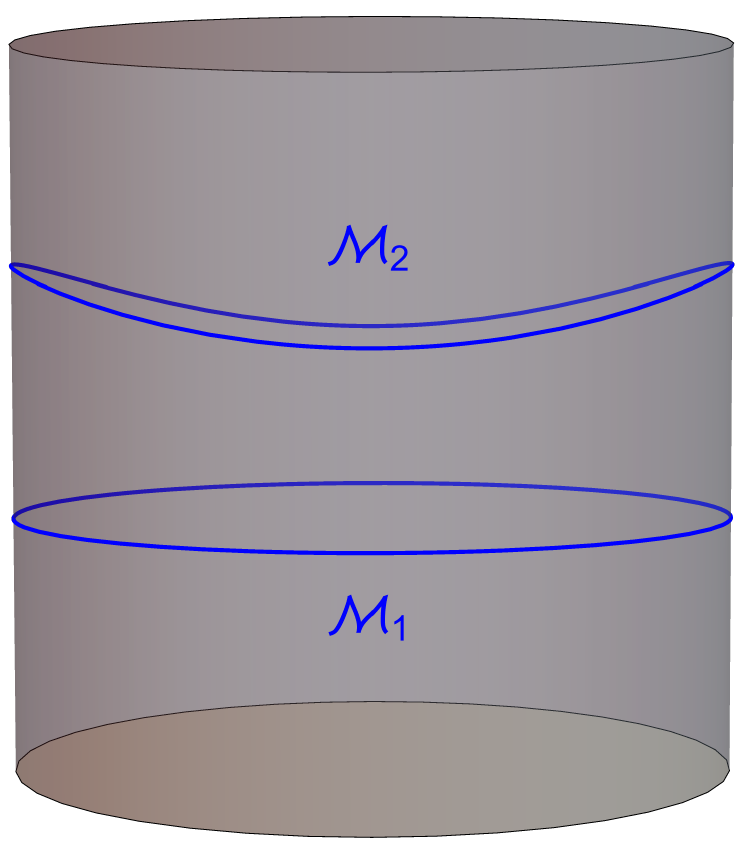}\qquad \qquad \qquad\qquad
\includegraphics[width=2.5cm]{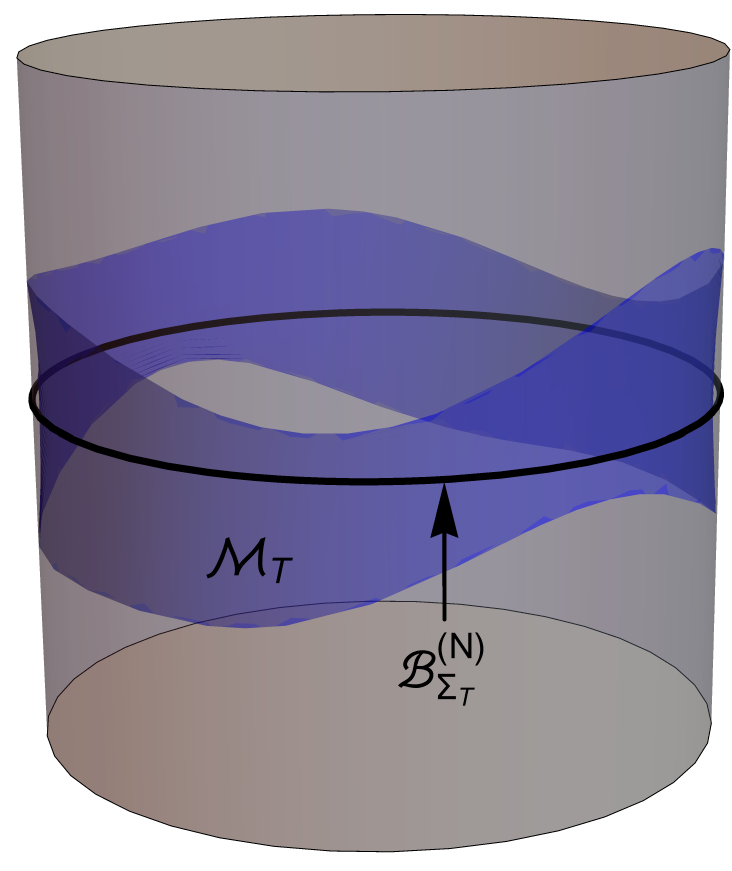} 
\caption[]{\small The cylinder represents the boundary spacetime. 
Left:  Algebras $\sM_1$ and $\sM_2$ generated by generalized free fields on two different Cauchy slices are inequivalent. Right: 
At finite $N$, the algebra associated with the time band indicated in the plot is equivalent to that on a single Cauchy slice. But in the large $N$ limit, it is inequivalent and is in general a proper subalgebra of the full algebra. }
\label{fig:caucy}
\end{center}
\end{figure}

The bulk theory described by $S_2 [\de \phi]$ can be regarded as an ordinary field theory in a curved spacetime with evolution of $\de \phi$ governed by Heisenberg equations of motions. In contrast, on the boundary,  correlation functions of operators in $\sA_\Psi$ are obtained by taking the large $N$ limit; there are no equations of motion governing the corresponding Gaussian fields. Thus $\sA_\Psi$ should be generated by  generalized free fields.\footnote{A generalized free field is a Gaussian field specified solely by its two-point correlation functions, with no equations of motion governing its evolution. For example, there is no Heisenberg equations governing the evolution of single-trace operators.} 
That there are no equations of motion governing generalized free fields has important implications 
for operator algebras:  operator algebras associated with different Cauchy slices on the boundary are inequivalent. This leads to many new emergent algebras in the large $N$ limit that are not present at finite $N$. See Fig.~\ref{fig:caucy}.

Another way to understand the independence of algebras at different times is to note that the boundary Hamiltonian $H$ does not survive the large-$N$ limit.  $H$ is defined as 
\be \label{eu000}
H = \int_{\sig} d^{d-1} x \, T^{00}, 
\ee
where $\sig$ is a boundary Cauchy slice.  $T^{\mu \nu}$ is the boundary stress tensor, which has the form 
\be \label{0nfs}
T^{\mu \nu} = N  \Tr (\cdots)  \ .
\ee
Due to the prefactor of $N$ in~\eqref{0nfs}, the stress tensor $T^{\mu \nu}$---and the Hamiltonian $H$---does not have a well-defined large-$N$ limit in {\it any} sector. As a result, single-trace operators at different times become independent in the large-$N$ limit. The rescaled operators
\be \label{ha100}
\hat T^{\mu \nu} \equiv {1 \ov N} T^{\mu \nu}, \quad \hat H \equiv {H \ov N} , 
\ee
do survive the large $N$ limit. However,  they do not generate finite time translations, 
\be \label{00haht}
i [\hat H, \sO(x)] = {1 \ov N} \p_t \sO (x) \ .
\ee

In a semiclassical state $\ket{\Psi}$ (such as the vacuum), where time-translation symmetry is present, there should exist a corresponding time-translation operator $\hat{h}_\Psi$ within that sector, which acts on single-trace operators as\footnote{
Below, $\sO(x)$ should be understood as $\pi_\Psi(\sO(x))$. For notational simplicity, we will suppress the symbol $\pi_\Psi$, as we will often do in later discussions.} 
\be \label{haht11}
i [\hat h_\Psi, \sO (x)] = \p_t \sO (x)  \ .
\ee
This is not in tension with the independence of operator algebras on different time slices. Unlike~\eqref{eu000}, the operator $\hat{h}\Psi$ cannot be expressed as the integral of a local operator over a {\it single} Cauchy slice. We will present an explicit example in the vacuum sector in the next subsection. On the gravity side, the corresponding bulk geometry $\phi_c$ should admit a timelike Killing vector, and $\hat{h}\Psi$ corresponds to the bulk generator associated with that symmetry.

Clearly, the above discussion applies to any symmetry in the large-$N$ limit, where the symmetry generator can no longer be expressed as an integral of local operators over a single time slice.

We now illustrate the general structure discussed above more explicitly using a few examples.



\subsection{The vacuum sector} \label{sec:vacuum}

For a CFT$_d$ on $\RR \times S_{d-1}$ (with $S_{d-1}$ having radius $R$), the vacuum state $\ket{\Om}$ is dual to empty global AdS, with metric 
 \be \label{gadsM}
 ds^2 = - \le(1 + {\rho^2 \ov R^2} \ri) dt^2 + {d \rho^2 \ov 1 + {\rho^2 \ov R^2}} + \rho^2 d \Om_{d-1}^2 , 
 \ee
together  with the bulk
vacuum state $\ket{0}_{\rm AdS} \in \sH_{\rm AdS}^{\rm (Fock)}$ that is invariant under AdS isometries
\be
 \ket{\Om} \quad  \lra  \quad  (\text{empty AdS}  , \ket{0}_{\rm AdS})  \ .
\ee
In this case, both the bulk Fock space and the boundary GNS Hilbert space can be constructed explicitly, and  the identification~\eqref{fockid} can be directly verified.

More explicitly, a bulk field $\phi (X)$  has the mode expansion
\be \label{bwa}
\phi (X) = \sum_k \le(u_k (X) a_k  + u_k^* (X) a_k^\da  \ri), \quad a_k \ket{0}_{\rm AdS} =0,
\ee
where $\{u_k (X)\}$ is the complete set of normalizable bulk wave functions in empty AdS. 
The Fock space $\sH_{\rm AdS}^{\rm (Fock)}$ is generated by acting $a_k^\da $ on $\ket{0}_{\rm AdS}$. 

On the boundary, {\it connected} vacuum correlation functions of single-trace operators have large-$N$ scalings of the form 
\be 
\vev{\sO} = 0 , \quad \vev{\sO_1 \sO_2}_c \sim O(N^0), \quad \vev{\sO_1 \cdots \sO_n}_c \sim N^{2-n} \ .
\ee
Thus, at the leading order of the $N \to \infty$ limit, 
\be\label{factoriza}
\vev{\sO_1 \cdots \sO_n} = \sum (\text{products of two-point functions}) \sim O(N^0) ,
\ee
defining a Gaussian theory in terms of two-point functions, which are known explicitly from conformal symmetry up to normalization constants.  Each single-trace operator can be viewed as a generalized free field, and accordingly the GNS Hilbert space $\sH_\Om^{(\rm GNS)}$ obtained from acting with the single-trace operator algebra~\eqref{SingOp} on $\ket{\Om}$
has the structure of a Fock space. Furthermore, it can be shown that 
\be \label{eunS}
\sB (\sH_\Om^{(\rm GNS)}) = (\pi_\Om (\sS))''  \ .
\ee
From the diagnostic discussed below~\eqref{Sejne}, $\om_\Om$ is a pure state for $\sS$, and 
 $\sS = \sA_\Om$, i.e., there is no other operator surviving the large $N$ limit in the vacuum sector.

 In $\sH_\Om^{(\rm GNS)}$, 
the representation of a single-trace operator $\sO$ can be expanded as 
\be \label{bdwa}
\pi_\Om (\sO (x) ) = \sum_k (v_k (x) b_k + v_k^* (x) b_k^\da), \quad b_k \ket{\bid}_\Om = 0 ,
\ee
where $v_k (x)$ is a complete set of basis functions on $\RR \times S_{d-1}$. 
 $\sH_\Om^{(\rm GNS)}$ is  generated by acting $b_k^\da$ on $ \ket{\bid}_\Om$. 
 For $\phi$ and $\sO$ dual to each other, it can be shown that~(with an appropriate choice of normalization for $\sO$ and basis functions) 
\be 
v_k (x) = \lim_{r \to \infty} r^\De u_k (X), \quad X = (r, x) ,
\ee
and~\eqref{exT} implies that we should identify 
\be \label{bbos}
a_k = b_k \ .
\ee
This establishes~\eqref{fockid}, i.e., we can identify 
\be\label{du7}
\sH_{\rm AdS}^{\rm (Fock)} = \sH_{\Om}^{(\rm GNS)} , \quad \ket{0}_{\rm AdS} = \ket{\bid}_{\Om} , \quad
\sB (\sH_{\rm AdS}^{\rm (Fock)}) = \sB (\sH_{\Om}^{(\rm GNS)} ) \ .
\ee

Equation~\eqref{bbos} can also be written in coordinate space as 
\be\label{hkll00}
\phi (X) = \int  d^d x\, K(X; x) \pi_\Om (\sO (x))  , 
\ee
where the kernel $K (X;x)$ satisfies the bulk equations of motion and~\eqref{exT}, and is  known as the global HKLL construction~\cite{BanDou98,Ben99,HamKab05,HamKab06}. 
From the fact that energies in a conformal representation are equally spaced with interval $2$, it can  be shown 
that~\cite{LeuLiu22} 
\be \label{miF}
 \sB \le(\sH_\Om^{\rm GNS} \ri) = \sY_{I_w} , \quad w \geq \pi R ,
\ee
where $I_w$ is a uniform time band on the boundary of width $w$. Thus in~\eqref{hkll00}, it is enough to restrict the 
$x$-integration to a time band of width $\pi R$. 

Quantum field theory in the geometry~\eqref{gadsM} possesses a time-translation global symmetry, with an associated bulk Hamiltonian $\hat{h}_\Omega$ given by
\be \label{bhamil} 
\hat h_\Om = \int_{\Sig_t}  d \rho d^{d-1} \Om \, \rho^{d-1}  \sT^t_t 
\ee
where $ \sT^t_t $ is the bulk stress tensor, which is quadratic in the bulk fields $\phi$ in the free theory limit. 
Plug~\eqref{bwa} into~\eqref{bhamil} we obtain a quadratic expression in terms of $a_k$, which can be viewed as a boundary operator with the identification~\eqref{bbos}. This is the boundary time-translation operator, and acts as~\eqref{haht11} by construction. It is quadratic in $\pi_\Omega(\sO)$ and involves integrals over the boundary spacetime across a time band of width $\pi R$, as implied by~\eqref{miF}.

For a semiclassical state $\ket{\Psi}$ whose bulk dual is a geometry with a single asymptotic boundary and no event horizon, the discussion parallels that of the vacuum sector. We next consider the simplest example that includes a horizon.

\subsection{Thermofield double state} \label{sec:TFD}

Now consider two copies of the boundary CFT, denoted as CFT$_R$ and CFT$_L$, whose Hilbert spaces and Hamiltonians are denoted respectively as $\sH_{R,L}$ and $H_{R,L}$. The thermofield double (TFD) state $\ket{\Psi_\b} \in \sH_R \otimes \sH_L$ is  defined at finite $N$ as 
\be \label{tfd0}
\ket{\Psi_\b} = {1 \ov \sqrt{Z_\b}} \sum_n e^{-{\b \ov 2} E_n} \ket{n}_R \ket{\Th  n}_L, \quad Z_\b = \sum_n e^{-\b E_n} \ . 
\ee
Here $\ket{n}$ denotes the energy eigenstates with eigenvalues $E_n$, and $\Th$ represents the $\sC \sR \sT$ operator. 
In the TFD state, tracing out one copy of the CFTs (say CFT$_L$) yields  the thermal density operator 
$\rho_\b = {1 \ov Z_\b} e^{-\b H_R}$. That is, from the perspective of CFT$_R$, the system is in a thermal state with 
 free energy $F(\b) = - {1 \ov \b} \log Z_\b$.  $\sB (\sH_R)$ is type I, and 
$\ket{\Psi_\b}$ is cyclic and separating with respect to it, with the modular operator given by
\be \label{each}
- \log \De_\b = \b (H_R - H_L) \ .
\ee


In the large $N$ limit, the system undergoes a first-order transition 
at a temperature $T_{\rm HP}$, known as the Hawking-Page temperature~\cite{HawPag83},  where the free energy $F(\b)$ jumps from $O(N^0)$ to $O(N^2)$ and the bulk dual transitions from thermal AdS to a black hole geometry~\cite{Wit98b}. 

\begin{figure}[H]
\begin{centering}
	\includegraphics[width=10cm]{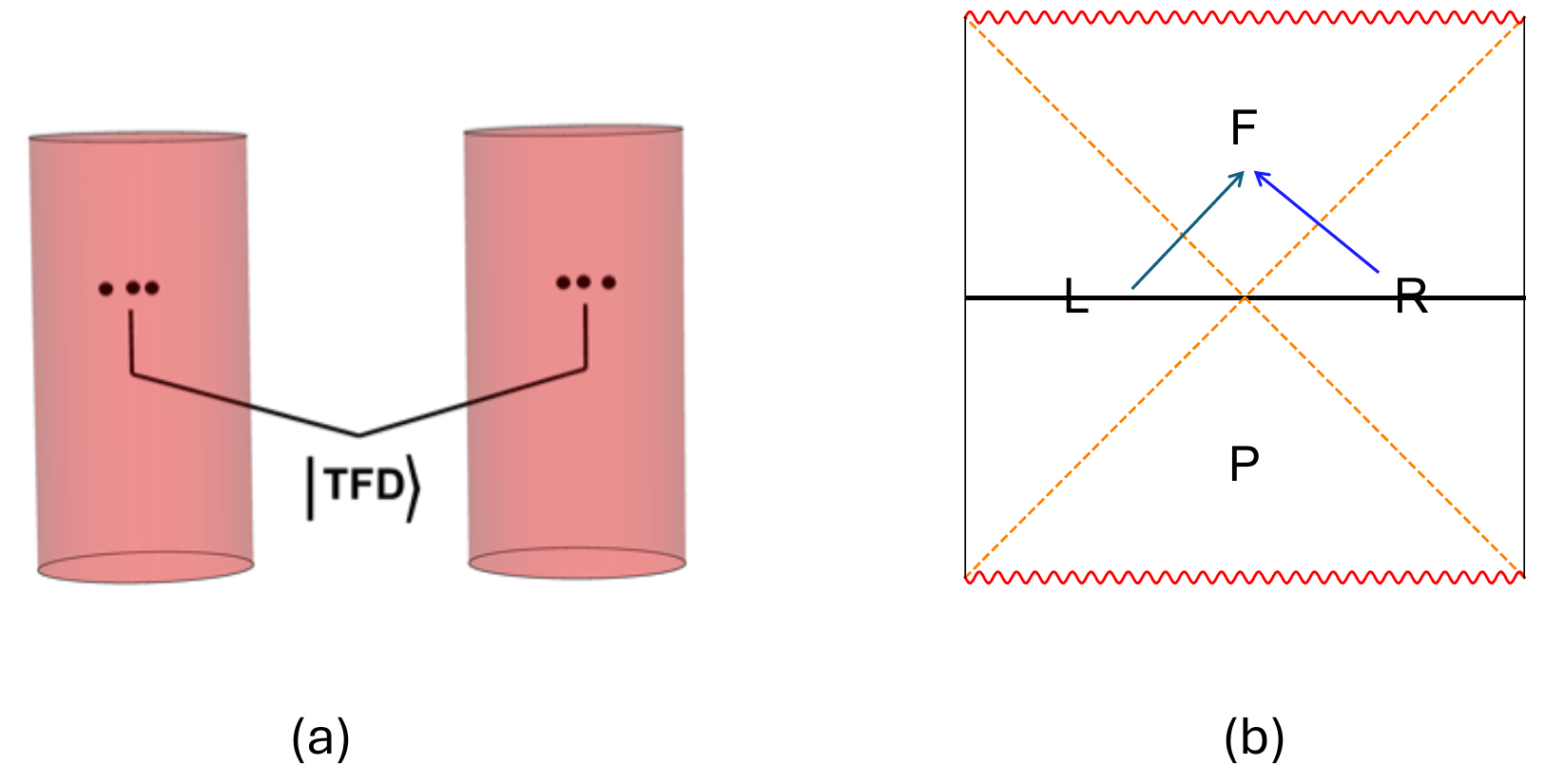} 
\par\end{centering}
\caption{\small (a)  For $T < T_{\rm HP}$,
$\ket{\Psi_\b}$ is dual to two disconnected global AdS spacetimes entangled in a thermofield double state.
(b) For $T > T_{\rm HP}$, $\ket{\Psi_\b}$ is dual to an eternal black hole in AdS.
The black horizontal line represents a Wilson line stretching from the left to the right boundary. Bulk degrees of freedom originating in the $R$ and $L$ regions of the black hole can cross the horizon---illustrated by arrows---and interact in the $F$-region.
}
\label{fig:tads}
\end{figure}

More explicitly, for $T < T_{\rm HP}$, the system has free energy of order $O(N^0)$, and the thermal ensemble is dominated by states with energies of order $O(N^0)$. From the perspective of the $R$-system, the bulk geometry is described by a thermal gas in global AdS, commonly referred to as thermal AdS.
The full $R$ and $L$ system can then be described by a {\it bulk} thermofield double state $\ket{\psi_\beta}$ living in two copies of global AdS~\cite{Mal01},
\begin{equation} \label{btfd}
\ket{\psi_\beta} = \frac{1}{\sqrt{\tilde Z_\beta}} \sum_n e^{- \frac{\beta}{2} \tilde E_n} \ket{\tilde n}_R \ket{\Theta \tilde n}_L, \quad \tilde Z_\beta = \sum_n e^{-\beta \tilde E_n} , ,
\end{equation}
where $\ket{\tilde n}$ and $\tilde E_n$ are the eigenstates and eigenvalues of the bulk Hamiltonian governing excitations around global AdS. See Fig.~\ref{fig:tads}(a).
Tracing out the $L$-system in~\eqref{btfd} then yields a thermal gas in global AdS for the $R$-system.

For $T > T_{\rm HP}$, the free energy $F(\beta)$ becomes $O(N^2)$, and the thermal ensemble is dominated by states with energies of order $O(N^2)$. The bulk dual is an eternal AdS black hole~\cite{Mal01}, described by the metric
\begin{equation} \label{adsBHM}
ds^2 = - f(\rho) dt^2 + \frac{d\rho^2}{f(\rho)} + \rho^2 d\Omega^2_{d-1}, \quad f(\rho) = 1 + \frac{\rho^2}{R^2} - \frac{\mu}{\rho^{d-2}} \ .
\end{equation}
See Fig.~\ref{fig:tads}(b). CFT$_R$ and CFT$_L$ live on the two asymptotic boundaries of this connected geometry. The thermal density matrix $\rho_\beta = \frac{1}{Z_\beta} e^{-\beta H_R}$ may be interpreted as dual to the $R$-region---i.e., the exterior of the black hole shown in Fig.~\ref{fig:tads}(b).

While the identification with the black hole geometry for $T > T_{\rm HP}$ had passed many nontrivial checks and was widely regarded as correct beyond reasonable doubt, the following conceptual puzzles had been noted:

\ben 

\item \textbf{Factorization puzzle}.
While the boundary system factorizes as a tensor product of CFT$_R$ and CFT$_L$, the bulk black hole is described by a connected spacetime  and does not admit a corresponding factorization. For instance, consider a Wilson line extending from the left boundary to the right in the bulk black hole geometry~\cite{Har15}~(see Fig.~\ref{fig:tads}(b)). There appears to be no way to represent the associated boundary operator as a product of operators in $\sB(\sH_R) \otimes \sB(\sH_L)$.

\item \textbf{Meeting-behind-the-horizon puzzle}. While there is no interaction between CFT$_R$ and CFT$_L$, the corresponding bulk degrees of freedom originating from the $R$ and $L$ regions of the black hole can, however,  ``meet''---and thus interact---after crossing the horizon~\cite{MarWal12}  (see Fig.~\ref{fig:tads}(b)). This puzzle is intimately connected to the boundary realization of time evolution generated by a Kruskal-like flow---i.e., evolution capable of traversing the horizon.

\een
These puzzles can be addressed by formulating the duality from an algebraic perspective. In this subsection, we focus on the factorization puzzle, deferring the discussion of the meeting-behind-the-horizon puzzle to Sec.~\ref{sec:krus}, where we explore the emergence of the bulk horizon from the boundary theory.

To set the stage, we first examine the $T < T_{\rm HP}$ phase more closely. In this regime, the large-$N$ limit of TFD state $\ket{\Psi_\beta}$ can be understood by taking the $N \to \infty$ limit directly in the sum of~\eqref{tfd0}. States with energies $E_n$ that scale nontrivially with $N$ drop out of the sum due to exponential suppression from the Boltzmann factor $e^{-\frac{1}{2} \beta E_n}$. The remaining contributions come entirely from states with $E_n \sim O(N^0)$, which, in the large-$N$ limit, can be interpreted as elements of the vacuum-sector GNS Hilbert spaces $\sH_\Omega^{R}$ and $\sH_\Omega^{L}$ of CFT$_R$ and CFT$_L$, respectively.

Denote the single-trace operator algebras from the CFT$_R$ and CFT$_L$ as $\sS_R$ and $\sS_L$. The GNS Hilbert space $\sH_{\Psi_\b}^{(\rm GNS)}$, obtained from acting with $\sS = \sS_R \otimes \sS_L$ on $\ket{\Psi_\b}$, factorizes as  
\bega \label{ejnn2}
\sH_{\Psi_\b}^{(\rm GNS)} =  \sH^R_\Om \otimes \sH_\Om^L \ . 
 \end{gather}
Denoting 
\be
\sY_{R} = (\pi_{\Psi_\b} (\sS_R))'', \quad \sY_{L} = (\pi_{\Psi_\b} (\sS_L))'' ,
\ee
we can further identify 
\bega
  \sY_R = \sB (\sH^R_\Om) , \quad \sY_L =  \sB (\sH^L_\Om)  \ .
   \label{ejnn3}
\end{gather}
Thus, at such temperatures, where the entanglement between CFT$_R$ and CFT$_L$ is of order $O(N^0)$,
$\sY_R$---the large $N$ limit of $\sB (\sH_R)$---remains type I. 

Given the identification $\sH_{\rm AdS}^{(\rm Fock)}= \sH_\Om^{(\rm GNS)}$, equation~\eqref{ejnn2} then implies that 
the gravity dual of $\sH_{\Psi_\beta}^{(\rm GNS)}$ corresponds to two copies of global AdS, with the identification 
\be
 \ket{\bid}_\b = \ket{\psi_\b} , \quad   \sY_R = \widetilde \sM_R , \quad  \sY_L =\widetilde \sM_L \ .
\ee
where $\wt \sM_R$ and $\wt \sM_L$ denote the bulk operator algebras associated with the right and left global AdS spacetimes, respectively.


Now consider $T > T_{\rm HP}$. With the thermal ensemble dominated by states of energies of order $O(N^2)$, the expression~\eqref{tfd0} becomes ill-defined in the $N \to \infty$ limit as neither the energies nor the associated eigenstates admit a well-defined limit. Moreover, a single-trace operator $\sO  = \Tr (\cdots) \in \sB(\sH_R)$ can have a divergent one-point function in the state $\ket{\Psi_\b}$,
\be \label{eh0o}
\vev{\Psi_\b|\sO |\Psi_\b} \sim O(N), \quad N \to \infty \ .
\ee
Nevertheless, we can define the $N \to \infty$ limit of the TFD state~\eqref{tfd0} 
 in terms of the limiting behavior of its correlation functions. 
To this end, we introduce  ``renormalized'' single-trace operators by subtracting their thermal expectation values
\be \label{sub0}
\hat \sO \equiv \sO -   \vev{\Psi_\b|\sO |\Psi_\b},  
\ee
and define the algebra $\sS_\b^{(R)}$ ($\sS_\b^{(L)}$) as being generated by such operators. 
From the large-$N$ behavior of thermal correlation functions, we have
\be \label{ttw0}
\vev{\Psi_\b|\hat \sO|\Psi_\b} =0, \quad 
 \vev{\Psi_\b \le |\hat \sO_1 (x_1) \hat \sO_2 (x_2) \ri |\Psi_\b} \sim O(N^0), \quad N \to \infty  \ .
\ee
Furthermore, higher-point functions factorize into products of two-point functions, as in~\eqref{factoriza}.

We can  build a GNS Hilbert space $\sH_{\Psi_\b}^{(\rm GNS)}$ using $\ket{\Psi_\b}$ and $\sS_\b \equiv \sS_\b^{(L)} \otimes \sS_\b^{(R)}$. 
Due to the large $N$ factorization, at the leading order, we have a generalized free field theory and the GNS Hilbert space has the structure of a Fock space. It can  be checked that the state $\om_\b$ defined by expectation values of $\sS_\b$ in $\ket{\Psi_\b}$ is pure with respect to $\sS_\b$, and thus  $\sA_{\Psi_\b}$ coincides with the set $\sS_\b$ generated by products of single-trace operators. Denoting 
\be
\sY_{R} = (\pi_{\Psi_\b} (\sS_{\b}^{(R)}))'', \quad \sY_{L} = (\pi_{\Psi_\b} (\sS_{\b}^{(L)}))'' ,
\ee
we have 
\be 
\sY_R' = \sY_L, \quad \sB ( \sH_{\Psi_\b}^{(\rm GNS)}) = \sY_R \lor \sY_L \ .
\ee

Now the nature of $\sH_{\Psi_\beta}^{(\rm GNS)}$, $\sY_R$, and $\sY_L$ is qualitatively different: the entanglement between CFT$_R$ and CFT$_L$ becomes of order $O(N^2)$. As a result, it has been argued that $\sY_R$---the large-$N$ limit of $\sB(\sH_R)$---becomes a type III$_1$ von Neumann algebra~\cite{LeuLiu21a,LeuLiu21b}, and the GNS Hilbert space $\sH_{\Psi_\beta}^{(\rm GNS)}$ no longer factorizes into separate components associated with CFT$_R$ and CFT$_L$.

On the gravity side, small perturbations around the black hole geometry can be described using the standard formalism of quantum field theory in a curved spacetime. 
 Quantizing these perturbations yields a Fock space $\sH_{\rm BH}^{\rm (Fock)}$, with the corresponding vacuum state given by the Hartle-Hawking vacuum $\ket{HH}$. We then have the identifications
\bega \label{ejnn}
\sH_{\Psi_\b}^{(\rm GNS)} = \sH_{\rm BH}^{\rm (Fock)}, \quad
 \ket{\bid}_\b = \ket{HH}  , \\
  \sY_R = \widetilde \sM_R, \quad \sY_L = \widetilde \sM_L ,
   \label{ejnn1}
\end{gather}
where $\wt \sM_R$ and $\wt \sM_L$ are respectively the bulk operator algebras associated with the $R$ and $L$ exterior regions
the black hole. The identifications~\eqref{ejnn1} can be made explicitly by matching the bulk QFT oscillators with those of the boundary generalized free field theory. 
Since $\widetilde{\sM}_R$ and $\widetilde{\sM}_L$ are operator algebras associated with subregions in a continuum QFT, they are of type III$_1$, matching the type of $\sY_R$ and $\sY_L$.

Thus, the non-factorization of the bulk dual description in terms of the black hole geometry can be directly understood as a consequence of the change in the large-$N$ limit of the boundary algebra, induced by the divergent entanglement in this limit.

We close this subsection with a few additional remarks: 

\ben 

\item Similar to~\eqref{bwa} and~\eqref{bdwa} in the vacuum sector, we can express 
a single-trace operator $\sO_{R,L}$  in CFT$_R$ or CFT$_L$ in terms of a mode expansion as 
\be 
\sO_\al (x) = \sum_k v_k^{(\al)} (x) a_k^{(\al)} , \quad \al = R, L , \quad k = (\om, \vk) , \quad
a_{k}^{(R)} \ket{\bid}_\b = e^{-{\b \om \ov 2} }  a_{-k}^{(L)} \ket{\bid}_\b , 
\ee
where $\om$ is the frequency and $\vk$ denotes the spatial momentum. 
From the identifications~\eqref{ejnn1}, the oscillators $a_k^{(R,L)}$ are identified with those   for the corresponding bulk field $\phi_{R, L}$ in the $R$ and $L$ regions,
\be \label{nuMo}
\phi_\al (X) = \sum_k u_k^{(\al)} (X) a_k^{(\al)} , \quad \al = R, L , \quad k = (\om, \vk)  \ .
\ee

\item Although the bulk dual of $\ket{\Psi_\beta}$ is already known, one could have deduced its key geometric features purely from the algebraic structure: the fact that $\sY_{R,L}$ are type I for $T < T_{\rm HP}$ and become type III$_1$ for $T > T_{\rm HP}$ implies that the bulk geometry must be disconnected in the former case and connected in the latter. This forms the basis of the algebraic formulation of ER=EPR, which we will elaborate on in Sec.~\ref{sec:EREPR}.

\item  For $T > T_{\rm HP}$, each temperature corresponds to a distinct sector with a non-overlapping GNS Hilbert space. This is because the states $\ket{\Psi_\beta}$ at different values of $\beta$ have entanglement entropies that differ by $O(N^2)$, and are thus separated by an infinite amount of entanglement in the large-$N$ limit.
In contrast, for $T < T_{\rm HP}$, there exists a single GNS Hilbert space common to all $\beta$, with the states $\ket{\bid}_\b$ representing different vectors within the same Hilbert space.

\item At any temperature, the CFT Hamiltonians $H_R$ and $H_L$ do not belong to the operator algebras in the large-$N$ limit, and therefore equation~\eqref{each} no longer holds. However, the modular operator $\Delta_\beta$ does survive the large-$N$ limit and continues to generate time translations on the two boundaries (in opposite directions).
For $T < T_{\rm HP}$, the modular operator factorizes as
\begin{equation}
- \log \Delta_\beta = \beta (\hat{h}_R - \hat{h}_L) , 
\end{equation}
where $\hat{h}$ denotes the bulk Hamiltonian of quantum fields in global AdS, and corresponds to the time-translation operator in the boundary generalized free field theory in the vacuum sector (recall the discussion around~\eqref{bhamil}).
For $T > T_{\rm HP}$, by contrast, the modular operator no longer factorizes, reflecting the emergent III$_1$ structure of boundary algebras.

\een

\subsection{General black holes} \label{sec:genBH0}

We now consider an example in which there exist operators, in addition to single-trace operators, that survive the large-$N$ limit; that is, the inclusion~\eqref{SingOp} is strict.



\begin{figure}[h]
\begin{centering}
	\includegraphics[width=8cm]{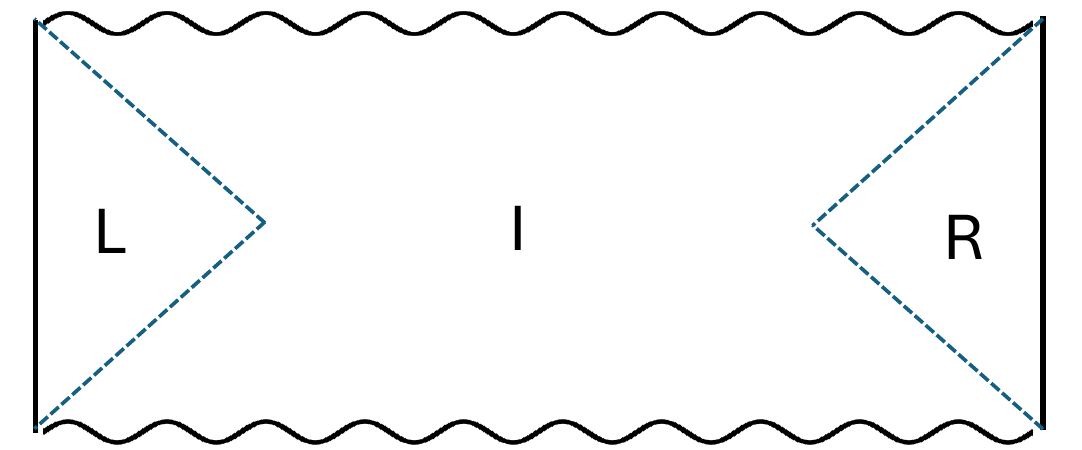}
\par\end{centering}
\caption{\small A long two-sided black hole.   The black hole interior region is labeled by $I$. 
}
\label{fig:longbh}
\end{figure}

The thermofield double state is special due to the presence of a bifurcating horizon. This is no longer the case for a general two-sided semi-classical state $\ket{\Psi}$---the large-$N$ limit of a certain entangled state in $\sH_R \otimes \sH_L$---which is dual to a ``long'' black hole, as illustrated in Fig.~\ref{fig:longbh}. For simplicity, we have taken the geometry to be time-reflection and left-right symmetric; the discussion readily generalizes to less symmetric cases.

As in the TFD state above the Hawking-Page temperature, we can define ``renormalized'' single-trace algebras $\sS^{(R)}_\Psi$ and $\sS^{(L)}_\Psi$ for the CFT$_R$ and CFT$_L$, respectively. In analogy with the identification~\eqref{ejnn1}, by studying the bulk QFT outside the horizon, we can establish
\be\label{ejnn21} 
\sY_R \equiv (\pi_\Psi (\sS^{(R)}_\Psi))'' = \widetilde \sM_R, \quad \sY_L \equiv (\pi_\Psi (\sS^{(L)}_\Psi))'' = \widetilde \sM_L ,
\ee
where $\wt \sM_R$ and $\wt \sM_L$ denote the bulk operator algebras associated with the $R$ and $L$ exterior regions of the black hole. However, unlike in the TFD case, the $R$ and $L$ regions no longer cover a full Cauchy slice. Bulk operators in the interior region $I$---as indicated in Fig.~\ref{fig:longbh}---should still correspond to boundary operators that survive the large-$N$ limit, but they do not belong to the algebra of single-trace operators.

Thus, in this case we must have
\be 
\sS \subset \sA_\Psi \ .
\ee
The additional operators required to describe bulk operators in region $I$ must be related in some way to single-trace operators, since a bulk field $\phi$ in region $R$ and in region $I$ represents the same underlying field.
{In Sec.~\ref{sec:ewr}, we will see that these extra operators can be related to single-trace operators via modular flow.}

\begin{figure}[H]
\begin{centering}
	\includegraphics[width=12cm]{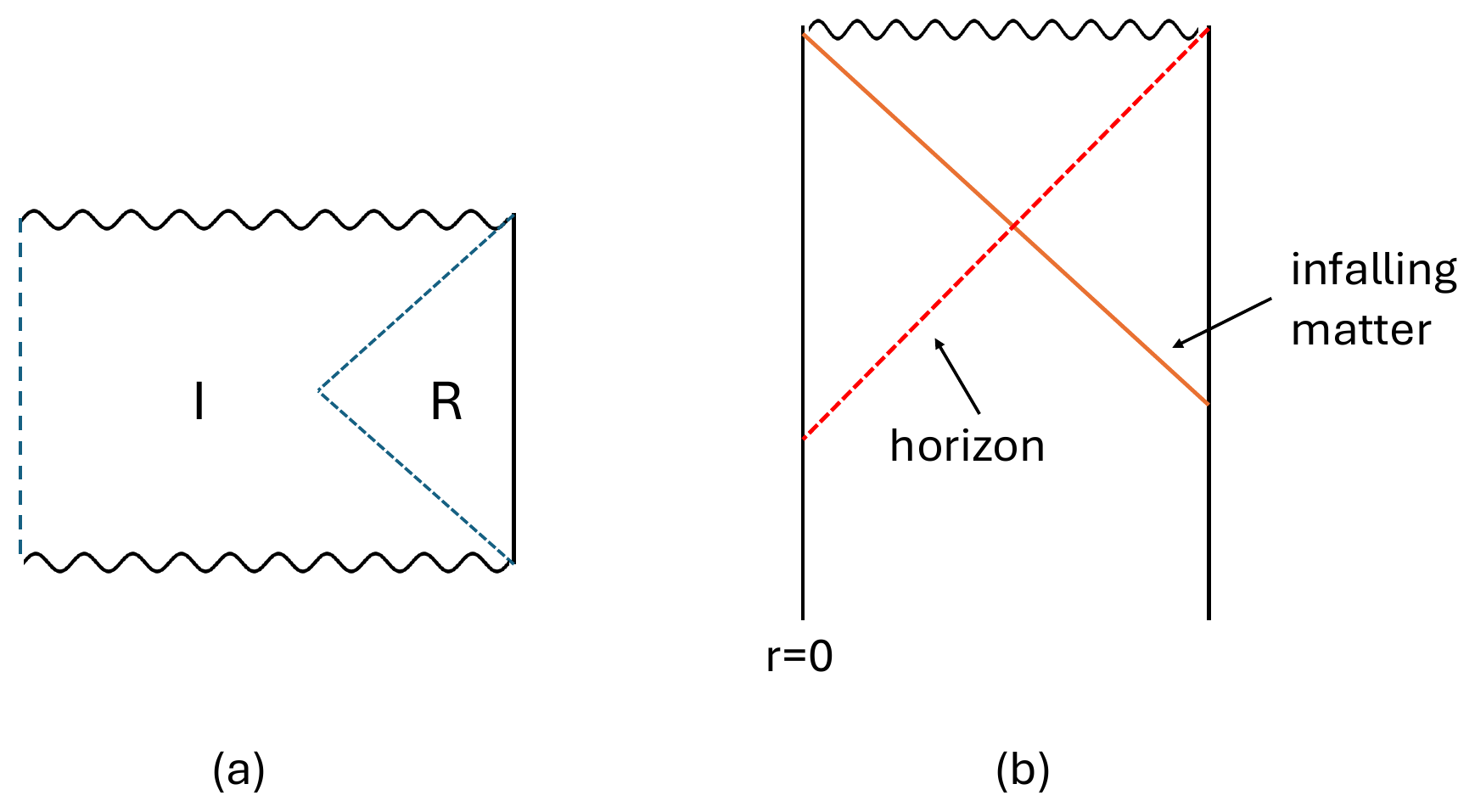}
\par\end{centering}
\caption{\small (a) A ``long'' single-sided black hole. The vertical dashed line marks the location where the spacetime either ends smoothly or terminates at an end-of-world brane.
(b) A black hole spacetime formed from gravitational collapse.
}
\label{fig:ssbh}
\end{figure}

A similar conclusion of $\sS \subset \sA_\Psi$ applies to a ``long'' single-sided black hole, illustrated in Fig.~\ref{fig:ssbh}(a): bulk operators in the interior region $I$ should correspond to boundary operators that survive the large-$N$ limit, but do not belong to the algebra of single-trace operators. However, the case of a single-sided black hole formed from gravitational collapse~(see 
Fig.~\ref{fig:ssbh}(b)) is different: as we will see later in Sec.~\ref{sec:hkll}, one finds that $\sA_\Psi = \sS$.

\subsection{A diagnostic of firewalls in generic highly excited states}\label{sec:firewall}


So far, we have considered semi-classical states---states that admit a semi-classical bulk description. Now, consider a generic highly excited state $\ket{\Psi} \in \sH$ with energy $E_\Psi \sim O(N^2)$. At such energies, the boundary theory is expected to be chaotic, and a generic state should exhibit thermal behavior in the large-$N$ limit. That is, correlation functions of single-trace operators are expected to satisfy
\be \label{eim0}
\vev{\Psi|\sO|\Psi} \approx \vev{\sO}_\b , \quad \vev{\Psi|\sO_1 \sO_2 \cdots \sO_n |\Psi} \approx 
\vev{\sO_1 \sO_2 \cdots \sO_n}_\b
\ee
to leading order in the $1/N$ expansion, where the right-hand sides represent thermal correlation functions at inverse temperature $\b$. The value of $\b$ is determined by energy matching: $E_\Psi = E_\b$, where $E_\b$ is the energy of the thermal density matrix at temperature $\b$.

Equations~\eqref{eim0} suggest that the state $\ket{\Psi}$ admits a geometric description corresponding to the exterior region of a black hole. But does it also admit a smooth geometric extension behind the horizon, as in the case of Fig.~\ref{fig:ssbh}? Alternatively, the interior geometry may not exist, in which case we say there is a firewall at the horizon. 

\begin{figure}[H]
\begin{centering}
	\includegraphics[width=6cm]{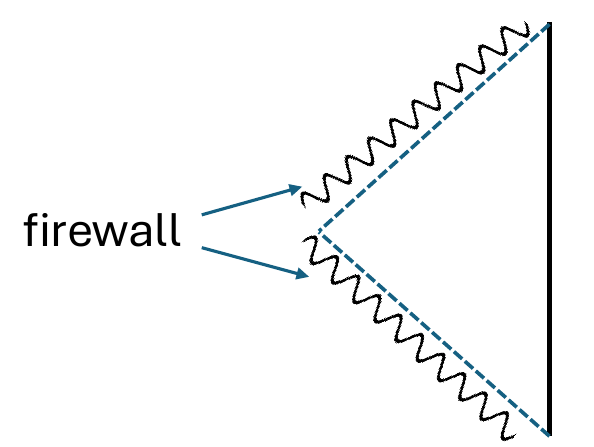} 
\par\end{centering}
\caption{\small Generic states do not have a smooth geometric beyond the horizon, i.e., there is a firewall at the horizon.}
\label{fig:nobh}
\end{figure}

As discussed in the previous subsection, describing the portion of a Cauchy slice behind the horizon requires the existence of operators in the large-$N$ limit that lie outside the single-trace operator algebra. 
 This is a highly nontrivial condition. If the state $\ket{\Psi}$ does not satisfy it, then a smooth bulk geometry behind the horizon cannot be defined, and there is a ``firewall''  at the horizon. See Fig.~\ref{fig:nobh}.


The diagnostic can be stated more formally as follows:

{\it Suppose a generic excited state $\ket{\Psi} \in \sH$ satisfies~\eqref{eim0} and $\sA_\Psi = \sS$. Then the bulk dual corresponds to a black hole with a ``firewall'' at the horizon.}

\noindent Note that in this case $\om_\Psi$ is a mixed state with respect to $\sA_\Psi$, in violation of~\eqref{pureS}.  

Similar statements apply to generic two-sided states in $\sH_R \otimes \sH_L$. Moreover, for a general two-sided state $\ket{\Psi}$, in addition to~\eqref{eim0} (where all operators there should be understood to belong to a single CFT), we also expect 
\be
\vev{\Psi|\sO_R \sO_L|\Psi}_c \to 0 \quad \text{as} \quad N \to \infty, \quad \sO_R \in \sS_\Psi^{(R)}, \ \sO_L \in \sS_\Psi^{(L)} \ ,
\ee
which is consistent with the interpretation that no smooth, connected geometry exists between the $L$ and $R$ boundaries.




\subsection{Perturbative $1/N$  corrections: conserved charges and bulk nonlocality} \label{sec:higher}


So far we have restricted to the leading order in the $G_N$ expansion.
On the gravity side, perturbative $G_N$ corrections can be obtained by including higher order terms in~\eqref{freac}--\eqref{0Inact}.  The bulk field theory is now interacting, with effects of interactions treated perturbatively. 
On the boundary,  three-point and higher-point functions are non-vanishing, suppressed by powers of $1/N$, resulting in a generalized field theory with interactions. 
To any finite order in the $1/N$ (or $G_N$) perturbative expansion, the leading-order characterization of the algebras remains valid, as the spectrum of the modular operator is dominated by its zeroth-order contribution.\footnote{This is analogous to the standard perturbative treatment of $\lambda \phi^4$ theory, where the Hilbert space at any finite order in perturbation theory remains that of the free theory.} For example, in the TFD state above the Hawking-Page temperature, $\sY_R$ and $\sY_L$ should still be type III$_1$. 

In this subsection we discuss two new features from $1/N$ (or $G_N$) perturbative corrections. The first deals with $1/N$ corrections to the relation~\eqref{hkll00} between bulk and boundary operators, while the second concerns boundary conserved quantities, which are closely tied to bulk nonlocal behavior arising from Gauss laws.
For definiteness, we will consider the vacuum sector. A similar discussion can be applied to other sectors. 

\subsubsection{$1/N$ corrections to bulk reconstruction} 

One immediate consequence of including higher order corrections in~\eqref{freac} is that the relation~\eqref{hkll00} between bulk and boundary operators becomes more complicated~\cite{KabLif11,KabLif13}. As an illustration here  we consider a simple example of a real scalar field $\phi$ dual to a boundary operator $\sO$.  Suppose $S_2$ and $S_3$ in~\eqref{freac}--\eqref{0Inact} 
for $\phi$ have the form 
\be 
S_2 =-\ha \int d^{d+1} X \, \sqrt{-g} \le((\p \phi)^2 + m^2 \phi^2 \ri), \quad  S_3 =- {1 \ov 3} \int d^{d+1} X \sqrt{-g} \,  \phi^3 + \cdots   
\ee
where $\cdots$ include couplings of $\phi$ to other fields, such as metric perturbations. 
The equation of motion for $\phi$ is then given by 
\be 
\p^2 \phi = m^2 \phi + \ka \phi^2  + \cdots  \ .
\ee
We can expand the above equation perturbatively in $\ka$ as 
\bega 
\phi = \phi^{(0)} + \ka \phi^{(1)} + \cdots \\
\label{perE}
(\p^2 - m^2) \phi^{(0)} = 0, \quad  (\p^2-m^2) \phi^{(1)} =  (\phi^{(0)})^2 ,
\end{gather} 
and $\phi^{(1)}$ can be solved as\footnote{The homogenous solution to the equation for $\phi^{(1)}$ can be set to zero by absorbing it into the definition of $\phi^{(0)}$, and $ (\phi^{(0)} (X') )^2$ should be understood as being normal ordered.} 
\be \label{perE1}
\phi^{(1)} (X) = \int d^{d+1} X'  \sqrt{-g} \, G (X, X')  (\phi^{(0)} (X') )^2, 
\ee
where $G (X,X')$ is a bulk propagator for $\phi$.  The zeroth order equation~\eqref{perE} leads to~\eqref{hkll00}, and~\eqref{perE1} then gives  
\bega 
\phi (X) = \int d^d x \, K(X; x) \sO^{(0)} (x) + \ka \int d^d x_1 d^d x_2 \, K(X; x_1, x_2) \sO^{(0)} (x_1)\sO^{(0)} (x_2) + \cdots, \\
 K(X; x_1, x_2)  \equiv \int d^{d+1} X'  \sqrt{-g} \, G (X, X') K (X'; x_1) K (X';x_2) ,
\end{gather} 
where we have added a superscript to $\sO^{(0)}$ to stress that it is the generalized free field dual to $\phi^{(0)}$. 
Including higher order corrections, we expect the following expansion in $1/N$, 
\ie \label{edu}
\phi (X) &= \int d^d x\, K (X; x) \sO^{(0)} (x) \cr
+ & \sum_{n=1}^\infty {1 \ov N^n} \int d^d x_{1} \cdots d^d x_{n+1} \, K_{i_1 \cdots i_{n+1}} (X; x_1, \cdots x_{n+1}) \sO_{i_1}^{(0)} (x_1) \cdots \sO_{i_{n+1}}^{(0)} (x_{n+1})  \ .
\fe
Applying the boundary limit of~\eqref{edu} to~\eqref{exT}, we then find that 
\be \label{edu1}
\sO (x) =  \sO^{(0)} (x)
+  \sum_{n=1}^\infty {1 \ov N^n} \int d^d x_{1} \cdots d^d x_{n+1} \, \tilde K_{i_1 \cdots i_{n+1}} (x; x_1, \cdots x_{n+1}) \sO_{i_1}^{(0)} (x_1) \cdots \sO_{i_{n+1}}^{(0)} (x_{n+1})  ,
\ee
where $ \tilde K_{i_1 \cdots i_{n+1}} (x; x_1, \cdots x_{n+1})$ is obtained from $K_{i_1 \cdots i_{n+1}} (X; x_1, \cdots x_{n+1})$ by taking $X$ to the boundary. Equation~\eqref{edu1} gives the expansion of the single-trace operator $\sO(x)$ perturbatively in terms of generalized free fields $\sO^{(0)}_i$.

\subsubsection{Conserved charges and bulk nonlocality}

We now discuss qualitative new features concerning conserved quantities when including $1/N$ corrections. 
We will first discuss the boundary story (emphasized in~\cite{Wit21b}), and then the corresponding bulk picture.

The boundary CFT possesses global symmetries, including conformal symmetries and potentially additional internal symmetries. 
As an illustration, we assume that the CFT has a $U(1)$ internal symmetry with a corresponding conserved current $J^\mu$ and conserved charge~($\sig$ is a boundary Cauchy slice) 
\be \label{eu00}
Q = \int_\sigma d^{d-1} x \, J^0(x) \ .
\ee
Acting on an operator $\sO(x)$ of charge $1$, we have 
\be \label{comY}
i [Q, \sO(x)] = \sO(x) \ .
\ee

In close parallel with the discussion of the Hamiltonian below~\eqref{eu000}, the conserved current 
$J^\mu$ takes the form $N  \Tr (\cdots)$. As a result, only the rescaled operators
\be \label{ha10}
\hat J^{\mu} \equiv \frac{1}{N} J^{\mu}, \quad \hat Q \equiv \frac{Q}{N}
\ee
survive in the large-$N$ limit. The commutator of $\hat Q$ with a single-trace operator is suppressed by $1/N$
\be\label{haht}
i [\hat Q, \sO(x)] = \frac{1}{N} \sO(x) \ .
\ee
In the boundary generalized free field theory, there should exist an operator $\hat q$ that generates charge rotations via
\be \label{haht1}
i [\hat q, \sO(x)] = \sO(x) \ .
\ee
As in the case of the Hamiltonian, $\hat q$ cannot be written as an integral of a local operator over a single Cauchy slice.

The $U(1)$ global symmetry on the boundary is dual to the asymptotic symmetry associated with a bulk $U(1)$ gauge symmetry. In empty AdS in the $G_N \to 0$ limit, this asymptotic symmetry extends to the full bulk as the global part of the gauge symmetry, and $\hat q$ corresponds to the Noether charge associated with this bulk global symmetry.


Now consider going beyond the leading order in the $G_N$ (or $1/N$) expansion. 
Equation~\eqref{haht} implies that the bulk theory should develop some {\it nonlocal} behavior. More explicitly, suppose $A_M = (A_z, A_\mu)$ is the bulk $U(1)$ gauge field dual to the boundary current $\hat J^\mu$, we have the extrapolation dictionary\footnote{For convenience we use the Poincare coordinates near the boundary ($z=0$).}  
\be \label{ex00}
\hat J^\mu (x) = \lim_{z \to 0} j^\mu (z, x), \quad  j^\mu (z,x) \equiv \sqrt{-g} F^{z \mu} (z, x),
\ee
where $F_{MN}$ is the field strength associated with $A_M$. Suppose $\phi$ is the charged scalar dual to $\sO$. From~\eqref{haht} and~\eqref{hkll00}, 
we have 
\be \label{bbco}
i  [\hat Q, \phi (z', x')] = {1 \ov N} \phi (z', x') \propto \ka \phi (z', x')  \ .
\ee
Taking $\sig$ in~\eqref{eu00} to be on the same bulk Cauchy slice as $(z',x')$, from~\eqref{ex00}, we find that~\eqref{bbco} requires nonzero  commutator between $F^{z\mu} (z,x)$ and $\phi (z',x')$ at {\it spacelike separations}. 

Equation~\eqref{bbco} can be understood on the gravity side as a consequence bulk gauge symmetries and the corresponding Gauss law constraints.
More explicitly, consider the lowest two orders of the bulk action of $\phi$ and $A_M$, which should have the form
\bega
S = S_2 + \ka S_3+ O(\ka^2), \quad 
S_2 = - \int d^{d+1} x \, \sqrt{-g} \, \le[{1 \ov 4} F_{MN} F^{MN} + |\p \phi|^2 
+ m^2 |\phi|^2   \ri], \\
S_3 = A_M J^M  + \cdots, \quad  J^M =  i \le( \phi^* \p^M \phi - \phi \p^M \phi^* \ri)  \ .
\label{cuac}
\end{gather} 
$J^M$ is the Noether current for the $U(1)$ global symmetry of $S_2$. In $S_3$ we suppressed other possible cubic terms, e.g., the interactions between $\phi$ and metric perturbations.  
Equations of motion for the gauge field are 
\bega \label{eomF} 
\nab_M F^{MN} = \ka J^N + O(\ka^2),
\end{gather} 
and gauge transformations have the form 
\be 
A_M \to A_M + \p_M \lam , \quad \phi \to \phi' = e^{i \ka \lam} \phi  = \phi + i \ka \lam \phi   + O(\ka^2) , 
\ee
where the gauge parameter $\lam (z, x)$ should go to zero approaching the boundary $z =0$.  $\phi (z, x)$ is not gauge invariant, but we can construct a gauge invariant operator by attaching to it a Wilson line going to the boundary, for example, 
\be 
\Phi (z,x) = e^{-i \ka V} \phi (z,x)  = \phi (z, x)-  i \ka V \phi + O(\ka^2) 
, \quad V  =\int_0^z dz' \, A_z (z', x) \ .
\ee
$\Phi(z, x)$ is known as a dressed observable, and it exhibits nonlocal commutation relations due to the inherent nonlocality of the Wilson line used in its construction.

Alternatively, we can fix a convenient gauge; in any gauge that completely fixes the gauge freedom, $\phi (z, x)$ corresponds to a gauge invariant observable. More explicitly, suppose a gauge has been fixed.  
Solving~\eqref{eomF} then leads to a solution of the form 
\be \label{exwp}
A_M (X) = A_M^{(0)} (X) + \ka \int  d^{d+1} X\, \sqrt{-g} \, G_{MN} (X, X')  J^N  (X') + \cdots
\ee
where $G_{MN} (X, X') $ is a propagator for the gauge field. 
The second term of~\eqref{exwp} can then lead to nonlocal commutation relation of the form~\eqref{bbco} (see e.g.~\cite{Hee12} for details). As in~\eqref{edu1}, the boundary limit of~\eqref{exwp} leads to a corrected extrapolation dictionary of the form 
\be \label{ejnu}
 \hat Q = \hat Q^{(0)} + {1 \ov N} \hat q + \cdots ,
\ee
where $\hat Q^{(0)}$ denotes the corresponding operator at the generalized free field level.


Parallel discussions can be given by time translations and other spacetime symmetries in which case we need to deal with bulk diffeomorphisms. We can either consider dressed observables by attaching to $\phi (z, x)$ a ``gravitational Wilson line'' going to the boundary, or work with a specific gauge. In the latter case, we have a rather parallel story to~\eqref{eomF} and~\eqref{exwp},  with~\eqref{eomF} replaced by the Einstein equations and $J^N$ by the bulk stress tensor of $\phi$. In particular, we expect the $\hat H$ (as defined in~\eqref{ha100}) to have the $1/N$ expansion  
\be \label{ejnu00}
\hat H = \hat H^{(0)} + {1 \ov N} \hat h_\Om + \cdots \ .
\ee

\section{Subregion-subalgebra duality}  \label{sec:sub}



 
 



The algebraic formulation of AdS/CFT in the large-$N$ limit, discussed in the previous section, leads directly to a simple but powerful principle: the subregion-subalgebra duality~\cite{LeuLiu22}, which \textit{equates} an arbitrary bulk spacetime subregion with an emergent type III$_1$ von Neumann subalgebra of the boundary CFT.

In this section, we first present the general formulation of the subregion-subalgebra duality, followed by several classes of examples---including a reformulation of entanglement wedge reconstruction. We then introduce a general theorem that provides a powerful method for identifying bulk subregions with boundary subalgebras. Finally, we argue that if the generalized entropy of a bulk region admits a finite-$N$ extension, then the corresponding subalgebra in the duality should  extend to a type I algebra at finite $N$.

The subregion-subalgebra duality offers a framework for addressing long-standing questions about the emergence of bulk locality and the reconstruction of general bulk regions. We explore its further implications for holography in Sec.~\ref{sec:Eme}.

\subsection{General formulation}

In Sec.~\ref{sec:genG}, we discussed how the boundary theory Hilbert space separates into disjoint sectors in the large $N$ limit, each corresponding to a distinct bulk geometry. The Fock space obtained by quantizing perturbations around a given bulk geometry can then be identified with the GNS Hilbert space of the corresponding boundary state, as expressed in~\eqref{fockid}.

\begin{figure}[H]
\begin{center}
\includegraphics[width=3cm]{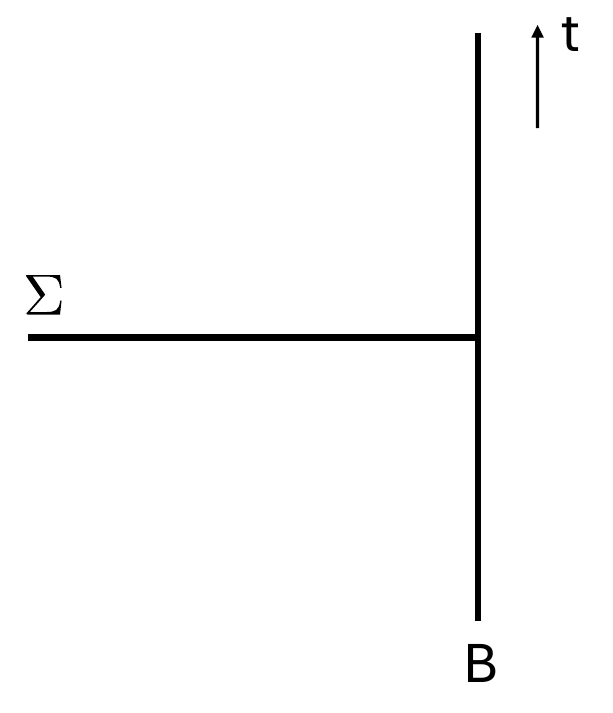}
\caption[]{\small For a general asymptotic AdS geometry, the bulk algebra $\sB (\sH_{\Psi}^{\rm (Fock)})$ can be defined on a single bulk Cauchy slice $\Sig$, while the boundary algebra $\sB (\sH_{\Psi}^{(\rm GNS)} )$ is defined for the full boundary spacetime. In the plot, the vertical line represents the boundary. 
}
\label{fig:mapping}
\end{center}
\end{figure}

The identification~\eqref{fockid} implies that 
\be \label{opide0}
\sB (\sH_{\Psi}^{\rm (Fock)}) = \sB (\sH_{\Psi}^{(\rm GNS)} )  \ .
\ee
 Note that $\sB (\sH_{\Psi}^{\rm (Fock)})$ is defined on a single bulk Cauchy slice, while $\sB (\sH_{\Psi}^{(\rm GNS)} )$ is defined for the full boundary spacetime (or at least a finite time band as in~\eqref{miF}), see Fig.~\ref{fig:mapping}.  
 Since the bulk has one more dimension than the boundary,  $\sB (\sH_{\Psi}^{\rm (Fock)})$ and $\sB (\sH_{\Psi}^{(\rm GNS)} )$ depend on the same number of dimensions.

 At the quadratic level, we can diagonalize bulk fields so that different fields decouple from one another. As a result, both $\sH_{\Psi}^{(\rm GNS)}$ and $\sB (\sH_{\Psi}^{(\rm GNS)} )$ exhibit a tensor product structure, decomposing into factors corresponding to these different fields (labeled by $i$) 
\bega \label{tesT}
\sH_{\Psi}^{(\rm GNS)} = \otimes_i  \sH_{\Psi, i}^{(\rm GNS)}, \quad 
 \sB (\sH_{\Psi}^{(\rm GNS)} )  = \otimes_i  \sB (\sH_{\Psi, i}^{(\rm GNS)}) ,  \\
 \sH_{\Psi, i}^{(\rm Fock)} =  \sH_{\Psi, i}^{(\rm GNS)}, \quad  \sB (\sH_{\Psi,i}^{\rm (Fock)}) = \sB (\sH_{\Psi,i}^{(\rm GNS)} )  \ .
\end{gather}


Equation~\eqref{opide0} further implies that operator subalgebras acting on $\sH_\Psi^{\rm (Fock)}$ and $ \sH_{\Psi}^{({\rm GNS})}$ must have one-to-one correspondence. In particular, consider an open subregion $\fb$ on a bulk Cauchy slice, and the von Neumann algebra\footnote{All the bulk subalgebras are denoted with a $\widetilde{}$ to distinguish them from  boundary algebras.}
$\wt \sM_\fb \subset \sB (\sH_\Psi^{\rm (Fock)})$ associated with the region. Then there must be a corresponding boundary subalgebra 
$\sM_\fb \subset\sB (\sH_{\Psi}^{({\rm GNS})})$ that is identified with it, i.e., 
\be \label{ssdual}
 \widetilde \sM_\fb = \sM_\fb  \ .
\ee
The equivalence~\eqref{ssdual} implies that the boundary subalgebra $\sM_\fb$ captures all the bulk operations which one can perform in $\fb$, and from our discussion of Sec.~\ref{sec:qft}, $\sM_\fb$ also encodes the causal structure, geometric information, as well as the entanglement structure of $\fb$. For example, entanglement of a bulk subregion $\fb$ with its causal complement is captured by eigenvalues of the modular operators (entanglement Hamiltonian), which can be probed by modular flows. The equivalence~\eqref{ssdual} identifies the bulk modular operator associated with $\wt \sM_\fb$ with the boundary modular operator for $\sM_\fb$. We can then obtain the bulk entanglement structure from modular flows of $\sM_\fb$.

 We therefore conclude that a bulk subregion $\fb$  can be fully ``reconstructed''  using the corresponding boundary algebra $\sM_\fb$. 
This is the {\bf subregion-subalgebra duality}. 

We have already encountered examples of subregion-subalgebra duality in Sec.\ref{sec:TFD} and Sec.\ref{sec:genBH0} (e.g.~\eqref{ejnn1} and~\eqref{ejnn21}), where the region outside a black hole was identified with the single-trace operator algebra on the corresponding boundary.

While the statement itself may appear simple, it provides a powerful framework to reconstruct the bulk system from the boundary. 
Furthermore, as we will see later, it may be used to ``define'' the bulk subregions and causal structure in the stringy regime when we take the boundary theory to be at finite 't Hooft coupling $\lam$. 
The subregion-subalgebra duality also leads to highly nontrivial requirements on the boundary system, implying that, in the large $N$ limit, $\sB (\sH_{\Psi}^{({\rm GNS})})$ must be rich enough to contain all the subalgebras associated to local bulk subregions.

Since at the leading order in the $G_N \to 0$ limit, the bulk is described by a quantum field theory in a curved spacetime, 
we assume that $\wt \sM_\fb$ satisfies various general properties discussed in Sec.~\ref{sec:qft}: 

\ben

\item It obeys the bulk Reeh-Schlieder theorem, i.e., the ``vacuum'' $\ket{0}_{\phi_c} \in \sH_\Psi^{\rm (Fock)}$ is cyclic and separating with respect to $\wt \sM_\fb$.  This implies that the GNS vacuum $\ket{\bid}_\Psi$ is cyclic and separating with respect to $\sM_\fb$.

\item The algebras for different regions respect (bulk) causality, i.e. 
\be \label{bcau}
 \widetilde \sM_\fb  = \wt \sM_{\hat b}
 \ee
where $\hat \fb$ is the bulk domain of dependence of $\fb$. Hence, the corresponding boundary subalgebra $\sM_\fb$ can in fact reconstruct $\hat \fb$.

\item $\wt \sM_\fb$ is type III$_1$, which implies that the corresponding boundary algebra $\sM_\fb$ must also be type III$_1$.

\item For topologically trivial regions  $\fb_1,~\fb_2$, the algebras are additive, 
\be\label{badd1}
\widetilde{\sM}_{\fb_1 \cup \fb_2} = \widetilde{\sM}_{\fb_1} \vee \widetilde{\sM}_{\fb_2}  \ .
\ee

\item The bulk algebras satisfy the Haag duality (for topologically trivial regions) 
\begin{align}
		\label{haagb1} 
	\widetilde{\sM}_{b'} &= \le(\widetilde{\sM}_{\fb}\ri)', 
\end{align}
where $\fb'$ denotes the complement of $\fb$ on its bulk Cauchy slice. 

\een
We will see later that  these properties have important boundary implications.

\subsection{Entanglement wedge reconstruction: algebraic reformulation} \label{sec:ewr}

Two immediate classes of examples of subregion-subalgebra duality are provided by the entanglement wedge and causal wedge reconstructions associated with a boundary region. The entanglement wedge reconstruction states that the physics in the entanglement wedge of a boundary subregion $A$ (on a Cauchy slice) can be fully described by that in $A$~\cite{Van09,CzeKar12,CzeKar12b,Wal12,HeaHub14,AlmDon14,JafSuh14,PasYos15,JafLew15,DonHar16,Har16,FauLew17,CotHay17}. Causal wedge reconstructions~\cite{BanDou98,Ben99,BalKraLaw98,HamKab05,HamKab06,KabLif11,Hee12,HeeMar12,PapRaj12,Mor14,Hub14,EngPen21a,Wit23}, also known as local HKLL reconstructions, express bulk local operators in the causal wedge of $A$ in terms of single-trace operators supported in the domain of dependence of $A$.

Below, we reformulate these statements in algebraic terms\footnote{The entanglement wedge reconstruction has also been formulated in algebraic terms within the framework of holographic quantum error correction~\cite{Har16, KanKol18, Fau20, KanKol21, GesKan21, FauLi22}, which we will comment on in Sec.~\ref{sec:QEC}.} as special cases of subregion-subalgebra duality~\cite{LeuLiu21b,LeuLiu22}, and highlight several new insights that emerge from this perspective in subsequent subsections. 

\subsubsection{Entanglement and causal wedge reconstructions} 

\begin{figure}[H]
\begin{center}
\includegraphics[width=4cm]{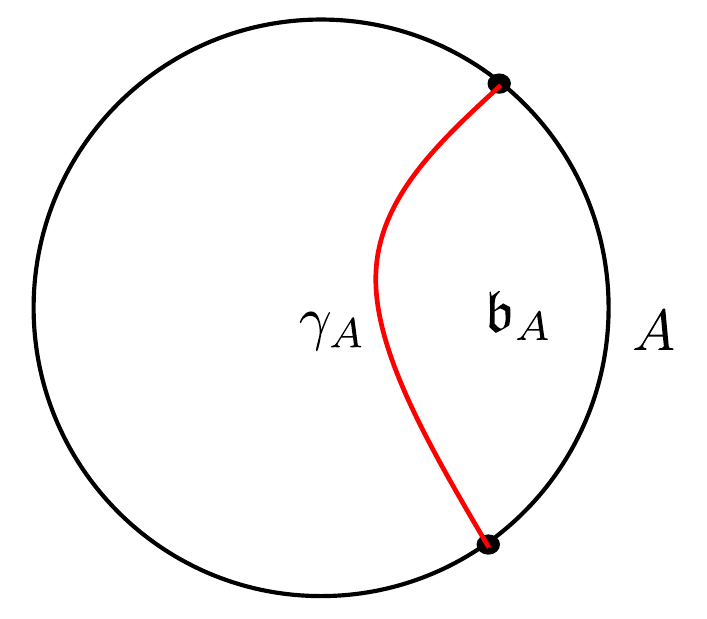}
\caption[]{
\small A constant-time Cauchy slice $\Sig$ of a static asymptotic AdS is shown, with the boundary indicated by the black circle. 
$\ga_A$ is the RT surface of the boundary subregion $A$ and the region between $\ga_A$ and $A$ is the bulk dual subregion $\fb_A$ of $A$. The entanglement wedge $\hat \fb_A$ (not shown in figure) is the (bulk) domain of dependence of $\fb_A$.
}
\label{fig:EW}
\end{center}
\end{figure}

For illustration, consider a {\it static} asymptotic AdS spacetime corresponding to a semi-classical state $\ket{\Psi}$. 
$A$ is a boundary subregion lying on a constant time bulk Cauchy slice $\Sig$. Denote $\fb_A$ as the bulk subregion between $A$ and the Ryu-Takayanagi (RT) surface $\ga_A$ on the Cauchy slice. See Fig.~\ref{fig:EW}.  The entanglement wedge of $A$ 
is the bulk domain of dependence of $\fb_A$, which we denote as $\hat \fb_A$. 


Since physics of a region can be fully described by the operator algebra in the region, the entanglement wedge reconstruction can be stated algebraically as
\be \label{enc0}
\sX_A =  \widetilde \sM_{\fb_A} = \widetilde \sM_{\hat \fb_A} , 
\ee
where $\sX_A \subset \sB(\sH_\Psi^{\rm (GNS)})$ denotes the boundary algebra in the region $A$ in the large $N$ limit. 

The definition of $\sX_A$ requires some elaboration.  At finite $N$, we have a von Neumann algebra $\sB_A^{(N)}$ of operators localized in $A$, which acts on $\sH_{\rm CFT}$, and satisfies various properties discussed in Sec.~\ref{sec:qft}. In particular, we have 
\be \label{xr0}
\sB_A^{(N)} =\sB_{\hat A}^{(N)} \ .
\ee
As $N \to \infty$, many operators in $\sB_A^{(N)}$ may not have a sensible limit and drop out of the large $N$ theory, so are not the right objects to consider. We define $\sX_A$ as~\cite{LeuLiu22} 
\bega \label{xr}
\sX_A = \le(\pi_\Psi( \lim_{N \to \infty, \Psi} \sB_A^{(N)})  \ri)''
= \sX_{\hat A}, 
\end{gather} 
where $\lim_{N \to \infty, \Psi} \sB_A^{(N)}$ is the collection of operators in $\sB_A^{(N)}$ that survive the large $N$ limit in 
the semi-classical state $\ket{\Psi}$.\footnote{$\lim_{N \to \infty, \Psi} \sB_A^{(N)}$ is the counterpart of $\sA_\Psi$, introduced in Sec.~\ref{sec:genG}, but restricted to the region $A$.} From~\eqref{xr0}, $\sX_A = \sX_{\tilde A} = \sX_{\hat A}$ where $\tilde A$ is any Cauchy slice of $\hat A$.

There is another algebra that can be naturally associated with $A$ in the large $N$ limit: $\sY_{\hat A}$~(recall~\eqref{Sejne}), which is the von Neumann algebra generated by single-trace operators localized in the domain of dependence $\hat A$ of $A$. 
Since the collection of single-trace operators restricted to $\hat A$ always survives the large $N$ limit, we have 
\be \label{Xsub0}
\sY_{\hat A} \subseteq \sX_A \ .
\ee

The causal wedge reconstruction states that bulk local operators in the causal wedge of $\hat A$ can be expressed in terms of single-trace operators supported in $\hat A$, i.e., elements of $\sY_{\hat A}$. For example, a bulk field $\Phi (X)$ in the causal wedge of $A$ can be expressed as 
\be \label{cauRE0} 
\Phi (X) = \int_{\hat A} d^d x\, \sK_A^{(C)} (X; x) \pi_\Psi(\sO (x))  , \quad X \in \sC_{\hat A}, \; x \in \hat A ,
\ee
where $\sO(x)$ is the boundary single-trace operator dual to $\Phi$, and $\sC_{\hat A} = \left(\tilde J^+ (\hat A) \cap \tilde J^- (\hat A)\right)''$ is the causal wedge of $\hat A$.\footnote{
Here $\tilde J^\pm (Y)$ denotes the causal future/past of a region $Y$ in the conformal completion
of the bulk spacetime, and the double prime denotes the (bulk) causal completion. Our definition corresponds to the causal completion of what is usually referred to as the causal wedge~\cite{HubRan12}.} $\sK^{(C)}_A(X; x)$ is a kernel, which should in general be understood as a distribution and be invertible when $X$ is restricted to a Cauchy slice of $\sC_{\hat A}$.
We can therefore identify
\be\label{CWR}
\widetilde{\sM}_{\sC_{\hat A}} = \sY_{\hat A}  \ .
\ee
Equations~\eqref{ejnn1} and~\eqref{ejnn21} are special cases of~\eqref{CWR}, with $A$ taken to be the full $R$ or $L$ boundary, respectively.

For highly symmetric boundary regions $A$, such as a half-space or a spherical region, the causal wedge reconstruction---and thus~\eqref{CWR}---can be demonstrated explicitly by solving bulk equations of motion with appropriate boundary conditions at 
$\hat A$~\cite{HamKab05,HamKab06,Mor14}. 
For more general regions $A$, it can be established using the time-like tube theorem~\cite{Bor61,Ara63,StrWit23b,Str00}, as shown in~\cite{Wit23}.

From~\eqref{enc0} and~\eqref{CWR}, equation~\eqref{Xsub0} is consistent with the geometric fact that
the causal wedge always lies inside the entanglement wedge~\cite{Wal12,HeaHub14},
\be \label{sbap}
\sC_{\hat A} \subseteq \hat \fb_A   \ .
\ee

\subsubsection{Modular flows and emergent type III$_1$ structure} 

Equation~\eqref{xr} provides an abstract definition that, by itself, does not specify which additional operators---beyond those in $\sY_{\hat A}$---are present, if any. On the gravity side, it is known that the causal wedge is generally a proper subregion of the entanglement wedge, except in special cases such as spherical boundary regions in the vacuum state. Therefore, based on their gravitational identifications, we expect that $\sY_{\hat A}$ is, in general, a proper subset of $\sX_A$.

\begin{figure}[H]
\begin{center}
\includegraphics[width=8cm]{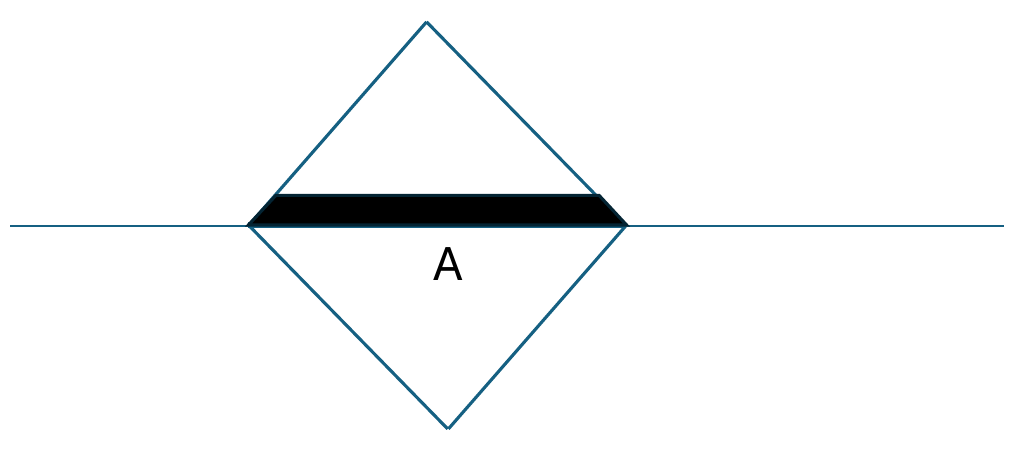}
\end{center}
\caption{\small Consider an infinitesimal time band in $\hat A$ of width $\ep$ around $A$, as shown by the shaded region, and denote the algebra generated by single-trace operators in this band as $\sY_A^\ep$. The semi-classical state under consideration is cyclic and separating for both $\sY_A^\ep \subset \sX_A$ and $\sX_A$. Lemma~\ref{le:erg} of Sec.~\ref{sec:modular} then implies that $\sX_A$ can be generated by acting with modular flows associated with $\sX_A$ on $\sY_A^\ep$, which gives  equation~\eqref{zws1}, where $\sO(\vx)$ should be understood as being smeared within the time band (local operators must be smeared to be well-defined).
}
 \label{fig:ewrR}
\end{figure}

Using Lemma~\ref{le:erg} stated at the end of Sec.~\ref{sec:modular}, we can in fact show that the full entanglement wedge---and therefore $\sX_A$---can be generated from single-trace operators via modular flow. More explicitly, since the entanglement wedge is a bulk local region, the state $\ket{\bid}_\Psi$ is cyclic and separating with respect to the algebra $\widetilde\sM_{\fb_A} = \sX_A$. Let the corresponding modular operator be denoted by $\Delta_{\sX_A}$.\footnote{Note from~\eqref{enc0}, we also have 
$\De_{\sX_A} = \wt \De_{\fb_A}$, where $\wt \De_{\fb_A}$ is the bulk modular operator for $\wt \sM_{\fb_A}$. \label{modID}}
From Lemma~\ref{le:erg}, $\sX_A$ can be generated by applying modular flow to any subalgebra of it for which $\ket{\bid}_\Psi$ is cyclic and separating. One choice is the causal wedge algebra $\sY_{\hat A}$. Alternatively, we may choose a smaller subalgebra $\sY_A^\epsilon$---the single-trace operator algebra supported in an infinitesimal time band of width $\epsilon$ within $\hat A$ centered around $A$, as illustrated in Fig.~\ref{fig:ewrR}. Then $\sX_A$ can be written as
\be \label{zws1}
\sX_A = \text{the algebra generated by} \; \{\sO (s, \vx) \equiv \De^{-is}_{\sX_A} \sO_\ep (\vx) \De_{\sX_A}^{i s}, \;\; \vx \in A, s \in \RR \} ,
\ee
where $\sO(\vx)$ denotes single-trace operators in $A$, 
and $\sO_\ep$ means that it should be smeared within a time band of width $\ep$ around $A$. 
For notational simplicity we have  suppressed $\pi_\Psi$ on the RHS of~\eqref{zws1}. 
Using~\eqref{zws1}, we can write~\eqref{enc0} more explicitly as 
\be \label{JLMS}
\Phi (X) = \int_{-\infty}^\infty ds \! \int_A d \vx \, \sK^{(E)}_A (X; s, \vx) \sO (s; \vx) , \quad X \in \hat \fb_A
\ee 
for some kernel $ \sK^{(E)}_A (X; s, \vx) $ which should be understood as a distribution. Equation~\eqref{JLMS} was first conjectured in~\cite{JafLew15}, with additional support provided in~\cite{FauLew17} based on arguments different from those presented here.

Denote as $\De_{A, \Psi}$ (with $N$-dependence suppressed) the modular operator for $\sB_A^{(N)}$ in state $\ket{\Psi}$. 
The modular flow generated by $\De_{\sX_A}$ can be understood as the large $N$ limit of 
the flow generated by $\De_{A, \Psi}$, i.e., 
\be \label{linMo1}
\pi_\Psi \le(\lim_{N \to \infty, \ket{\Psi}} \De_{A, \Psi}^{-i s} \sO (x)  \De_{A, \Psi}^{i s} \ri) = \De^{-is}_{\sX_A}\pi_\Psi ( \sO(x) ) \De_{\sX_A}^{i s}  \ .
\ee
Equation~\eqref{linMo1} was first conjectured in~\cite{LeuLiu22}, and more support will be given in upcoming work~\cite{BerLiu25}.

Equation~\eqref{linMo1} implies that operators in $\sX_A$ that do not belong to $\sY_{\hat A}$ arise from the large $N$ limit 
of modular-flowed operators. 
From equations~\eqref{zws1} and~\eqref{linMo1}, when the modular flows generated by $\Delta_{A, \Psi}$ are geometric---i.e., they act as pointwise transformations---the set~\eqref{zws1} should coincide with $\sY_{\hat A}$, as such geometric flows simply fill the domain $\hat A$. In this case, we have
\be \label{eqce}
\sX_A = \sY_{\hat A} \quad \text{and} \quad \hat \fb_A = \sC_{\hat A} \ .
\ee
This situation arises when $A$ is a half-space or a spherical region in the vacuum state $\ket{\Omega}$, or when $A$ is the full $R$ boundary in the thermofield double state.
In more general situations, however, modular flows are not geometric, and the inclusion $\sY_{\hat A} \subset \sX_A$ is proper.
 

Now consider a general boundary subregion $A$ in the vacuum state $\ket{\Omega}$, for which $\sY_{\hat A} \subset \sX_A$. The operators in $\sX_A$ but not in $\sY_{\hat A}$---or equivalently, those supported in $\hat \fb_A$ but not in $\sC_A$---can be generated by applying modular flow to single-trace operators in $A$. Since in the vacuum state, $\sS = \sA_\Omega$, these operators must be expressible in terms of single-trace operators, and therefore must involve single-trace operators supported \textit{outside} $\hat A$. That is, the large-$N$ limit of operators in $\sB_A^{(N)}$ may include contributions from single-trace operators outside the domain of dependence $\hat A$. This does not contradict causality, as there are no equations of motion that relate single-trace operators at different times.

The equivalences~\eqref{enc0} and~\eqref{CWR} imply that both $\sX_A$ and $\sY_{\hat A}$ should be type III$_1$. We stress that these type III$_1$ structures have nothing to do with the type III$_1$ of $\sB_A^{(N)}$ at finite $N$. Consider putting the boundary theory on a lattice, in which case $\sB_A^{(N)}$ becomes type I, but $\sX_A$ and $\sY_{\hat A}$ are still type III$_1$. That is, these type III$_1$ structures are emergent in the large $N$ limit. 

\begin{figure}[H]
\begin{center}
\includegraphics[width=4cm]{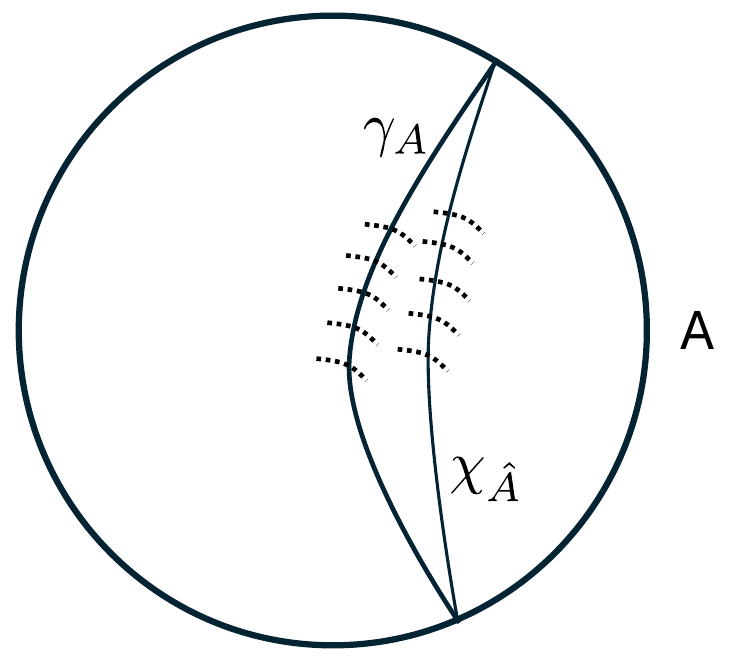}
\caption[]{
\small Shown in the plot is a constant-time bulk Cauchy slice $\Sigma$, with $\chi_{\hat A} = \partial \sC_{\hat A} \cap \Sigma$. 
The infinite long-range entanglement between $A$ and its complement in the large-$N$ limit---responsible for the type III$_1$ nature of $\sX_A$ and $\sY_{\hat A}$---manifests in the bulk as infinite short-range entanglement localized near $\gamma_A$ and $\chi_{\hat A}$, respectively, as depicted in the plot by short dotted lines.
}
\label{fig:bulkEN}
\end{center}
\end{figure}

In the continuum limit, the type III$_1$ nature of $\sB_A^{(N)}$ at finite $N$ arises from an infinite amount of short-range entanglement near $\p A$, stemming from the infinite number of short-distance degrees of freedom present there in the CFT. In the bulk, this is reflected in the infinite proper size of AdS near the boundary. By contrast, the type III$_1$ nature of $\sX_A$ and $\sY_{\hat A}$ originates from an infinite amount of long-range entanglement between $A$ and its complement, which emerges in the $N \to \infty$ limit (as illustrated in Fig.~\ref{fig:newSt}). In the bulk, this infinite entanglement is mirrored by an infinite amount of short-range entanglement localized near $\ga_A$ or the edge of $\sC_{\hat A}$ (denoted $\chi_{\hat A}$), as illustrated in Fig.~\ref{fig:bulkEN}.

The algebraic reformulation~\eqref{enc0} of the entanglement wedge reconstruction also offers an alternative way to 
characterize the RT-surface $\ga_A$ for a boundary region $A$. 
From the discussion of Sec.~\ref{sec:qftS}, near the RT surface $\ga_A$, bulk modular flows associated with $\fb_A$ should act geometrically as local boosts, and $\ga_A$ is the invariant submanifold 
of the local boosts. 
 Given the identification $\De_{\sX_A} = \tilde \De_{\fb_A}$ (see footnote~\ref{modID}), $\ga_A$ can then be understood as an {\it asymptotic invariant submanifold} of the boundary modular flow,\footnote{The word ``asymptotic'' is important: since $\ga_A$ lies  on the boundary of $\fb_A$, operators on $\ga_A$ do not belong to the local subalgebra of $\fb_A$. $\ga_A$ can only be obtained as a limit of points in $\fb_A$.} which can be used as an alternative definition of $\ga_A$. In fact, 
the derivation~\cite{LewMal13} of the RT-formula can be reinterpreted and adapted to show that the bulk asymptotic invariant submanifold of boundary modular flows is indeed the minimal surface anchored on the boundary at $A$~\cite{BerLiu25}. 

\subsubsection{Entanglement and causal wedges for extended gravitational systems}  \label{sec:ewri}

The algebraic formulation of entanglement and causal wedge algebras provides a natural generalization of these concepts that does not rely on geometric descriptions. For instance, instead of specifying a subsystem of the boundary CFT at finite $N$ by a geometric subregion, we may specify it by a von Neumann algebra $\sM$. The corresponding entanglement wedge algebra $\sX_\sM$ can then be defined in direct analogy with~\eqref{xr}. In general, such an $\sX_\sM$ need not admit a bulk dual described by a geometric region. Indeed, even when $\sM$ corresponds to a geometric boundary region $A$, a bulk geometric description of $\sX_A$ does not always exist, when the entanglement between $A$ and its complement is of order $O(N^0)$ (see~\cite{Liu25} for an example). Nevertheless, the algebraic definition always applies.

We may also consider extended gravitational systems in which a gravitational sector with a boundary dual is entangled or coupled to a non-gravitational sector. Denote the gravitational sector (together with its boundary dual) by $B$, and the non-gravitational sector by $R$, so that the full system is $B \cup R$. When focusing only on $R$, we can assume its quantum information is fully accessible. In contrast, for questions involving $B$, two levels of description arise: a fully quantum one, which is not precisely understood, and a semi-classical one. When $B$ is treated semi-classically, consistency requires that the $R$ sector be handled at the same level of approximation. For convenience, we will refer to this regime as the $G_N \to 0$ limit, even though $R$ itself does not possess a gravitational coupling.

Now consider a subsystem $Q$ of the full system $B \cup R$. $Q$ may be a subregion of the boundary CFT, a subsystem of $R$, or {\it a union of both}. Denote the operator algebra of $Q$ at the fully quantum level (i.e., finite $G_N$) by $\sB_Q$, and define its $G_N \to 0$ limit as
\be \label{linRa}
\sX_Q = \lim_{G_N \to 0, \ket{\Psi}} \sB_Q ,
\ee
where $\ket{\Psi}$ is the state under consideration. Equation~\eqref{linRa} should be understood in the same sense as~\eqref{xr}. We interpret $\sX_Q$ as the entanglement wedge algebra of $Q$, whether or not a geometric description exists. Whenever such a description is available, the algebraic and geometric definitions should be regarded as equivalent.

We may also define a causal wedge algebra $\sY_Q$ associated with $Q$, namely the algebra appearing in the low-energy effective description of $Q$. While somewhat abstract in general, this definition becomes concrete in specific settings. For example, consider an evaporating black hole, where the full system consists of the black hole plus its radiation. Taking $Q$ to be the radiation subsystem, $\sY_Q$ corresponds to the algebra of Hawking quanta as described in the low-energy effective theory, while $\sX_Q$ arises as the $G_N \to 0$ limit of the exact radiation algebra.




\subsection{Quantum informational aspects of EWR from algebraic perspective}

In this subsection, we discuss various quantum informational aspects of EWR from the perspective of its algebraic formulation.

\subsubsection{Entanglement wedge nesting and superadditivity of boundary algebras}

An important geometric property of entanglement wedges, which follows from the extremality of the RT/HRT surface $\gamma_A$, is entanglement wedge nesting~\cite{Wal12}: for two boundary regions $A_1 \subseteq A_2$, the corresponding entanglement wedges satisfy $\hat{\fb}_{A_1} \subseteq \hat{\fb}_{A_2}$. This property can be naturally understood from the boundary perspective via the identification~\eqref{enc0}: since $A_1 \subseteq A_2$, it follows by definition that $\sX_{A_1} \subseteq \sX_{A_2}$, which in turn implies $\wt{\sM}_{\hat{\fb}_{A_1}} \subseteq \wt{\sM}_{\hat{\fb}_{A_2}}$.

Entanglement wedge nesting has an immediate corollary: the superadditivity of entanglement wedges. We now elaborate on this property and make its boundary origin more explicit. For simplicity of illustration, we will take the bulk spacetime to be static and the boundary regions to lie on constant-time slices.

\begin{figure}[H]
\begin{center}
\includegraphics[width=10cm]{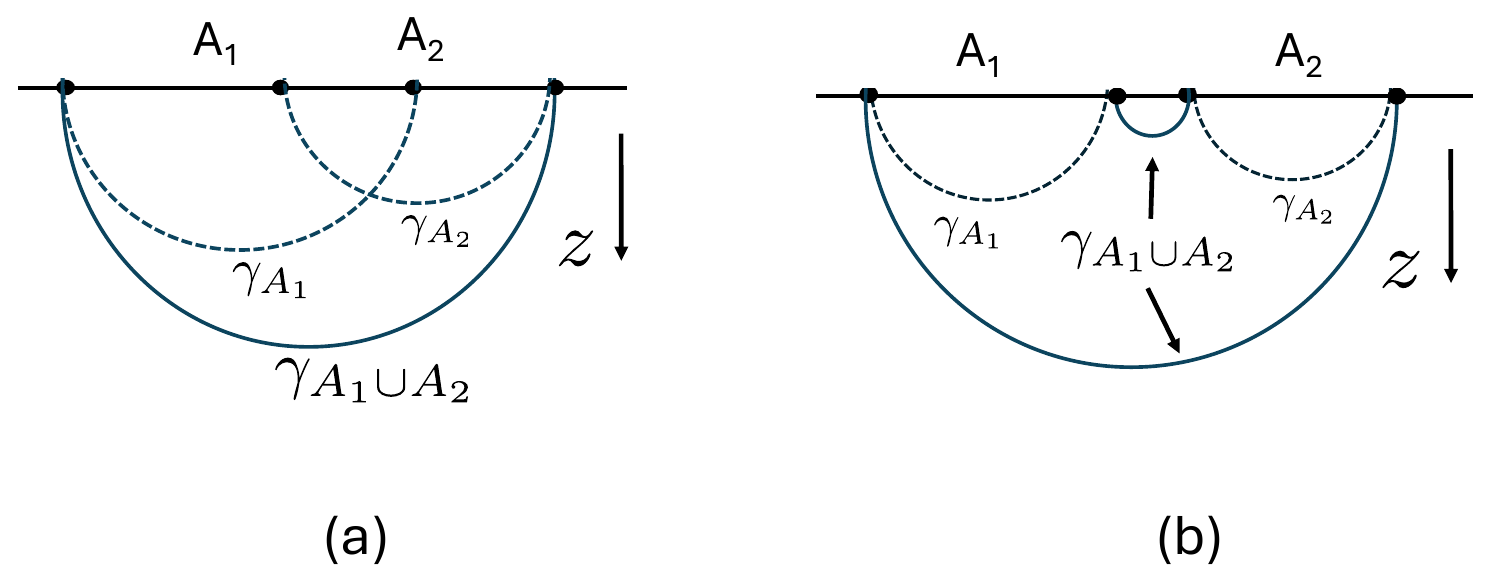}
\caption[]{
\small Examples of superadditivity of entanglement wedges~\eqref{Aoo2} in the AdS$_3$/CFT$_2$ duality, shown on a constant-time bulk Cauchy slice in the Poincare patch: (a) two intersecting intervals $A_1$ and $A_2$; (b) two disjoint intervals $A_1$ and $A_2$ placed sufficiently close together. The RT surfaces  $\gamma_{A_1 \cup A_2}$ are shown as solid lines.
}
\label{fig:addIntBulk}
\end{center}
\end{figure}

Consider two boundary spatial regions $A_1$ and $A_2$ on a constant-time slice. Entanglement wedge nesting implies that 
\be \label{Aoo2}
\fb_{A_1} \cup \fb_{2} \subseteq \fb_{A_1 \cup A_2}, \quad   \fb_{A_1 \cap A_2}  \subseteq \fb_{A_1} \cap \fb_{A_2}  \ .
\ee
Equations~\eqref{Aoo2} imply that the entanglement wedge of the union $A_1 \cup A_2$ is no smaller than the union of the entanglement wedges for $A_1$ and $A_2$. In fact, the inclusions in~\eqref{Aoo2} are generically proper, i.e., the entanglement wedges for boundary subregions are {\it superadditive}. See Fig.~\ref{fig:addIntBulk} for  simple examples.

From the identification~\eqref{enc0}, equations~\eqref{Aoo2} imply that the corresponding boundary subalgebras should be superadditive, i.e.,   
\be \label{hen3}
\sX_{A_1} \lor \sX_{A_2} \subseteq \sX_{A_1 \cup A_2}, \qquad \sX_{A_1 \cap A_2}   \subseteq  \sX_{A_1} \land \sX_{A_2} \ . 
\ee
However, recall from Sec.~\ref{sec:ceRQ} that for a general RQFT---that is, at finite 
$N$---the algebras associated with topologically trivial spatial regions are expected to be additive, namely,
\be \label{add}
\sB_{A_1}^{(N)} \lor \sB_{A_2}^{(N)} = \sB_{A_1 \cup A_2}^{(N)} , \qquad
\sB_{A_1}^{(N)} \land \sB_{A_2}^{(N)} = \sB_{A_1 \cap A_2}^{(N)} \ .
\ee

\begin{figure}[h]
\begin{centering}
\includegraphics[width=10.5cm]{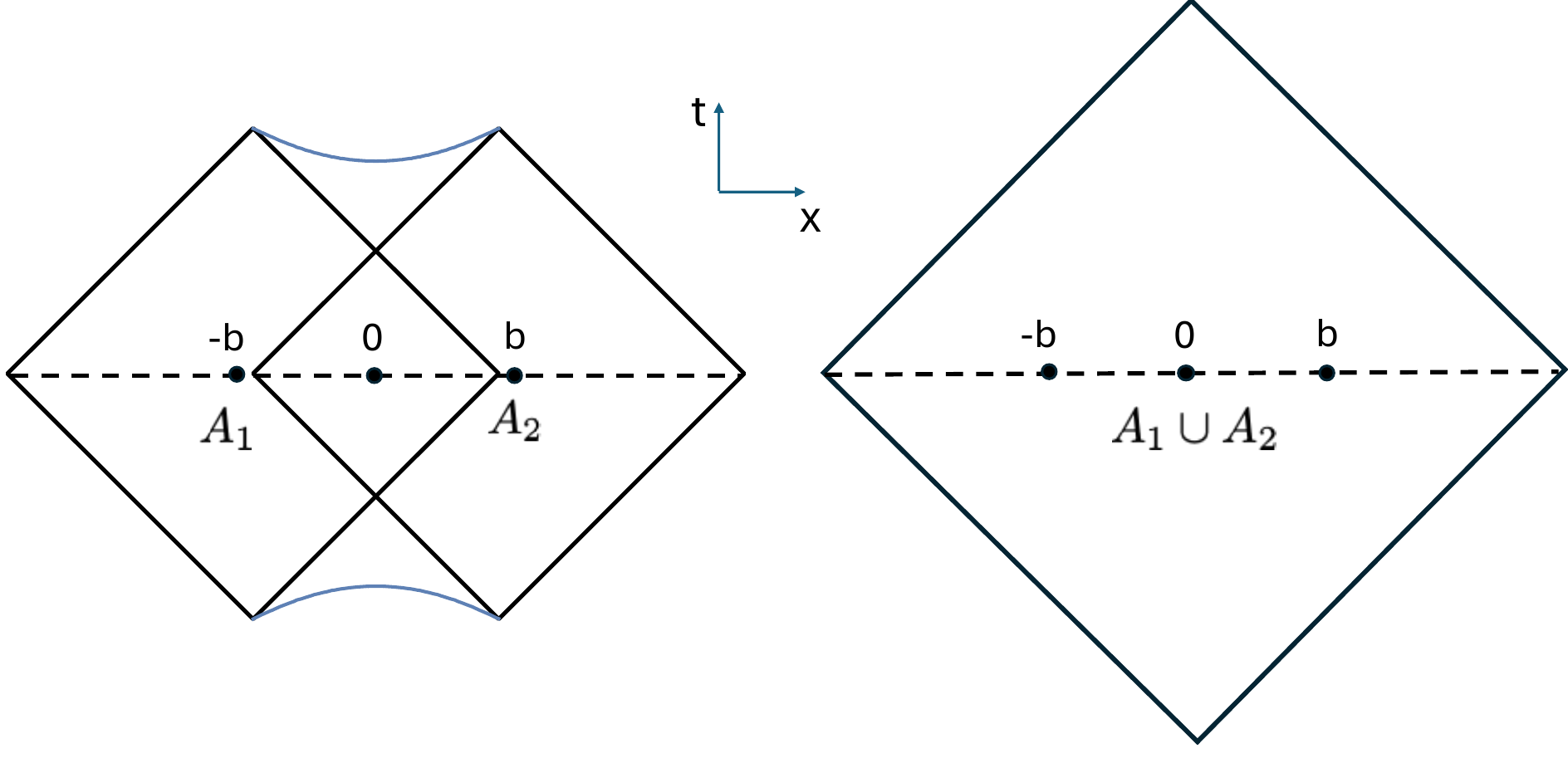}
\end{centering}
\caption[]{\justifying\small Consider the example of Fig.~\ref{fig:addIntBulk}(a), where we take the two intervals $A_{1,2}$ to each have half-width $a$, respectively centered at $x = \pm b$, with $b < a$. 
For an interval $A$,  $\sX_A = \sY_{\hat A}$~(recall~\eqref{eqce}), 
the algebra generated by single-trace operators in $\hat A$. Hence, $\sX_{A_1 \cup A_2} = \sY_{\wh{A_1 \cup A_2}}$, i.e., the algebra generated by single-trace operators in the right plot. Determining  $ \sX_{A_1} \lor \sX_{A_2}$ is nontrivial. Fortunately, it was worked out in an old result of Araki~\cite{Ara63}, and is given by the algebra of single-trace operators in the region shown in the left plot, where the curves at the top and bottom are given by 
     $t = \pm \sqrt{x^2 + a^2 -b^2}$ for $x \in (-b, b)$.
     We thus find $\sX_{A_1} \lor \sX_{A_2} \subset \sX_{A_1\cup A_2}$. It can be further checked that the causal wedge of the region in the left plot is precisely $\widehat{\fb_{A_1} \cup \fb_{A_2}}$~\cite{LeuLiu22}.  
}
\label{fig:twoDiam}
\end{figure}

Thus, for consistency with~\eqref{hen3}, the additivity property~\eqref{add} must be violated in the large $N$ limit. This violation can indeed be demonstrated explicitly on the boundary side for the example of Fig.~\ref{fig:addIntBulk}(a)~\cite{LeuLiu22,LeuLiu24}. See  Fig.~\ref{fig:twoDiam} for illustrations.\footnote{It is worth emphasizing that the field-theoretic discussion in Fig.~\ref{fig:twoDiam} apply to any CFT$_2$ in the large central charge limit, independent of the coupling strength.} Thus, the superadditivity of entanglement wedges originates from, and can be viewed as, a bulk geometric manifestation of the superadditivity of boundary subalgebras in the large 
$N$ limit.


Recall from the discussion of Sec.\ref{sec:qft} that the additivity property~\eqref{add} can be understood from the fact that the algebras $\sB^{(N)}_{A_1}, \sB^{(N)}_{A_2}$ are locally generated. Equations~\eqref{hen3} imply that in the large $N$ limit, the algebras associated with local regions are no longer locally generated. This has important implications for quantum informational properties of the boundary system~\cite{LeuLiu22,LeuLiu24}. For example, it can be leveraged to provide a boundary explanation of, and generalize, the enhanced quantum tasks on the boundary enabled by bulk operations~\cite{May19,MayPen19,May21,MaySor22}.
Below we will discuss its connection to quantum error correction.

At finite $N$, the boundary subalgebra for a region $A$ is also expected to satisfy the Haag duality~\eqref{nal}, i.e., 
\be \label{haagD}
	\sB_{\bar{A}}^{(N)} = (\sB_A^{(N)})'  \ .
\ee
With the assumption that the bulk algebras satisfy the Haag duality, we can deduce that 
\be \label{haags}
 \sX_A'  = \sX_{\bar A}  ,
 \ee
 i.e., the Haag duality survives the large $N$ limit. More explicitly, from the bulk Haag duality~\eqref{haagb1}, 
\be \label{1hassj}
\wt \sM_{\fb_A}' = \wt \sM_{\overline{\fb_A}}  =  \wt \sM_{\fb_{\bar A}}, 
\ee
where in the second equality we have used that for the system in a pure state, $A$ and $\bar A$ share the same RT surface, and thus 
\be 
\overline{\fb_A} = \fb_{\bar A}  \ .
\ee 
Equation~\eqref{haags} then follows from~\eqref{1hassj} upon using the identification~\eqref{enc0}.

In contrast, the algebra of single-trace operators  associated with a spacetime region is by definition additive, but does {\it not} satisfy the Haag duality.

\subsubsection{Explanation of quantum error correcting features} \label{sec:QEC} 

\begin{figure}[H]
\begin{center}
\includegraphics[width=3.5cm]{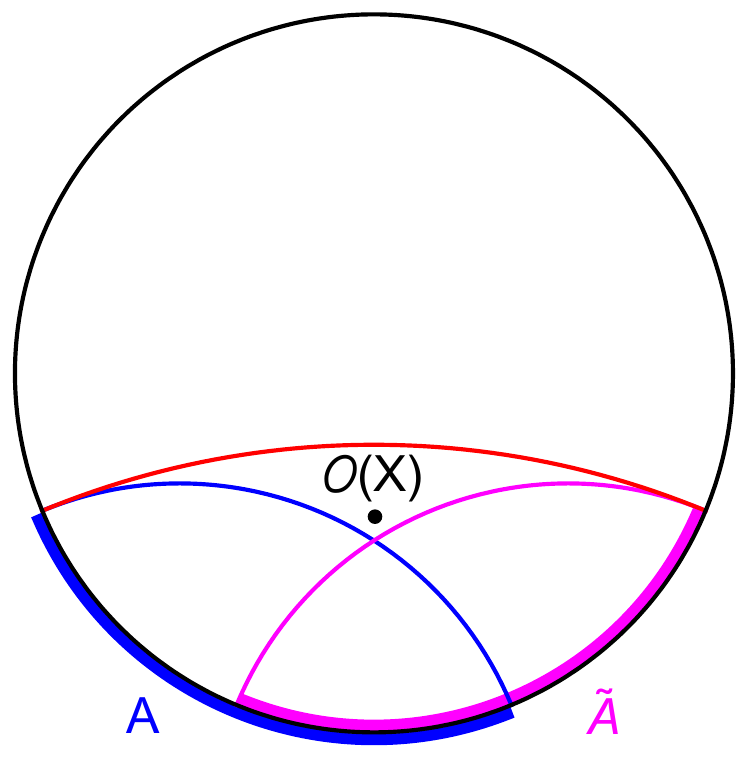} \qquad \qquad \qquad
\includegraphics[width=3.5cm]{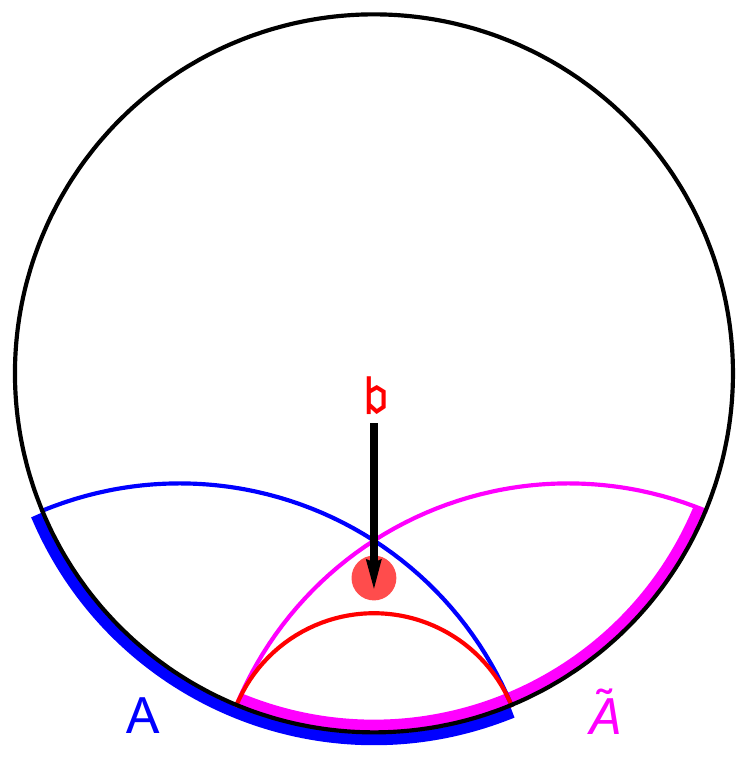}
\caption{\small 
Superadditivity of entanglement wedges~\eqref{Aoo2} can be interpreted in terms of quantum error correcting properties. 
 Here the figure should be understood as representing a single time slice in AdS with the circle denoting the boundary while the interior of the circle is the bulk of AdS.
(a) That $\fb_A \cup \fb_{\tilde A} \subsetneq \fb_{A\cup \tilde A}$ implies that there exist operations, e.g. the operator $O(X)$, that cannot be reconstructed from those in either $A$ or $\tilde A$, but can be reconstructed from those $A \cup \tilde A$.
  (b)  Quantum operations in a local subregion $\fb$ are encoded in the boundary system in a highly redundant way as 
they can be contained in the entanglement wedges of many possible boundary subregions, such as $A$ and $\tilde A$, 
and in principle infinitely many others. Here it is important $\fb$ lies in the intersection of entanglement wedges of $A$ and $\tilde A$, but not in the entanglement wedge of $A \cap \tilde A$. 
}
\label{fig:redu}
\end{center}
\end{figure}

Entanglement wedge reconstruction has been interpreted in terms of quantum error correction~\cite{AlmDon14}. 
Heuristically, the ability to recover all the quantum information in a bulk subregion $\fb_A$ with knowledge of only the subregion $A$ on the boundary implies that the quantum information in $\fb_A$ is robust even if we ``make errors'' in or even completely ``erase''  the complement, $\bar A$, of $A$. Furthermore, superadditivity of entanglement wedges~\eqref{Aoo2} implies that the union of two subsystems $A_1$ and $A_2$ contains more quantum information than the ``combination'' of the respective quantum information for two subsystems, and quantum information in a bulk subregion is ``stored'' in the boundary system in a highly redundant way. See Fig.~\ref{fig:redu} for an illustration. 

 The quantum error correction perspective has provided 
an important guiding principle to construct toy models of holographic duality using systems of finite dimensional Hilbert spaces, such as various types of tensor networks (see e.g.~\cite{PasYos15, HayNez16}),  as well as an interpretation of the RT formula~\cite{Har16}.\footnote{See also~\cite{KanKol18,Fau20, KanKol21, GesKan21, FauLi22} for further development using von Neumann algebras. 
}




The identification of subsystems~\eqref{enc0} in terms of subalgebras, together with the superadditivity of subalgebras~\eqref{hen3} in the large $N$ limit, provides a boundary explanation for the origin of quantum error correcting properties. The quantum error correcting features illustrated in Fig.~\ref{fig:redu} can be attributed to the fact that boundary subalgebras associated with boundary subregions are non-locally generated in the large $N$ limit. 

Building on these considerations, let us now examine a bulk operator $\Phi(X)$ with $X$ lying in region $b$ of Fig.~\ref{fig:redu}(b). 
It can be reconstructed from both regions $A$ and $\tilde A$, but lies outside the entanglement wedge of $A \cap \tilde A$. For definiteness, suppose we are in the vacuum sector, and that for $A$ and $\tilde A$ the entanglement and causal wedges coincide.\footnote{The discussion below can be readily extended to more general situations using~\eqref{JLMS}--\eqref{linMo1}.} From~\eqref{cauRE0}, we then have  
\bega \label{eun000}
\Phi (X) 
= \pi_\Om (\sO_{A} (X)) = \pi_\Om (\sO_{\tilde A} (X)) , \\ 
\sO_{A} (X) \equiv \int_{\hat A} d^{d} x \, K_A^{(C)} \le(X; x\ri)  \sO (x), 
\label{eun001}
\end{gather} 
with $\sO_{\tilde A}(X)$ defined by replacing $A$ with $\tilde A$ in~\eqref{eun001}.
$\sO_A(X)$ and $\sO_{\tilde A}(X)$ are completely different boundary operators, involving distinct boundary degrees of freedom\footnote{We stress that $X$ does not lie in the entanglement wedge of $A \cap \tilde A$.}, even though their representations on the GNS Hilbert space $\sH^{\mathrm{(GNS)}}_\Omega$ coincide. By considering other regions whose entanglement wedge encloses $X$, there is an infinite number of ways to describe $\Phi(X)$, all of which coincide in their representations on $\sH^{\mathrm{(GNS)}}_\Omega$.

This implies that we cannot associate $\Phi(X)$---that is, any bulk degrees of freedom in region $b$---to $A$, to $\tilde A$, or to any boundary region whose entanglement wedge encloses $b$. This further elucidates the emergent nature of bulk degrees of freedom: they are collectively encoded in the boundary system and can, in principle, be reconstructed in infinitely many ways, each involving distinct sets of boundary degrees of freedom.





The algebraic formulation of entanglement wedge reconstruction also clarifies the nature of holographic quantum error correction codes discussed in~\cite{AlmDon14,PasYos15,HayNez16}. In that setup, the Hilbert space describing bulk physics in the semi-classical regime, $\sH_{\rm bulk}$, is embedded into the full CFT Hilbert space as a ``code'' subspace, i.e., $\sH_{\rm bulk} \simeq \sH_{\rm code} \subset \sH_{\rm CFT}$. Since the full boundary CFT Hilbert space is involved, this formulation is only meaningful at finite $N$. However, the bulk Hilbert space $\sH_{\rm bulk}$ in the semi-classical regime can be precisely defined only in the $N \to \infty$ limit.\footnote{As discussed in Sec.~\ref{sec:genG}, $\sH_{\rm bulk}$ should be identified with the GNS Hilbert space on the boundary.}

Thus, holographic quantum error correction codes should be understood not as a literal implementation of the holographic dictionary, but rather as a framework for understanding how semi-classical bulk physics connects to the finite-$N$ quantum regime. While in the original toy models~\cite{AlmDon14,PasYos15,HayNez16}, the codes were isometric and exact, various arguments suggest~\cite{Kel16,Fau20} that realistic codes must be approximate.\footnote{For example, there are tensions between exact quantum error correction and the Reeh-Schlieder theorem. We note that~\cite{HayPen18} also argues for the necessity of approximate codes, though for different reasons: their argument is based on considering ``large'' code spaces that include black hole states with dimension $O(N^2)$---a situation we do not consider here.} A formulation of holographic quantum error correction that can apply to realistic holographic systems (beyond toy models) was proposed in~\cite{FauLi22}. There, the authors introduce asymptotically isometric codes---codes that become isometric only in the $N \to \infty$ limit---within a von Neumann algebraic framework. This asymptotic setup provides a powerful way to relate semi-classical bulk physics to finite-$N$ boundary descriptions for realistic holographic systems.

\subsection{An algebraic formulation of entanglement islands}

Recent significant progress has been made in understanding the emergence of the Page curve for an evaporating black hole within the AdS/CFT duality~\cite{Pen19,AlmEng19}. A striking implication of this discussion is the ``entanglement island'' phenomenon~\cite{Pen19,AlmEng19,AlmMah19a}: specifically, after the Page time $t_P$, the black hole interior becomes part of the radiation system and is referred to as an entanglement island of the radiation. Despite its unintuitive and still mysterious nature, the entanglement island phenomenon appears to be an inevitable consequence of the self-consistency of entanglement wedge reconstruction. In this subsection, we provide an algebraic formulation of entanglement islands through the lens of subregion-subalgebra duality. We present a general definition of an island associated with a subsystem, formulated in terms of operator algebras, which reveals that it is a common feature of holographic systems. Moreover, this algebraic definition provides an intrinsic boundary diagnostic for identifying its presence.

\begin{figure}[H]
\begin{center}
 \includegraphics[width=8cm]{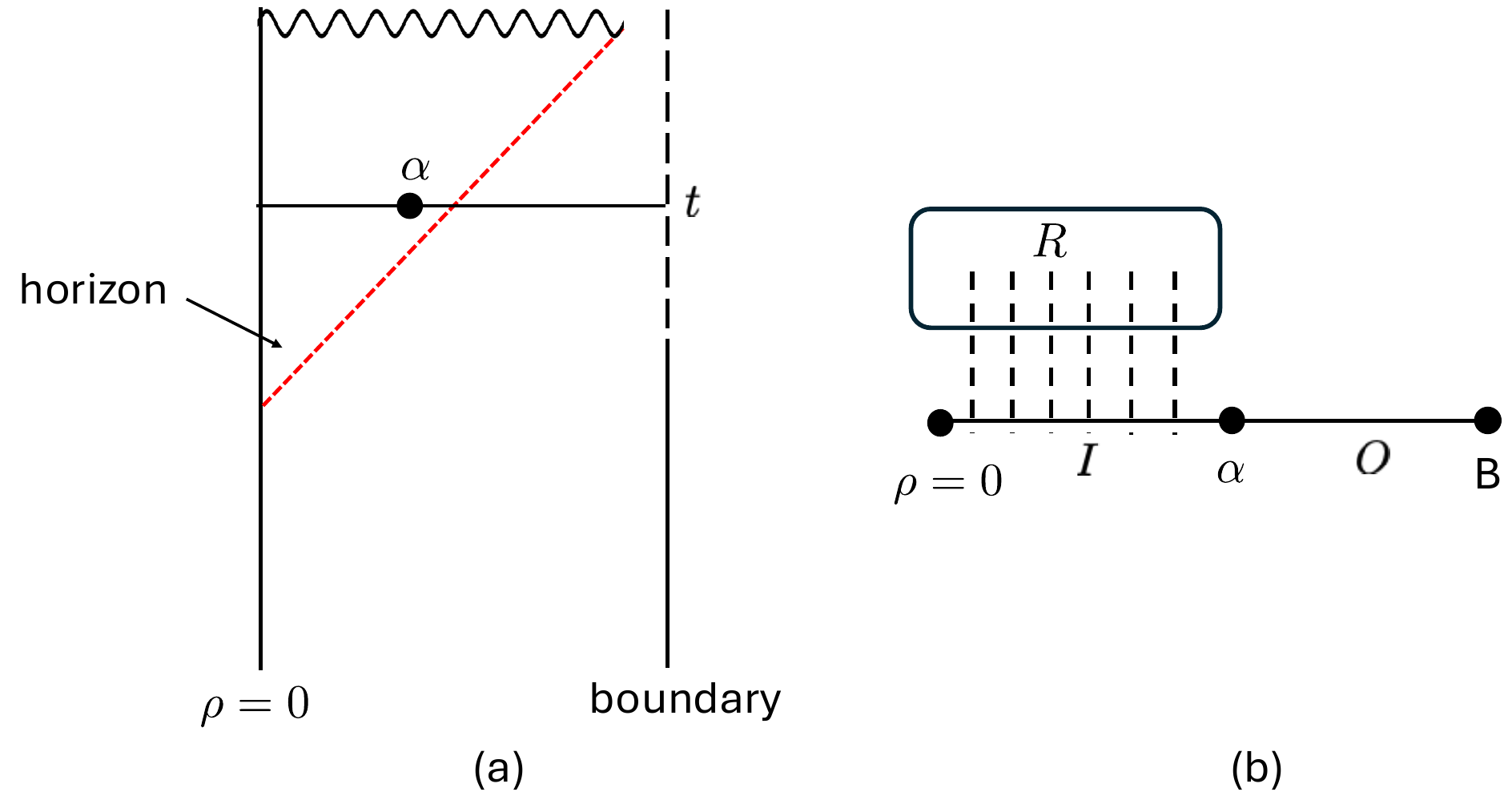}
\caption{\small An evaporating black hole in AdS. (a) The Penrose diagram of an evaporating black hole in AdS, with dashed vertical line denoting transparent boundary where the Hawking radiation can pass through. A Cauchy slice at time $t$ is shown as a horizontal line, with two competing QES: one being the empty set (i.e., at $\rho=0$), and the other, denoted by $\al$, lying slightly inside the horizon. (b) Cartoon of the Cauchy slice shown in (a), with $B$ denoting the boundary. $O$ is the region between $\al$ and the boundary, and $I$ is the region interior to $\al$.
$R$ denotes the radiation system and dashed lines represent entanglement between the black hole interior and the radiation. 
}
\label{fig:EWBH}
\end{center}
\end{figure}

Now consider the example of an evaporating black hole, as illustrated in Fig.~\ref{fig:EWBH}.
The full system separates into the black hole system ($B$) which is described by a boundary CFT and the emitted Hawking radiation ($R$).  Before the Page time $t_P$, the minimal quantum extremal surface (QES) associated with the boundary subsystem $B$ is the empty set, with the entanglement wedge $\hat \fb_B$ of the boundary including a full bulk Cauchy slice, i.e.,  $I \cup O$ in Fig.~\ref{fig:EWBH}. For $t> t_P$, the minimal QES is given by $\al$ in Fig.~\ref{fig:EWBH}, and $\hat \fb_B$ includes only the part of a Cauchy slice exterior to $\al$, i.e., region $O$ in Fig.~\ref{fig:EWBH}.
From entanglement wedge reconstruction, the interior of the black hole is reconstructible from the boundary CFT for $t < t_P$, but not for $t > t_P$.  From~\eqref{enc0}, we have 
\be \label{LinRa2}
\sX_B = \bca \wt \sM_O \lor \wt \sM_I & t < t_P \cr
\wt \sM_O & t > t_P 
\eca ,
 \ee
 where $\wt \sM_O $ and $\wt \sM_I$ denote the operator algebras of bulk fields on a Cauchy slice exterior and interior to the horizon, respectively.  After the Page time ($t > t_P$),  the black hole interior $\wt \sM_I$ is no longer part of $B$. Since $B \cup R$ is the full system, it must then belong to $R$.\footnote{As emphasized below~\eqref{eun000}, the bulk degrees of freedom are collectively encoded in the boundary. The statement that $\widetilde{\sM}_I$ belongs to $R$ should not be understood to mean that the degrees of freedom constituting $\widetilde{\sM}_I$ form a subset of those of $R$.}

By using the entanglement wedge algebra $\sX_R$ and the causal wedge algebra $\sY_R$ defined for non-gravitational systems such as $R$ in Sec.~\ref{sec:ewri}, the inclusion of the black hole interior within $R$ can be expressed in a more precise mathematical language as
\be\label{LinRa1}
\sX_{R} = \bca \sY_{R}   & t < t_P \cr
 \sY_{R}  \lor \wt \sM_{I} & t > t_P 
 \eca ,
 \ee
where the causal wedge algebra $\sY_{R}$ denotes the  algebra associated with the  Hawking quanta in the low-energy effective field theory of the joint gravity-radiation system.  
Equation~\eqref{LinRa1}---compared with~\eqref{LinRa2}---provides a precise way to describe 
the ``transfer'' of the black hole interior from $B$ to $R$ at $t_P$~\cite{EngLiu23} . 

From $\sX_R$ and $\sY_{R} $, we can determine existence of an island from the radiation system alone, by considering 
\be \label{isddef}
\sI_{R}  \equiv \sY_{R} ' \cap \sX_{R}   \ .
\ee
From~\eqref{LinRa1}, $\sI_{R} = \wt \sM_I$ for $t > t_P$. If $\sI_{R}$ is non-empty, there is an island---that is, there exist operators independent of those in the low-energy effective description of the radiation. As in~\eqref{zws1}, the island algebra $\sI_{R}$ can be obtained from $\sY_{R}$ using modular flows associated with $\sX_{R}$.

We can apply the definition of an island~\eqref{isddef} to any subsystem $Q$ of an extended holographic system discussed around~\eqref{linRa} as 
\be \label{islaD0}
\sI_Q = \sY_{Q}' \cap \sX_Q,
\ee
where $\sY_{Q}$ is the ``causal wedge'' algebra of $Q$, i.e., the algebra associated with the low-energy effective description of $Q$. {\it An island for a subsystem $Q$ thus corresponds to the part which survives the semi-classical limit but does not belong to the low-energy description of $Q$ itself.}

\begin{figure}[H]
\begin{center}
\includegraphics[width=4cm]{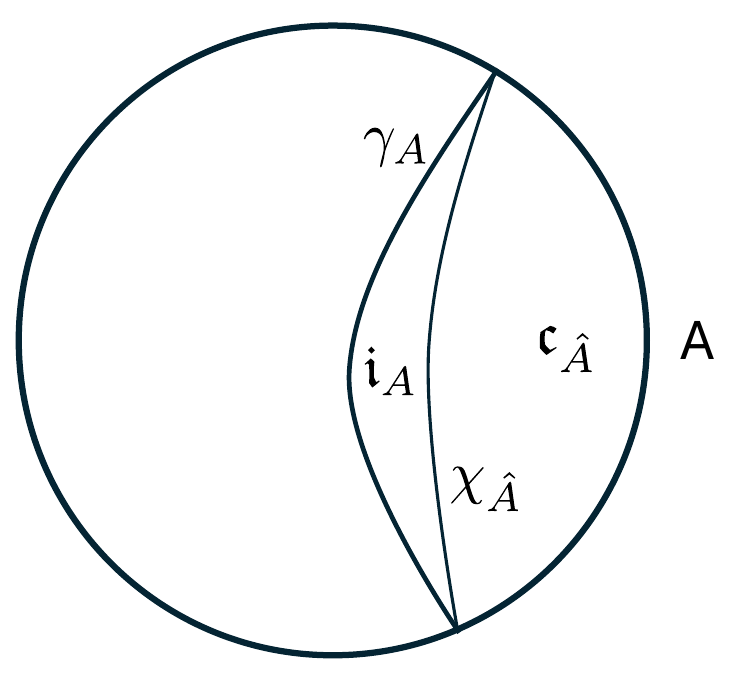}
\caption{\small A constant-time bulk Cauchy slice with $\chi_{\hat A}$ being the edge of the causal wedge, and the RT surface $\chi_A$ being edge of the entanglement wedge. $\fc_{\hat A}$ is the part between $\chi_{\hat A}$ and $A$, and is a Cauchy slice for the causal wedge. 
$\fri_A$ is region between $\chi_{\hat A}$ and $\ga_A$. The regions $\fri_A$ and $\fc_{\hat A}$ are highly entangled through short-distance entanglement near $\chi_{\hat A}$
}
\label{fig:islanD}
\end{center}
\end{figure}

Now consider the case where $Q$ is taken to be a boundary subregion $A$. We can identify $\sY_{Q}$ with $\sY_{\hat A}$, the algebra generated by single-trace operators in $\hat A$. In this case, the island algebra $\sI_A$ consists of operators generated by modular flows from $\sY_{\hat A}$ that do not belong to $\sY_{\hat A}$. In the bulk, these are operators that lie outside the causal wedge of $A$ but within its entanglement wedge.
Geometrically, $\fb_A = \fc_{\hat A} \cup \fri_A$, where $\fc_{\hat A}$ and $\fri_A$ are illustrated in Fig.~\ref{fig:islanD}, and thus
\be \label{intdec0}
\wt \sM_{\fb_A} = \wt \sM_{\fc_{\hat A}} \lor \wt \sM_{\fri_A}  \quad \Ra\quad 
\sX_A = \sY_{\hat A} \lor \wt \sM_{\fri_A}  , \quad \sI_A = \wt \sM_{\fri_A} ,
\ee
which identified the region $\fri_A$ (or equivalently its algebra $\wt \sM_{\fri_A}$) as the island for $A$. 


While identifying region $\fri_A$ as the island for $A$ may appear surprising at first sight, this situation is not fundamentally different from that of the radiation. In both cases, the island part is geometrically separated from the subsystem and is highly entangled with the low-energy part, except that for radiation, the entanglement between $\sY_{R}$ and $\wt \sM_I$ (as illustrated in Fig.~\ref{fig:EWBH}(b)) is non-local, while for a boundary region $A$, the entanglement between $\sY_{\hat A}$ and $\wt \sM_{\fri_A}$ occurs through bulk short-distance entanglement near $\chi_{\hat A}$.

\begin{figure}[H]
    \centering
  \includegraphics[width=4cm]{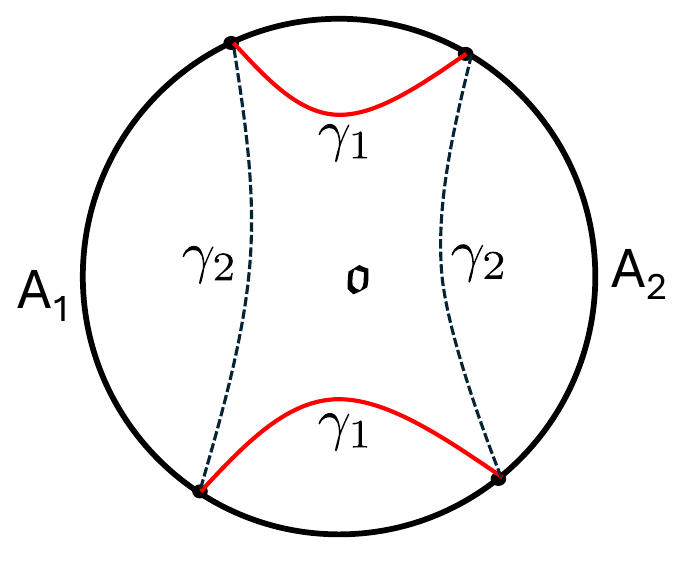}
    \caption{\small Transfer of an island between $A = A_1 \cup A_2$ (consisting of two intervals) and $\bar A$ in the vacuum state of a boundary CFT$_2$, as the size of $A$ is changed. 
 When $|A| = \pi + \ep$ where $|A|$ denotes the length of $A$, the RT surface is given by $\ga_1$, giving rise to a connected entanglement wedge for $A$. In particular, the middle region labeled by $\fo$ can be identified as an island of $A$ from the discussion around~\eqref{intdec0}.  
 For $|A| = \pi -\ep$, the RT surface is given by $\ga_2$. Region $\fo$ is now identified as an island of $\bar A$. 
As we shrink the size of $A$ from $\pi + \ep$ to $\pi - \ep$, the island algebra $\wt \sM_\fo$  is transferred from $\sX_{A}$ to $\sX_{\overline{A}}$. 
    }
    \label{fig:twointervals}
\end{figure}

The transfer of $\wt \sM_I$ at $t_P$ from~\eqref{LinRa2} to~\eqref{LinRa1} can now be interpreted as a transfer of an island between different subsystems. This is, in fact, a common phenomenon. An example is  presented in Fig.~\ref{fig:twointervals}, 
where the subalgebra $\wt \sM_\fo$ of the middle region $\fo$---now interpreted as an island---is transferred from $A$ to $\oA$ when $A$ becomes smaller than $\oA$. In both the black hole evaporation example and that of Fig.~\ref{fig:twointervals}, the transfer occurs when the entropies of a system and its complement become equal, and there is an exchange of dominance of QES. Similar transfer of an island is expected whenever there is such an exchange of dominance~\cite{EngLiu23}.

Note that the semiclassical description of a (sub)system is expected to vary continuously under any continuous change of the system's parameters; therefore, a sharp transition can only occur in the island part. Since the island part of an algebra $\sX_A$ can be generated from the causal wedge algebra $\sY_A$ by modular flows, the transfers of ``islands'' in an evaporating black hole at the Page time and in the example of Fig.~\ref{fig:twointervals} can be attributed to a transition in the properties of modular flows as the state of the system is varied with respect to some parameter.





\subsection{Boundary description of a bulk causal diamond and generalized causal wedge reconstruction} \label{sec:hkll}

In this subsection we discuss the boundary subalgebras dual to bulk diamond-like regions which do not touch the boundary. We start with some simple examples and then give a general theorem which includes a large class of examples. 

\begin{figure}[H]
\begin{center}
\includegraphics[width=6cm]{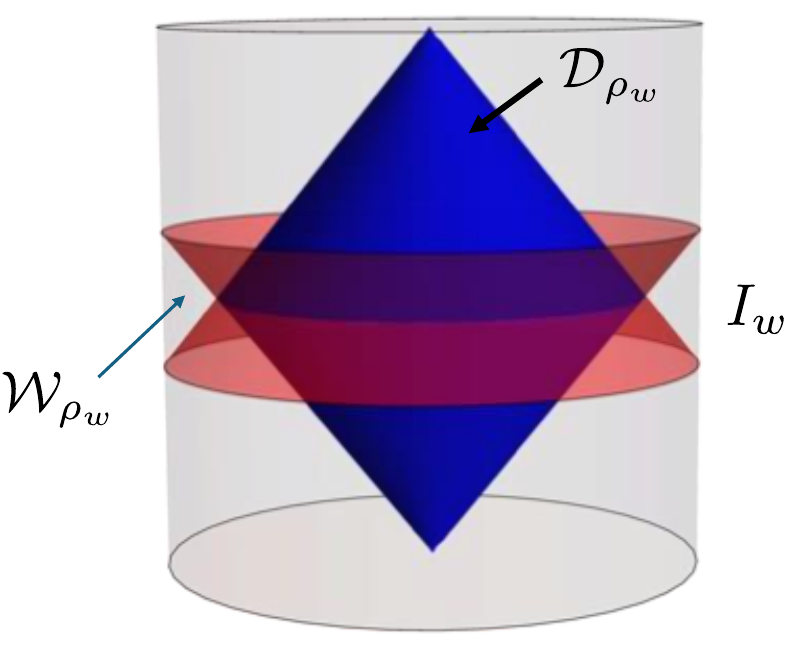}
\caption{\small  $\sW_{\rho_w}$ and $\sD_{\rho_w}$ are, respectively, the bulk duals of the time-band algebra $\sY_{I_w}$ and its commutant $\sY_{I_w}'$ in the vacuum sector.
}
\label{fig:bulkD}
\end{center}
\end{figure}

  Consider the vacuum state $\ket{\Omega}$, which is dual to empty global AdS as given in~\eqref{gadsM}. Let $I_w$ denote a time band on the boundary of width $w < \pi R$, i.e., the spacetime region with $t \in \left(-\frac{w}{2}, \frac{w}{2}\right)$.
 The ``causal wedge'' of this region $\sW_{\rho_w} \equiv \tilde J^+ (I_w) \cap \tilde J^- (I_w)$ 
 is a ``spherical Rindler'' region with  ``radius'' $\rho_w$ given by 
\be \label{timebad}
 \rho_w =R \tan \le({\pi \ov 2} - {w \ov 2R} \ri) , 
 \ee
as illustrated in Fig.~\ref{fig:bulkD}. Note that $\sW_{\rho_w} = \sW_{\rho_w}''$,  is a bulk domain of dependence. 
Extending the causal wedge reconstruction to this case, we find the identification~\cite{BanBry16,LeuLiu22}: 
\be \label{timeBI}
 \wt \sM_{\sW_{\rho_w}} = \sY_{I_w}  \ .
\ee
That is, the bulk spherical Rindler region $\sW_{\rho_w}$ is dual to a boundary time-band algebra $\sY_{I_w}$. 

For $w \geq \pi R$, $\sW_{\rho_w}$ covers an entire bulk Cauchy slice, and 
$\sW_{\rho_w}''$ becomes the full global AdS spacetime. From~\eqref{miF}, $\sY_{I_w} = \sY_B$, the full boundary algebra.   In this case, equation~\eqref{timeBI} is simply replaced by  the last expression of~\eqref{eunS}.

Taking commutant on both sides of~\eqref{timeBI} and using~\eqref{haagb1}, we find 
\be 
\wt \sM_{\sD_{\rho_w}} = \sY_{I_w}' 
\ee
where $\sD_{\rho_w} = \sW_{\rho_w}'$ is the spherical diamond region in the center of global AdS, shown in Fig.~\ref{fig:bulkD}.
We then have the correspondence~\cite{LeuLiu22}: 
\be 
\sD_{\rho_w} \quad \lra \quad  \sY_{I_w}'  \ .
\ee

We find that the boundary algebra describing the bulk diamond region $\sD_{\rho_w}$ does not have a boundary geometric description: it is defined algebraically through the commutant structure of the boundary system. By considering increasing values of $w < \pi R$, we can use $\sY_{I_w}'$ to describe progressively smaller diamonds in the bulk. As $w$ approaches $\pi R$, the diamond can become arbitrarily small, and its local physics should approach that of flat spacetime.

This is rather intuitive: for a larger time band, there are more operators in the algebra $\sY_{I_w}$, and the commutant should become smaller. Physically, as we increase the size $w$ of the time band, physical operations described by operators in $\sY_{I_w}$ can probe longer boundary time scales. The commutant $\sY_{I_w}'$ consists of physical operations that are ``independent'' of those in $\sY_{I_w}$, and may heuristically be interpreted as operators probing even longer time scales than those in $\sY_{I_w}$. It is such operations that correspond to those in the bulk diamond region $\sD_{\rho_w}$.
Such a relation is consistent with, and may be viewed as a precise operator-algebraic realization of, the usual IR/UV connection~\cite{SusWit98,PeePol98}.



\begin{figure}[H]
\begin{center}
\includegraphics[width=11cm]{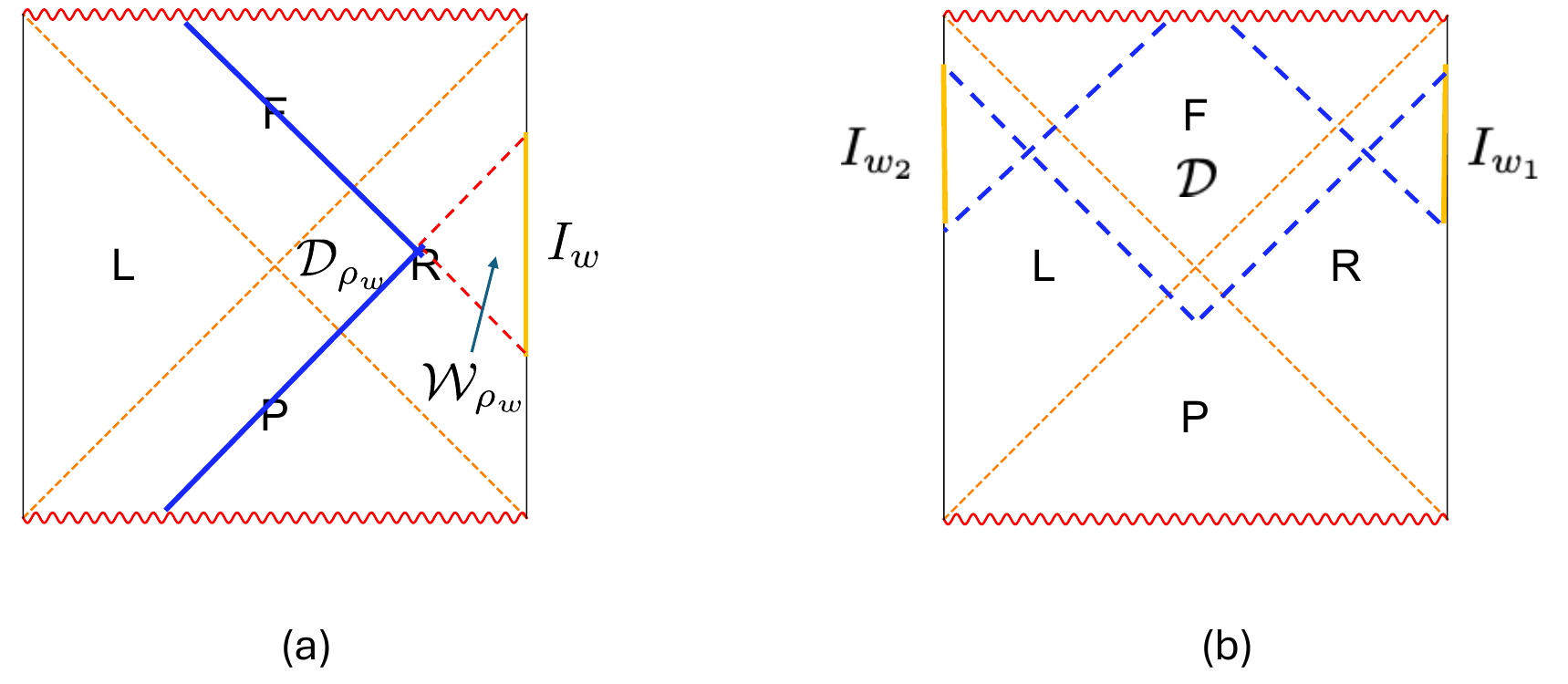}
\caption{\small 
(a) In the plot, $\sW_{\rho_w}$ and $\sD_{\rho_w}$ are, respectively, the bulk duals of the time-band algebra $\sY^{(R)}_{I_w}$ in the thermofield double state and its relative commutant $(\sY^{(R)}_{I_w})' \cap \sY_R$. $\wt \sD_{\rho_w}$ is the bulk region to the left of the two thick blue solid lines, including the $L$ region as well as parts of the $F$ and $P$ regions.
(b) The bulk region $\sD$ (enclosed by dashed blue lines) is dual to the commutant $(\sY_{I_{w_1}}^{(R)} \cup \sY_{I_{w_2}}^{(L)})'$.}
\label{fig:bulkDBH}
\end{center}
\end{figure}

Now consider $Y = I_w^{(R)}$, a time band of width $w$ on the right boundary of the thermofield double state $\ket{\Psi_\b}$ at $T > T_{\rm HP}$, with the bulk metric given by~\eqref{adsBHM}. 
We then have 
\be \label{sphWB}
\wt \sM_{\sW_{\rho_w}} = \sY^{(R)}_{I_w} , \quad w = 2 \int_{\rho_w}^\infty {d \rho \ov f (\rho)}, \quad \forall w > 0 ,
\ee
where $\sW_{\rho_w}$ is the spherical wedge region in the $R$-region of the black hole, shown in Fig.~\ref{fig:bulkDBH}(a).

The existence of a horizon  implies that for any choice of $w$, $\sW_{\rho_w}$ covers only a proper subset of the $t=0$ Cauchy slice of the $R$-region of the black hole, i.e., the $R$-region can never be ``filled up'' by  spherical wedge regions, no matter how large $w$ is.

 Taking the commutant on both sides of~\eqref{sphWB}, we find 
\be 
  \wt \sM_{\wt \sD_{\rho_w} }  = (\sY^{(R)}_{I_w})' , \qquad   \wt \sM_{\sD_{\rho_w} } = (\sY^{(R)}_{I_w})' \cap \sY_R \ .
 \ee
See Fig.~\ref{fig:bulkDBH}(a). 
The geometric statement that $\sW_{\rho_w}$ is a proper subset of the $R$-region for any $w$---equivalently, the existence of the horizon---translates to the algebraic condition 
\be \label{docoN}
(\sY^{(R)}_{I_w})' \cap \sY_R \neq \emptyset, \quad  \forall w \ .
\ee
Thus, the existence of a horizon in the bulk can be diagnosed on the boundary through commutants of time band algebras.
See Fig.~\ref{fig:bulkDBH}(b) for the bulk region dual to the commutant of the algebra associated with the union of time-bands on $L$ and $R$.

Now consider a semi-classical state $\ket{\Psi}$ corresponding to a single-sided black hole geometry formed via a homogenous gravitational collapse (recall Fig.~\ref{fig:ssbh}(b)). On the boundary, the collapse process can be interpreted as the thermalization of $\ket{\Psi}$, with the
black hole describing the macroscopic equilibrium state at late times.\footnote{In the Heisenberg picture, the state does not evolve, but correlation functions of single-trace operators approach those of a thermal state.}

\begin{figure}[H]
\begin{center}
\includegraphics[width=5cm]{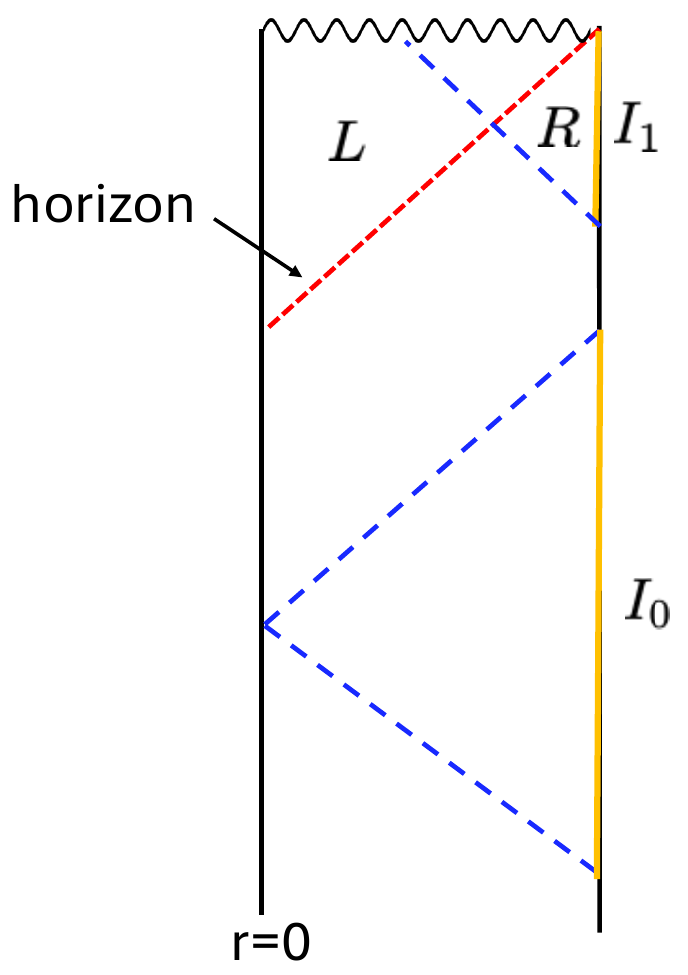}
\caption{\small 
Bulk duals of boundary time bands and their commutants in a single-sided collapsing black hole. For clarity, the collapsing matter is not shown.}
\label{fig:timeBH}
\end{center}
\end{figure}

Consider the time band $I_0$ on the boundary of Fig.~\ref{fig:timeBH}. Since the causal wedge of $I_0$ covers an entire Cauchy slice in the bulk, the causal wedge reconstruction gives 
\be \label{i0bul}
\sY_{I_0} = \sB (\sH_{\rm bulk})
\ee 
where  $\sH_{\rm bulk}$ denotes the bulk Hilbert space. 
The bulk algebra $\wt \sM_R$ for the region $R$ in Fig.~\ref{fig:timeBH} long after the black hole has formed can be identified with the algebra $\sY_{I_1}$ generated by single-trace operators localized in the semi-infinite time band $I_1$ on the boundary. The bulk algebra $\wt \sM_L$ in the region labeled by $L$ can be identified with the commutant $\sY_{I_1}'$.
This gives a thermofield double structure for a single-sided black hole at times long after the black hole has formed.\footnote{$\sY_I'$ provides a precise definition for the ``mirror operators'' envisioned in~\cite{PapRaj12,PapRaj13b}.} 
From~\eqref{i0bul}, $\sY_{I_0}$ is type I, while $\sY_{I_1}$  is a proper subalgebra of $\sY_{I_0}$ and is type III$_1$. In fact we have 
\be \label{sitiM}
\sY_{I_0} = \sY_{I_1} \lor \sY_{I_1}'  = \wt \sM_R \lor \wt \sM_L \ .
\ee


At early times, the algebra of a sufficiently large time band is of type $I$. However, as the system approaches thermal equilibrium, the algebra associated with a time band becomes type III$_1$, regardless of how large the band is (as long as its lower end is fixed). This change in the type of the algebra can be interpreted as a signature of horizon formation in the bulk and thermalization on the boundary (see Sec.~\ref{sec:causalS} for further discussion).

We find that the commutants of time-band algebras provide a powerful framework for describing bulk causally complete regions, including those that do not intersect the boundary. Thus far, our discussion has been limited to uniform time bands---homogeneous in the boundary spatial directions---and to spherically symmetric bulk geometries. 

Extending causal wedge reconstruction to a more general boundary region $Y$ is, however, more intricate. Domains of dependence discussed in Sec.~\ref{sec:ewr} and uniform time bands in spherically symmetric spacetimes are special cases in which the causal wedge intersects the boundary exactly at the corresponding boundary region. This property does not hold in general, owing to the well-known focusing of null geodesics: a null congruence fired from the boundary into the bulk will typically not retrace a congruence fired from the bulk to the boundary, as caustics develop along the null generators.

On the boundary side, this is reflected in the fact that the algebra $\sY_Y$ generated by single-trace operators restricted to a boundary region $Y$ is, in general, not a von Neumann algebra: taking the double commutant $\sY_Y''$ can involve single-trace operators supported on a larger region. For instance, as illustrated in Fig.~\ref{fig:twoDiam}(a), the algebra $\sY_{A_1} \lor \sY_{A_2}$ extends beyond $A_1 \cup A_2$. Moreover, $\sY_Y''$ depends on the chosen GNS sector. For example, in the thermofield double state, only at infinite temperature, does one find $\sY_{A_1} \lor \sY_{A_2} = \sY_{A_1 \cup A_2}$~\cite{LeuLiu24}.

In~\cite{EngLiu25},  a general theorem is established: The algebra $\sY_Y$ of single-trace operators associated with a boundary spacetime region $Y$ admits a standard reconstruction of the causal wedge and is a von Neumann algebra if and only if the following condition is true:  (1) it is causally convex; (2) denoting its generalized causal wedge as
\be \label{dedsC}
\sC_Y = \le(\tilde J^{+}[Y]\cap \tilde J^{-}[Y] \ri)'' ,
\ee
and the boundary manifold as $B$, we have 
\be \label{keycon}
\sC_Y \cap B=Y \ .
\ee
For such a $Y$, it can be shown that the causal wedge reconstruction holds 
\be \label{gcwr}
\sY_Y  = \wt \sM_{\sC_Y} 
\ .
\ee
Taking commutant on both sides of~\eqref{gcwr} gives
\be \label{cdiaD}
\sY_Y'  = \wt \sM_{\sC_Y}' = \wt \sM_{\sC_Y'} ,
\ee
where in the second equality we have used~\eqref{haagb1}. Equation~\eqref{cdiaD} identifies the bulk causal complement  of $\sC_Y$ with the commutant  of $\sY_Y$. 

For a boundary region $Y$ not satisfying the above conditions, we have
\be 
Y \subset Y_{\rm max} \equiv \sC_Y \cap B
\ee
and we expect $\sY_Y'' = \sY_{Y_{\rm max}}$.

The algebra $\sY_{Y}$ in a generalized causal wedge reconstruction~\eqref{gcwr} and its commutant $\sY_{Y}'$ are emergent structures that arise in the large-$N$ limit. At finite $N$, arbitrary products of single-trace operators can, in principle, generate the entire operator algebra. In particular, when $Y$ covers a full boundary Cauchy slice, by the Heisenberg equations of motion, the resulting algebra on any Cauchy slice becomes equivalent to the full algebra $\sB(\sH_{\rm CFT})$. As a result, for any time band, the operator algebra becomes indistinguishable from $\sB(\sH_{\rm CFT})$, whose commutant is trivial.

This emergent nature of $\sY_{Y}$ and $\sY_{Y}'$ for a time-band region $Y$ is consistent with the expectation that sharply defined bulk subregions only exist in the $G_N \to 0$ limit.

In the next subsection, we will argue that $\sY_{Y}$ may nevertheless admit a finite-$N$ extension that remains strictly smaller than $\sB(\sH_{\rm CFT})$ and possesses a nontrivial commutant. Such finite-$N$ extensions of $\sY_{Y}$ and $\sY_{Y}'$ could provide an algebraic framework for defining bulk subsystems in the quantum gravitational regime.

\subsection{Generalized entropy and subregion-subalgebra duality at finite $N$} 
\label{sec:finiteN}

We discussed earlier in Sec.~\ref{sec:higher} that, at any finite order in the $G_N$ perturbation theory, the leading order type III$_1$ characterization of the algebras suffices.  
The story should change significantly when $G_N$ is treated non-perturbatively. For example, the operator algebra of CFT$_R$ in the thermofield double state should become type I and be state-independent. 
Another example is the boundary algebra describing the bulk diamond region $\fb$ in Fig.~\ref{fig:bulkD}; the argument for its existence breaks down at finite $N$ and it is not clear whether an algebra can still be defined.

Treating $N$ non-perturbatively in a precise manner is currently out of reach. However,  the expected behavior of generalized gravitational entropy gives an indication that the 
subregion-subalgebra duality may still be defined at finite $N$~\cite{LeuLiu25,EngLiu25}, despite that geometrically a bulk subregion can no longer be sharply defined due to spacetime fluctuations. See also~\cite{BahBel22,JenRaj24} for other discussions of time band algebras at finite $N$.


Consider two copies of the boundary CFT  in the TFD state, at a temperature above the Hawking-Page transition. 
As discussed in Sec.~\ref{sec:TFD}, the algebra of observables for CFT$_R$ is given by $\sY_R = \wt \sM_R$ in the large $N$ limit, and by $\sB (\sH_R)$ at finite $N$. 
At finite $N$, the algebra $\sB (\sH_R)$ 
has a well-defined entropy $S_R$---the entanglement entropy between CFT$_R$ and CFT$_L$. 
In the large $N$ limit, $S_R$ can be identified with the generalized entropy of the dual black hole, defined as 
\be \label{genE}
S_{\rm gen} \equiv {A_{\rm hor} \ov 4 G_N (\ep)} + S_{\rm bulk} (\ep) \ .
\ee
Here, $A_{\rm hor}$ denotes the area of the horizon, $S_{\rm bulk}(\epsilon)$ is the bulk entanglement entropy of the $R$ region, and $G_N(\epsilon)$ is the bare Newton's constant, with $\epsilon$ representing a short-distance cutoff.



In the $G_N \to 0$ limit,\footnote{Throughout the discussion below, $G_N$ should be understood as the renormalized coupling, while the bare coupling is denoted by $G_N (\ep)$, with explicit dependence on the cutoff parameter $\ep$.} the bulk algebra $\wt \sM_R$ for the $R$-region is type III$_1$, for which $S_{\rm bulk}$ cannot be defined. In practice, one introduces a bulk short-distance cutoff $\ep$ to regularize the algebra, turning it into a type I algebra $\wt \sM_R^\ep$, whose entropy has leading divergence of the form 
\be\label{div1}
S_{\rm bulk} (\ep) = a  {A_{\rm hor}  \over \ep^{d-1}} + \cdots,  \quad \ep \to 0,
\ee
where $a$ is some constant and $\cdots$ denotes less divergent and finite terms. The first term in~\eqref{genE} also exhibits bulk UV divergences through the bare coupling $G_N(\epsilon)$. 
Therefore, in the $G_N \to 0$ limit, where the bulk is described by an effective field theory, neither of the terms on the right-hand side of~\eqref{genE} is individually well-defined. 

However,  there are indications that the UV divergences of the two terms in~\eqref{genE} cancel each other, and the sum is in fact UV finite~\cite{SusUgl94}, i.e, 
\be \label{genE1}
S_{\rm gen} = \lim_{\ep \to 0} \le({A_{\rm hor} \ov 4 G_N (\ep)} + S_{\rm bulk} (\ep)  \ri)
\ee
has a well-defined limit. 
 The cancellation is an indication that the generalized entropy~\eqref{genE} can be extended to finite $G_N$, which is of course required if it is to be identified with the boundary entanglement entropy $S_R$. 
 

Equations~\eqref{genE}--\eqref{genE1} also give a hint on how type III$_1$ algebra $\sY_R$ of the large $N$ limit becomes a type~I algebra $\sB (\sH_R)$ at finite $N$, as follows. We first treat $G_N$ perturbatively, with the short-distance cutoff $\ep$ independent of the renormalized coupling $G_N$, which is implicit in the limit~\eqref{genE1}. 
From the subregion-subalgebra duality, there should exist a type I boundary algebra $\sY^\ep_R$, 
\be \label{sub1}
\wt \sM_R^\ep = \sY_R^\ep \ .
\ee
That is, there exists a regularization on the boundary in the $N \to \infty$ limit that turns $\sY_R$ into a type I algebra $\sY_R^\ep$. While  we do not currently know how to describe the regularization explicitly for a general boundary theory, the duality implies that it should exist.  

Now consider a small but nonzero $G_N$ (i.e., with $G_N$ treated non-perturbatively). We can no longer take $\ep$ all the way to $0$. For $\ep \sim \ell_p$, where $\ell_p$ is the Planck length, quantum gravitational effects should become important and we should interpret~\eqref{genE} in the full quantum gravitational theory. 
In this regime, it is no longer possible to have a clean separation between the $1/G_N$ term in~\eqref{genE} and the term~\eqref{div1} in $S_{\rm bulk} (\ep)$ as ${1\ov G_N} \sim {1 \ov \ell_p^{d-1}} \sim {1 \ov \ep^{d-1}}$.  
 Therefore, now it is more sensible to interpret the total contribution $S_{\rm gen}$ as the entropy for $\wt \sM_R^{\ep} = \sY_R^\ep$. Now recall that at finite $N$, $S_{\rm gen}$ is the entropy of $\sB (\sH_R)$, which leads us to identify 
 \be 
 \wt \sM_R^{\ep} = \sY_R^\ep = \sB (\sH_R), \quad \ep \sim \ell_p \ .
 \ee
 In other words, as $\ep$ is decreased all the way to the Planck length $\ell_p$, we expect that throughout the process, there exists a type I algebra $\wt \sM_R^{\ep} = \sY_R^\ep$, which eventually becomes $\sB (\sH_R)$. 

A similar discussion applies to the entanglement wedge $\hat \fb_A$ of a boundary region $A$. In this case, according to the RT formula~\cite{RyuTak06,HubRan07,FauLew13}, the generalized entropy of the bulk subregion $\fb_A$ can be identified with the boundary entanglement entropy $S_A$\footnote{A boundary short-distance cutoff is always assumed.} of $A$,
\be
S_{\rm gen}(\fb_A) = S_A,
\ee
and therefore should admit a finite $G_N$ extension. The same argument as above suggests that a regularized $\sX_A^\ep$ should become 
$\sB_A^{(N)}$ when $\ep \sim \ell_p$.




The generalized entropy~\eqref{genE} can be extended to a general spacetime region. For example, consider 
the spherical Rindler region $\sW_{\rho_w}$ in Fig.~\ref{fig:bulkD}, which is dual to the time band algebra $\sY_{I_w}$ on the boundary. 
The generalized entropy for $\sW_{\rho_w}$ is defined as 
\be \label{genB} 
S_{\rm gen} [\sW_{\rho_w}] = \lim_{\ep \to 0} \left( {{\rm Area} (\sW_{\rho_w}) \over 4 G_N (\ep)} + \wt S_{\sW_{\rho_w}} (\ep) \right)  \ .
\ee
Here $\wt S_{\sW_{\rho_w}} (\ep)$ is the bulk entanglement entropy between $\sW_{\rho_w}$ and its complement $\sD_{\rho_w}$. As in~\eqref{genE}, while each individual term on the right hand side of~\eqref{genB} suffers from UV divergences, the sum is believed to be finite (see e.g., a review in~\cite{BouFis15a}). 
By using the same argument as above, this can be used as an indication of: 

\ben[(a)] 

\item $S_{\rm gen} [\sW_{\rho_w}] $ can be extended to finite $G_N$.  

\item The algebra $\sY_{I_w}$ admits a finite-$N$ extension to a type I subalgebra $\sB_{I_w}$, whose entropy is given by the generalized entropy $S_{\rm gen}[\sW_{\rho_w}]$. Importantly, $\sB_{I_w}$ cannot coincide with the full boundary algebra $\sB(\sH_{\rm CFT})$, since the latter has zero entropy in the vacuum state.

Similarly, the commutant $\sY_{I_w}'$, which is dual to the diamond region, should also possess a finite-$N$ extension, denoted $\sB_{\sD_{\rho_w}}$.

Note that at finite $N$, the bulk regions $\sW_{\rho_w}$ and $\sD_{\rho_w}$ can no longer be sharply defined due to spacetime fluctuations. Nevertheless, the above discussion suggests that there exist well-defined operator algebras that can be taken as the natural definitions of the corresponding subsystems in the quantum gravitational regime.\footnote{See also discussions in~\cite{ColDon23} from path integral perspectives.}  





\een

\section{Emergence of bulk geometric concepts} \label{sec:Eme}


The subregion-subalgebra duality identifies an arbitrary bulk subregion $\fb$ with a type III$_1$ boundary subalgebra $\sM_\fb$. 
The subalgebra $\sM_\fb$ in general does not have a geometric origin itself, and can be precisely defined only in the large $N$ limit, which reflects that the bulk subregion can only be sharply defined in the perturbative $G_N$ expansion. 
An example is the commutant of the time band algebra $\sY_I$ discussed in Sec.~\ref{sec:hkll}, describing a bulk diamond region that does not touch the boundary. 

As discussed in Sec.~\ref{sec:qft}, to describe a QFT in a curved spacetime, instead of using the action of the QFT with a background metric, we can use the collection of local subalgebras associated with open subregions. 
In contrast to the conventional action-based formulation, the local algebraic description makes manifest local physics, the causal structure, and the entanglement properties of the system---all of which are hidden in the action approach.
In this section, we  illustrate how the subregion-subalgebra duality can be used to understand the emergence of these aspects from properties of boundary algebras.

\subsection{Algebraic characterization of bulk causal structure}\label{sec:causalS}

In this subsection we discuss how  the subregion-subalgebra duality can be used to characterize the bulk causal structure using  boundary algebras, including an algebraic characterization of bulk horizons. 

Boundary operators automatically commute when they are spacelike separated on the boundary. 
However, time-like separated operators may also commute. Their commutation relations depend on the representations of boundary operators on $\sH_\Psi^{(\rm GNS)}$, which are in turn specified by two-point functions of 
boundary operators in $\ket{\Psi}$. The commutant structure of boundary subalgebras can be much richer than what is required of the boundary causal structure. In fact, from the subregion-subalgebra duality, it captures the {\it bulk causal structure}. 

We first describe how the commutant structure of boundary algebras can be used to define a causal depth parameter, which quantifies the ``depth'' of the bulk interior~\cite{GesLiu24}.

\begin{figure}[H]
\centering
\includegraphics[width=10cm]{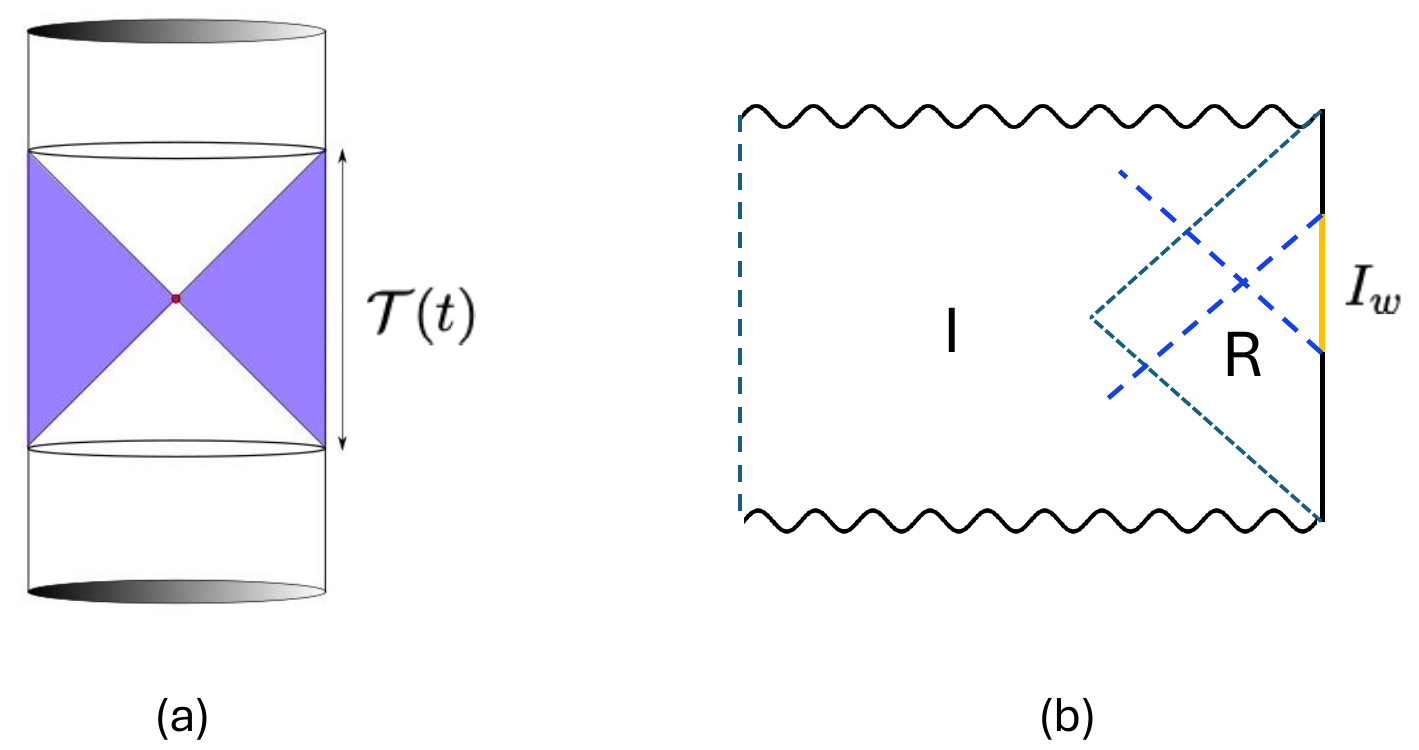}
\caption{\small  
(a) The depth parameter $\mathcal{T} (t)$ can be seen as 
the minimal width of a time band centered around time $t$ such that the ``causal wedge'' of the time band can cover a full Cauchy slice in the bulk. For an asymptotic AdS geometry without a horizon, the $\sT$ is expected to be finite.  See~\cite{Eng16} for development of this line of reasoning. (b) For a single-sided black hole, $\sY_{I_w}$ always has a nontrivial commutant for any finite value of $w$, and thus $\sT (t)= \infty$ for any $t$.
}
\label{fig:depth}
\end{figure}

 
 Recall from the discussion around~\eqref{miF}, in empty AdS, the algebra $\sY_{I_w}$ associated with a time band $I_w$ of width $w$ becomes the full algebra when $w \geq \pi R$. Geometrically, $w = \pi R$ is the minimal width of the band such that light rays from the band can cover a full  bulk Cauchy slice. It can thus be  used to characterize the ``depth'' of the radial direction probed in terms of the causal structure, see Fig.~\ref{fig:depth}(a). Motivated by this, a causal depth parameter can be defined in the boundary system as follows.

\begin{defn}\label{def:h1}
The  causal depth parameter $\mathcal{T} (t)$ of a semi-classical  state $\ket{\Psi}$ at time $t$ is the largest value of $w$ for which the algebra $\sY_{I_w (t)}$ for the time band $I_w (t)= (t-w/2, t +w/2)$ has a nontrivial commutant. 
\end{defn}

For empty AdS, we have $\sT (t) = \pi R$, being  time-independent. For a general asymptotic AdS geometry, $\sT (t)$ is in general time-dependent, but if the bulk geometry does not contain a black hole, we expect $\sT (t)$ to be finite for all time. 

For a single-sided black hole, from Fig.~\ref{fig:depth}(b), $\sT (t) = \infty$ for all time, which provides  an intrinsic boundary characterization of the horizon. By considering  time bands $I_w$ with increasingly larger width $w$ on the boundary, we can probe 
progressively larger subregions of the exterior of the black hole using $\sY_{I_w}$; the full exterior region is accessible  only in the  limit $w \to \infty$.\footnote{In~\cite{NebQi23}, a quantum circuit model following the same line of thought to probe the bulk horizon was constructed.} 

For a single-sided black hole formed from collapse, from Fig.~\ref{fig:timeBH} and the discussion above~\eqref{sitiM}, $\sT (t)$ is finite for any $t$, but increases monotonically to $\infty$ as $t \to \infty$. 
This divergence of $\sT$ as $t \to \infty$ can be considered as a signature of the horizon formation.

Definition~\ref{def:h1} applies to a single-sided geometry. For a two-sided state, we can take the time band $I_w$ to be on one of the boundaries, say the $R$-boundary. We can then define a depth parameter $\sT_R (t)$ for the causal wedge of the $R$-boundary by using the relative commutant $\sY_{I_w}' \cap \sY_R$ of $\sY_{I_w}$ in the single-trace operator algebra $\sY_R$ for the $R$-boundary.

\begin{defn}\label{def:h2} 
The  causal depth parameter $\mathcal{T}_R (t)$ for the right causal wedge of a two-sided semi-classical state $\ket{\Psi}$ at time $t$ is defined as the largest value of $w$ for which the algebra $\sY_{I_w (t)}$, associated with the time band $I_w (t) = (t-w/2, t +w/2)$ on the right boundary, has a nontrivial relative commutant in $\sY_R$. 
\end{defn}

Consider, for example, the thermofield double state. For $T < T_{\rm HP}$, with the bulk geometry given by two disconnected copies of AdS, with the causal wedge of the $R$-boundary given by the right AdS. 
$\sT_R (t)$ simply reduces to that of empty AdS, given by $\pi R$. For $T > T_{\rm HP}$, the bulk geometry is given by the eternal black hole. From Fig.~\ref{fig:bulkDBH}(a) and discussion around~\eqref{docoN}, $\sT = \infty$ for all time, which provides a boundary characterization of the bulk bifurcating horizon.


From the perspective of the boundary CFT,  the depth parameter $\sT$ is the minimal width of a time interval $I_w$ such that the knowledge of $\sY_{I_w}$ is enough to determine the full (single-trace) algebra of the system~(or the $R$-system in the case of a two-sided state).
For example, in the case of empty AdS,  $\sY_{I_w} = \sB (\sH_\Om^{(\rm GNS)})$ for $w > \pi R$, i.e., we already have access to all the operations in the system given such an $\sY_{I_w}$. But in the case of the eternal black hole, it is not possible to have access to the full right algebra $\sY_R$ for any finite $w$. 

For the theory at finite $N$, access to the operator algebra on a single Cauchy slice suffices 
to determine the full algebra. In the large $N$ limit, this property breaks down as there are no equations of motion.
Even though operators on different Cauchy slices cannot be expressed in terms of each other, there are still~(state-dependent) relations among them (encoded in the structure of two-point functions that specifies the generalized free field theory). Such relations make it possible to recover the full algebra by having access to the operator algebra for a time band with a width $w > \sT$. 
For a black hole geometry, 
$\sT=\infty$ implies that the relations among operators at different Cauchy slices are so weak that the full algebra cannot be recovered by having access to operator algebras of arbitrary finite time bands.\footnote{In fact, it is not enough to have access to the algebra of a semi-infinite time band of the form $I = (t_0, \infty)$.}

The causal depth parameter $\mathcal{T}$ can thus also be viewed as quantifying the level of determinism in the boundary theory.  In a theory with equations of motion, there is full determinism and $\sT=0$.  
Whenever $\sT$ is nonzero, there is a loss of determinism, and when it is infinite, the determinism is lost completely. 
Under this interpretation, the emergence of the radial direction is tied to loss of determinism on the boundary, and 
a black hole represents a complete loss of determinism.
 It should also be mentioned that $\sT =0$ by itself does not imply full determinism from a single time slice. 

The causal depth parameter can be determined by the spectral function of the boundary theory, and it is equal to an invariant known in harmonic analysis as the~\textit{exponential type}. For simplicity, we will restrict to the time-translationally invariant case where $\sT$ is time-independent.  Consider the commutator\footnote{In the discussion below, for notational simplicity we will suppress spatial dependence which does not play a significant role.}
\be \label{com1}
 \rho (t-t') \equiv  \vev{\Psi|[\sO (t), \sO (t')]| \Psi} , 
\ee
whose Fourier transform $\rho (\om)$  is called the spectral function. 
Operators in a time-band algebra $\sY_{I}$ can be generated by
\be 
\sO (f_I) = \int dt \, \sO (t) f_I (t), \quad {\rm supp} \, f_I \in I, 
\ee
with ${\rm supp} \, f_I$ denoting the support of function $f_I$. 
The (relative) commutant of $\sY_{I}$ can be shown to be generated by $\sO (g) = \int dt \, \sO (t) g(t)$, for $g$ real-valued,  satisfying  
\be \label{defg} 
[\sO (g), \sO (f_I)] =0 \quad \to \quad (g, f_I) \equiv \int dt dt' \, g(t) \rho(t-t') f_I (t') = 0  \ .
\ee
Equation~\eqref{defg} can also be stated as that $g(t)$ lies in the symplectic complement of the subspace spanned by $f_I$, since $(g,f)$, as defined by~\eqref{defg}, is a symplectic product. Equation~\eqref{defg} can be written in frequency space as
\be 
\int {d \om  \ov 2 \pi} g^* (\om) \rho (\om) f_I (\om) =0  \ .
\ee
Finding $\sT$, then becomes: 


\textbf{Exponential type problem:} Given a measure $\rho(\omega)$, what is the smallest value of $\sT$ for which the Fourier transforms of compactly supported distributions on $I = (-\sT/2, \sT/2)$ become dense in $L^2(\rho)$?

The exponential type problem was first introduced in the work of Kolmogorov and others on chaos in classical dynamical systems (see for example \cite{Kre45}), where they asked for how much time is necessary to observe a system in order to be able to predict its whole evolution. This is closely related to the loss of determinism in a large $N$ system described above. 

The value of the causal depth parameter $\mathcal{T}$ can be obtained  
from the largest value of $a$ for which an \textit{$a$-uniform, or $a$-regular}, sequence can be embedded into the spectral function $\rho (\om)$~\cite{PolT13}.  Roughly speaking,  it is a sequence $\{\lambda_n\}$
such that
\begin{align}\label{eihg}
\rho (\lam_n) \neq 0, \quad \lam_n \approx \frac{n}{a} + c_{\pm}, \quad n \to \pm\infty \ .
 \end{align} 
See Appendix C of~\cite{GesLiu24} for a precise description. We stress that only the behavior at large frequencies matters when deciding whether a sequence is $a$-uniform or not, and thus $\sT$ is  solely encoded in the large-frequency behavior of the boundary spectral function. This may seem to be in tension with the IR/UV connection of the duality, as large-frequency behavior is typically associated with bulk physics near the boundary, whereas 
$\sT$ clearly serves as a parameter probing the interior (or the global structure) of the bulk spacetime.
 We note that,  it is not how $\rho (\om)$ grows or decays with a large $\om$---usually characterized as the UV behavior---that is important here. Rather, it is the average spacing between the points in the support of $\rho$ that is required to determine $\mathcal{T}$.

\subsection{Emergence of Kruskal-like time and causal structure of an eternal black hole} \label{sec:krus}

In the last subsection, we introduced a causal depth parameter $\sT$, which quantifies the depth of the bulk as probed by light rays. In particular, $\sT = \infty$ can be viewed as an indication of the presence of an event horizon. In this subsection, using the eternal black hole as an illustrative example, we show that a much finer understanding of bulk causal structure can be obtained. This includes the emergence of a Kruskal-like time, sharp boundary signatures of the horizon crossing, and probes of the black hole interior~\cite{LeuLiu21a,LeuLiu21b}.
In particular, we will provide a resolution of the meeting-behind-the-horizon puzzle mentioned in Sec.~\ref{sec:TFD}. 

Consider the boundary system in the thermofield double state. In Sec.~\ref{sec:TFD}, we saw that in the large $N$ limit, the boundary algebra for CFT$_R$ becomes type III$_1$ for $T > T_{\rm HP}$, while for $T < T_{\rm HP}$, the algebra is of type I.
These cases correspond, respectively, to the bulk eternal black hole and thermal AdS geometries. Recall Fig.~\ref{fig:tads}. From the bulk perspective, a key difference between the two is that the $L$ and $R$ boundaries are connected in the black hole geometry, but disconnected in thermal AdS.

This connection between the $L$ and $R$ regions in the black hole geometry has important implications for the causal structure: it should be possible for time evolution to take an operator from the $R$ system across the horizon into the $F$ region, where it becomes causally connected to operators in the $L$ system and underlies the ``meeting-behind-the-horizon puzzle'' discussed in Sec.~\ref{sec:TFD}. 
In contrast, for two disconnected copies of AdS, such causal connection through time evolution is not possible.



\begin{figure}[H]
\begin{centering}
	\includegraphics[width=2in]{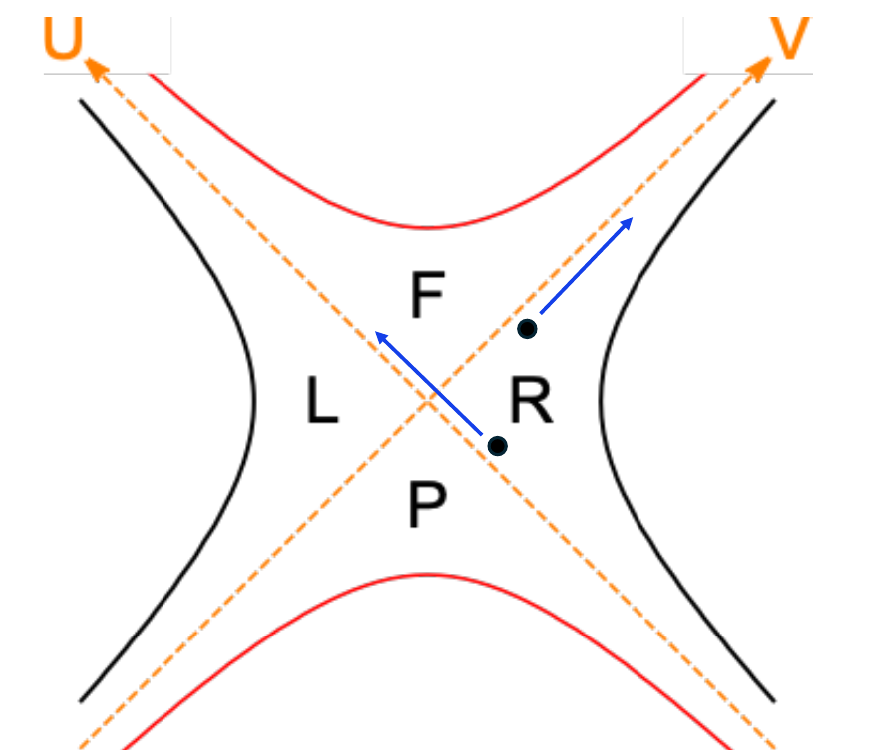}
	\par\end{centering}
\caption{\small Kruskal diagram for an eternal black hole, with the Kruskal time oriented upward and $U,V$ labeling the Kruskal null coordinates. Transverse directions $x_\perp$ are suppressed. 
The dashed lines are event horizons ($U =0$ or $V=0$), the solid red lines are the singularities ($UV = 1$), and the solid black lines are the boundaries~($UV =-1$).
We also show the translation of a point near the past horizon ($V \ll 1$) or the future horizon ($|U| \ll 1$) along the $U$ or $V$ direction, respectively.
}
\label{fig:cauC}
\end{figure}


In the AdS Schwarzschild geometry, there are two distinct types of time, one is the Schwarzschild time, which maps the $R$ ($L$) region to itself, and is an isometry. Another is the Kruskal time which can take a point in the $R$ (or $L$) region to the $F$ or $P$ regions, see Fig.~\ref{fig:cauC}.  Approximating the near-horizon region by a locally flat Minkowski patch, the Kruskal time coincides with the Minkowski time, while the Schwarschild time coincides with the Rindler time. In particular, the Kruskal null coordinates $U, V$ coincide with the light-cone coordinates $x^-, x^+$ of Fig.~\ref{fig:shiftedWedge}.
The Kruskal time is not an isometry, and in the Hartle-Hawking state the black hole behaves as a time-dependent system in terms of its evolution. We will refer to a time evolution that takes the Cauchy slice at $t=0$ to one across the horizon as a Kruskal-like evolution. 


The Schwarzschild time reduces to the boundary time as the boundary is approached and thus can be naturally understood in terms of the boundary. In fact, they can be directly identified based on equivalence of bulk and boundary algebras. 
The Schwarzschild time is the modular time of the bulk algebra $\wt \sM_R$, while in the TFD state, 
from~\eqref{each}, the boundary time coincides with the modular time for $\sY_R$.\footnote{Modular flows act in opposite directions on $\sY_R$ and $\sY_L$  and we take their boundary times to also go in opposite directions.} Given the identification of $\wt \sM_R = \sY_R$, the corresponding modular times are thus identified.

The boundary interpretation of a Kruskal-like time evolution is much more mysterious, as the boundary time approaches $\pm \infty$ at the horizon and thus stops there. There appears to be no time evolution in the boundary that can be identified with it. We will now show that, in the large $N$ limit, the emergent type III$_1$ structure of the boundary algebra $\sY_R$ gives rise to additional boundary times that can be identified in the bulk as Kruskal-like times.

%

More explicitly, in Sec.~\ref{sec:halfs} we discussed that given a \vNa\ $\sM$ with a cyclic and separating vector $\ket{\Psi}$, if there is a subalgebra $\sN \subset \sM$ for which $\ket{\Psi}$ is also cyclic, then there exists a positive operator,
\be \label{krG}
G = K_\sM - K_\sN \geq 0, \quad G \ket{\Psi}=0, 
\ee
which can be used to generate new time flows that leave $\ket{\Psi}$ invariant. 
Furthermore, if $\sN$ satisfies the half-sided modular inclusion condition~\eqref{ghb} (or~\eqref{ghb1}), then $G, K_\sM, K_\sN$ form a closed algebra, and the flow generated by $G$ has additional nice properties discussed in detail in Sec.~\ref{sec:halfs}. 

\begin{figure}[H]

\begin{centering}
	\includegraphics[width=5cm]{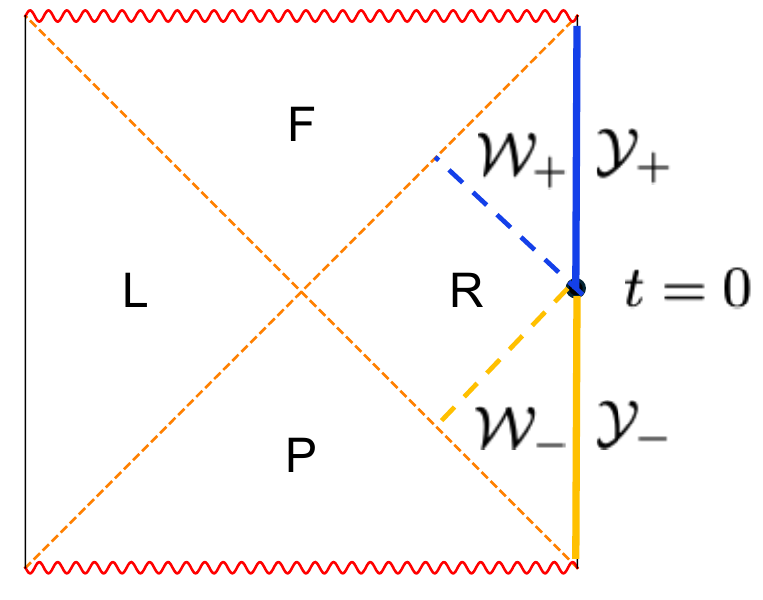}
	\par\end{centering}
\caption{\small 
The algebras $\mathcal{Y}_+$ and $\mathcal{Y}_-$, which are supported on the semi-infinite time intervals $t > 0$ and $t < 0$, are dual to the
regions  $\sW_+$ and $\sW_-$ in the bulk, respectively. 
}
\label{fig:half}
\end{figure}

We can now construct Kruskal-like times from the boundary as follows. Take $\sM$ to be $\sY_R$ with $\ket{\Psi_\b}$ being the cyclic and separating vector. We take $\sN$ to be the subalgebra $\sY_{-}$ associated with the semi-infinite time band $t < 0$ (i.e., 
generated by single-trace operators with $t < 0$). From the discussion of Sec.~\ref{sec:hkll}, $\sY_-$ is dual to the bulk subregion $\sW_-$ indicated in Fig.~\ref{fig:half}. Thus it is also type III$_1$, and $\ket{\Psi_\b}$ is cyclic and separating with respect to it. 

Since the modular flow of $\sY_R$ with respect to $\ket{\Psi_\b}$ is simply time translation, $\sY_-$ satisfies the half-sided modular inclusion condition~\eqref{ghb}. The corresponding operator defined by~\eqref{krG}, which we denote as $G_+$, then generates a new time flow for $\sY_R$. Via the identification $\wt \sM_R = \sY_R$, this also induces a new flow for bulk operators in the $R$-region.

Similarly, we can consider the subalgebra $\sY_+$ associated with the semi-infinite time band $t > 0$ (see Fig.~\ref{fig:half}), with the corresponding operator defined by~\eqref{krG} denoted as $G_-$. The operator $G_-$ generates an another time flow for $\sY_R = \wt \sM_R$.

How do these flows act on a bulk operator? We can in fact get a qualitative picture without doing any explicit calculation. 
Since $\sY_\pm$ are equivalent to the bulk algebras in the regions $\sW_\pm$ of Fig.~\ref{fig:half}, we recover near the horizon---where the $R$-region of the black hole geometry reduces to Rindler spacetime---the situation discussed in Fig.~\ref{fig:shiftedWedge} of Sec.~\ref{sec:halfs}. Therefore, near the horizon, the flows generated by $G_\pm$ should correspond, respectively, to null shifts in $U$ and $V$ (see Fig.~\ref{fig:cauC}), i.e., 
\bega \label{horsh}
e^{i G_+ s} \phi (X) e^{- i G_+ s} = \phi (X_s), \quad X \equiv (U, V, x_\perp), \quad X_s = (U+s, V, x_\perp), \quad V \ll 1,  \\
e^{i G_- s} \phi (X) e^{- i G_- s} = \phi (X_s),  \quad  X_s = (U, V +s , x_\perp), \quad U \ll 1 
 \ .
  \label{horsh1}
\end{gather} 
This shows that flow generated by $G_\pm$ can take an operator localized in the $R$ region to the $F$ and $P$ regions. 
We can also define 
\be \label{spatime}
p = G_- - G_+, \quad h  = G_+ + G_-
\ee
which generate, respectively, horizontal (spatial) and vertical (Kruskal-time) translations in the near-horizon region of Fig.~\ref{fig:cauC}.

This discussion resolves the meeting-behind-the-horizon puzzle: although the system Hamiltonian $H = H_R + H_L$ contains no interaction between the $R$ and $L$ CFTs, in the large-$N$ limit the entanglement structure of the state gives rise to emergent Hamiltonians $G_\pm$. These couple the $R$ and $L$ systems and generate translations that move operators from the $R$ and $L$ regions behind the horizon.

Understanding how $G_\pm$ act explicitly on a bulk local operator at general locations requires more elaborate technical  machinery (see~\cite{LeuLiu21b} for details). Denoting $\Phi (X, s) \equiv e^{i G_+ s} \phi (X) e^{- i G_+ s}$, we 
summarize here the explicit results for the BTZ  black hole in AdS$_3$:

\ben 

\begin{figure}[H]
\begin{centering}
	\includegraphics[width=10cm]{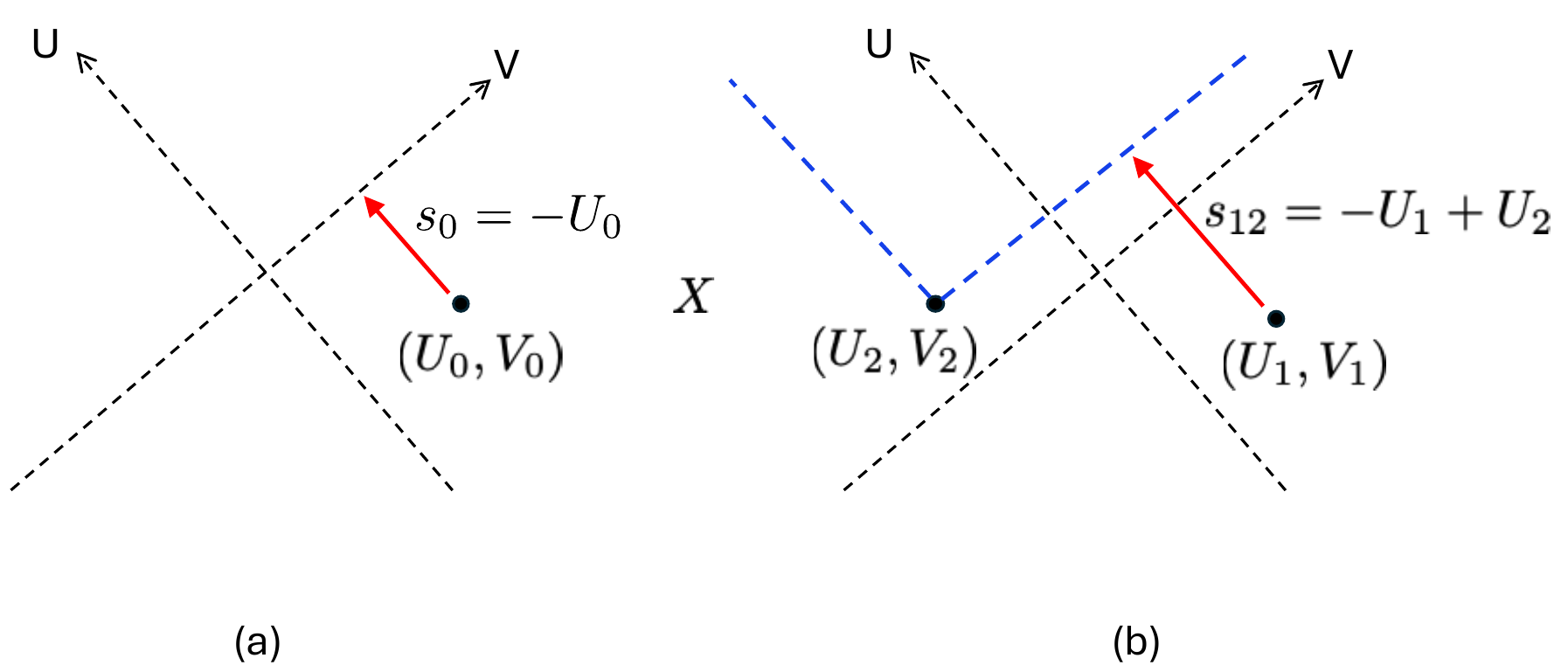}
	\par\end{centering}
\caption{\small The boundary and horizon are suppressed in both Kruskal diagrams for clarity.  (a) Translating $X = (U_0, V_0)$ by $s > s_0 \equiv -U_0$ along the $U$ direction moves the point into the interior of the horizon. There, the operator comes into causal contact with the $L$ region, and consequently $\Phi(X; s)$ must include operators in $\widetilde{\sM}_L$.  (b) Translation by $s > s_{12} \equiv -U_1 + U_2$ along the $U$ direction moves $X_1 = (U_1, V_1)$ into the light cone of $X_2 = (U_2, V_2)$, at which point the commutator~\eqref{ocab} becomes nonzero.
}
\label{fig:shift}
\end{figure}

\item $\Phi (X, s)$ is not a bulk local operator, but may be understood as $\phi$ smeared over a certain spacetime region. 
With $X = (U_0, V_0, x_\perp)$ in terms of Kruskal null coordinates,  we find 
$\Phi (X; s)$ is supported only for $U < U_0 + s$.  In particular, for $s < s_0 \equiv -U_0$, $\Phi(X; s) \in \widetilde{\sM}_R$, while for $s > s_0$, $\Phi(X; s)$ also involves operators in $\widetilde{\sM}_L$. This is exactly what one expects when $\Phi(X)$ crosses the horizon along the $U$ direction (see Fig.~\ref{fig:shift}(a)).


We stress that the existence of a finite $s_0$ at which operators in $\wt \sM_L$ ``suddenly'' appear in $\Phi (X,s)$ is a highly nontrivial feature. In general, if $G_+$ couples $R$ and $L$ operators in a nontrivial way, then the action of $e^{i G_+s}$ will generate operators in $\wt \sM_L$ for {\it any nonzero} $s$.

\item For general points $X_1 \in R$ and $X_2 \in L$,  
we find that 
\be \label{ocab}
[  \Phi (X_1,s) ,  \phi (X_2)] =0
\ee
for $s < s_{12} \equiv -U_1 + U_2$, but the commutator becomes nonzero when $s >s_{12}$, precisely reproducing the casual structure expected from the black hole geometry.  See Fig.~\ref{fig:shift}(b).

We see that while the evolution is nonlocal, it still respects the sharp causal structure.

\begin{figure}[H]
\begin{centering}
	\includegraphics[width=10.0cm]{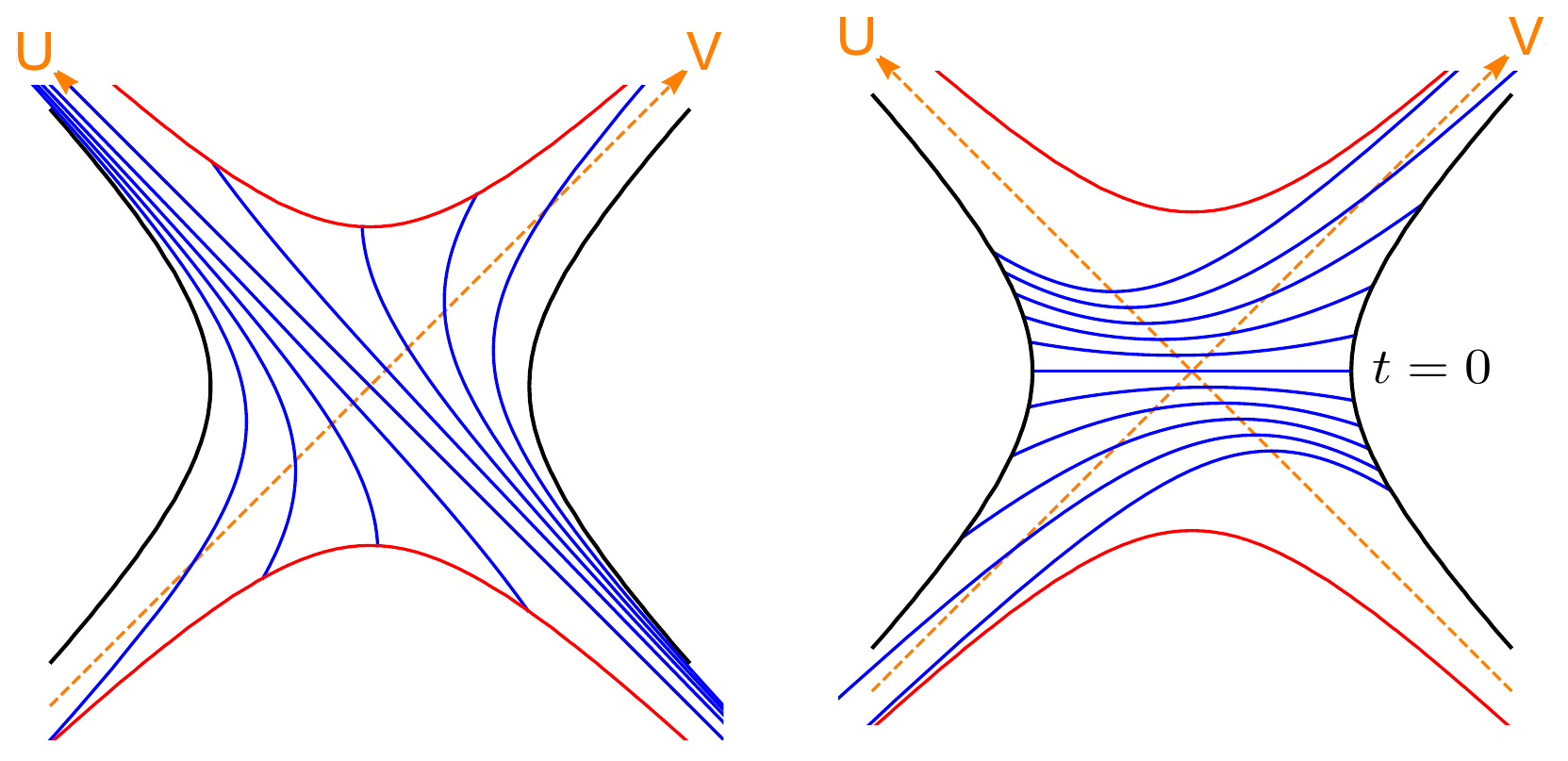}
\par\end{centering}
\caption{\small The left plot gives trajectories of~\eqref{heb}. The right plot gives constant $s$ surfaces evolved from the $t =0$ slice.  
The orange dashed lines are the event horizons, black solid lines are the boundaries, while the red solid lines are the singularities.   }
\label{fig:uEvolvedObserver}
\end{figure}

\item In the large $\Delta$ limit, for operators that are smeared uniformly in the boundary spatial direction,  
 the transformation turns out to be local 
\bega  \label{heb}
\Phi (X,s) =\lam_X  (s) \phi (X_s), \quad X_s = (U_s, V_s), \\ U_s = U_0+ s, \quad
V_s 
= {V_0 \ov 1 - s V_0}  ,   \quad
\lam_X = \sqrt{1- {2 s V_0 \ov 1 - U_0 V_0}  } \ .
\end{gather} 
The trajectories following from~\eqref{heb} are shown in 
Fig.~\ref{fig:uEvolvedObserver}. In particular, for $s \to s_1 \equiv {1-U_0 V_0 \ov 2 V_0}$, $X_s$ approaches the black hole singularity\footnote{More explicitly, at $s=s_1$ we have $U_{s_1} = {1+U_0 V_0 \ov 2 V_0}$, $V_{s_1} = {2 V_0 \ov 1+U_0 V_0 }$
and $U_{s_1} V_{s_1} = 1$.}, where  $\lam_X \to 0$.

\een  

Parallel statements can be made for evolution generated by $G_-$ with the roles of $U$ and $V$ exchanged. 
In the large $\De$ limit, it can be shown that  $G_\pm$ and $K$ form an $SL(2, R)$ algebra~\cite{CuiLiu25}.
From the actions of $G_\pm$, we can also obtain the actions generated by~\eqref{spatime}.
For a black hole in AdS$_2$, the algebra can be constructed explicitly for any operator and shown to be $SL(2, R)$~\cite{LasLeu24}.

For $T < T_{\rm HP}$, the algebra $\sY_R$ is type I, for which there does not exist the half-sided modular inclusion structure. This is consistent with the bulk picture that the two boundaries are disconnected.

\subsection{Emergent spacetime connectivity: algebraic ER=EPR} \label{sec:EREPR}

In the thermofield double state, the bulk dual is given by the eternal black hole for $T > T_{\rm HP}$. For  $T < T_{\rm HP}$, we find two disconnected spacetimes with entangled quantum fields. This has been elevated into a general ER=EPR principle~\cite{MalSus13, Van13, VerVer13a}: any 
entangled quantum gravitational systems are connected by some kind of Einstein-Rosen (ER) bridge. However, there are 
some significant weaknesses in the proposal:

\bi 

\item When $T < T_{HP}$, with left and right systems have entanglement of order $\sO(G_N^0)$, 
the ER ``bridge'' has to be some kind of quantum wormhole. However, no independent definition 
of such a wormhole was given. Therefore, the proposal is more of a slogan than an equivalence in this case. 

\item One may be tempted to refine the ER=EPR proposal by saying that there is an Einstein-Rosen (ER) bridge when there is  
$O(1/G_N)$ entanglement. However, a counterexample was identified in~\cite{EngFol22} in the context of an evaporating black hole, showing that the statement cannot hold in general. In particular, it was pointed out that at some time $t < t_P$, before the Page time, the black hole system is classically disconnected from the radiation, yet there exists $O(1/G_N)$ entanglement between them.

\ei

In Sec.~\ref{sec:TFD}, we saw that classical connectivity of two systems in the thermofield double state  implies that  the operator algebra of each system is a type III$_1$ von Neumann algebra in the large $N$ limit. In other words, classical spacetime connectivity in this context requires a very specific entanglement structure (as characterized by type III$_1$ von Neumann algebras) in the corresponding boundary system. Furthermore, we discussed earlier in Sec.~\ref{sec:krus} that the type III$_1$ structure plays a crucial role for the emergence of an event horizon and associated causal structure.

Motivated by the weakness mentioned earlier and these observations, in~\cite{EngLiu23}, a more refined description of spacetime connectivity, called {\it algebraic ER=EPR}, was proposed, which associates bulk spacetime connectivity/disconnectivity with the operator algebraic structure of a quantum gravity system.\footnote{See~\cite{NogBan21,BanDor22, BanMor23} for a different algebraic perspective.} 
The operator algebraic structure not only includes information on the amount of entanglement, but also more importantly {\it the structure of entanglement}. The proposal gives a definition of quantum wormhole, which, as we will see momentarily, does not include the example of the TFD state at $T < T_{\rm HP}$. 


More explicitly, the algebraic ER=EPR proposal for a two-sided state can be stated as follows~\cite{EngLiu23}. 

{\it 
Consider two entangled systems $R_1$ and $R_2$ in a pure semi-classical state $\ket{\psi_{R_1 R_2}}$ with a gravitational bulk dual spacetime $\sW_{R_1 R_2}$,
in the $G_N \to 0$ limit. Assume that every part of $\sW_{R_1 R_2}$ is connected to at least one of the boundaries.\footnote{This excludes situations involving baby universes, for which the proposal requires further refinement.}
Let $\sM_{R_1}$ and $\sM_{R_2}$ be the operator algebras of $R_1$ and $R_2$ in the large $N$ limit.  Then,  $\sW_{R_1 R_2}$
\begin{enumerate}[I.]
 \item is disconnected if  and only if ${\cal M}_{R_{1}}$ and ${\cal M}_{R_{2}}$ are both type I.  
 	\item has a classical wormhole connecting $R_{1}$ to $R_{2}$ if and only if ${\cal M}_{R_{1}}$ and ${\cal M}_{R_{2}}$ are both type III$_{1}$ and $\sW_{R_1 R_2}$  is classical. 
 	\item  has a quantum wormhole connecting $R_{1}$ to $R_{2}$ 
	if and only if  $\sW_{R_1 R_2}$ is quantum volatile and $\sM_{R_1}, \sM_{R_2}$ are not type I. 
\end{enumerate}}

The proposal uses the notion of a quantum volatile spacetime, which is {\it not} classical in the usual sense even in the $G_N \to 0$ limit. More precisely, we say that a spacetime $\sW_{R_1 R_2}$ is classical\footnote{We assume there exists a family of Cauchy slices with boundary time ranges of order $O(G_N^0)$, and a coordinate basis on each member of this family, in which the metric components are all of order ${\cal O}(G_N^0)$.}  if in the $G_N \to 0$ limit, (i) fluctuations of diffeomorphism invariants go to zero as ${\cal O}(G_{N}^{a})$ for some $a>0$;  (ii) spacetime diffeomorphism invariants such as volumes, areas, and lengths  do not scale as $O(G_{N}^{-a})$ with $a > 0$. A spacetime is  {\it quantum volatile}  
when either or both of the two conditions are violated. Examples of a quantum volatile spacetime include an evaporating black hole  around the Page time where the interior of the black hole has a proper length of order $O(G_N^{-1})$. We will see other examples in Sec.~\ref{sec:diff}.

For the TFD state at $T < T_{\rm HP}$, the boundary algebras are type I, and thus no classical or quantum wormhole connects the two boundaries. Now consider the evaporating black hole illustrated in Fig.~\ref{fig:EWBH}. At times $t < t_P$ (but still of order $O(1/G_N)$), the entanglement wedge of the black hole is quantum-volatile, with an interior of proper length $O(1/G_N)$. In this regime, the system is connected to the radiation by a quantum wormhole. For $t > t_P$, the entanglement wedge of the black hole becomes classical and connects to the radiation through a classical wormhole anchored at the QES $\alpha$.\footnote{The boundary algebra is of type III$_1$ both before and after the Page time. The distinction between the two regimes can be analyzed using the modular depth parameter, which generalizes definitions~\ref{def:h1}--\ref{def:h2} by employing modular time bands~(i.e., intervals defined in modular time). It was argued in~\cite{GesLiu24} that the modular depth parameter of the boundary algebra is finite for $t < t_P$, but diverges for $t > t_P$.}


The  bipartite algebraic ER=EPR can also be generalized to multipartite case, using the bulk dual of canonical purification~\cite{EngWal18}, see~\cite{EngLiu23} for details.  


We stress that the above algebraic ER=EPR proposal applies to the $\apr \to 0$ limit. In the stringy regime, geometric concepts used in the formulation of the proposal do not apply, and type III$_1$ alone is not enough for connectivity (even in the absence of quantum volatility). We will discuss this issue further in Sec.~\ref{sec:stringyH}.


\subsection{Stringy geometry and stringy black holes} \label{sec:stringy}

Our discussions so far have been restricted to the {\it semi-classical} regime, where the bulk is described by the Einstein gravity theory coupled to matter fields. In the example of the $\sN =4$ SYM theory, this requires not only taking $N \to \infty$, but also the limit $\lam \to \infty$, which corresponds to the $\apr \to 0$ limit in the bulk (recall~\eqref{du3}). 
In this subsection, we discuss  how to use the boundary algebras to probe the bulk gravity system in the stringy regime~\cite{GesLiu24}.


\subsubsection{Subregions in the stringy regime and causal structure?} 

On the bulk side, at finite $\apr$, an infinite tower of stringy fields with mass $m \sim {1 \ov \sqrt{\apr}}$ appear, and moreover the number of such fields exhibits Hagedorn growth, i.e. growing with mass as $e^{O(m \sqrt{\apr})}$.  The bulk can no longer be considered as a quantum field theory in a curved spacetime, but rather a classical string theory. It is not precisely known how to generalize the usual geometric concepts such as spacetime subregions and causal structure 
to this regime or whether they can still be defined. 
Bulk stringy fields are dual to single-trace operators on the boundary with scaling dimensions proportional to $\lambda^{1/4}$. At finite $\lambda$, these operators enter the spectrum as dynamical degrees of freedom.

At the leading order in the small $G_N$ expansion\footnote{Equivalently $g_s \to 0$ limit, where $g_s$ is the dimensionless string coupling.}, the bulk theory is still a free theory. In particular, if we consider a specific spacetime field corresponding to a stringy excitation, it should still be described by a free quantum field theory in a curved spacetime. 
Similarly, on the boundary each single-trace operator is still described by a generalized free field, and 
the discussion of operator algebras associated with each single-trace operator goes exactly as that  
given before. 
The boundary (bulk) operator algebra still has the tensor product structure of~\eqref{tesT}. It is just that now the label $i$ needs to go over many more fields.

Starting with the equivalence~\eqref{ssdual} at infinite $\lam$, and extending the corresponding boundary algebra $\sM_b$ 
to finite $\lam$, we can then use~\eqref{ssdual} as an {\it algebraic definition} of the bulk spacetime subregion $\fb$ at finite $\apr$. In particular, the commutant structure of $\sM_\fb$ can be used to {\it define} the causal structure of $\fb$ in the stringy regime. 

Then there appears no difference between the previous Einstein gravity regime and the stringy regime? 
There is, however, an important catch here. From the bulk perspective, what is special about the Einstein gravity 
regime is the universal nature of how matter fields couple to gravity: except in some special situations, matter fields ``see'' the same background geometry, which may be viewed as a version of the equivalence principle. When we include stringy corrections, which at low energies take the form of higher derivative corrections, such universality is broken. For example, consider a bulk scalar field $\phi$, 
\be 
\sL = -\ha g^{\mu \nu} \p_\mu \phi \p_\nu \phi + c_1 R^{\mu \nu} \p_\mu  \phi \p_\nu \phi + \cdots 
\ee
where for illustration we have written only one possible higher derivative corrections explicitly. Different scalars may have different coefficients $c_1$, which means that different scalars ``see'' different effective metrics. 
Similarly, when including higher derivative terms such as the Gauss-Bonnet term to the Einstein action, different components of metric perturbations ``see'' different effective metrics (see~\cite{BriLiu07} for an explicit example).

It is thus possible that, in the stringy regime, the definition of a bulk subregion---and the associated causal structure---may differ from field to field, even though, at leading order in the $G_N$ expansion, these notions may still be sharply defined for each individual field.

\begin{figure}[H]
\centering
\includegraphics[scale=0.3]{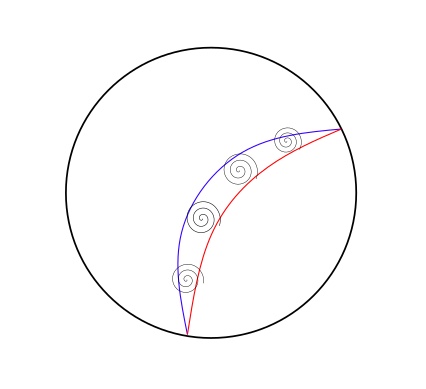}
\caption{\small A cartoon illustrating how the notion of a subregion may break down in the stringy regime. The bulk region corresponding to the subalgebra generated by a boundary generalized free field in $\sM_\fb$ may depend on the field. Two such regions, corresponding to two different generalized free fields, are depicted in blue and red. If these fields interact, the distinction between the corresponding bulk geometries may become blurred.}
\label{fig:blurry}
\end{figure}

This implies that, at finite $\lambda$, while the boundary algebra $\sM_\fb$ continues to define a subsystem, that subsystem may no longer correspond to a single, well-defined bulk subregion. Instead, the subalgebra associated with each generalized field in $\sM_\fb$ may correspond to a different subregion. When interactions between different fields are included (e.g., perturbatively in $G_N$), spacetime and its causal structure will appear ``fuzzy,'' as there is no longer a universal geometric description valid for all fields. See Fig.~\ref{fig:blurry} for an illustration. From the boundary perspective, the equivalence principle in the Einstein gravity regime requires that the boundary algebras associated with different single-trace operators should ``conspire'' to give the same bulk subregion.


The causal depth parameter introduced in Sec.~\ref{sec:causalS} can be used to probe such differences~\cite{GesLiu24}. For each single-trace operator $\sO$, we can define such a causal depth parameter $\sT_\sO (t)$. In the large $\lam$ limit, universality of bulk geometry for all fields means that $\sT_\sO (t)$ should be independent of $\sO$. At finite $\lam$, $\sT_\sO (t)$ may depend on $\sO$, and such dependence can be used to probe the violation of the equivalence principle and possible definition of a stringy geometry.

The above discussion does not imply that subregions cease to exist entirely in the stringy regime. In highly symmetric situations---for example, the boundary theory in the vacuum state, which corresponds to empty AdS---the behavior of two-point functions is governed by conformal symmetry. In such cases, it is possible that various geometric notions, including certain subregions, may still be well-defined.

Another crucial element in the stringy regime is the Hagedorn growth of fields, a feature that significantly impacts the short-distance structure and thermodynamic behavior of the system. The subregion-subalgebra duality formulated earlier was based on the correspondence between individual bulk fields and boundary single-trace operators.  
It is natural to question whether qualitative properties of modular flows as well as the type of algebras could be modified in the presence of Hagedorn growth of density of states.

\subsubsection{Stringy black hole and half-sided modular inclusions} \label{sec:stringyH}

An important question for systems in the stringy regime is whether, despite the absence of a universal geometric bulk description, the notion of a black hole---or a sharp stringy horizon---can still be defined in a universal way across all fields.
Consider, for definiteness, the boundary system (such as the $\sN =4$ SYM theory) in the thermofield double state at a temperature $T > T_{\rm HP}$ at finite $\lam$.\footnote{It can be argued that the Hawking-Page transition is due to the large $N$ limit, and is present at all values of $\lam$~\cite{AhaMar03}.}
Is there a precise way to formulate the notion that  the bulk system is still described by a black hole with a horizon? 




In Sec.~\ref{sec:causalS}, we discussed that in the semi-classical regime, the commutant structure of boundary time band algebras  can be used to diagnose existence of event horizons. 
In particular, for a thermofield double state, existence of the bifurcating horizon is reflected by the causal depth parameter $\sT = \infty$. The definitions~\ref{def:h1} and~\ref{def:h2} of the causal depth parameter can be readily extended to finite $\lam$. Can we still use $\sT=\infty$ as a diagnostic of an event horizon for the bulk system in the stringy regime? Beyond the geometric picture of the semi-classical regime, the situation becomes more intricate.


%

%

In the semi-classical regime, while an event horizon is defined through the causal structure, it has also many other properties. For example, it is a hypersurface of infinite redshift, a property that plays an essential role for the existence of Hawking radiation. In particular, this implies that quantum fields outside the black hole must have spectral support at arbitrarily small frequencies (with respect to Schwarzschild time). In the context of AdS/CFT duality, this further implies that boundary operators in the TFD state must also have spectral support at arbitrarily small frequencies. 
We would like to have a definition for a stringy black hole that still retains this property. 
However, in the stringy regime, it is unclear whether a definition based solely on the commutant structure---used as a proxy for stringy causal structure---can fully capture all aspects of black holes as described in the semi-classical regime.

From the boundary perspective, at $\lam=\infty$, duality with the bulk system implies a ``conspiracy'' that $\sT =\infty$ should come together with spectral support at arbitrarily small frequencies. At finite $\lam$, such conspiracy may no longer happen. 
 Indeed, there exist systems with $\sT = \infty$, but that do not have spectral support at arbitrarily small frequencies.\footnote{As discussed around~\eqref{eihg}, the value of 
$\sT$, including whether it is infinite, depends solely on the asymptotic large-frequency behavior of the spectral function of a boundary operator and is therefore generally independent of its small-frequency behavior.} A simple example is the 
vacuum spectral function of a free scalar field of mass $m$  on $\RR^{1,d-1}$, 
\be \label{frema}
\rho (\om, \vk) = f(-k^2) \th (-k^2 - m^2) (\th (\om)  - \th (-\om)), \quad k^2 \equiv -\om^2 + \vk^2  \ .
\ee
The spectral function has a spectral gap at $\om =0$ for any $\vk$, but has a continuous spectrum beyond the gap. 
The corresponding depth parameter is infinite. 

Therefore, we need a stronger condition for defining a black hole at finite $\lam$. Nevertheless, that $\sT = \infty$ still says something about the bulk string dual. In~\cite{GesLiu24}, it was proposed that it can be used to define a notion of connectedness in the stringy regime: 

\begin{defn}The two sides of the thermofield double state are \textit{stringy-connected} if the algebra $\sY_{(-t_0,t_0)}$ is a strict subalgebra of $\sY_{R}$ for any finite value of $t_0$.\label{def0}\end{defn}

This definition of stringy connectivity is stronger than that given in Sec.~\ref{sec:EREPR} which only requires $\sY_R$ to be type III$_1$.\footnote{Note that here we consider the TFD state above the Hawking-Page temperature, so $\sY_R$ is already type III$_1$.} One can easily find examples of type III$_1$ algebras that have a finite $\sT$ (including $\sT =0$). 

To search for a new definition of a black hole in the stringy regime, note that in the semi-classical regime, an important property associated with a  bifurcation surface, is half-sided modular inclusions along the future or past horizon, see Fig.~\ref{fig:half}. 
The corresponding bulk regions, labelled by $\sW_-$ and $\sW_+$ in the figure, are respectively dual to boundary algebras $\sY_-$ and $\sY_+$ supported on past and future semi-infinite time bands. 
 The existence of a half-sided modular inclusion then requires that the algebra of a semi-infinite time band is a {\it strict subalgebra} 
of $\sY_R$. This a strictly stronger condition than $\sT =\infty$, as it involves semi-infinite intervals and not only finite intervals. While it may not be immediately intuitive, it can be shown that the emergence of a half-sided inclusion ensures that the spectral support of boundary operators~(and the corresponding bulk fields)  
includes arbitrarily small frequencies~\cite{GesLiu24}: 
\begin{Prop}
In a (0+1)-D generalized free field theory at finite temperature carrying a half-sided modular inclusion, the spectral function cannot vanish on any open interval.
\label{prop:hsmi}
\end{Prop}
This  motivates the following definition for the existence of a stringy horizon~\cite{GesLiu24}:
\begin{defn}
There is a stringy horizon in the thermofield double state if the subalgebra $\sY_{(0,\infty)}$ is a strict subalgebra of $\sY_{R}$.
\label{def2}
\end{defn}
Note that, by invariance under time translation and time reversal, this definition implies that $\sY_{(t_0, \infty)}$ is a strict subalgebra of $\sY_R$, and similarly for $\sY_{(-\infty, t_0)}$, for all values of $t_0$. As a result, each inclusion satisfies the conditions for a half-sided modular inclusion in its respective direction. One can further show that the converse of Proposition~\ref{prop:hsmi} is also true~\cite{GesLiu24}: 
\begin{Prop}
If the spectral density $\rho (\om)$  is a continuous function that vanishes only at zero, is differentiable at $0$ with continuous nonzero first derivative, and decays at most polynomially, then $\mathcal{T}=\infty$ and there is a stringy horizon.
\end{Prop}

See also~\cite{LasLeu24} for elaborations on stringy horizon and ergodic hierarchy in the dual generalized free field theory.  

For the $\sN=4$ SYM theory at finite $\lam$ above the Hawking-Page temperature, it has been argued that, a generic boundary operator has spectral support on the full frequency axis~\cite{FesLiu06}, and thus half-sided modular inclusions should exist. 
A bifurcate horizon should then be present for generic bulk fields. However, do horizons for different fields coincide? At the moment, we do not have a direct way to answer this question, and will just make some general remarks. The thermofield double state has a left-right reflection symmetry, which means that the left and right horizons for each field should coincide. Therefore, if there is a notion of stringy geometry for all fields (although the effective metric may be different for different fields), then the horizons should sit in the middle of that spacetime, and thus coincide.

\section{Algebraic Approaches to Quantum-Gravity Regimes} \label{sec:diff}

%


We saw in Sec.~\ref{sec:higher} that once we go beyond leading order in the $G_N$
expansion, bulk operators need to be ``dressed'' to maintain diffeomorphism invariance---a procedure that inevitably introduces nonlocality. In practice, this gravitational dressing is implemented perturbatively by expanding in metric fluctuations around the background.

In this section, we first introduce a simple model in which the dressing is decoupled from the $G_N$
expansion, allowing us to study its effects nonperturbatively even {\it at the leading order of the $G_N \to 0$ limit}. 
The basic idea is to introduce static observers by hand into the system, and to dress bulk observables by attaching them to one of these observers~\cite{ChaLon22}. This simple model can be adapted to a variety of gravitational settings---black holes~\cite{ChaPen22,KudLeu23}, de Sitter space~\cite{ChaLon22,Gom22,Gom23,Gom23a,Gom23c,Wit23b,KolLiu24,KudLeu24}, or arbitrary local regions~\cite{JenSor23,KliLei23}---offering new insights into nonlocality from gravitational constraints, horizon structure, generalized entropies, and the operational meaning of bulk observables.

We will see that the crossed product construction of Sec.~\ref{sec:crossed} emerges naturally from the dressing---and that it often involves taking the crossed product by the modular group.\footnote{See also~\cite{KliLei23a,AliChe24a,AliChe24,AliKli24,FewJan24,DeVEcc24b,DeVEcc24,DeVEcc24a} for more general discussions of crossed product and gravitational constraints.} This procedure converts the original type III algebra into a type II algebra, reflecting a dramatic shift in the entanglement structure introduced by the dressing. In line with the general principle that entanglement encodes bulk geometry, we will see that this algebraic transition induces a modification of the bulk spacetime structure,  yielding the quantum-volatile geometries described in Sec.~\ref{sec:EREPR}.

The Type II structure produced by the crossed product admits a  von Neumann entropy. As explained in Sec.~\ref{sec:crossed}, this Type II entropy corresponds to the relative entropy 
 of the original Type III algebra. In the black-hole setting, that relative entropy can be related to the black hole's generalized entropy. In de Sitter space, it yields an explicit interpretation of the de Sitter horizon entropy as entanglement entropy. Moreover, when applied to a local spacetime region, this construction sheds  light on the meaning of generalized gravitational entropies.

In a fully dynamical theory, observers should not be treated as external static reference systems; rather, they must themselves be dynamical and part of the system.\footnote{See~\cite{ChePen24} for a discussion in which observers arise dynamically.} We then consider a model of dynamical observers in de Sitter space~\cite{KolLiu24}, for which the gravitational dressing becomes more intricate and is no longer described by a crossed product.

We finally turn to Jackiw-Teitelboim (JT) gravity with matter, where the operator algebras of quantum gravity (i.e., with finite $G_N$) can be constructed directly from boundary algebras, without resorting to crossed products or external observers. In this setting, the algebras are found to be of type II~\cite{PenWit23,Kol23,PenWit24,Gao24}. Such low-dimensional models provide rare cases in which the algebraic structure of quantum gravity can be analyzed in detail.






\subsection{A model of gravitational dressing from observers} \label{sec:toy}


Consider a quantum field theory in a certain spacetime with Hilbert space $\sH_Q$. Let $\sM$ be the algebra of a  region on a Cauchy slice and $\sM'$ its commutant, with $\sM \lor \sM' = \sB (\sH_Q)$. Both $\sM$ and $\sM'$ are type III$_1$. Suppose $\ket{\Psi} \in \sH_Q$ is cyclic and separating with respect to $\sM$ and denote $K$ the corresponding modular Hamiltonian. 
As discussed in Sec.~\ref{sec:modular}, $K$ can be used to generate an ``internal'' time flow in $\sM$ (and $\sM'$), 
\be 
A (t) = e^{i K t} A  e^{- i K t}, \quad A' (t) = e^{- i K t} A'  e^{i Kt}, \quad A \in \sM, A' \in \sM , t \in \RR \ .
\ee
The discussion below will not depend on whether $K$ has a geometric description, but in physical examples of interest it often does.

We now introduce a pair of $R$ and $L$ observers~\cite{Wit21b,ChaLon22},\footnote{The introduction of an observer can also be viewed as introducing a quantum reference frame~\cite{FewJan24,DeVEcc24b,DeVEcc24}.} with respective Hamiltonians $\hat q_R$ and $\hat q_L$. For now we will assume that $\hat q_R, \hat q_L$  have spectra $(-\infty, +\infty)$, with conjugates $\hat p_R, \hat p_L$.
The eigenstates 
$\ket{\tau}_R$ and $\ket{\tau}_L$ of $\hat p_R, \hat p_L$ can serve as clock states for the 
$R$ and $L$ observers, respectively---i.e., $\ket{\tau}_R$ represents the 
$R$-observer's clock reading $\tau$. We denote the Hilbert spaces of $R$ and $L$ observers as $\sH_R$ and $\sH_L$, respectively.

We require the full system---the QFT plus $R$ and $L$ observer systems---to be invariant under time translations generated by 
\be \label{Heum}
H = K + \hat q_R - \hat q_L  \ .
\ee
This system can be viewed as a toy model for a time-diffeomorphism-invariant theory, with 
$H$ playing the role of an analogue of the Hamiltonian constraint in general relativity.

Naively, we can construct gauge invariant  states through projection\footnote{The discussion below of physical Hilbert space and dressed operators follows the presentation of~\cite{HoeSmi19,DeVEcc24a}.} 
\bega \label{phy1}
\ket{\psi}_P = \Pi  \ket{\psi} , \quad  \ket{\psi} \in \sH \equiv \sH_R \otimes \sH_L \otimes \sH_Q , \\
\Pi \equiv \int_{-\infty}^\infty dt \, U (t) , \quad  U(t) = e^{- i H t} , \quad t \in \RR \ .
\end{gather} 
However, since the integration range of $t$ is uncompact, such states are not normalizable as 
\be 
\Pi^2 = \int_{-\infty}^\infty dt dt' e^{- i  H(t+t')} = \Pi  \int_{-\infty}^\infty dt  \ .
\ee
We can avoid the problem by regularizing the integration range of $t$. Alternatively, we can suitably modify  
the definitions of gauge invariant states and their inner products as follows. 

The gauge invariant states are defined to be equivalence classes $[\psi]$ where states in an equivalence class satisfying 
\be 
\Pi \ket{\psi_1} = \Pi \ket{\psi_2} = \Pi \ket{\psi}, \quad \ket{\psi_1}, \ket{\psi_2} \in [\psi] \ .
\ee 
We denote the  equivalence relation as $ \ket{\psi_1} \sim  \ket{\psi_2} \sim  \ket{\psi}$. 
Clearly, states related by the action of $e^{i Ht}$ are  equivalent. We define inner products among physical states as 
\be 
(\psi|\phi) \equiv \vev{\psi|\Pi|\phi}  \ .
\ee
which can be readily checked to depend only on the equivalence classes $[\psi]$ and $[\phi]$.

Now given a state $\ket{\psi} \in \sH$, we can always ``gauge fix'' the $L$-system to be in an eigenstate of $\hat p_L$ with eigenvalue to be any desired value $\tau$. That is, 
there exists a state $\ket{\psi_\tau}_{RQ} \in \sH_R \otimes \sH_Q$ such that 
\be \label{gxLC}
\ket{\psi} \sim \ket{\tau}_L  \otimes \ket{\psi_\tau}_{RQ} , \quad \text{i.e.,} \quad
\Pi \ket{\psi} = \Pi  \le(\ket{\tau}_L\otimes \ket{\psi_\tau}_{RQ} \ri)
\ .
\ee
To see this, expand $\ket{\psi}$ as 
\be
\ket{\psi} = \int d \tau  d \tau' \, \sum_n \psi_n (\tau, \tau') \ket{\tau}_R \otimes \ket{\tau'}_L \otimes \ket{n} ,
\ee
where $\{\ket{n}\}$ denotes a complete basis for $\sH_Q$. From the action of $\Pi$, we can readily verify that~\eqref{gxLC} is satisfied for 
\be \label{nye1}
\ket{\psi_\tau}_{RQ} = {_L} \vev{\tau|\Pi |\psi}  \in \sH_R \otimes \sH_Q  \ .
\ee
$\ket{\psi_\tau}_{RQ}$ as defined in the above equation depends only on the equivalence class of $\ket{\psi}$ and we can view~\eqref{nye1} as a map from physical Hilbert space $\sH_{\rm phys} $ to $ \sH_R  \otimes \sH_Q$. It can be shown that it is an isometry.  More explicitly, given two states $\ket{\psi}, \ket{\phi}$, 
\bega 
(\phi|\psi) = \vev{\psi|\Pi|\psi} =  \vev{\phi|\Pi|\tau}_L  {_L} \vev{\tau|\Pi|\psi}
= {_{RQ}}\vev{\phi_\tau|\psi_\tau}_{RQ},
\end{gather} 
where in the second equality we have used the second equation of~\eqref{gxLC} and~\eqref{nye1}. 

In the gauge~\eqref{gxLC}, $\hat q_L$ is no longer dynamical. We can set~\eqref{Heum} to zero to ``solve'' it in terms of $\hat q_R$ and $K$, viewing it as a ``composite'' operator 
\be \label{yeh11}
\hat q_L = \hat q_R+  K  \ .
\ee
Indeed, under the mapping~\eqref{nye1}, we have  
\be 
e^{- i \hat q_L \tau_1} \ket{\psi} \to e^{- i (\hat q_R + K ) \tau_1} \ket{\psi_\tau}_{RQ}  \ .
\ee

Similarly, we can choose a gauge by fixing the clock of the $R$-observer, 
\be  \label{gfx1}
\ket{\psi} \sim \ket{\tau}_R \otimes \ket{\psi_\tau}_{LQ}, \quad \ket{\psi_\tau}_{LQ} = {_R} \vev{\tau|\Pi |\psi}  \in \sH_L \otimes \sH_Q \ . 
\ee
In this gauge, we have 
\be \label{yeh12}
\hat q_R = \hat q_L - K  \ .
\ee

Now we turn to construction of gauge invariant observables, which are required to commute with~\eqref{Heum}. 
 For operators $A \in \sM$ or $A' \in \sM'$, we can 
construct gauge invariant operators by dressing them to $R$ or $L$ observers. For example, we can dress elements of $\sM$ to 
the $R$-observer and elements of $\sM'$ to the $L$-observer, obtaining 
\bega \label{dre1}
\hat A_R (\tau) = \int_{-\infty}^\infty dt \, e^{ i H t} \le(\ket{\tau} \bra{\tau}_R \otimes A \ri)e^{ - i H t}  
 = A_R (\hat p_R - \tau)
= e^{i K \hat p_R} A (-\tau) e^{-i K \hat p_R} 
,  \\
\label{dre3}
\hat A'_L (\tau) = \int_{-\infty}^\infty dt \, e^{-i H t} \le(\ket{\tau} \bra{\tau}_L \otimes A' \ri)e^{i H t}  
= A'_L (\hat p_L - \tau)
= e^{-i K \hat p_L} A' (-\tau) e^{-i K \hat p_L}  \ .
\end{gather} 
By construction, these operators commute with $H$. $\hat A_{R} (\tau)$ 
are the physical operators corresponding to $A$ dressed to the $R$ observers at its clock time $\tau$. 
Similarly with $A'_{L} (\tau)$. They can be translated by  $\hat q_R$ and $\hat q_L$, respectively, 
\be \label{tieb}
\hat A_R (\tau) = e^{i \hat q_R \tau} \hat A_R  e^{-i \hat q_R \tau} , \quad
\hat A'_L (\tau ) = e^{i \hat q_L \tau} \hat A'_L   e^{-i \hat q_L \tau} ,
\ee
where $\hat A_{R} \equiv \hat A_{R} (\tau=0)$ and $\hat A'_{L} \equiv \hat A'_{L} (\tau=0)$.

It is important to stress that the time $\tau$ in $\hat A_R (\tau)$ and $\hat A'_L (\tau )$ is now a quantum number, rather than a coordinate.
In a general state that is not an eigenstate of $\hat p_R$ or $\hat p_L$, its value can fluctuate.  In other words, the operators $\hat A_R$ and $\hat A'_L$ live in a ``quantum'' spacetime. 


The algebra of gauge invariant operators associated with $\sM$ and $\sM'$ can then written as 
\bega \label{cro1}
\wh \sM = \{\hat A_R  = e^{i K \hat p_R} A e^{- i K \hat p_R}  , e^{i \hat q_R s} |A \in \sM, s \in \RR\}'',  \\
\wh \sM' = \{\hat A'_L = e^{-i K \hat p_L} A' e^{i K \hat p_L}  , e^{i \hat q_L s} |A' \in \sM', s \in \RR\}'' \ .
\label{cros4}
\end{gather} 
where from~\eqref{tieb}, $e^{i \hat q_R s}$ and $e^{i \hat q_L s}$ generate ``time'' evolution of $\hat A_R$ and $\hat A'_L$, respectively.\footnote{Note $\wh{\sM'} = (\wh \sM)'$.} 
 

Now consider the gauge~\eqref{gxLC} with $\tau =0$, where $\hat A'_L =  e^{-i K \hat p_L} A'  e^{-i K \hat p_L} = A'$ as the system is in the eigenstate of $\hat p_L$ with eigenvalue $\tau=0$. $\wh \sM' $ can then be written as 
\be\label{cro2}
\wh \sM' = \{A'  , e^{i \hat q_L s} = e^{i (K + \hat q_R) s} |A' \in \sM', s \in \RR\}'' ,
\ee
where we have also used~\eqref{yeh11}. Equations~\eqref{cro1}--\eqref{cro2} are precisely the crossed product algebras~\eqref{n00} and~\eqref{n01} discussed in Sec.~\ref{sec:crossed1} with $\hat q$ there identified with $\hat q_R$. 
In the gauge~\eqref{gfx1}, we have $\hat A_R =  e^{i K \hat p_R} A (-\tau) e^{-i K \hat p_R}  = A$ and~\eqref{yeh12},
giving 
\be\label{cros3}
\wh \sM = \{A , e^{i \hat q_R s} =  e^{i (\hat q_L-K) s}|A \in \sM, s \in \RR\}'', 
\ee
which together with~\eqref{cros4} are precisely~\eqref{n1}--\eqref{n2} with $\hat q$ there identified with $\hat q_L$. 
We see that the two representations of the crossed product algebras of Sec.~\ref{sec:crossed1} correspond to two choices of gauge fixings here.


From the discussion of Sec.~\ref{sec:crossed}, given that $K$ is the modular Hamiltonian for $\sM$ and $\sM'$ for some vector $\ket{\Psi}$, the algebras $\wh \sM$ and $\wh \sM'$ discussed above are type II$_\infty$ algebras, for which density operators and entropies can be defined. We will now consider explicit applications of the above abstract models to various physical contexts.

\subsection{Application I: Quantum volatile black hole spacetime and its entropy} \label{sec:BHEN}

We now apply the model of Sec.~\ref{sec:toy} to an eternal AdS black hole. We take the QFT of Sec.~\ref{sec:toy} to be the bulk quantum field theory in the black hole geometry in the $G_N \to 0$ limit, with $\ket{\Psi}$ being the Hartle-Hawking vacuum $\ket{HH}$, and $\sM = \wt \sM_R$, $\sM' = \wt \sM_L$.  The corresponding modular operator $K$ generates the Schwarzschild time translation. 
In this case, the presence of two asymptotic boundaries makes it natural to identify the $R$ and $L$ ``observers'' with the $R$ and $L$ boundaries, respectively. Correspondingly, the eigenvalues $p_R$ and $p_L$ of the conjugate operators $\hat p_R$ and $\hat p_L$ can be understood as the boundary times of the $R$ and $L$ regions.



Various physical interpretations can be given to $\hat q_R$ and $\hat q_L$. In~\cite{ChaPen22}, the authors considered a black hole dual to a microcanonical version of the TFD state, 
\be\label{micTFD}
\ket{\tilde \Psi_{\rm TFD}} = \sum_n f (E_n - E_0) \ket{E_n}_L \ket{E_n}_R
\ee
where the sum is centered around some energy $E_0 \sim O(N^2)$ with the support of function $f (E-E_0)$ within the energy range of $O(N^0)$.
In the bulk, $\hat q_R$ is identified with the  perturbation of the ADM Hamiltonian around the background value, which on the boundary corresponds to the Hamiltonian $H_R$ of the right CFT with the expectation value subtracted, i.e., $\hat q_R = H_R - \vev{H_R}$ where $\vev{\cdot}$ denotes the expectation value in~\eqref{micTFD}.   Similarly for $\hat q_L$. 

In~\cite{KudLeu23} (see also~\cite{DeVEcc24}), where the standard eternal AdS black hole (i.e., dual to the TFD state in the canonical ensemble) is considered, $\hat q_R$ and $\hat q_L$ are identified as the second order charges associated with the metric perturbations, and~\eqref{Heum} follows from the Hamiltonian constraint corresponding to the Killing vector associated with the bulk Schwarzschild time. $\hat q_R, \hat q_L$ also arise naturally in a mini-superspace canonical quantization of quantum gravity in the black hole geometry~\cite{Kuc94}. 

In all these interpretations, $\hat q_R, \hat q_L$ are treated as independent (bulk) degrees of freedom from the QFT, and 
the leading order $G_N$ expansion is considered. The spectra of  $\hat q_R$ and $\hat q_L$  lie in $(-\infty, \infty)$ as 
they represent small energy perturbations which can go in both directions.

From the discussion of Sec.~\ref{sec:toy}, we can choose to gauge fix the right observer time to coincide with Schwarzschild time $t$, for which the bulk algebras for the $R$ and $L$ regions are given by~\eqref{cros3} and~\eqref{cros4}.  
Alternatively, we can fix the left observer time which results in~\eqref{cro1} and~\eqref{cro2}.

\begin{figure}[H]
\begin{centering}
	\includegraphics[width=4cm]{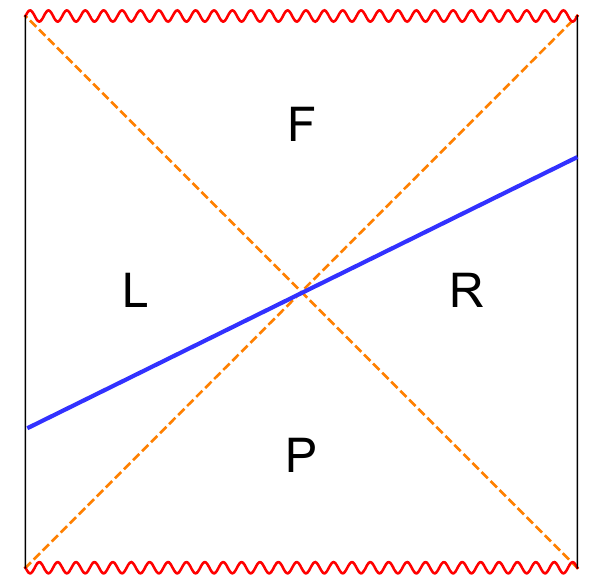}
\par\end{centering}
\caption{\small  Geometrically, the expectation value of $\hat p_R + \hat p_L$ corresponds to the time separation between the endpoints of a geodesic running from the $R$ to the $L$ boundary.}
\label{fig:timeshift}
\end{figure}

The crossed product algebras~\eqref{cro1}--\eqref{cros4} are type II$_\infty$, signaling change in the entanglement structure of the system. This is reflected in  changes in the spacetime structure. Firstly, the dressed observables in~\eqref{cros4} and~\eqref{cro1} may be interpreted as that bulk quantum fields live on a spacetime where time is quantum. 
In a general state, the time difference between $R$ and $L$ boundaries---that is, the expectation value of $\hat p_R + \hat p_L$, which is gauge invariant\footnote{$\hat p_R + \hat p_L$ commutes with~\eqref{Heum}.}---has fluctuations of $O(1)$ in the $G_N \to 0$ limit.
Geometrically, this time difference can be represented by a geodesic going from the $R$ to $L$ boundaries as indicated in Fig.~\ref{fig:timeshift}. 
We thus have an example of ``quantum volatile geometry'' mentioned in Sec.~\ref{sec:EREPR}: the $R$ and $L$ boundaries are connected by a ``quantum wormhole.'' This is also intutively  natural in the case of microcanonical TFD state~\eqref{micTFD}, as energy uncertainties of $O(1)$ should lead to time uncertainties of $O(1)$.

Now consider a general excited state $\ket{\Phi}$ in the bulk Hilbert space $\sH_Q$ and consider a state 
in $\sH_Q \otimes L^2 (\RR)$ of the form
\be \label{secSi}
\ket{\hat \Phi} = \ket{\Phi}  \otimes \ket{g}, \quad 
\ket{\Phi} \in \sH_Q, \quad  \ket{g} \in L^2 (\RR)  , 
\ee
where $\ket{g}$ has coordinate space wave function $g(q)$ which is taken to be nonzero everywhere and normalized as 
\be 
\int dq \, |g (q)|^2 = 1 \ .
\ee
Here we assume that  the gauge choice of~\eqref{gxLC} or~\eqref{gfx1} has been chosen, with $q$ here denoting the eigenvalues of (the remaining) $\hat q_R$ or $\hat q_L$.

We can now consider the type II entropy for $\wh \sM_R$ in the state~\eqref{secSi}, which as we discussed in Sec.~\ref{sec:crossed} can be identified with the relative entropy of the original type III algebra $\sM_R$ 
 \be 
S_R^{(\hat \Phi)} = - S (\Phi|\Psi)  + {\rm const}  \ .
\ee
Here $\ket{\Psi}$ is taken to be the Hartle-Hawking vacuum of the QFT in the black hole geometry, and the constant does not depend 
on the state $\ket{\Phi}$. 
It has been shown in~\cite{Wal11} that up to a state-independent (i.e. $\ket{\Phi}$-independent) constant, the relative entropy can in fact be identified with 
the negative of the generalized entropy in the state $\ket{\Phi}$ 
\be
S (\Phi|\Psi)  =  - S_{\rm gen}^{(\Phi)}  + {\rm const} 
\ee
with the assumption that the system settles into the equilibrium state $\ket{\Psi}$ as $t \to +\infty$. 
It then follows that 
\be 
S_R^{(\hat \Phi)} =   S_{\rm gen}^{(\Phi)}+ {\rm const}  
\ee
i.e., $S_R^{(\hat \Phi)}$ gives the generalized entropy of the black hole, up to an additive constant which is independent of $\ket{\Phi}$ (i.e., the bulk state of matter fields in the black hole geometry). The generalized entropy includes both the entanglement entropy of the matter fields outside the black hole and the change of the black hole entropy (i.e., the horizon area) due to the backreaction of matter fields on the black hole geometry.

\subsection{Application II: de Sitter entropy} \label{sec:CLPW}

In this section we apply the model of Sec.~\ref{sec:toy} to de Sitter spacetime to shed light on the operator algebras in de Sitter and de Sitter entropy~\cite{ChaLon22,Wit23b}.\footnote{See also~\cite{NarVer23} for a proposal of de Sitter holography motivated from such a model of observers.}

\begin{figure}[H]
	\centering
	\includegraphics[width=5cm]{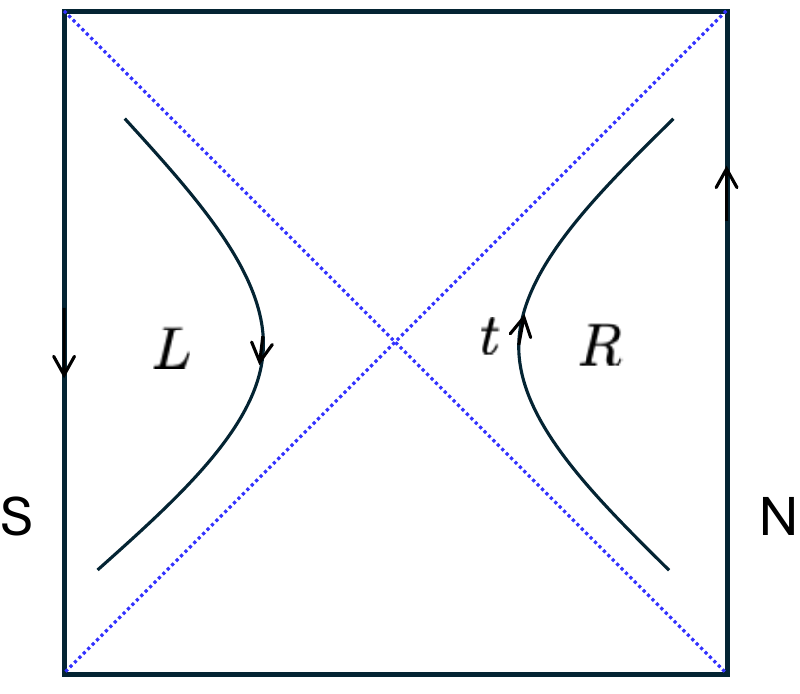}
		\caption{\small Static observers  located at the north and south poles of de Sitter space (labeled $N$ and $S$)
		can only access $R$ and $L$ regions, respectively. 
	 Dotted lines are their cosmological horizons. Translation in $t$ in an isometry 
	and the corresponding Killing vector acts in opposite directions in $R$ and $L$ static patches. 
	}
	\label{fig:desitter}
\end{figure}

The Penrose diagram of global de Sitter is given in Fig.~\ref{fig:desitter}, whose Cauchy slice is given by a sphere. 
A key geometric feature of de Sitter is that a static observer can only access part of the 
full spacetime. In Fig.~\ref{fig:desitter}, a static observer sitting at the north/south pole of the spatial manifold can only access the $R/L$ region of the spacetime, bounded by its past and future horizon.  $R$ and $L$ regions are known as static patches. Their definition as well as the corresponding cosmological horizons are observer-dependent. The metric for a static patch can be written as
\be \label{emdS}
ds^2 = - f_0 dt^2 + {1 \ov f_0} dr^2 + r^2 d \Om_{d-2}^2 , \quad f_0 = 1- {r^2 \ov R^2} 
\ee
where $R$ is the de Sitter radius, related to the cosmological constant $\Lam$ by $2 \Lam = {(d-1) (d-2) \ov R^2}$. 
The observer sits at $r=0$, with the cosmological horizon at $r= R$. Translation in $t$ is an isometry and the corresponding Killing vector acts on $t$ in the $R$ and $L$ patches in opposite directions~(see Fig.~\ref{fig:desitter}). 

In contrast to quantum gravity in AdS spacetime, formulating quantum gravity for a closed universe like de Sitter space presents far deeper conceptual challenges. The absence of a boundary exacerbates the ``problem of time''---it is unclear how to define time evolution in an unambiguous manner. Compounding this issue are the difficulties in constructing  physical observables and the Hilbert space of physical states.

An extensively discussed subject about de Sitter space is the physical interpretation of the entropy associated with the cosmological horizon of a static patch~\cite{GibHaw77a}
\be \label{dsEN}
 S_{\rm dS}  \equiv {A^{(0)}_{\rm hor} \ov 4 G_N} ={R^{d-1} \om_{d-2} \ov 4 G_N} ,
\ee
where $A^{(0)}_{\rm hor}$ is the area of the cosmological horizon for empty de Sitter~\eqref{emdS} and $\om_{d-2}$ is the volume of a $(d-2)$-dimensional unit sphere. The nature of the entropy in~\eqref{dsEN} has inspired much speculation---for instance, whether its exponential might correspond to the dimension of the Hilbert space in de Sitter spacetime (see e.g.~\cite{Ban00}). However, concrete progress on this issue has been limited.

Now consider the generalized entropy for the 
$R$-static patch for a general asymptotic de Sitter spacetime
\be \label{dsEN1}
S_{\rm gen} = {A_{\rm hor} \ov 4 G_N}  + \tilde S_R  ,
\ee
where $A_{\rm hor}$ is the area of the cosmological horizon and $\tilde S_R$ is the entropy of quantum fields in the $R$-patch. One  ``peculiar'' aspect of the generalized de Sitter entropy~\eqref{dsEN1}---which distinguishes it from black hole entropy---is that exciting matter in the $R$-patch, while increasing $\tilde S_R$,  {\it decreases} the total entropy~\eqref{dsEN1} through the associated reduction of $A_{\rm hor}$~\cite{MaeKoi98}. This can be seen, for example, by considering a black hole in the $R$-patch (imagine we excite enough matter that they collapse to form a black hole), with the metric 
\be \label{dSschM}
ds^2 = - f dt^2 + {1 \ov f} dr^2 + r^2 d \Om_{d-2}^2 , \quad f = 1- {\mu \ov r^{d-3} }- {r^2 \ov R^2}  \ .
\ee
For $\mu$ not too large, there is a the black hole horizon (with radius $r_b$) inside the cosmological horizon (with radius $r_c$), i.e., $r_b < r_c$. In this case, we can use the black hole entropy to represent the leading contribution to $\tilde S_R$, giving 
\be \label{gendE}
S_{\rm gen} = \om_{d-2} {r_c^{d-2} + r_b^{d-2} \ov 4 G_N} < S_{\rm dS}   ,
\ee
 where the last inequality in~\eqref{gendE} can be readily checked explicitly. It thus appears that
 \be 
 S_{\rm gen}^{(\rm max)} = S_{\rm dS} \ .
 \ee


Interpreting $S_{\rm gen}$ as the entanglement entropy between an ``$R$-system'' (associated with the $R$-patch) and an ``$L$-system'' (associated with the $L$-patch), one immediately recognizes the parallel with the entangled spin example at $\theta = {\pi \over 4}$ discussed in Sec.~\ref{sec:inf}.
In that example, with the system being in a maximally entangled state, any perturbations of the system will decrease the total entanglement entropy.  The discussion of Sec.~\ref{sec:II} tells us that in the large $N$ limit, the operator algebra for the $R$-system at $\theta = {\pi \over 4}$ is given by a type II$_1$ algebra, and the (type-II) entropy of the corresponding algebra in a perturbed state can be interpreted as the difference with the maximal entanglement entropy. 
Now suppose we interpret empty de Sitter space as describing the maximally entangled state between two quantum systems in the $G_N \to 0$ limit, can we push the parallel further to see that the operator algebra in the right static patch is given by a type II$_1$ algebra? 

It turns out this can indeed be achieved using the model of Sec.~\ref{sec:toy} with a small tweak. The key idea to introduce 
an ``external'' observer equipped with a clock for each static patch~\cite{ChaLon22}\footnote{This is an old idea. See for example~\cite{Mon78,Mon79}.}. Observables can then be dressed to the observers and a nontrivial time-diffeomorphism algebra can be constructed.
 We take $\sH_Q$ of Sec.~\ref{sec:toy} to be the QFT Hilbert space of matter fields in global de Sitter spacetime, with $\ket{\Psi}$ identified as the Bunch-Davies vacuum $\ket{0}_{\rm BD}$, and $\sM, \sM'$ as the algebras in the $R$- and $L$-patches, respectively.
 The corresponding modular Hamiltonian is simply the boost generator $K$ of $t$-translations.
 We take $\hat q_R, \hat q_L$ to be associated with $R$ and $L$ static observers sitting respectively at the north and south poles of the sphere.  




The model of Sec.~\ref{sec:toy} then offers a simple toy model of ``quantum'' de Sitter spacetime which 
is invariant under time diffeomorphisms associated with the boosts.\footnote{The model can also be motivated from linear instability, see~\cite{DeVEcc24,KapMar24,KudLeu23}.}
 The discussion of Sec.~\ref{sec:toy} then leads to diffeomorphism invariant operator algebras $\wh \sM_R$ and $\wh \sM_L = \wh \sM_R'$ for $R$ and $L$ patches. 
The algebras~\eqref{cro1},  \eqref{cro2} and~\eqref{cros3}, \eqref{cros4} correspond to gauge fixing the $L$ and $R$ clocks, respectively. 

Compared with the black hole system of Sec.~\ref{sec:BHEN}, the new element here is that we now impose that $\hat q_R, \hat q_L$ are bounded from below, with their eigenvalues non-negative. 
As discussed around~\eqref{tyii1} in Sec.~\ref{sec:IIcro}, with $\hat q$ that is bounded from below, the crossed product algebras $\wh \sM_R$ and $\wh \sM_L$ become type II$_1$. The corresponding type II$_1$ entropy is negative, and can be interpreted as the difference from that of the maximally entangled state, represented by $\ket{\hat \Psi_T}$~\eqref{tyitr}. In the current context, $\ket{\hat \Psi_T}$ has the form 
\begin{equation}
\ket{\hat \Psi_T} = e^{-\ha q} \otimes \ket{0}_{\rm BD}, 
\label{eq:hh}
\ee
with the trace on $\wh \sM_R$ given by 
\begin{equation}
\tr( \cdots ) \equiv  \vev{\hat \Psi_T | \cdots |\hat \Psi_T } \ .
\end{equation}


One may object that if empty de Sitter is described by a maximally entangled state, then it should have an infinite temperature. However, for an observer in a static patch using time $t$ of~\eqref{emdS}, there is a finite de Sitter temperature~\cite{FigHoe75,GibHaw77a}, given by 
$T_{\rm dS} = {1 \ov 2 \pi R}$. We note that the state $\ket{\hat \Psi_T}$ has a density operator for the $R$-algebra $\rho = e^{- K}$, i.e., it is a thermal state with dimensionless temperature $1$. In terms of units of $t$ of~\eqref{emdS}, it becomes the de Sitter temperature. 

To summarize, a model based on external observers of positive-definite Hamiltonians can give a natural explanation of features of the de Sitter entropy and provides support for empty de Sitter corresponding to a maximally entangled state.

\subsection{Application III: Generalized gravitational entropy for a local spacetime region} \label{sec:region}

To gain insight into the nature of generalized gravitational entropy and its possible finite $N$ interpretation (see Sec.~\ref{sec:finiteN}), we next apply the model of Sec.~\ref{sec:toy} to more general spacetime regions. The AdS eternal black hole (Sec.~\ref{sec:BHEN}) and de Sitter space (Sec.~\ref{sec:CLPW}) are special cases, admitting states whose modular Hamiltonians coincide with generators of isometric time flows. For generic regions in general spacetimes, such states do not exist, so the use of the model is more involved. 



\begin{figure}[H]
	\centering
	  \includegraphics[width=10cm]{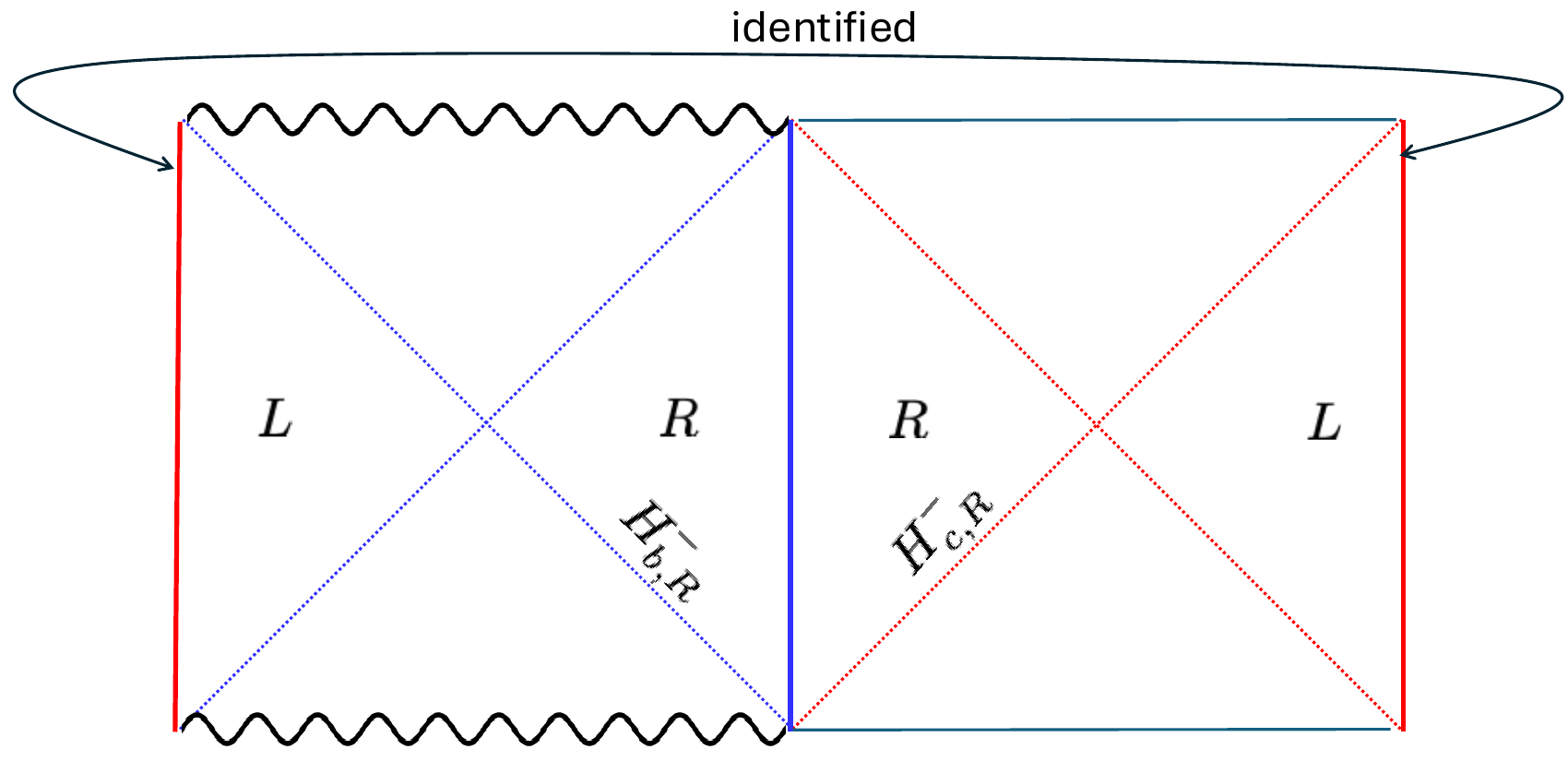}   
	\caption{\small  Penrose diagram of Schwarzschild-de Sitter spacetime. The red vertical lines at the left and right edges of the diagram are identified. Each point in the diagram represents a $(d-2)$-dimensional sphere, so a spatial slice has topology $S_1 \times S_{d-2}$. Blue dotted lines indicate black hole event horizons (at radius $r_b$), while red dotted lines are cosmological horizons (at radius $r_c > r_b$). The trajectories of the $R$ and $L$ observers are shown as solid blue and red vertical lines, and the regions accessible to them are labeled $R$ and $L$, respectively. We can choose a Cauchy slice $\Sigma$ of the $R$ region as the union of the past black hole and cosmological horizons, $\Sigma = H_{b,R}^- \cup H_{c,R}^-$.}
	\label{fig:dSsch}
\end{figure}

An interesting example is provided by Schwarzschild de Sitter~\eqref{dSschM}, whose Penrose diagram is given by Fig.~\ref{fig:dSsch}. 
We are interested in the operator algebra in region $R$ and $L$, i.e., we take $\sM = \sM_R$ and $\sM' = \sM_{L}$. 
Region $R$, whose radius coordinate lies in the range $(r_b, r_c)$ is bounded by a black hole horizon at $r_b$ and a cosmological horizon at $r_c$. In general, the two horizons have different temperatures, and as a result there does not exist a state whose modular Hamiltonian coincides with the generator of time $t$ in~\eqref{dSschM}. Thus the modular flow of $\sM$ is no longer geometric. 

This case is still special, since the $R$ region is bounded by Killing horizons: we can choose a Cauchy slice for the $R$ region as $\Sigma = H_{b,R}^- \cup H_{c,R}^-$, the union of the past black hole and cosmological horizons (see Fig.~\ref{fig:dSsch}). The key idea of~\cite{KudLeu23} for circumventing the absence of a global geometric modular flow is to decompose the algebra $\sM_R$ as $\sM_R = \sM_c \otimes \sM_b$, where $\sM_c$ and $\sM_b$ are the algebras associated with $H_{c,R}^-$ and $H_{b,R}^-$, respectively. One can then select a state\footnote{The state is defined as the vacuum for initial data on the past black hole and cosmological horizons.}, for which the modular flows of the horizon algebras are geometric, coinciding with the boost symmetries of the corresponding horizons.

We now introduce two sets of observer Hamiltonians, $\hat q_{R,c}, \hat q_{L,c}$ and $\hat q_{R,b}, \hat q_{L,b}$, for the cosmological and black hole horizons, respectively, and impose the constraints\footnote{These constraints can be motivated from the gravitational analysis of the Hamiltonian constraints~\cite{KudLeu23}.}
\be
K_c + \hat q_{R,c} = \hat q_{L,c}, \quad K_b + \hat q_{R,b} = \hat q_{L,b} ,
\ee
after which the previous discussion applies. Interestingly, it was found in~\cite{KudLeu23} that the algebra becomes type II$_\infty$, rather than type II$_1$, even when the observer Hamiltonians are bounded from below. The associated type II entropy can again be identified with the $O(G_N^0)$ contribution to the generalized entropy,
\be
S_{\rm gen}^{(R)} = {A_{\rm COS} \over 4 G_N} + {A_{\rm BH} \over 4 G_N} + \tilde S_R \ .
\ee
The type II$_\infty$ answer is natural, since the type II entropy corresponds to the $O(G_N^0)$ part of $S_{\rm gen}^{(R)}$, i.e., the perturbation around the leading $O(1/G_N)$ contribution ${A_{\rm COS} \ov 4 G_N} + {A_{\rm BH} \ov 4 G_N}$, which is no longer the maximal entropy (as discussed around~\eqref{gendE}). Consequently, the state under consideration is not a maximal entropy state, as would be expected for a type II$_1$ algebra.

Extending the discussion to a general spacetime region such as $\sD_{\rho_w}$ in Fig.~\ref{fig:bulkD} is even more challenging, as no Killing horizon is present. In~\cite{JenSor23}, the authors conjectured that although no state exists in which the modular flow of the algebra coincides globally with a geometric time flow in the QFT (see~\cite{Sor24} for further investigations), there nevertheless exists a state for which the corresponding modular flow is geometric \emph{instantaneously} on a single Cauchy slice of that region.
They further showed that, under this ``instantaneously geometric modular flow'' conjecture, the model of Sec.~\ref{sec:toy} can be applied, and that the generalized gravitational entropy of a region can thereby be obtained as an entanglement entropy. A peculiarity arises if one imposes that the observer Hamiltonian be bounded: in that case, the algebra becomes of type II$_1$, which conflicts with the general expectation that no maximally entangled state should exist in such a setting. This tension may indicate either that the bounded-Hamiltonian assumption is not justified, or that there is a subtlety in the geometric modular flow conjecture itself.



\subsection{A model of dynamical observers and emergence of cosmological horizon}

In the model of Sec.~\ref{sec:toy} and subsequent applications of Sec.~\ref{sec:BHEN}--\ref{sec:region}, the observers themselves are not dynamical, only carrying  dynamical ``clocks'' described by $L^2 (\RR)$.  
In a quantum gravity system, observers should be dynamical. Here we briefly discuss a model in two-dimensional de Sitter spacetime (dS$_2$) with a pair of dynamical observers of finite mass~\cite{KolLiu24}. 
The regime of static observers corresponds to taking the mass of the observers to infinity.

Consider the standard action of a free massive particle,
\begin{equation}
S_{\text{particle}} = -m \int d\tau \sqrt{-g_{\tau \tau}},
\label{eq:sparticle}
\end{equation}
where $\tau$ is a parameter on the observer's worldline, and $g_{\tau \tau}$ is the pull-back of the background metric to the worldline. We can obtain the action of an observer with a dynamical clock by  promoting the mass parameter $m$ to a dynamical variable $Q(\tau)$, with conjugate momentum $P(\tau)$~\cite{Wit23b}, 
\begin{equation}
S_{\text{observer}} = \int d\tau \, \left[ P \partial_\tau Q - Q \sqrt{-g_{\tau \tau}}  \right] \ .
\end{equation}
The Hilbert space $\sH_{\rm obs}$ of an observer in dS$_2$ is obtained from that of a free quantum particle by augmenting it with an additional quantum number corresponding to the mass.
 
Now  consider a free scalar field $\vp$ in dS$_2$ with Hilbert space $\sH_Q$ and two observers, $R$ and $L$, and impose invariance under the isometry group of dS$_2$, treating this symmetry group as gauged. We denote the Hamiltonians for the observers as $H_R$ and $H_L$ respectively. 

The physical states are obtained by projecting the pre-Hilbert space $\calhobsR \otimes \calh_{Q} \otimes \sH_{\rm obs, L}$ onto the singlet sector of the isometry group. The invariant states are again not normalizable, and, as in the discussion of Sec.~\ref{sec:toy}, a new inner product must be introduced to define the physical Hilbert space $\sH$. The construction is, however, considerably more involved, and we refer the reader to~\cite{KolLiu24} for details. A simple example is 
\be \label{bhdhh} 
\ket{\Psi_{HH}} = \ket{0}_{\rm BD} \otimes \ket{f_{HH}} 
\ee
where $\ket{0}_{\rm BD}$ is the Bunch-Davies vacuum in the theory of $\vp$ and $\ket{f_{HH}}$ is a singlet (Hartle-Hawking) state in the two observers' Hilbert space $\calhobsR \otimes \sH_{\rm obs,L}$, prepared using a Euclidean path integral.

Gauge invariant operators can be obtained by dressing the scalar field $\varphi$ to one of the observers.  For example, operator $\phi_R$ obtained by dressing $\vp$ to the $R$-observer can be defined as 
\begin{equation}
	\vev{ n_R^\prime, s'_R| \phi_R |  n_R,s_R} \equiv \int d^2 x \, \sqrt{-g} \left(\Psi^{s_R'}_{n_R^\prime}(x)\right)^* \vp (x) \Psi^{s_R}_{n_R}(x), 
	\label{eq:4.4}
\end{equation}
where $\ket{n_R, s_R}$ denotes a basis of $\calhobsR$ ($n_R$ labels the quantum numbers on dS$_2$, and $s_R$ specifies the observer's mass), $\Psi^{s_R}_{n_R}(x)$ is the corresponding wave function on dS$_2$, and the integration on the right-hand side is over the full dS$_2$. Equation~\eqref{eq:4.4} defines $\phi_R$ 
as an operator in $\calhobsR \otimes \calh_{Q} \otimes \sH_{\rm obs, L}$ 
in terms of matrix elements on $\calhobsR$ that is valued in the space of operators on $\calh_Q$. 
It can be shown that $\phi_R$ maps singlets of  $\calhobsR \otimes \calh_{Q} \otimes \sH_{\rm obs, L}$ to singlets and thus 
gives a well-defined operator on $\calh$. We again refer to~\cite{KolLiu24} for details. We can ``evolve'' $\phi_R$ using $H_R$ 
\be 
\phi_R (\tau) = e^{i H_R \tau} \phi_R e^{- i H_R \tau} \ .
\ee
Similarly, we can define operators dressed to the $L$-observer. 

 We define the R-observer's algebra $\cala_R$ to be the von Neumann algebra generated by all the $R$-dressed operators together with $H_R$. 
 A key result of~\cite{KolLiu24} is that  $\cala_R$ is a direct integral of type I$_\infty$ factors.\footnote{The center of $\cala_R$ is the Hamiltonian $H_L$ of the $L$-observer.} 
  Intuitively, this is related to that when the observer is fully dynamical, it can change trajectory, and thus there is no proper subregion in which the algebra dressed to it is confined. Thus, local operators dressed to an observer are effectively smeared over an entire Cauchy slice in a state-dependent way. In a QFT on global de Sitter, the algebra of observables associated to a Cauchy slice is type I.  
 Since there is no cosmological horizon, entropy associated with the type I algebra $\sA_R$
 should not be interpreted as the generalized entropy of a horizon; it simply counts the gauge-invariant states in $\calh$.

The static-observer limit is achieved by writing $Q = \Lambda + q$ and taking $\Lam \to \infty$, and the results of Sec.~\ref{sec:CLPW} can be recovered by taking the limit appropriately. Note that the static-observer limit does not commute with taking the double-commutant of the generating algebra of $\cala_R$. If we take the $\Lam \to \infty$ limit after the double-commutant, we obtain a type I algebra. In contrast, if the $\Lam \to \infty$ limit is taken first, then a type II algebra is recovered, reflecting the emergence of the observer's cosmological horizon. As discussed in Sec.~\ref{sec:CLPW}, the trace on this type II algebra is given by the expectation value in the state~\eqref{eq:hh}. Away from the $\Lam \to \infty$ limit, however, one can check that this state does not define a trace on the observer's algebra; it only becomes cyclic in the limit. Thus the trace on the type II algebra is also emergent---and so is the associated Gibbons-Hawking entropy.

When an observer's mass is not strictly infinite, they can in principle have access to the entire spacetime.
With a large but finite mass $\Lam$ for the observer, an uncertainty $1/\Lam$ in its location at $\tau=0$ is magnified to ${1 \ov \Lam} e^{{2 \pi \ov \b_{\rm dS}} \tau} $ after time $\tau$ due to the exponential expansion of de Sitter, where $\b_{\rm dS}$ is the inverse dS temperature. This motivates the definition of a ``scrambling time''\footnote{See also the discussion in~\cite{ABan23} for a different perspective.}
\be\label{dsSc} 
 \tau_s \equiv {\b_{\rm dS} \ov 2 \pi} \log  \Lambda
\ee
after which the observer has an $O(1)$ probability to escape the original static patch. 

The scrambling time~\eqref{dsSc} can be probed using an out-of-time-ordered correlator (OTOC). Although $\varphi$ is a free field and
does not interact directly with the observer, the gravitational constraints (arising from gauging the isometry group) nevertheless ``couple'' them in a nontrivial way. With the observers being dynamical, rich physics can emerge from these couplings. For instance, inserting a matter operator $\phi_R(\tau)$ at time $\tau$ along an observer's worldline induces a recoil of its trajectory (see the left plot of Fig.~\ref{fig:recoil}), leading to nontrivial OTOCs.


More explicitly, consider two free scalar fields labeled $A$ and $B$, and 
\be
G_4 (\tau) = \vev{\Psi_{HH} \le| \phi^B_R \le(\tfrac{\tau}{2} \ri) \,  \phi^A_R \le(-\tfrac{\tau}{2}\ri) \, \phi^B_R \le(\tfrac{\tau}{2}+u_2\ri) \, \phi^A_R \le(-\tfrac{\tau}{2}+u_1\ri) \ri|\Psi_{HH}}
\label{OTOC1}
\ee
where $\ket{\Psi_{HH}}$ was defined in~\eqref{bhdhh}, and $u_1, u_2$ are taken to be small. 
 
The calculation of~\eqref{OTOC1} was discussed in detail in~\cite{KolLiu24}. Here we briefly review its qualitative behavior. When the mass of the observer is fixed, $G_4$ decays to zero for large $\tau$, reflecting the loss of ``coherence'' between the two $\phi^A$  due to recoil effects induced by the insertion of $\phi^B$. In the opposite limit, with $\tau$ fixed but the observer mass taken to 
infinity~(corresponding to the static-observer limit with no recoil), one finds (with the subscript ${\rm BD}$ suppressed)
\be
G_4 = G_4^{(0)} \equiv \vev{0 \le| \varphi_R^B \le(\tfrac{\tau}{2} \ri) \, \varphi_R^B \le(\tfrac{\tau}{2} + u_2 \ri) \ri|0}
\vev{0 \le| \varphi_R^A \le(-\tfrac{\tau}{2} \ri) \, \varphi_R^A \le(-\tfrac{\tau}{2}+u_1 \ri) \ri|0} \ .
\ee
With a large but finite observer mass $\Lam$, we find 
\be \label{lyaP}
G_4 - G_4^{(0)} \propto {e^{{4 \pi \ov \b_{\rm dS}} \tau} \ov \Lam^2} \propto e^{{4 \pi \ov \b_{\rm dS}} (\tau - \tau_s)}  \ .
\ee
That is, we find quantum chaotic behavior with a Lyapunov exponent ${4 \pi \ov \b_{\rm dS}}$.\footnote{This is twice the chaos bound~\cite{MalShe15}. This same exponent was observed in~\cite{AalShi20} for dS$_3$, but due to a different physical mechanism.
However, the result of~\cite{AalShi20}  has since been disputed~\cite{Kol25}.} 
Here the scrambling time $\tau_s$ gives the time scale characterizing the loss of coherence between the two $\phi^A$ insertions due to recoil effects. See the right plot of Fig.~\ref{fig:recoil} for an illustration of the physical origin of the behavior~\eqref{lyaP}.

  
 \begin{figure}[H]
	\centering
	\includegraphics[width=3cm]{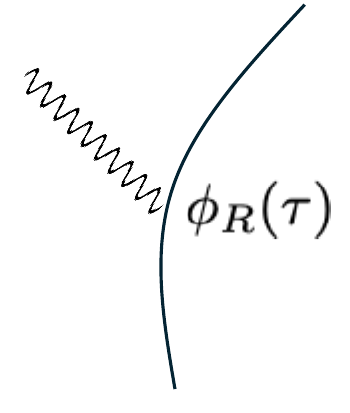} \qquad \qquad 
	\includegraphics[width=0.4\textwidth]{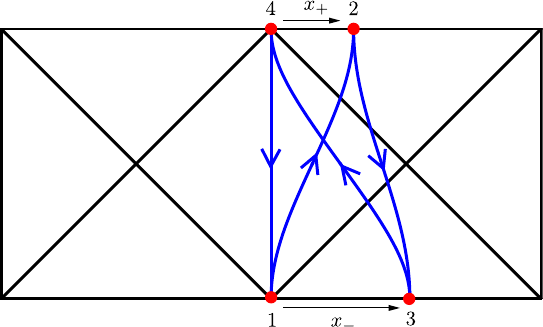}
		\caption{\small Left: Inserting a matter operator $\phi_R$ (wavy line) along the observer's worldline (solid line) induces a ``recoil'' of its trajectory.
Right: Recoil effects leading to~\eqref{lyaP}. Shown is the Penrose diagram for dS$_2$, with the left and right edges identified. The central vertical line connecting points 1 and 4 corresponds to the North Pole, where the $R$-observer originally resides. Each red dot denotes one of the four operators in the OTOC, labeled 1-4 from right to left. For simplicity, we take $\tau$ to be very large (though still smaller than $\tau_s$), such that operators 1 and 3 are inserted essentially at past infinity, and operators 2 and 4 at future infinity.
When the observer's mass $\Lam$ is not strictly infinite, acting with a dressed operator causes recoil. After operator 1 (i.e., $\phi^A_R(-\tfrac{\tau}{2}+u_1)$) acts, the observer's late-time location at $\tfrac{\tau}{2}$ is given by $x_+ \propto e^{\tfrac{2\pi}{\b_{\rm dS}}\tau}/\Lam$. Likewise, after operator 2 (i.e., $\phi_R^B(\tfrac{\tau}{2}+u_2)$) acts, the trajectory is further deflected, shifting its location on past infinity to $x_-$. As the OTOC is computed by evolving along the observer's worldline, the observer travels from point 1 back to point 1 following the blue arrows.
	}
	\label{fig:recoil}
\end{figure}

\subsection{Operator algebras of JT gravity} 

We now turn to the operator algebras of JT gravity~\cite{Tei83,Jackiw,AlmPol14}  coupled to matter, which is a two-dimensional gravity theory (see~\cite{MerTur22} for a review). It serves as a valuable toy model for exploring aspects of finite-$G_N$ physics in higher-dimensional gravity theories.  The full analysis is technically involved and rather intricate. Here we review only the main results.\footnote{Here we consider only bosonic JT gravity. For the story of super-JT gravity, see~\cite{PenWit24}.}

 The action for JT gravity coupled to matter can be written as
\be\label{ljt} 
\sI = {1 \ov 16 \pi G_N} \int_M d^2 x \, \sqrt{-g} \, \Phi \, (\sR +2 ) +  {1 \ov 8 \pi G_N} \int_{\p M} dt \, \sqrt{-\ga} \, \Phi\, (K -1)  + \sI_{\rm matter} [g, \phi] ,
\ee
where $M$ denotes a two-dimensional spacetime with metric $g_{\mu \nu}$, $\mathcal{R}$ is its Ricci scalar, $K$ the extrinsic curvature of the boundary $\partial M$ with induced metric $\ga$, $\Phi$ a scalar field commonly referred to as the dilaton, and $\phi$ collectively denotes the matter fields, which do not couple directly to $\Phi$. 

The equations of motion for $\Phi$ lead to\footnote{This holds also at the quantum level; integrating over $\Phi$ in the path integral leads to a delta function enforcing~\eqref{ads2}.}
\be\label{ads2}
\mathcal{R} + 2 = 0 \ ,
\ee
which implies that the spacetime manifold is locally isomorphic to AdS$_2$, with metric
\be
ds^2 = - \cosh^2 \sigma  du^2 + d\sigma^2 , \qquad \sigma, u \in (-\infty, +\infty) \ .
\ee
Note that AdS$_2$ has two disconnected boundaries at $\sigma = \pm \infty$.
To make the boundary terms from the variations of~\eqref{ljt} vanish, boundary conditions should be imposed on $\Phi$ and the metric. A convenient choice is~\cite{AlmPol14,MalSta16a} 
\be \label{bdjt}
\ga_{tt}|_{\p M} = - { 1\ov \ep^2} , \quad \Phi_{\p M} = {\phi_b \ov \ep} ,
\ee
where we parameterize the boundary $\p M$ using a coordinate $t$ and $\ga_{tt}$ is the induced metric on the boundary, and $\ep$ is an infinitesimal parameter.  $\phi_b$ is a constant that may be regarded as a proxy for $1/G_N$, since only the combination $\Phi/G_N$ appears in the action. The geometry resulting from the boundary condition~\eqref{bdjt} is that of a two-sided black hole, as illustrated in Fig.~\ref{fig:ads2}(a). One can also consider the Euclidean theory, where AdS$_2$ is replaced by a hyperbolic disk (with a single boundary), as shown in Fig.~\ref{fig:ads2}(b).

In the absence of matter, the dynamics of JT gravity reduces to boundary dynamics, as the bulk is frozen by~\eqref{ads2}. In Lorentzian signature, the $R$ and $L$ boundaries can be parameterized by $(u_R(t), \sigma_R(t))$ and $(u_L(t), \sigma_L(t))$, respectively. The first condition in~\eqref{bdjt} allows $\sigma_R(t)$ and $\sigma_L(t)$ to be expressed in terms of $u_R(t)$ and $u_L(t)$. The action~\eqref{ljt} then reduces to the boundary actions $I_R$ and $I_L$ for $u_R(t)$ and $u_L(t)$, so that
\be
\sI = I_R [u_R (t)] + I_L [u_L (t)] \ .
\ee
From $I_R$ and $I_L$, one can in turn derive the boundary Hamiltonians $H_R$ and $H_L$ for the $R$ and $L$ boundaries.

$u_R(t)$ and $u_L(t)$ are not independent, since one must still quotient by the isometry group of AdS$_2$ (a gauge symmetry). This quotient identifies the boundary Hamiltonians, giving $H_L = H_R \equiv H$, and reduces the classical phase space to two dimensions~\cite{HarJaf18}. The phase space can be parametrized either by $(H, \delta t)$, where $\delta t$ is the time shift between the two boundary trajectories, or by $(\ell, p_\ell)$, where $\ell$ is the (renormalized) geodesic distance between the $L$ and $R$ boundaries and $p_\ell$ is its conjugate momentum.

The theory can then be canonically quantized, with Hilbert space $L^2(\mathbb{R})$. An (overcomplete) basis $\{\ket{\b}, \b \in \RR_+\}$ for this Hilbert space can be constructed using the Euclidean path integral, as illustrated in Fig.~\ref{fig:ads2}(b).

 \begin{figure}[H]
	\centering
	\includegraphics[width=12cm]{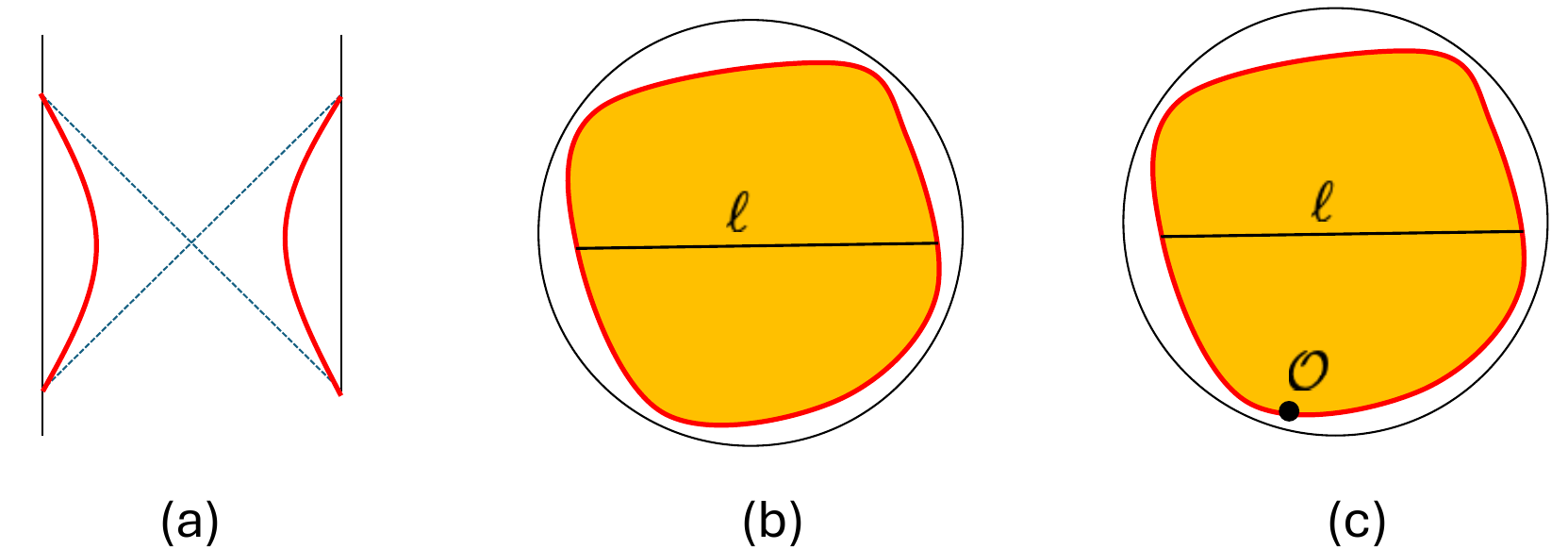} 
		\caption{\small (a) Lorentzian geometry of JT gravity: the thick red lines denote the dynamical boundaries of the spacetime.
(b) Euclidean description: given by a hyperbolic disk, with the boundary (represented by thick red line) topologically a circle. The path integral over the lower half-disk with boundary length ${\beta \over 2}$ prepares the wave function $\psi_\beta(\ell)$ for the state $\ket{\beta}$.
(c) In the presence of matter, a basis of the bulk Hilbert space $\sH$ can be obtained from the path integral over the lower half-disk with matter operators inserted on the lower boundary.}
	\label{fig:ads2}
\end{figure}

Now let us include matter fields. For illustration, consider a scalar field $\phi$, which may be interacting. Since matter fields do not couple to the dilaton $\Phi$, the spacetime geometry remains AdS$_2$. We denote the boundary limit of $\phi$ as $\sO$, which defines a single-trace operator in the dual boundary quantum mechanical system, as in~\eqref{exT}.
With matter present, $H_R$ and $H_L$ become independent, as they now include contributions from the matter fields. Upon quantization, the full Hilbert space takes the form $\sH = L^2 (\RR) \otimes \sH_{\rm matt}$. 
An (overcomplete) basis for $\sH$ can again be constructed via Euclidean path integrals, now with insertions of the matter operator $\sO$ on the Euclidean boundary~\cite{PenWit23,Kol23}, as illustrated in Fig.~\ref{fig:ads2}(c).
It is important to emphasize that, in contrast with the discussion in Sec.~\ref{sec:TFD}, even at finite $G_N$ there is only a single Hilbert space $\sH$; in particular, it cannot be factorized into separate Hilbert spaces for the $R$ and $L$ systems.

Denote the algebras generated by the single-trace operator $\sO$ on the $R$ and $L$ boundaries as $\sA_R^0$ and $\sA_L^0$, respectively. In the $G_N \to 0$ (or equivalently $\phi_b \to \infty$) limit, we obtain a quantum field theory in ``classical'' AdS$_2$ at leading order. The situation is analogous to that in Sec.~\ref{sec:TFD} and Sec.~\ref{sec:krus} for an AdS eternal black hole in higher dimensions: $\sA_R^0$ and $\sA_L^0$ are type III$_1$ algebras, dual to the bulk algebras in the $R$ and $L$ regions of AdS$_2$, where the causal and entanglement wedges coincide. 


We can further consider the model of Sec.~\ref{sec:BHEN} by introducing ``static'' boundary observers with Hamiltonians $\hat q_R $ and $\hat q_L$. As discussed there, the algebra of gauge-invariant operators is obtained via a crossed-product construction, and the resulting algebras are of type II$_\infty$. In this case, although $G_N \to 0$, the timeshift between the two boundaries exhibits $O(1)$ fluctuations, and the spacetime becomes quantum and volatile.

Now finally consider finite $G_N$ (or finite $\phi_b$). In this regime, the boundary algebra $\sA_R^0$ should be enlarged to $\sA_R$ by including $H_R$, and similarly $\sA_L^0$ to $\sA_L$ by including $H_L$. The spacetime is now fully quantum: the $R$ and $L$ boundaries become fully dynamical; horizons can no longer be sharply defined; and bulk subregions such as entanglement and causal wedges are no longer well defined. As a result, it is not clear how to separate the bulk spacetime into $R$ and $L$ regions.
Interestingly, the operator algebra $\sB(\sH)$ of the bulk quantum gravity system can still be ``factorized'' into $R$ and $L$ algebras~\cite{PenWit23,Kol23}: 
\be 
\sB(\sH) = \sA_R \lor \sA_L, \qquad \sA_R' = \sA_L \ .
 \ee 
 Moreover, $\sA_{R}$ and $\sA_{L}$ are type II$_\infty$ factors, with the trace defined as 
 \be \label{triiJT}
 \tr a = \lim_{\beta \to 0} \vev{\beta | a | \beta}, \quad a \in \sA_R 
 \ee 
 where $\ket{\beta}$ denotes the state prepared in Fig.~\ref{fig:ads2}(b). Thus, even though no ``geometric'' entanglement wedge exists at the fully quantum level, there nevertheless remains an ``algebraic'' entanglement wedge.

In the $G_N \to 0$ (or $\phi_b \to \infty$) limit, $\ket{\b}$ becomes non-normalizable and the trace~\eqref{triiJT} is no longer defined,  consistent with the emergent type III$_1$ structure in that limit.

 
\section{Conclusions and discussions}  \label{sec:diss}

In this review, we have explored how entanglement in systems with infinitely many degrees of freedom can be described through the framework of operator algebras, and how emergent algebraic structures in the $N \to \infty$ limit of the boundary theory shed light on the emergence of spacetime in quantum gravity in the $G_N \to 0$ limit. Below we first summarize some of the main features discussed  and then offer speculative remarks on how operator algebras may provide a pathway toward uncovering the mathematical structure of quantum gravity.

\subsection{Summary}

Here we summarize the main lessons discussed in this article:

\bi

\item In the $G_N \to 0$ limit, a general bulk subregion is defined by an operator algebra in the boundary theory, which typically lacks a direct geometric description. The corresponding algebra exhibits an emergent type III$_1$ structure, arising from the infinite long-range entanglement generated in the large-$N$ limit of the boundary system. 

The boundary algebras describing bulk regions can be extended  readily to the stringy regime, potentially providing an algebraic definition of stringy subregions.

\item Bulk causal structure is encoded in the commutant structure of boundary algebras. While boundary algebras encode boundary causality by definition (with spacelike separated operators commuting), timelike separated boundary operators may also commute with each other. This extended commutant structure reflects bulk causality.

Quantitative measures such as causal depth parameters can be defined in terms of boundary algebras to characterize the global causal structure of bulk spacetime, including the presence of horizons. Moreover, these measures extend naturally to the stringy regime, potentially providing boundary definitions of stringy horizons.

\item Boundary modular flows and half-sided modular flows can be used to characterize the emergence of time in bulk spacetime. We showed how these structures elucidate the appearance of Kruskal-like flows in black hole geometries and provide probes of the black hole interior. In addition, the structure of boundary algebras can be employed to characterize bulk connectivity, motivating an algebraic formulation of the ER=EPR proposal.



\item The algebras arising in the large $N$ limit can be separated into two types: 

\ben 

\item An entanglement wedge algebra $\sX$ is defined as the large-$N$ limit (in a semi-classical state) of an algebra $\sB$ in a finite-$N$ theory. By construction, $\sX$ admits a finite-$N$ extension $\sB$.

\item  In contrast to the entanglement wedge algebra $\sX$, a causal wedge algebra $\sY$ is intrinsically a semi-classical construct, defined in terms of single-trace operators specified by semi-classical bulk fields through the extrapolation dictionary~\eqref{exT}.

\een
Neither type of algebra is guaranteed to have a geometric interpretation in the bulk, particularly once we move beyond the Einstein regime.

Both types of algebras can also be defined for a quantum system that lacks a gravity dual of its own but is coupled to, or entangled with, a holographic system that does. Such a system need not possess a parameter $N$; rather, the notions of large-$N$, finite-$N$, and the semi-classical regime are defined with respect to the holographic partner.

\item We also presented several simple models of quantum gravity constructed from operator algebras. The model with static observers, though simple, already helps clarify the physical origin of generalized gravitational entropies---in black holes, the static patch of de Sitter spacetime, and arbitrary bulk regions---in terms of entanglement entropies. The model with dynamical observers in de Sitter spacetime illustrates the emergence of cosmological horizons and offers hints toward a possible holographic description of de Sitter space. Finally, operator algebras in JT gravity provide an example of a quantum gravity system in which the Hilbert space does not factorize, while the operator algebra does.



\ei

\subsection{The mathematical structure of quantum gravity?}  
  
 In the usual formulation of quantum mechanics, Hilbert space plays a fundamental role. The states of the system correspond to density operators in a Hilbert space, with observables represented by Hermitian operators. There are strong indications that in quantum gravity 
 the role of Hilbert space  is much less fundamental. 
 It is generally believed that it is not possible in quantum gravity to have physical processes that change the asymptotic structure of a spacetime. For example, in type IIB string theory, 
it should not be possible to have a physical process transitioning from an asymptotic AdS$_5 \times S_5$ spacetime 
to AdS$_3 \times S_3 \times K3$, or from either of these spacetimes to ten-dimensional Minkowski spacetime. 
Therefore, IIB string theory in each of these spacetimes should have a separate Hilbert space. Indeed, IIB string theory in 
asymptotic AdS$_5 \times S_5$ and AdS$_3 \times S_3 \times K_3$ spacetimes are described by 
a four-dimensional and a two-dimensional
CFT, respectively. Each of the CFTs has its own Hilbert space; there is no physical process that can take a state
in one CFT to any state of the other. Yet AdS$_5 \times S_5$,  AdS$_3 \times S_3 \times K_3$, and ten-dimensional Minkowski spacetime are considered to be different vacua within the same quantum gravity theory described by type II superstring theory. 

Without a single global Hilbert space encompassing all its physical states, how should we characterize the mathematical structure of a quantum gravity theory?


We can gain insights from quantum field theory (QFT) in curved spacetimes (see, e.g., \cite{HolWal08,HolWal14} for reviews), which describes the semiclassical $G_N \to 0$ limit of a quantum gravity system.
As an example, consider a free scalar field theory in Minkowski spacetime. The theory can be quantized using either Minkowski time or Rindler time. Both quantizations are physically sensible and describe different physical situations. However, they lead to inequivalent Hilbert spaces and to quantum systems that are unitarily inequivalent.
One could treat these different quantizations as distinct theories, but this is conceptually unsatisfactory, since there is a clear sense in which they represent the same free scalar theory, merely quantized with a different choice of time. More generally, one may consider the same scalar field theory defined in different spacetimes---for instance, in Minkowski spacetime, in a Schwarzschild black hole background, or in de Sitter spacetime. Again, the resulting Hilbert spaces are inequivalent, but should they really be regarded as entirely different quantum systems?

There is a unified way to treat Minkowski and Rindler quantizations by defining the theory in terms of the algebras of the scalar field. In this framework, different quantizations---and the resulting Hilbert spaces---can be understood as different representation spaces of the same underlying abstract algebra.


This is the basic idea of algebraic QFT~(AQFT), which we already briefly mentioned in Sec.~\ref{sec:qft}. 
In AQFT, algebras of operators play a fundamental role, with Hilbert space  being a derived concept. More explicitly, in the AQFT approach, a QFT on a spacetime manifold $M$ can be viewed as being defined by a pair $(\sA, \sS)$, where 
$\sA$ is the algebra of quantum fields on $M$ (which we can take to be a $C^*$ algebra), and $\sS$ is the space of allowed states.
Here states are not states in a Hilbert space, but defined abstractly as  linear, positive functionals on the algebra\footnote{Recall the definition of Sec.~\ref{sec:states}.}.
 Given a state $\om$ on $\sA$, we can then construct  a Hilbert space $\sH_\om$ using the GNS procedure of Sec.~\ref{sec:GNS}, and obtain a representation $\pi_\om (\sA)$ of $\sA$ on $\sH_\om$.
 Some states in $\sS$ may belong to $\sH_\om$, but other states may lead to distinct Hilbert spaces. Each such Hilbert space provides a representation space for the algebra $\sA$. 
$\sS$ does not have the structure of a Hilbert space, as in general it is not possible to define an inner product for all states in it. 




Such a conceptual framework can be naturally generalized to a quantum gravitational system~\cite{Liu25}. We can postulate that  a quantum gravity system (with a finite $G_N$) is specified by an abstract $*$-algebra $\mathcal{A}$, and some collection of allowed states. Spacetimes of differing asymptotic structure correspond to distinct states. Given a state $\omega$ that specifies an asymptotic structure, the GNS construction again furnishes a Hilbert space $\sH_\omega$, within which the full physics of spacetimes sharing that asymptotic structure is realized. The Hilbert space $\sH_\omega$ carries a representation of the algebra $\pi_\omega(\sA)$.

For example, suppose type IIB superstring theory is described by an algebra $\sA_{\rm IIB}$. A background such as $\mathrm{AdS}_5 \times S_5$ could then be interpreted as corresponding to a state $\omega_{\mathrm{AdS}_5 \times S_5}$ on $\sA_{\rm IIB}$, with the associated GNS Hilbert space identified with the Hilbert space of $\mathcal{N}=4$ Super-Yang-Mills (SYM) theory. Similarly, $\mathrm{AdS}_3 \times S_3 \times K3$ would correspond to another state, with the resulting GNS Hilbert space identified with that of the (deformed) symmetric orbifold theory. Each realization of the AdS/CFT correspondence involving type IIB superstring theory would thus correspond to a distinct state $\om$ on the same algebra $\sA_{\rm IIB}$, with different boundary CFTs providing different representations $\pi_\om (\sA_{\rm IIB})$ of this algebra. Ten-dimensional Minkowski spacetime and potential de Sitter vacua may be interpreted in the same way, although we still lack a holographic description for them.
In this framework, all solutions of type IIB string theory could be understood in a unified manner, independent of their asymptotic structure.


At present, we do not have a background-independent definition of $\sA_{\mathrm{IIB}}$, nor a complete characterization of the allowed states. String field theory (see, e.g.~\cite{Erb21,SenZwi24} for recent reviews) may play an important role in identifying such a formulation. For a chosen background state $\omega$, string field theory provides the Hilbert space $\sH_\omega$ together with the background-dependent algebra $\pi_\omega(\sA_{\mathrm{IIB}})$. Moreover, its equations of motion can be used to systematically identify further consistent states and their associated algebras. See also~\cite{Wit23b} for a recent proposal to define a background-independent algebra in quantum gravity using an algebra of operators along an observer's worldline.

Our current understanding of quantum gravity has been shaped by the asymptotic structure of spacetime, with AdS/CFT providing the sharpest formulation. Yet this reliance on asymptotics presents a fundamental barrier to extending holographic ideas to closed universes or asymptotically flat spacetimes. The algebraic approach discussed above~\cite{Liu25} offers a way forward: it treats asymptotic structure and the cosmological constant as state-dependent data rather than defining ingredients, allowing concepts such as Hilbert space and degrees of freedom to emerge as derived, state-dependent notions.\footnote{See~\cite{Liu25} for examples of how a closed universe in AdS/CFT can be described in this manner.} We regard this as a promising step toward a unified and more general formulation of quantum gravity.

\vspace{0.2in}   \centerline{\bf{Acknowledgements}} \vspace{0.2in}
We would like to thank Justin Berman, Wentao Cui, Elliott Gesteau, Netta Engelhardt, David Kolchmeyer, Sam Leutheusser, and Manu Srivastava for collaborations and many discussions. We also thank  David Kolchmeyer, Nima Lashkari, Sam Leutheusser, 
Jon Sorce,  Edward Witten, and in particular Justin Berman for comments on a draft and helpful suggestions, and Thomas Faulkner, Ping Gao, Antony Speranza, and Jakob Ygnason for clarifications. This work is supported by the Office of High Energy Physics of U.S. Department of Energy under grant Contract Number  DE-SC0012567 and DE-SC0020360 (MIT contract \# 578218), and was made possible through the support of grant \#63670 from the John Templeton Foundation.

\appendix

\section{Examples of infinite entanglement with factorizable Hilbert space} \label{app:counter}

Here we give two simple examples of factorized Hilbert space with states of infinite entanglement. 

The first example involves two entangled harmonic oscillators (denoted by $L$ and $R$) in the state\footnote{We thank Antony Speranza for suggestion and discussion on this example.} 
\be \label{cosS} 
\ket{\Psi} = \sum_{n=0}^\infty a_n \ket{n}_L \otimes \ket{n}_R, \quad |a_n|^2 \sim  {1 \ov n \log^2 n}, \; \; n \to \infty \ .
\ee 
With such a falloff, $\ket{\Psi}$ is normalizable, and so is the reduced density operator $\rho_R$ for $R$. 
The entanglement entropy 
\be 
S_R = - \Tr \rho_R \log \rho_R = - \sum_{n=1}^\infty |a_n|^2 \log |a_n|^2 
\sim \sum_{n \to \infty} {1 \ov n \log n} = \infty \ .
\ee
As expected, such a state $\ket{\Psi}$ has an infinite energy ($H_R$ is the Hamiltonian of the $R$-oscillator)
\be 
E_R = \vev{\Psi|H_R|\Psi} \propto \sum_{n=0}^\infty n |a_n|^2 \sim \sum_{n \to \infty} {1 \ov \log^2 n} = \infty  \ .
\ee
The physical origin of infinite entanglement in this example---arising from a slow tail---is clearly very different from that in Sec.~\ref{sec:inf}, which results from an infinite number of degrees of freedom.

The second example involves the entangled spin system discussed in Sec.~\ref{sec:inf} and~\ref{sec:ensR}, by allowing the $i$-th pair to have an $i$-dependent $\th_i$. It can be shown that the algebra $\sM_R$ is type I if $\th_i$ satisfy the condition~\cite{AraWoo68}, 
\be \label{tyIC}
\sum_{i=1}^\infty \lam_i < \infty , \quad \lam_i = \sin^2 \th_i  \ .
\ee 
Since $\sM_R$ is type I, the Hilbert space can be factorized. The total entanglement entropy between $L$ and $R$ is given by 
\be 
S = - \sum_{i=1}^\infty (\lam_i \log \lam_i + (1-\lam_i) \log (1-\lam_i)< - \sum_{i=1}^\infty \lam_i \log \lam_i \ .
\ee
Choosing $\lam_i \sim {1 \ov i \log^2 i}$ for large $i$ as in~\eqref{cosS}, equation~\eqref{tyIC} is satisfied, but $S =\infty$.

\section{Calculation of modular operator} \label{app:mod}


Here we give an explicit derivation that~\eqref{neso1} solves equation~\eqref{neso}. We will be slightly more general, considering  $\ket{\hat \Psi}$ to have the form 
\be 
\ket{\hat \Psi}= \ket{\Psi} \otimes g^\ha (q) , \quad \Phi \in \sH, 
\ee
where $g^\ha (q)$ is a real coordinate space wave function in $L^2 (\RR)$. For~\eqref{haSa}, $g(q) =1$. 

It is convenient to use the form~\eqref{n1}, and 
we can take $\hat A, \hat B$ to be basis vectors $\hat A = A e^{i s (K - \hat q )}, \hat B = B e^{i t (K- \hat q)}$ and the above equation becomes 
\be 
\vev{\hat  \Psi|A e^{i s (K -\hat q )} B e^{i t (K - \hat q)}|\hat \Psi} 
= \vev{\hat \Psi|B e^{i t (K- \hat q)} \hat \De A e^{i s (K - \hat q )} |\hat  \Psi} 
\ee
which after using $K \ket{\Psi} = 0$ can be written as 
\be 
\vev{\hat \Psi|A e^{i s K} B e^{-i (t +s) \hat q}|\hat  \Psi} 
= \vev{\hat  \Psi|B e^{i t (K-\hat q)} \hat \De A e^{- i s  \hat q} |\hat  \Psi}  \ .
\ee
The LHS then can be written as (below $B_s \equiv e^{is K} B e^{-is K}$)
\bea 
{\rm LHS} &=& \int {dq} \, g(q) e^{-i (t +s) q} \vev{\Psi| A e^{is K} B|\Psi} 
= \tilde g (t+s) \vev{\Psi| A B_{s} |\Psi}  \cr
&=& \tilde g (t+s) \vev{\Psi |B_{s} \De_\Psi A|\Psi} 
= \tilde g (t+s) \vev{\Psi |B  \De_\Psi e^{- i s K } A |\Psi}  ,
\label{khs0}
\eea
where we have used the KMS relation~\eqref{kmsm}, and  
\be 
\tilde g (s) = \int {dq } \, g (q) e^{-i s q}, \quad g(q) = \int {ds \ov 2 \pi} \, \tilde g(s) e^{i sp} \ .
\ee
The RHS can be written as 
\bega 
{\rm RHS} =\int {d q}  \int {d q'}  g^\ha (q)  g^\ha (q') e^{- i t q - i s q'} 
\vev{\Psi, q|B e^{i t K} \hat \De A  |\Psi, q'} 
\label{khs1}
\end{gather} 
where $\ket{\Psi,q} \equiv \ket{\Psi} \otimes \ket{q}$. 

Comparing~\eqref{khs0} and~\eqref{khs1}, we consider the following ansatz 
\be
\hat \De= \De_\Psi \, g (\hat q- K) g^{-1} (\hat q) = {\De_\Psi \ov g(\hat q)} \int {ds' \ov 2 \pi} \, e^{-i s' (K-\hat q)} \tilde g(s')  
\ee
which gives 
\bea
{\rm RHS} &=& \int d q  \int {ds' \ov 2 \pi} \, \tilde g(s') e^{- i q (t+s) + i q s'} \vev{\Psi|B e^{i (t -s')K } \De A  |\Psi} \cr
&=& \tilde g(s+t) \vev{\Psi|B  \De_\Psi e^{- i s K } A  |\Psi} = {\rm LHS}  \ . 
\eea

$\hat \De$ can be written in a factorized form 
\bega
\hat \De = e^{-K} g (\hat q- K) g^{-1} (\hat q) 
= {e^{-\hat q} \ov g (\hat q) }  e^{\hat q -K} g (\hat q- K) = \rho'^{-1}  \rho , \\
\rho =  e^{\hat q -K} g (\hat q- K)  , \quad  \rho'  = g (\hat q) e^{\hat q}   \ .
\end{gather}

For the special case $g=1$,  we have 
\be
\ket{\hat \Psi } = \ket{\Psi} \otimes \ket{p=0}  , \quad \hat \De = e^{-K}  = \De, \quad \rho = e^{\hat q -K} , \quad  \rho' = e^{\hat q}  \ .
\ee

\bibliographystyle{jhep}
\bibliography{all}

\end{document}

%% file: HLdefs.tex
\usepackage[usenames, dvipsnames]{color}

\newcommand{\hl}[1]{\color{magenta}}

\usepackage[
	  pagebackref=false,
	  colorlinks=true,
      linkcolor=blue,
      urlcolor=blue,
      filecolor=black,
      citecolor=red,
      pdfstartview=FitV,
      pdftitle={},
        pdfauthor={},
        pdfsubject={},
        pdfkeywords={},
        pdfpagemode=None,
        bookmarksopen=true
      ]{hyperref}

\usepackage[T1]{fontenc}
\usepackage[latin9]{inputenc}
\usepackage{textcomp}
\usepackage{slashed} 
 \usepackage{xcolor}

\usepackage[normalem]{ulem}
\usepackage{amsmath}
\usepackage{enumerate}
\usepackage{amsfonts}
\usepackage{epsfig}
\usepackage{mathbbol}
\usepackage{cancel}
\usepackage{mathrsfs}
\usepackage{geometry}
\usepackage{esint}
\usepackage{amsthm}
\usepackage{amssymb}

\usepackage{array}
\usepackage{float}
\usepackage{multirow}

\usepackage{bm}

\def\Tr{\mathop{\rm Tr}}
\def\tr{\mathop{\rm tr}}

\newcommand\half{{\ensuremath{\frac{1}{2}}}}
\newcommand\p{\ensuremath{\partial}}

\newcommand\field[1]{{\ensuremath{\mathbb{{#1}}}}}

\newcommand\vev[1]{{\ensuremath{\left\langle{#1}\right\rangle}}}

\newcommand\ket[1]{\ensuremath{\lvert{#1}\rangle}}
\newcommand\bra[1]{\ensuremath{\langle{#1}\rvert}}

\newcommand{\CC}{\field{C}}

\newcommand{\NN}{\field{N}}

\newcommand{\RR}{\field{R}}

\newcommand{\ZZ}{\field{Z}}

\newcommand{\be}{\begin{equation}}
\newcommand{\ee}{\end{equation}}
\newcommand{\bea}{\begin{eqnarray}}
\newcommand{\eea}{\end{eqnarray}}
\newcommand{\bega}{\begin{gather}}
\newcommand{\eega}{\end{gather}}

\newcommand{\bi}{\begin{itemize}}
\newcommand{\ei}{\end{itemize}}
\newcommand{\ben}{\begin{enumerate}}
\newcommand{\een}{\end{enumerate}}
\newcommand{\bca}{\begin{cases}}
\newcommand{\eca}{\end{cases}}
\newcommand{\bln}{\begin{align}}
\newcommand{\eln}{\end{align}}
\newcommand{\bst}{\begin{split}}
\newcommand{\est}{\end{split}}
\def\ie{\begin{equation}\begin{aligned}}
\def\fe{\end{aligned}\end{equation}}
\newcommand{\bma}{\le(\begin{matrix}}
\newcommand{\ema}{\end{matrix}\ri)}
\newcommand{\bfl}{\noindent\begin{flushleft}}
\newcommand{\efl}{\par\end{flushleft}}

\newcommand\al{{\alpha}}
\newcommand\ep{\epsilon}
\newcommand\sig{\sigma}
\newcommand\Sig{\Sigma}
\newcommand\lam{\lambda}
\newcommand\Lam{\Lambda}
\newcommand\om{\omega}
\newcommand\Om{\Omega}

\newcommand\ga{{\ensuremath{{\gamma}}}}
\newcommand\Ga{{\ensuremath{{\Gamma}}}}
\newcommand\de{{\ensuremath{{\delta}}}}
\newcommand\De{{\ensuremath{{\Delta}}}}
\newcommand\vp{\varphi}
\newcommand\ka{\kappa}

\newcommand\da{{\dagger}}
\newcommand\nab{{\nabla}}
\newcommand\Th{{\Theta}}
\def\th{{\theta}}

\newcommand{\lra}{{\leftrightarrow}}

\newcommand\Ra{{\Rightarrow}}

\newcommand\ov{\over}
\newcommand\ha{{\half}}

\newcommand\apr{{\ensuremath{{\alpha'}}}}
\def\le{\left}
\def\ri{\right}

\newcommand\sA{{\ensuremath{{\mathcal A}}}}
\newcommand\sB{{\ensuremath{{\mathcal B}}}}
\newcommand\sC{{\ensuremath{{\mathcal C}}}}
\newcommand\sD{{\ensuremath{{\mathcal D}}}}

\newcommand\sH{{\ensuremath{{\mathcal H}}}}
\newcommand\sI{{\ensuremath{{\mathcal I}}}}
\newcommand\sJ{{\mathcal J}}
\newcommand\sK{{\ensuremath{{\mathcal K}}}}
\newcommand\sL{{\ensuremath{{\mathcal L}}}}
\newcommand\sM{{\ensuremath{{\mathcal M}}}}
\newcommand\sN{{\ensuremath{{\mathcal N}}}}
\newcommand\sO{{\ensuremath{{\mathcal O}}}}
\newcommand\sP{{\ensuremath{{\mathcal P}}}}

\newcommand\sR{{\mathcal R}}
\newcommand\sS{{\mathcal S}}
\newcommand\sT{{\mathcal T}}

\newcommand\sW{{\mathcal W}}
\newcommand\sX{{\mathcal X}}
\newcommand\sY{{\mathcal Y}}
\newcommand\sZ{{\mathcal Z}}

\newcommand\vx{{\vec x}}
\newcommand\vk{{\vec k}}

\newcommand{\fa}{{\mathfrak{a}}}

\newcommand{\fb}{{\mathfrak{b}}}

\newcommand{\fo}{{\mathfrak o}}

\newcommand{\rt}{{\rm t}}

\newcommand{\oA}{{\overline{A}}}

\global\long\def\b{\beta}

\newcommand{\wh}{\widehat}

\newcommand{\bid}{{\mathbf 1}}

%% file: lectures6.bbl
\providecommand{\href}[2]{#2}\begingroup\raggedright\begin{thebibliography}{100}

\bibitem{EinPod35}
A.~Einstein, B.~Podolsky, and N.~Rosen, {\it Can quantum-mechanical description
  of physical reality be considered complete?},  {\em Phys. Rev.} {\bf 47}
  (May, 1935) 777--780.

\bibitem{Sch35}
E.~Schrödinger, {\it Discussion of probability relations between separated
  systems},  {\em Mathematical Proceedings of the Cambridge Philosophical
  Society} {\bf 31} (1935), no.~4 555--563.

\bibitem{Mal97}
J.~Maldacena, {\it The large {$N$} limit of superconformal field theories and
  supergravity},  {\em Adv. Theor. Math. Phys.} {\bf 2} (1998) 231,
  [\href{http://arxiv.org/abs/hep-th/9711200}{{\tt hep-th/9711200}}].

\bibitem{GubKle98}
S.~S. Gubser, I.~R. Klebanov, and A.~M. Polyakov, {\it Gauge theory correlators
  from noncritical string theory},  {\em Phys. Lett.} {\bf B428} (1998) 105,
  [\href{http://arxiv.org/abs/hep-th/9802109}{{\tt hep-th/9802109}}].

\bibitem{Wit98}
E.~Witten, {\it {Anti-de Sitter space and holography}},  {\em Adv. Theor. Math.
  Phys.} {\bf 2} (1998) 253--291,
  [\href{http://arxiv.org/abs/hep-th/9802150}{{\tt hep-th/9802150}}].

\bibitem{RyuTak06}
S.~Ryu and T.~Takayanagi, {\it {Holographic derivation of entanglement entropy
  from AdS/CFT}},  {\em Phys.Rev.Lett.} {\bf 96} (2006) 181602,
  [\href{http://arxiv.org/abs/hep-th/0603001}{{\tt hep-th/0603001}}].

\bibitem{RyuTak06-2}
S.~Ryu and T.~Takayanagi, {\it {Aspects of Holographic Entanglement Entropy}},
  {\em JHEP} {\bf 0608} (2006) 045,
  [\href{http://arxiv.org/abs/hep-th/0605073}{{\tt hep-th/0605073}}].

\bibitem{HubRan07}
V.~E. Hubeny, M.~Rangamani, and T.~Takayanagi, {\it {A Covariant holographic
  entanglement entropy proposal}},  {\em JHEP} {\bf 0707} (2007) 062,
  [\href{http://arxiv.org/abs/0705.0016}{{\tt arXiv:0705.0016}}].

\bibitem{Wal11}
A.~C. Wall, {\it {A proof of the generalized second law for rapidly changing
  fields and arbitrary horizon slices}},  {\em Phys.Rev.} {\bf D85} (2012),
  no.~6 104049, [\href{http://arxiv.org/abs/1105.3445}{{\tt arXiv:1105.3445}}].

\bibitem{EngWal14}
N.~Engelhardt and A.~C. Wall, {\it {Quantum Extremal Surfaces: Holographic
  Entanglement Entropy beyond the Classical Regime}},  {\em JHEP} {\bf 01}
  (2015) 073, [\href{http://arxiv.org/abs/1408.3203}{{\tt arXiv:1408.3203}}].

\bibitem{LewMal13}
A.~Lewkowycz and J.~Maldacena, {\it {Generalized gravitational entropy}},  {\em
  JHEP} {\bf 1308} (2013) 090, [\href{http://arxiv.org/abs/1304.4926}{{\tt
  arXiv:1304.4926}}].

\bibitem{FauLew13}
T.~Faulkner, A.~Lewkowycz, and J.~Maldacena, {\it {Quantum corrections to
  holographic entanglement entropy}},  {\em JHEP} {\bf 1311} (2013) 074,
  [\href{http://arxiv.org/abs/1307.2892}{{\tt arXiv:1307.2892}}].

\bibitem{LeuLiu21a}
S.~Leutheusser and H.~Liu, {\it {Causal connectability between quantum systems
  and the black hole interior in holographic duality}},
  \href{http://arxiv.org/abs/2110.05497}{{\tt arXiv:2110.05497}}.

\bibitem{LeuLiu21b}
S.~Leutheusser and H.~Liu, {\it {Emergent times in holographic duality}},
  \href{http://arxiv.org/abs/2112.12156}{{\tt arXiv:2112.12156}}.

\bibitem{LeuLiu22}
S.~Leutheusser and H.~Liu, {\it {Subalgebra-subregion duality: emergence of
  space and time in holography}},  \href{http://arxiv.org/abs/2212.13266}{{\tt
  arXiv:2212.13266}}.

\bibitem{Neu30}
J.~von Neumann, {\it Zur algebra der funktionaloperationen und theorie der
  normalen operatoren},  {\em Mathematische Annalen} {\bf 102} (1930) 370--427.

\bibitem{Neu36}
J.~v.~Neumann, {\it On a certain topology for rings of operators},  {\em Annals
  of Mathematics} {\bf 37} (1936), no.~1 111--115.

\bibitem{MurNeu36}
F.~J. Murray and J.~v.~Neumann, {\it On rings of operators},  {\em Annals of
  Mathematics} {\bf 37} (1936), no.~1 116--229.

\bibitem{MurNeu37}
F.~J. Murray and J.~von Neumann, {\it On rings of operators. ii},  {\em
  Transactions of the American Mathematical Society} {\bf 41} (1937), no.~2
  208--248.

\bibitem{GesLiu24}
E.~Gesteau and H.~Liu, {\it {Toward stringy horizons}},
  \href{http://arxiv.org/abs/2408.12642}{{\tt arXiv:2408.12642}}.

\bibitem{HerKud25}
A.~Herderschee and J.~Kudler-Flam, {\it {Stringy algebras, stretched horizons,
  and quantum-connected wormholes}},
  \href{http://arxiv.org/abs/2510.01556}{{\tt arXiv:2510.01556}}.

\bibitem{EngLiu23}
N.~Engelhardt and H.~Liu, {\it {Algebraic ER=EPR and Complexity Transfer}},
  \href{http://arxiv.org/abs/2311.04281}{{\tt arXiv:2311.04281}}.

\bibitem{Van09}
M.~Van~Raamsdonk, {\it {Comments on quantum gravity and entanglement}},
  \href{http://arxiv.org/abs/0907.2939}{{\tt arXiv:0907.2939}}.

\bibitem{CzeKar12}
B.~Czech, J.~L. Karczmarek, F.~Nogueira, and M.~Van~Raamsdonk, {\it {The
  Gravity Dual of a Density Matrix}},  {\em Class.Quant.Grav.} {\bf 29} (2012)
  155009, [\href{http://arxiv.org/abs/1204.1330}{{\tt arXiv:1204.1330}}].

\bibitem{CzeKar12b}
B.~Czech, J.~L. Karczmarek, F.~Nogueira, and M.~Van~Raamsdonk, {\it {Rindler
  Quantum Gravity}},  {\em Class. Quant. Grav.} {\bf 29} (2012) 235025,
  [\href{http://arxiv.org/abs/1206.1323}{{\tt arXiv:1206.1323}}].

\bibitem{Wal12}
A.~C. Wall, {\it {Maximin Surfaces, and the Strong Subadditivity of the
  Covariant Holographic Entanglement Entropy}},  {\em Class.Quant.Grav.} {\bf
  31} (2014), no.~22 225007, [\href{http://arxiv.org/abs/1211.3494}{{\tt
  arXiv:1211.3494}}].

\bibitem{HeaHub14}
M.~Headrick, V.~E. Hubeny, A.~Lawrence, and M.~Rangamani, {\it {Causality \&
  holographic entanglement entropy}},  {\em JHEP} {\bf 12} (2014) 162,
  [\href{http://arxiv.org/abs/1408.6300}{{\tt arXiv:1408.6300}}].

\bibitem{AlmDon14}
A.~Almheiri, X.~Dong, and D.~Harlow, {\it {Bulk Locality and Quantum Error
  Correction in AdS/CFT}},  {\em JHEP} {\bf 04} (2015) 163,
  [\href{http://arxiv.org/abs/1411.7041}{{\tt arXiv:1411.7041}}].

\bibitem{JafSuh14}
D.~L. Jafferis and S.~J. Suh, {\it {The Gravity Duals of Modular
  Hamiltonians}},  \href{http://arxiv.org/abs/1412.8465}{{\tt
  arXiv:1412.8465}}.

\bibitem{PasYos15}
F.~Pastawski, B.~Yoshida, D.~Harlow, and J.~Preskill, {\it {Holographic quantum
  error-correcting codes: Toy models for the bulk/boundary correspondence}},
  {\em JHEP} {\bf 06} (2015) 149, [\href{http://arxiv.org/abs/1503.06237}{{\tt
  arXiv:1503.06237}}].

\bibitem{JafLew15}
D.~L. Jafferis, A.~Lewkowycz, J.~Maldacena, and S.~J. Suh, {\it {Relative
  entropy equals bulk relative entropy}},
  \href{http://arxiv.org/abs/1512.06431}{{\tt arXiv:1512.06431}}.

\bibitem{DonHar16}
X.~Dong, D.~Harlow, and A.~C. Wall, {\it {Reconstruction of Bulk Operators
  within the Entanglement Wedge in Gauge-Gravity Duality}},  {\em Phys. Rev.
  Lett.} {\bf 117} (2016), no.~2 021601,
  [\href{http://arxiv.org/abs/1601.05416}{{\tt arXiv:1601.05416}}].

\bibitem{Har16}
D.~Harlow, {\it {The Ryu-Takayanagi Formula from Quantum Error Correction}},
  {\em Commun. Math. Phys.} {\bf 354} (2017), no.~3 865--912,
  [\href{http://arxiv.org/abs/1607.03901}{{\tt arXiv:1607.03901}}].

\bibitem{FauLew17}
T.~Faulkner and A.~Lewkowycz, {\it {Bulk locality from modular flow}},  {\em
  JHEP} {\bf 07} (2017) 151, [\href{http://arxiv.org/abs/1704.05464}{{\tt
  arXiv:1704.05464}}].

\bibitem{CotHay17}
J.~Cotler, P.~Hayden, G.~Penington, G.~Salton, B.~Swingle, and M.~Walter, {\it
  {Entanglement Wedge Reconstruction via Universal Recovery Channels}},  {\em
  Phys. Rev. X} {\bf 9} (2019), no.~3 031011,
  [\href{http://arxiv.org/abs/1704.05839}{{\tt arXiv:1704.05839}}].

\bibitem{BanDou98}
T.~Banks, M.~R. Douglas, G.~T. Horowitz, and E.~J. Martinec, {\it {AdS dynamics
  from conformal field theory}},
  \href{http://arxiv.org/abs/hep-th/9808016}{{\tt hep-th/9808016}}.

\bibitem{Ben99}
I.~Bena, {\it {On the construction of local fields in the bulk of AdS(5) and
  other spaces}},  {\em Phys. Rev.} {\bf D62} (2000) 066007,
  [\href{http://arxiv.org/abs/hep-th/9905186}{{\tt hep-th/9905186}}].

\bibitem{BalKraLaw98}
V.~Balasubramanian, P.~Kraus, and A.~E. Lawrence, {\it {Bulk versus boundary
  dynamics in anti-de Sitter space-time}},  {\em Phys. Rev.} {\bf D59} (1999)
  046003, [\href{http://arxiv.org/abs/hep-th/9805171}{{\tt hep-th/9805171}}].

\bibitem{HamKab05}
A.~Hamilton, D.~N. Kabat, G.~Lifschytz, and D.~A. Lowe, {\it {Local bulk
  operators in AdS/CFT: A Boundary view of horizons and locality}},  {\em
  Phys.Rev.} {\bf D73} (2006) 086003,
  [\href{http://arxiv.org/abs/hep-th/0506118}{{\tt hep-th/0506118}}].

\bibitem{HamKab06}
A.~Hamilton, D.~N. Kabat, G.~Lifschytz, and D.~A. Lowe, {\it {Holographic
  representation of local bulk operators}},  {\em Phys.Rev.} {\bf D74} (2006)
  066009, [\href{http://arxiv.org/abs/hep-th/0606141}{{\tt hep-th/0606141}}].

\bibitem{KabLif11}
D.~Kabat, G.~Lifschytz, and D.~A. Lowe, {\it {Constructing local bulk
  observables in interacting AdS/CFT}},  {\em Phys. Rev. D} {\bf 83} (2011)
  106009, [\href{http://arxiv.org/abs/1102.2910}{{\tt arXiv:1102.2910}}].

\bibitem{Hee12}
I.~Heemskerk, {\it {Construction of Bulk Fields with Gauge Redundancy}},  {\em
  JHEP} {\bf 09} (2012) 106, [\href{http://arxiv.org/abs/1201.3666}{{\tt
  arXiv:1201.3666}}].

\bibitem{HeeMar12}
I.~Heemskerk, D.~Marolf, J.~Polchinski, and J.~Sully, {\it {Bulk and
  Transhorizon Measurements in AdS/CFT}},  {\em JHEP} {\bf 10} (2012) 165,
  [\href{http://arxiv.org/abs/1201.3664}{{\tt arXiv:1201.3664}}].

\bibitem{PapRaj12}
K.~Papadodimas and S.~Raju, {\it {An Infalling Observer in AdS/CFT}},
  \href{http://arxiv.org/abs/1211.6767}{{\tt arXiv:1211.6767}}.

\bibitem{Mor14}
I.~A. Morrison, {\it Boundary-to-bulk maps for ads causal wedges and the
  reeh-schlieder property in holography},  {\em Journal of High Energy Physics}
  {\bf 2014} (May, 2014).

\bibitem{Hub14}
V.~E. Hubeny, {\it {Covariant Residual Entropy}},  {\em JHEP} {\bf 09} (2014)
  156, [\href{http://arxiv.org/abs/1406.4611}{{\tt arXiv:1406.4611}}].

\bibitem{EngPen21a}
N.~Engelhardt, G.~Penington, and A.~Shahbazi-Moghaddam, {\it {A World without
  Pythons would be so Simple}},  \href{http://arxiv.org/abs/2102.07774}{{\tt
  arXiv:2102.07774}}.

\bibitem{Wit23}
E.~Witten, {\it {Algebras, Regions, and Observers}},
  \href{http://arxiv.org/abs/2303.02837}{{\tt arXiv:2303.02837}}.

\bibitem{EngLiu25}
N.~Engelhardt and H.~Liu, {\it {The Making of von Neumann Algebras from Bulk
  Focusing}},  \href{http://arxiv.org/abs/2509.05413}{{\tt arXiv:2509.05413}}.

\bibitem{LeuLiu24}
S.~Leutheusser and H.~Liu, {\it {Superadditivity in large $N$ field theories
  and performance of quantum tasks}},
  \href{http://arxiv.org/abs/2411.04183}{{\tt arXiv:2411.04183}}.

\bibitem{AlmMah19a}
A.~Almheiri, R.~Mahajan, J.~Maldacena, and Y.~Zhao, {\it {The Page curve of
  Hawking radiation from semiclassical geometry}},  {\em JHEP} {\bf 03} (2020)
  149, [\href{http://arxiv.org/abs/1908.10996}{{\tt arXiv:1908.10996}}].

\bibitem{Con95}
A.~Connes, {\em Noncommutative Geometry}.
\newblock Elsevier Science, 1995.

\bibitem{KanKol18}
M.~J. Kang and D.~K. Kolchmeyer, {\it {Holographic Relative Entropy in
  Infinite-Dimensional Hilbert Spaces}},  {\em Commun. Math. Phys.} {\bf 400}
  (2023), no.~3 1665--1695, [\href{http://arxiv.org/abs/1811.05482}{{\tt
  arXiv:1811.05482}}].

\bibitem{Fau20}
T.~Faulkner, {\it {The holographic map as a conditional expectation}},
  \href{http://arxiv.org/abs/2008.04810}{{\tt arXiv:2008.04810}}.

\bibitem{KanKol21}
M.~J. Kang and D.~K. Kolchmeyer, {\it {Entanglement wedge reconstruction of
  infinite-dimensional von Neumann algebras using tensor networks}},  {\em
  Phys. Rev. D} {\bf 103} (2021), no.~12 126018,
  [\href{http://arxiv.org/abs/1910.06328}{{\tt arXiv:1910.06328}}].

\bibitem{GesKan21}
E.~Gesteau and M.~J. Kang, {\it {Nonperturbative gravity corrections to bulk
  reconstruction}},  {\em J. Phys. A} {\bf 56} (2023), no.~38 385401,
  [\href{http://arxiv.org/abs/2112.12789}{{\tt arXiv:2112.12789}}].

\bibitem{FauLi22}
T.~Faulkner and M.~Li, {\it {Asymptotically isometric codes for holography}},
  \href{http://arxiv.org/abs/2211.12439}{{\tt arXiv:2211.12439}}.

\bibitem{PenWit23}
G.~Penington and E.~Witten, {\it {Algebras and States in JT Gravity}},
  \href{http://arxiv.org/abs/2301.07257}{{\tt arXiv:2301.07257}}.

\bibitem{Kol23}
D.~K. Kolchmeyer, {\it {von Neumann algebras in JT gravity}},  {\em JHEP} {\bf
  06} (2023) 067, [\href{http://arxiv.org/abs/2303.04701}{{\tt
  arXiv:2303.04701}}].

\bibitem{Lin22}
H.~W. Lin, {\it {The bulk Hilbert space of double scaled SYK}},  {\em JHEP}
  {\bf 11} (2022) 060, [\href{http://arxiv.org/abs/2208.07032}{{\tt
  arXiv:2208.07032}}].

\bibitem{Gao24}
P.~Gao, {\it {Modular flow in JT gravity and entanglement wedge
  reconstruction}},  \href{http://arxiv.org/abs/2402.18655}{{\tt
  arXiv:2402.18655}}.

\bibitem{PenWit24}
G.~Penington and E.~Witten, {\it {Algebras and states in super-JT gravity}},
  \href{http://arxiv.org/abs/2412.15549}{{\tt arXiv:2412.15549}}.

\bibitem{Ban25j}
T.~Banks, {\it {Note on Type $III_1$ Algebras in $ c= 1$ String Theory and Bulk
  Causal Diamonds}},  \href{http://arxiv.org/abs/2504.15076}{{\tt
  arXiv:2504.15076}}.

\bibitem{Tei83}
C.~Teitelboim, {\it {Gravitation and Hamiltonian Structure in Two Space-Time
  Dimensions}},  {\em Phys. Lett. B} {\bf 126} (1983) 41--45.

\bibitem{Jackiw}
R.~Jackiw, {\it {Lower Dimensional Gravity}},  {\em Nucl. Phys. B} {\bf 252}
  (1985) 343--356.

\bibitem{AlmPol14}
A.~Almheiri and J.~Polchinski, {\it {Models of AdS$_{2}$ backreaction and
  holography}},  {\em JHEP} {\bf 11} (2015) 014,
  [\href{http://arxiv.org/abs/1402.6334}{{\tt arXiv:1402.6334}}].

\bibitem{Wit21b}
E.~Witten, {\it {Gravity and the crossed product}},  {\em JHEP} {\bf 10} (2022)
  008, [\href{http://arxiv.org/abs/2112.12828}{{\tt arXiv:2112.12828}}].

\bibitem{ChaLon22}
V.~Chandrasekaran, R.~Longo, G.~Penington, and E.~Witten, {\it {An algebra of
  observables for de Sitter space}},  {\em JHEP} {\bf 02} (2023) 082,
  [\href{http://arxiv.org/abs/2206.10780}{{\tt arXiv:2206.10780}}].

\bibitem{Wit23b}
E.~Witten, {\it {A background-independent algebra in quantum gravity}},  {\em
  JHEP} {\bf 03} (2024) 077, [\href{http://arxiv.org/abs/2308.03663}{{\tt
  arXiv:2308.03663}}].

\bibitem{KolLiu24}
D.~K. Kolchmeyer and H.~Liu, {\it {Chaos and the Emergence of the Cosmological
  Horizon}},  \href{http://arxiv.org/abs/2411.08090}{{\tt arXiv:2411.08090}}.

\bibitem{AliJef23}
S.~Ali~Ahmad and R.~Jefferson, {\it {Crossed product algebras and generalized
  entropy for subregions}},  \href{http://arxiv.org/abs/2306.07323}{{\tt
  arXiv:2306.07323}}.

\bibitem{KliLei23}
M.~S. Klinger and R.~G. Leigh, {\it {Crossed Products, Extended Phase Spaces
  and the Resolution of Entanglement Singularities}},
  \href{http://arxiv.org/abs/2306.09314}{{\tt arXiv:2306.09314}}.

\bibitem{KliLei23a}
M.~S. Klinger and R.~G. Leigh, {\it {Crossed Products, Conditional Expectations
  and Constraint Quantization}},  \href{http://arxiv.org/abs/2312.16678}{{\tt
  arXiv:2312.16678}}.

\bibitem{AliChe24a}
S.~Ali~Ahmad, W.~Chemissany, M.~S. Klinger, and R.~G. Leigh, {\it {Quantum
  reference frames from top-down crossed products}},  {\em Phys. Rev. D} {\bf
  110} (2024), no.~6 065003, [\href{http://arxiv.org/abs/2405.13884}{{\tt
  arXiv:2405.13884}}].

\bibitem{AliChe24}
S.~Ali~Ahmad, W.~Chemissany, M.~S. Klinger, and R.~G. Leigh, {\it {Relational
  quantum geometry}},  {\em Nucl. Phys. B} {\bf 1015} (2025) 116911,
  [\href{http://arxiv.org/abs/2410.11029}{{\tt arXiv:2410.11029}}].

\bibitem{AliKli24}
S.~Ali~Ahmad, M.~S. Klinger, and S.~Lin, {\it {Semifinite von Neumann algebras
  in gauge theory and gravity}},  {\em Phys. Rev. D} {\bf 111} (2025), no.~4
  045006, [\href{http://arxiv.org/abs/2407.01695}{{\tt arXiv:2407.01695}}].

\bibitem{Gom22}
C.~Gomez, {\it {Cosmology as a Crossed Product}},
  \href{http://arxiv.org/abs/2207.06704}{{\tt arXiv:2207.06704}}.

\bibitem{Gom23}
C.~Gomez, {\it {Entanglement, Observers and Cosmology: a view from von Neumann
  Algebras}},  \href{http://arxiv.org/abs/2302.14747}{{\tt arXiv:2302.14747}}.

\bibitem{Gom23a}
C.~Gomez, {\it {Traces and Time: a de Sitter Black Hole correspondence}},
  \href{http://arxiv.org/abs/2307.01841}{{\tt arXiv:2307.01841}}.

\bibitem{Gom23c}
C.~Gomez, {\it {On the algebraic meaning of quantum gravity for closed
  Universes}},  \href{http://arxiv.org/abs/2311.01952}{{\tt arXiv:2311.01952}}.

\bibitem{Fio23}
B.~Fiol, {\it {From von Neumann algebras to quantum de Sitter}},  {\em PoS}
  {\bf CORFU2022} (2023) 133.

\bibitem{MerTap25}
T.~G. Mertens, T.~Tappeiner, and B.~de~S.~L.~Torres, {\it {Fiducial observers
  and the thermal atmosphere in the black hole quantum throat}},
  \href{http://arxiv.org/abs/2507.20983}{{\tt arXiv:2507.20983}}.

\bibitem{ChaPen22}
V.~Chandrasekaran, G.~Penington, and E.~Witten, {\it {Large N algebras and
  generalized entropy}},  \href{http://arxiv.org/abs/2209.10454}{{\tt
  arXiv:2209.10454}}.

\bibitem{JenSor23}
K.~Jensen, J.~Sorce, and A.~Speranza, {\it {Generalized entropy for general
  subregions in quantum gravity}},  \href{http://arxiv.org/abs/2306.01837}{{\tt
  arXiv:2306.01837}}.

\bibitem{KudLeu23}
J.~Kudler-Flam, S.~Leutheusser, and G.~Satishchandran, {\it {Generalized Black
  Hole Entropy is von Neumann Entropy}},
  \href{http://arxiv.org/abs/2309.15897}{{\tt arXiv:2309.15897}}.

\bibitem{KudLeu24}
J.~Kudler-Flam, S.~Leutheusser, and G.~Satishchandran, {\it {Algebraic
  Observational Cosmology}},  \href{http://arxiv.org/abs/2406.01669}{{\tt
  arXiv:2406.01669}}.

\bibitem{FewJan24}
J.~C. Fewster, D.~W. Janssen, L.~D. Loveridge, K.~Rejzner, and J.~Waldron, {\it
  {Quantum Reference Frames, Measurement Schemes and the Type of Local Algebras
  in Quantum Field Theory}},  {\em Commun. Math. Phys.} {\bf 406} (2025), no.~1
  19, [\href{http://arxiv.org/abs/2403.11973}{{\tt arXiv:2403.11973}}].

\bibitem{vanVer24}
J.~van~der Heijden and E.~Verlinde, {\it {An operator algebraic approach to
  black hole information}},  {\em JHEP} {\bf 02} (2025) 207,
  [\href{http://arxiv.org/abs/2408.00071}{{\tt arXiv:2408.00071}}].

\bibitem{DeVEcc24b}
J.~De~Vuyst, S.~Eccles, P.~A. Hoehn, and J.~Kirklin, {\it {Gravitational
  entropy is observer-dependent}},  {\em JHEP} {\bf 07} (2025) 146,
  [\href{http://arxiv.org/abs/2405.00114}{{\tt arXiv:2405.00114}}].

\bibitem{DeVEcc24}
J.~De~Vuyst, S.~Eccles, P.~A. Hoehn, and J.~Kirklin, {\it {Linearization
  (in)stabilities and crossed products}},
  \href{http://arxiv.org/abs/2411.19931}{{\tt arXiv:2411.19931}}.

\bibitem{DeVEcc24a}
J.~De~Vuyst, S.~Eccles, P.~A. Hoehn, and J.~Kirklin, {\it {Crossed products and
  quantum reference frames: on the observer-dependence of gravitational
  entropy}},  {\em JHEP} {\bf 07} (2025) 063,
  [\href{http://arxiv.org/abs/2412.15502}{{\tt arXiv:2412.15502}}].

\bibitem{CheJun25}
Y.~Chen, M.~Junge, and N.~Lashkari, {\it {Operator Algebras and Third
  Quantization}},  \href{http://arxiv.org/abs/2509.02293}{{\tt
  arXiv:2509.02293}}.

\bibitem{Liu25}
H.~Liu, {\it {Towards a holographic description of closed universes}},
  \href{http://arxiv.org/abs/2509.14327}{{\tt arXiv:2509.14327}}.

\bibitem{KudWit25}
J.~Kudler-Flam and E.~Witten, {\it Emergent mixed states for baby universes and
  black holes},  \href{http://arxiv.org/abs/to appear}{{\tt to appear}}.

\bibitem{LeuLiu25}
S.~Leutheusser and H.~Liu, {\it {Volume as an index of a subalgebra}},
  \href{http://arxiv.org/abs/2508.00056}{{\tt arXiv:2508.00056}}.

\bibitem{BahBel22}
E.~Bahiru, A.~Belin, K.~Papadodimas, G.~Sarosi, and N.~Vardian, {\it
  {State-dressed local operators in the AdS/CFT correspondence}},  {\em Phys.
  Rev. D} {\bf 108} (2023), no.~8 086035,
  [\href{http://arxiv.org/abs/2209.06845}{{\tt arXiv:2209.06845}}].

\bibitem{NogBan21}
F.~S. Nogueira, S.~Banerjee, M.~Dorband, R.~Meyer, J.~v.~d. Brink, and
  J.~Erdmenger, {\it {Geometric phases distinguish entangled states in wormhole
  quantum mechanics}},  {\em Phys. Rev. D} {\bf 105} (2022), no.~8 L081903,
  [\href{http://arxiv.org/abs/2109.06190}{{\tt arXiv:2109.06190}}].

\bibitem{BanDor22}
S.~Banerjee, M.~Dorband, J.~Erdmenger, R.~Meyer, and A.-L. Weigel, {\it {Berry
  phases, wormholes and factorization in AdS/CFT}},  {\em JHEP} {\bf 08} (2022)
  162, [\href{http://arxiv.org/abs/2202.11717}{{\tt arXiv:2202.11717}}].

\bibitem{BanMor23}
S.~Banerjee, M.~Dorband, J.~Erdmenger, and A.-L. Weigel, {\it {Geometric phases
  characterise operator algebras and missing information}},  {\em JHEP} {\bf
  10} (2023) 026, [\href{http://arxiv.org/abs/2306.00055}{{\tt
  arXiv:2306.00055}}].

\bibitem{BurDas23}
V.~Burman, S.~Das, and C.~Krishnan, {\it {A smooth horizon without a smooth
  horizon}},  {\em JHEP} {\bf 03} (2024) 014,
  [\href{http://arxiv.org/abs/2312.14108}{{\tt arXiv:2312.14108}}].

\bibitem{KriMoh23}
C.~Krishnan and V.~Mohan, {\it {State-independent Black Hole Interiors from the
  Crossed Product}},  \href{http://arxiv.org/abs/2310.05912}{{\tt
  arXiv:2310.05912}}.

\bibitem{BanVos23}
S.~Banerjee and G.~Vos, {\it {Behind-the-horizon excitations from a single 2d
  CFT}},  \href{http://arxiv.org/abs/2401.00890}{{\tt arXiv:2401.00890}}.

\bibitem{JenRaj24}
K.~Jensen, S.~Raju, and A.~J. Speranza, {\it {Holographic observers for
  time-band algebras}},  {\em JHEP} {\bf 06} (2025) 242,
  [\href{http://arxiv.org/abs/2412.21185}{{\tt arXiv:2412.21185}}].

\bibitem{Bah25}
E.~Bahiru, {\it {Algebraic traversable wormholes}},
  \href{http://arxiv.org/abs/2508.13283}{{\tt arXiv:2508.13283}}.

\bibitem{PenTab25}
G.~Penington and E.~Tabor, {\it {The algebraic structure of gravitational
  scrambling}},  \href{http://arxiv.org/abs/2508.21062}{{\tt
  arXiv:2508.21062}}.

\bibitem{Sia25}
B.~S. Sia, {\it {Principle of Diminishing Potentialities in Large N Algebras}},
   \href{http://arxiv.org/abs/2508.11688}{{\tt arXiv:2508.11688}}.

\bibitem{GenJia25}
H.~Geng, Y.~Jiang, and J.~Xu, {\it {Algebras, Entanglement Islands, and
  Observers}},  \href{http://arxiv.org/abs/2506.12127}{{\tt arXiv:2506.12127}}.

\bibitem{deBBah25}
J.~de~Boer, B.~Najian, J.~van~der Heijden, and C.~Zukowski, {\it {Modular
  chaos, operator algebras, and the Berry phase}},  {\em JHEP} {\bf 09} (2025)
  086, [\href{http://arxiv.org/abs/2505.04682}{{\tt arXiv:2505.04682}}].

\bibitem{DiGDor25}
G.~Di~Giulio, M.~Dorband, J.~Erdmenger, and H.~Scheppach, {\it {Modular theory
  and symmetry resolution in hyperfinite von Neumann algebras}},
  \href{http://arxiv.org/abs/2510.02441}{{\tt arXiv:2510.02441}}.

\bibitem{Haa92}
R.~Haag, {\em {Local Quantum Physics: Fields, Particles, Algebras}}.
\newblock Springer Verlag, Berlin, 1992.

\bibitem{Ara99}
H.~Araki, {\em {Quantum Theory}}.
\newblock Oxford University Press, 10, 1999.

\bibitem{Bor00}
H.~J. Borchers, {\it {On revolutionizing quantum field theory with Tomita's
  modular theory}},  {\em J. Math. Phys.} {\bf 41} (2000) 3604--3673.

\bibitem{Yng04}
J.~Yngvason, {\it {The Role of type III factors in quantum field theory}},
  {\em Rept. Math. Phys.} {\bf 55} (2005) 135--147,
  [\href{http://arxiv.org/abs/math-ph/0411058}{{\tt math-ph/0411058}}].

\bibitem{HolWal08}
S.~Hollands and R.~M. Wald, {\it {Axiomatic quantum field theory in curved
  spacetime}},  {\em Commun. Math. Phys.} {\bf 293} (2010) 85--125,
  [\href{http://arxiv.org/abs/0803.2003}{{\tt arXiv:0803.2003}}].

\bibitem{HolWal14}
S.~Hollands and R.~M. Wald, {\it {Quantum fields in curved spacetime}},  {\em
  Phys. Rept.} {\bf 574} (2015) 1--35,
  [\href{http://arxiv.org/abs/1401.2026}{{\tt arXiv:1401.2026}}].

\bibitem{Yng14}
J.~Yngvason, {\it {Localization and Entanglement in Relativistic Quantum
  Physics}},  {\em Lect. Notes Phys.} {\bf 899} (2015) 325--348,
  [\href{http://arxiv.org/abs/1401.2652}{{\tt arXiv:1401.2652}}].

\bibitem{Wit18}
E.~Witten, {\it {APS Medal for Exceptional Achievement in Research: Invited
  article on entanglement properties of quantum field theory}},  {\em Rev. Mod.
  Phys.} {\bf 90} (2018), no.~4 045003,
  [\href{http://arxiv.org/abs/1803.04993}{{\tt arXiv:1803.04993}}].

\bibitem{Wit21a}
E.~Witten, {\it {Why Does Quantum Field Theory In Curved Spacetime Make Sense?
  And What Happens To The Algebra of Observables In The Thermodynamic Limit?}},
   \href{http://arxiv.org/abs/2112.11614}{{\tt arXiv:2112.11614}}.

\bibitem{BraRobV2}
O.~Bratteli and D.~Robinson, {\em Operator Algebras and Quantum Statistical
  Mechanics}.
\newblock No.~v. 2 in Operator Algebras and Quantum Statistical Mechanics.
  Springer, 1979.

\bibitem{Sew02}
G.~L. Sewell, {\em Quantum Mechanics and Its Emergent Macrophysics}.
\newblock Princeton University Press, 2002.

\bibitem{OhyPet}
M.~Ohya and D.~Petz, {\em Quantum Entropy and Its Use}.
\newblock 1993.

\bibitem{CasTes17}
H.~Casini, E.~Teste, and G.~Torroba, {\it {Modular Hamiltonians on the null
  plane and the Markov property of the vacuum state}},  {\em J. Phys. A} {\bf
  50} (2017), no.~36 364001, [\href{http://arxiv.org/abs/1703.10656}{{\tt
  arXiv:1703.10656}}].

\bibitem{FauLei16}
T.~Faulkner, R.~G. Leigh, O.~Parrikar, and H.~Wang, {\it {Modular Hamiltonians
  for Deformed Half-Spaces and the Averaged Null Energy Condition}},  {\em
  JHEP} {\bf 09} (2016) 038, [\href{http://arxiv.org/abs/1605.08072}{{\tt
  arXiv:1605.08072}}].

\bibitem{BalFau17}
S.~Balakrishnan, T.~Faulkner, Z.~U. Khandker, and H.~Wang, {\it {A General
  Proof of the Quantum Null Energy Condition}},
  \href{http://arxiv.org/abs/1706.09432}{{\tt arXiv:1706.09432}}.

\bibitem{CeyFau18}
F.~Ceyhan and T.~Faulkner, {\it {Recovering the QNEC from the ANEC}},  {\em
  Commun. Math. Phys.} {\bf 377} (2020), no.~2 999--1045,
  [\href{http://arxiv.org/abs/1812.04683}{{\tt arXiv:1812.04683}}].

\bibitem{FauLi18}
T.~Faulkner, M.~Li, and H.~Wang, {\it {A modular toolkit for bulk
  reconstruction}},  {\em JHEP} {\bf 04} (2019) 119,
  [\href{http://arxiv.org/abs/1806.10560}{{\tt arXiv:1806.10560}}].

\bibitem{HolLon25}
S.~Hollands and R.~Longo, {\it {A New Proof of the QNEC}},
  \href{http://arxiv.org/abs/2503.04651}{{\tt arXiv:2503.04651}}.

\bibitem{Ges23b}
E.~Gesteau, {\it {Emergent spacetime and the ergodic hierarchy}},
  \href{http://arxiv.org/abs/2310.13733}{{\tt arXiv:2310.13733}}.

\bibitem{FurLas23}
K.~Furuya, N.~Lashkari, M.~Moosa, and S.~Ouseph, {\it {Information loss, mixing
  and emergent type III$_1$ factors}},
  \href{http://arxiv.org/abs/2305.16028}{{\tt arXiv:2305.16028}}.

\bibitem{OusFur23}
S.~Ouseph, K.~Furuya, N.~Lashkari, K.~L. Leung, and M.~Moosa, {\it {Local
  Poincar{\'e} algebra from quantum chaos}},  {\em JHEP} {\bf 01} (2024) 112,
  [\href{http://arxiv.org/abs/2310.13736}{{\tt arXiv:2310.13736}}].

\bibitem{GesSan24}
E.~Gesteau and L.~Santilli, {\it {Explicit large $N$ von Neumann algebras from
  matrix models}},  \href{http://arxiv.org/abs/2402.10262}{{\tt
  arXiv:2402.10262}}.

\bibitem{Reh99a}
K.-H. Rehren, {\it Algebraic holography},  {\em Annales Henri Poincar\'e} {\bf
  1} (2000) 607--623,
  [\href{http://arxiv.org/abs/http://arXiv.org/abs/hep-th/9905179}{{\tt
  http://arXiv.org/abs/hep-th/9905179}}].

\bibitem{Reh00}
K.-H. Rehren, {\it Local quantum observables in the anti-de{S}itter - conformal
  {QFT} correspondence},  {\em Phys. Lett.} {\bf B493} (2000) 383--388,
  [\href{http://arxiv.org/abs/http://arXiv.org/abs/hep-th/0003120}{{\tt
  http://arXiv.org/abs/hep-th/0003120}}].

\bibitem{DueReh02a}
M.~Duetsch and K.-H. Rehren, {\it {A Comment on the dual field in the scalar
  AdS / CFT correspondence}},  {\em Lett. Math. Phys.} {\bf 62} (2002)
  171--184, [\href{http://arxiv.org/abs/hep-th/0204123}{{\tt hep-th/0204123}}].

\bibitem{DueReh02b}
M.~Duetsch and K.-H. Rehren, {\it {Generalized free fields and the AdS - CFT
  correspondence}},  {\em Annales Henri Poincare} {\bf 4} (2003) 613--635,
  [\href{http://arxiv.org/abs/math-ph/0209035}{{\tt math-ph/0209035}}].

\bibitem{PapRaj13b}
K.~Papadodimas and S.~Raju, {\it {State-Dependent Bulk-Boundary Maps and Black
  Hole Complementarity}},  {\em Phys.Rev.} {\bf D89} (2014) 086010,
  [\href{http://arxiv.org/abs/1310.6335}{{\tt arXiv:1310.6335}}].

\bibitem{Jef18}
R.~Jefferson, {\it {Comments on black hole interiors and modular inclusions}},
  {\em SciPost Phys.} {\bf 6} (2019), no.~4 042,
  [\href{http://arxiv.org/abs/1811.08900}{{\tt arXiv:1811.08900}}].

\bibitem{BraRobV1}
O.~Bratteli and D.~Robinson, {\em Operator Algebras and Quantum Statistical
  Mechanics 1: C*- and W*-Algebras. Symmetry Groups. Decomposition of States}.
\newblock Operator Algebras and Quantum Statistical Mechanics. Springer, 1987.

\bibitem{TakV1}
M.~Takesaki, {\em Theory of Operator Algebras I}.
\newblock Encyclopaedia of Mathematical Sciences. Springer Berlin Heidelberg,
  2001.

\bibitem{TakV2}
M.~Takesaki, {\em Theory of Operator Algebras II}.
\newblock Encyclopaedia of Mathematical Sciences. Springer Berlin Heidelberg,
  2002.

\bibitem{TakV3}
M.~Takesaki, {\em Theory of Operator Algebras III}.
\newblock Encyclopaedia of Mathematical Sciences. Springer Berlin Heidelberg,
  2002.

\bibitem{Sor23a}
J.~Sorce, {\it {Notes on the type classification of von Neumann algebras}},
  {\em Rev. Math. Phys.} {\bf 36} (2024), no.~02 2430002,
  [\href{http://arxiv.org/abs/2302.01958}{{\tt arXiv:2302.01958}}].

\bibitem{CasHue13}
H.~Casini, M.~Huerta, and J.~A. Rosabal, {\it {Remarks on entanglement entropy
  for gauge fields}},  {\em Phys. Rev. D} {\bf 89} (2014), no.~8 085012,
  [\href{http://arxiv.org/abs/1312.1183}{{\tt arXiv:1312.1183}}].

\bibitem{Pow67}
R.~T. Powers, {\it Representations of uniformly hyperfinite algebras and their
  associated von neumann rings},  {\em Annals of Mathematics} {\bf 86} (1967),
  no.~1 138--171.

\bibitem{AraWoo68}
H.~Araki and E.~J. Woods, {\it A classification of factors},  {\em Publications
  of The Research Institute for Mathematical Sciences} {\bf 4} (1968) 51--130.

\bibitem{Ara64}
H.~Araki, {\it {Type of von Neumann Algebra Associated with Free Field}},  {\em
  Progress of Theoretical Physics} {\bf 32} (12, 1964) 956--965.

\bibitem{Lon82}
R.~Longo, {\it {Algebraic and modular structure of von Neumann algebras of
  physics}},  {\em Commun. Math. Phys.} {\bf 38} (1982) 551.

\bibitem{Fre84}
K.~Fredenhagen, {\it {On the Modular Structure of Local Algebras of
  Observables}},  {\em Commun. Math. Phys.} {\bf 97} (1985) 79.

\bibitem{LiHal08}
H.~Li and F.~D.~M. Haldane, {\it Entanglement spectrum as a generalization of
  entanglement entropy: Identification of topological order in non-abelian
  fractional quantum hall effect states},  {\em Phys. Rev. Lett.} {\bf 101}
  (Jul, 2008) 010504.

\bibitem{Sor23b}
J.~Sorce, {\it {An intuitive construction of modular flow}},  {\em JHEP} {\bf
  12} (2023) 079, [\href{http://arxiv.org/abs/2309.16766}{{\tt
  arXiv:2309.16766}}].

\bibitem{Fau25}
T.~Faulkner, {\it {to appear}}, .

\bibitem{BisWic76}
J.~Bisognano and E.~Wichmann, {\it {On the Duality Condition for Quantum
  Fields}},  {\em J.Math.Phys.} {\bf 17} (1976) 303--321.

\bibitem{Unr76}
W.~G. Unruh, {\it Notes on black hole evaporation},  {\em Phys. Rev. D} {\bf
  14} (1976) 870.

\bibitem{Buc74}
D.~Buchholz, {\it {PRODUCT STATES FOR LOCAL ALGEBRAS}},  {\em Commun. Math.
  Phys.} {\bf 36} (1974) 287--304.

\bibitem{DopLon84}
S.~Doplicher and R.~Longo, {\it {Standard and split inclusions of von Neumann
  algebras}},  {\em Invent. Math.} {\bf 75} (1984) 493--536.

\bibitem{BucWic86}
D.~Buchholz and E.~H. Wichmann, {\it {Causal Independence and the Energy Level
  Density of States in Local Quantum Field Theory}},  {\em Commun. Math. Phys.}
  {\bf 106} (1986) 321.

\bibitem{BucFre87}
D.~Buchholz, C.~D'Antoni, and K.~Fredenhagen, {\it {The Universal Structure of
  Local Algebras}},  {\em Commun. Math. Phys.} {\bf 111} (1987) 123.

\bibitem{Few16}
C.~J. Fewster, {\it The split property for quantum field theories in flat and
  curved spacetimes},  \href{http://arxiv.org/abs/1601.06936}{{\tt
  arXiv:1601.06936}}.

\bibitem{Haa87}
U.~Haagerup, {\it {Connes bicentralizer problem and uniqueness of the injective
  factor of type III1}},  {\em Acta Mathematica} {\bf 158} (1987), no.~none 95
  -- 148.

\bibitem{BucDop86}
D.~Buchholz, S.~Doplicher, and R.~Longo, {\it On noether's theorem in quantum
  field theory},  {\em Annals of Physics} {\bf 170} (1986), no.~1 1--17.

\bibitem{ConSto78}
A.~Connes and E.~StÃžrmer, {\it Homogeneity of the state space of factors of
  type iii1},  {\em Journal of Functional Analysis} {\bf 28} (1978), no.~2
  187--196.

\bibitem{CasHue19}
H.~Casini, M.~Huerta, J.~M. Mag\'an, and D.~Pontello, {\it {Entanglement
  entropy and superselection sectors. Part I. Global symmetries}},  {\em JHEP}
  {\bf 02} (2020) 014, [\href{http://arxiv.org/abs/1905.10487}{{\tt
  arXiv:1905.10487}}].

\bibitem{CasHue20}
H.~Casini, M.~Huerta, J.~M. Magan, and D.~Pontello, {\it {Entropic order
  parameters for the phases of QFT}},  {\em JHEP} {\bf 04} (2021) 277,
  [\href{http://arxiv.org/abs/2008.11748}{{\tt arXiv:2008.11748}}].

\bibitem{Bor92}
H.~J. Borchers, {\it The cpt-theorem in two-dimensional theories of local
  observables},  {\em Communications in Mathematical Physics} {\bf 143} (1992)
  315--332.

\bibitem{Bor93}
H.~J. {Borchers}, {\it {On modular inclusion and spectrum condition}},  {\em
  Lett Math Phys} {\bf 27} (1993) 311--324.

\bibitem{Wie93a}
H.-W. {Wiesbrock}, {\it {Symmetries and half-sided modular inclusions of von
  Neumann algebras}},  {\em Letters in Mathematical Physics} {\bf 28} (June,
  1993) 107--114.

\bibitem{Wie93b}
H.~W. Wiesbrock, {\it {Half sided modular inclusions of von Neumann algebras}},
   {\em Commun. Math. Phys.} {\bf 157} (1993) 83--92. [Erratum:
  Commun.Math.Phys. 184, 683--685 (1997)].

\bibitem{Bor96}
H.~Borchers, {\it Half-sided modular inclusion and the construction of the
  poincar{\'e} group},  {\em Communications in mathematical physics} {\bf 179}
  (1996) 703--723.

\bibitem{Bor98}
H.~J. {Borchers}, {\it {Half-Sided Translations and the Type of von Neumann
  Algebras}},  {\em Lett Math Phys} {\bf 44} (1998) 283?290.

\bibitem{Ara05}
H.~Araki and L.~Zsid{\'o}, {\it Extension of the structure theorem of borchers
  and its application to half-sided modular inclusions},  {\em Reviews in
  Mathematical Physics} {\bf 17} (2005), no.~05 491--543.

\bibitem{Lon84}
R.~Longo, {\it Solution of the factorial stone-weierstrass conjecture. an
  application of the theory of standard split w*-inclusions.},  {\em
  Inventiones mathematicae} {\bf 76} (1984) 145--156.

\bibitem{Lon87}
R.~Longo, {\it {Simple and rigid injective subfactors}},  {\em Adv. Math.} {\bf
  63} (1987) 152--171.

\bibitem{Bor99}
H.~J. {Borchers}, {\it {On the Embedding of von Neumann Subalgebras}},  {\em
  Communications in Mathematical Physics} {\bf 205} (Aug., 1999) 69--79.

\bibitem{Dae78}
A.~v. Daele, {\em Continuous Crossed Products and Type III Von Neumann
  Algebras}.
\newblock London Mathematical Society Lecture Note Series. Cambridge University
  Press, 1978.

\bibitem{Tak73}
M.~Takesaki, {\it Duality for crossed products and the structure of von neumann
  algebras of type iii},  {\em Acta Math.} {\bf 131} (1973) 249--310.

\bibitem{Mal01}
J.~M. Maldacena, {\it {Eternal black holes in anti-de Sitter}},  {\em JHEP}
  {\bf 04} (2003) 021, [\href{http://arxiv.org/abs/hep-th/0106112}{{\tt
  hep-th/0106112}}].

\bibitem{SchWit22}
J.-M. Schlenker and E.~Witten, {\it {No ensemble averaging below the black hole
  threshold}},  {\em JHEP} {\bf 07} (2022) 143,
  [\href{http://arxiv.org/abs/2202.01372}{{\tt arXiv:2202.01372}}].

\bibitem{HawPag83}
S.~Hawking and D.~N. Page, {\it {Thermodynamics of Black Holes in anti-De
  Sitter Space}},  {\em Commun.Math.Phys.} {\bf 87} (1983) 577.

\bibitem{Wit98b}
E.~Witten, {\it {A}nti-de~{S}itter space, thermal phase transition, and
  confinement in gauge theories},  {\em Adv. Theor. Math. Phys.} {\bf 2} (1998)
  505, [\href{http://arxiv.org/abs/hep-th/9803131}{{\tt hep-th/9803131}}].

\bibitem{Har15}
D.~Harlow, {\it {Wormholes, Emergent Gauge Fields, and the Weak Gravity
  Conjecture}},  {\em JHEP} {\bf 01} (2016) 122,
  [\href{http://arxiv.org/abs/1510.07911}{{\tt arXiv:1510.07911}}].

\bibitem{MarWal12}
D.~Marolf and A.~C. Wall, {\it {Eternal Black Holes and Superselection in
  AdS/CFT}},  {\em Class.Quant.Grav.} {\bf 30} (2013) 025001,
  [\href{http://arxiv.org/abs/1210.3590}{{\tt arXiv:1210.3590}}].

\bibitem{KabLif13}
D.~Kabat and G.~Lifschytz, {\it {Decoding the hologram: Scalar fields
  interacting with gravity}},  {\em Phys. Rev.} {\bf D89} (2014), no.~6 066010,
  [\href{http://arxiv.org/abs/1311.3020}{{\tt arXiv:1311.3020}}].

\bibitem{HubRan12}
V.~E. Hubeny and M.~Rangamani, {\it {Causal Holographic Information}},  {\em
  JHEP} {\bf 1206} (2012) 114, [\href{http://arxiv.org/abs/1204.1698}{{\tt
  arXiv:1204.1698}}].

\bibitem{Bor61}
H.~J. Borchers, {\it Über die vollständigkeit lorentzinvarianter felder in
  einer zeitartigen röhre},  {\em Il Nuovo Cimento} {\bf 19} (1961) 787.

\bibitem{Ara63}
H.~Araki, {\it A generalization of borchers theorem},  {\em Helvetica Physica
  Acta (Switzerland)}.

\bibitem{StrWit23b}
A.~Strohmaier and E.~Witten, {\it {The Timelike Tube Theorem in Curved
  Spacetime}},  \href{http://arxiv.org/abs/2303.16380}{{\tt arXiv:2303.16380}}.

\bibitem{Str00}
A.~Strohmaier, {\it {On the local structure of the Klein-Gordon field on curved
  space-times}},  {\em Lett. Math. Phys.} {\bf 54} (2000) 249--261,
  [\href{http://arxiv.org/abs/math-ph/0008043}{{\tt math-ph/0008043}}].

\bibitem{BerLiu25}
J.~Berman and H.~Liu, {\it {to appear}}, .

\bibitem{May19}
A.~May, {\it {Quantum tasks in holography}},  {\em JHEP} {\bf 10} (2019) 233,
  [\href{http://arxiv.org/abs/1902.06845}{{\tt arXiv:1902.06845}}]. [Erratum:
  JHEP 01, 080 (2020)].

\bibitem{MayPen19}
A.~May, G.~Penington, and J.~Sorce, {\it {Holographic scattering requires a
  connected entanglement wedge}},  {\em JHEP} {\bf 08} (2020) 132,
  [\href{http://arxiv.org/abs/1912.05649}{{\tt arXiv:1912.05649}}].

\bibitem{May21}
A.~May, {\it {Holographic quantum tasks with input and output regions}},  {\em
  JHEP} {\bf 08} (2021) 055, [\href{http://arxiv.org/abs/2101.08855}{{\tt
  arXiv:2101.08855}}].

\bibitem{MaySor22}
A.~May, J.~Sorce, and B.~Yoshida, {\it {The connected wedge theorem and its
  consequences}},  {\em JHEP} {\bf 11} (2022) 153,
  [\href{http://arxiv.org/abs/2210.00018}{{\tt arXiv:2210.00018}}].

\bibitem{HayNez16}
P.~Hayden, S.~Nezami, X.-L. Qi, N.~Thomas, M.~Walter, and Z.~Yang, {\it
  {Holographic duality from random tensor networks}},  {\em JHEP} {\bf 11}
  (2016) 009, [\href{http://arxiv.org/abs/1601.01694}{{\tt arXiv:1601.01694}}].

\bibitem{Kel16}
W.~R. Kelly, {\it {Bulk Locality and Entanglement Swapping in AdS/CFT}},  {\em
  JHEP} {\bf 03} (2017) 153, [\href{http://arxiv.org/abs/1610.00669}{{\tt
  arXiv:1610.00669}}].

\bibitem{HayPen18}
P.~Hayden and G.~Penington, {\it {Learning the Alpha-bits of Black Holes}},
  {\em JHEP} {\bf 12} (2019) 007, [\href{http://arxiv.org/abs/1807.06041}{{\tt
  arXiv:1807.06041}}].

\bibitem{Pen19}
G.~Penington, {\it {Entanglement Wedge Reconstruction and the Information
  Paradox}},  \href{http://arxiv.org/abs/1905.08255}{{\tt arXiv:1905.08255}}.

\bibitem{AlmEng19}
A.~Almheiri, N.~Engelhardt, D.~Marolf, and H.~Maxfield, {\it {The entropy of
  bulk quantum fields and the entanglement wedge of an evaporating black
  hole}},  {\em JHEP} {\bf 12} (2019) 063,
  [\href{http://arxiv.org/abs/1905.08762}{{\tt arXiv:1905.08762}}].

\bibitem{BanBry16}
S.~Banerjee, J.-W. Bryan, K.~Papadodimas, and S.~Raju, {\it {A toy model of
  black hole complementarity}},  {\em JHEP} {\bf 05} (2016) 004,
  [\href{http://arxiv.org/abs/1603.02812}{{\tt arXiv:1603.02812}}].

\bibitem{SusWit98}
L.~Susskind and E.~Witten, {\it The holographic bound in {A}nti-de~{S}itter
  space},  \href{http://arxiv.org/abs/{h}ep-th/9805114}{{\tt
  {h}ep-th/9805114}}.

\bibitem{PeePol98}
A.~W. Peet and J.~Polchinski, {\it {UV/IR} relations in {AdS} dynamics},  {\em
  Phys. Rev. D} {\bf 59} (1999) 065011,
  [\href{http://arxiv.org/abs/http://arXiv.org/abs/hep-th/9809022}{{\tt
  http://arXiv.org/abs/hep-th/9809022}}].

\bibitem{SusUgl94}
L.~Susskind and J.~Uglum, {\it Black hole entropy in canonical quantum gravity
  and superstring theory},  {\em Phys. Rev. D} {\bf 50} (1994) 2700--2711,
  [\href{http://arxiv.org/abs/hep-th/9401070}{{\tt hep-th/9401070}}].

\bibitem{BouFis15a}
R.~Bousso, Z.~Fisher, S.~Leichenauer, and A.~C. Wall, {\it {Quantum focusing
  conjecture}},  {\em Phys. Rev.} {\bf D93} (2016), no.~6 064044,
  [\href{http://arxiv.org/abs/1506.02669}{{\tt arXiv:1506.02669}}].

\bibitem{ColDon23}
E.~Colafranceschi, X.~Dong, D.~Marolf, and Z.~Wang, {\it {Algebras and Hilbert
  spaces from gravitational path integrals: Understanding Ryu-Takayanagi/HRT as
  entropy without invoking holography}},
  \href{http://arxiv.org/abs/2310.02189}{{\tt arXiv:2310.02189}}.

\bibitem{NebQi23}
T.~M. Nebabu and X.~Qi, {\it {Bulk Reconstruction from Generalized Free
  Fields}},  \href{http://arxiv.org/abs/2306.16687}{{\tt arXiv:2306.16687}}.

\bibitem{Eng16}
N.~Engelhardt, {\it Exploring the bulk in $\mathrm{AdS}/\mathrm{CFT}$: A
  covariant approach},  {\em Phys. Rev. D} {\bf 95} (Mar, 2017) 066005.

\bibitem{Kre45}
M.~Krein, {\it On a problem of extrapolation of a.n. kolmogorov},  {\em Dokl.
  Akadl. Vauk SSSR} {\bf 46} (1945) 306--309.

\bibitem{PolT13}
A.~Poltoratski, {\it A problem on completeness of exponentials},  {\em Annals
  of Mathematics} (2013) 983--1016.

\bibitem{CuiLiu25}
W.~Cui and H.~Liu, {\it to appear}, .

\bibitem{LasLeu24}
N.~Lashkari, K.~L. Leung, M.~Moosa, and S.~Ouseph, {\it {Modular Intersections,
  Time Interval Algebras and Emergent AdS$_2$}},
  \href{http://arxiv.org/abs/2412.19882}{{\tt arXiv:2412.19882}}.

\bibitem{MalSus13}
J.~Maldacena and L.~Susskind, {\it {Cool horizons for entangled black holes}},
  \href{http://arxiv.org/abs/1306.0533}{{\tt arXiv:1306.0533}}.

\bibitem{Van13}
M.~Van~Raamsdonk, {\it {Evaporating Firewalls}},  {\em JHEP} {\bf 11} (2014)
  038, [\href{http://arxiv.org/abs/1307.1796}{{\tt arXiv:1307.1796}}].

\bibitem{VerVer13a}
E.~Verlinde and H.~Verlinde, {\it {Passing through the Firewall}},
  \href{http://arxiv.org/abs/1306.0515}{{\tt arXiv:1306.0515}}.

\bibitem{EngFol22}
N.~Engelhardt and r.~Folkestad, {\it {Canonical purification of evaporating
  black holes}},  {\em Phys. Rev. D} {\bf 105} (2022), no.~8 086010,
  [\href{http://arxiv.org/abs/2201.08395}{{\tt arXiv:2201.08395}}].

\bibitem{EngWal18}
N.~Engelhardt and A.~C. Wall, {\it {Coarse Graining Holographic Black Holes}},
  {\em JHEP} {\bf 05} (2019) 160, [\href{http://arxiv.org/abs/1806.01281}{{\tt
  arXiv:1806.01281}}].

\bibitem{BriLiu07}
M.~Brigante, H.~Liu, R.~C. Myers, S.~Shenker, and S.~Yaida, {\it {Viscosity
  Bound Violation in Higher Derivative Gravity}},  {\em Phys. Rev. D} {\bf 77}
  (2008) 126006, [\href{http://arxiv.org/abs/0712.0805}{{\tt
  arXiv:0712.0805}}].

\bibitem{AhaMar03}
O.~Aharony, J.~Marsano, S.~Minwalla, K.~Papadodimas, and M.~Van~Raamsdonk, {\it
  {The Hagedorn - deconfinement phase transition in weakly coupled large N
  gauge theories}},  {\em Adv. Theor. Math. Phys.} {\bf 8} (2004) 603--696,
  [\href{http://arxiv.org/abs/hep-th/0310285}{{\tt hep-th/0310285}}].

\bibitem{FesLiu06}
G.~Festuccia and H.~Liu, {\it {The Arrow of time, black holes, and quantum
  mixing of large N Yang-Mills theories}},  {\em JHEP} {\bf 12} (2007) 027,
  [\href{http://arxiv.org/abs/hep-th/0611098}{{\tt hep-th/0611098}}].

\bibitem{ChePen24}
C.-H. Chen and G.~Penington, {\it {A clock is just a way to tell the time:
  gravitational algebras in cosmological spacetimes}},
  \href{http://arxiv.org/abs/2406.02116}{{\tt arXiv:2406.02116}}.

\bibitem{HoeSmi19}
P.~A. Hoehn, A.~R.~H. Smith, and M.~P.~E. Lock, {\it {Trinity of relational
  quantum dynamics}},  {\em Phys. Rev. D} {\bf 104} (2021), no.~6 066001,
  [\href{http://arxiv.org/abs/1912.00033}{{\tt arXiv:1912.00033}}].

\bibitem{Kuc94}
K.~V. Kuchar, {\it {Geometrodynamics of Schwarzschild black holes}},  {\em
  Phys. Rev. D} {\bf 50} (1994) 3961--3981,
  [\href{http://arxiv.org/abs/gr-qc/9403003}{{\tt gr-qc/9403003}}].

\bibitem{NarVer23}
V.~Narovlansky and H.~Verlinde, {\it {Double-scaled SYK and de Sitter
  holography}},  {\em JHEP} {\bf 05} (2025) 032,
  [\href{http://arxiv.org/abs/2310.16994}{{\tt arXiv:2310.16994}}].

\bibitem{GibHaw77a}
G.~W. Gibbons and S.~W. Hawking, {\it Cosmological event horizons,
  thermodynamics, and particle creation},  {\em Phys. Rev. D} {\bf 15} (1977)
  2738--2751.

\bibitem{Ban00}
T.~Banks, {\it Cosmological breaking of supersymmetry or little {L}ambda goes
  back to the future {II}},  \href{http://arxiv.org/abs/hep-th/0007146}{{\tt
  hep-th/0007146}}.

\bibitem{MaeKoi98}
K.~Maeda, T.~Koike, M.~Narita, and A.~Ishibashi, {\it Upper bound for entropy
  in asymptotically de~{S}itter space-time},  {\em Phys. Rev. D} {\bf 57}
  (1998) 3503--3508, [\href{http://arxiv.org/abs/gr-qc/9712029}{{\tt
  gr-qc/9712029}}].

\bibitem{Mon78}
V.~Moncrief, {\it {Invariant States and Quantized Gravitational
  Perturbations}},  {\em Phys. Rev. D} {\bf 18} (1978) 983--989.

\bibitem{Mon79}
V.~Moncrief, {\it {QUANTUM LINEARIZATION INSTABILITIES}},  {\em Gen. Rel.
  Grav.} {\bf 10} (1979) 93--97.

\bibitem{KapMar24}
M.~Kaplan, D.~Marolf, X.~Yu, and Y.~Zhao, {\it {De Sitter quantum gravity and
  the emergence of local algebras}},  {\em JHEP} {\bf 04} (2025) 171,
  [\href{http://arxiv.org/abs/2410.00111}{{\tt arXiv:2410.00111}}].

\bibitem{FigHoe75}
R.~Figari, R.~Hoegh-Krohn, and C.~R. Nappi, {\it Interacting relativistic boson
  fields in the de {S}itter universe with two space-time dimensions},  {\em
  Commun. Math. Phys.} {\bf 44} (1975) 265.

\bibitem{Sor24}
J.~Sorce, {\it {Analyticity and the Unruh effect: a study of local modular
  flow}},  {\em JHEP} {\bf 24} (2024) 040,
  [\href{http://arxiv.org/abs/2403.18937}{{\tt arXiv:2403.18937}}].

\bibitem{ABan23}
S.~A, T.~Banks, and W.~Fischler, {\it {Quantum theory of three-dimensional de
  Sitter space}},  {\em Phys. Rev. D} {\bf 109} (2024), no.~2 025011,
  [\href{http://arxiv.org/abs/2306.05264}{{\tt arXiv:2306.05264}}].

\bibitem{MalShe15}
J.~Maldacena, S.~H. Shenker, and D.~Stanford, {\it {A bound on chaos}},  {\em
  JHEP} {\bf 08} (2016) 106, [\href{http://arxiv.org/abs/1503.01409}{{\tt
  arXiv:1503.01409}}].

\bibitem{AalShi20}
L.~Aalsma and G.~Shiu, {\it {Chaos and complementarity in de Sitter space}},
  {\em JHEP} {\bf 05} (2020) 152, [\href{http://arxiv.org/abs/2002.01326}{{\tt
  arXiv:2002.01326}}].

\bibitem{Kol25}
D.~K. Kolchmeyer, {\it {Bottom-up de Sitter Holography}},
  \href{http://arxiv.org/abs/talk at IAS, Sept. 19, 2025}{{\tt talk at IAS,
  Sept. 19, 2025}}.

\bibitem{MerTur22}
T.~G. Mertens and G.~J. Turiaci, {\it {Solvable models of quantum black holes:
  a review on Jackiw{\textendash}Teitelboim gravity}},  {\em Living Rev. Rel.}
  {\bf 26} (2023), no.~1 4, [\href{http://arxiv.org/abs/2210.10846}{{\tt
  arXiv:2210.10846}}].

\bibitem{MalSta16a}
J.~Maldacena, D.~Stanford, and Z.~Yang, {\it {Conformal symmetry and its
  breaking in two dimensional Nearly Anti-de-Sitter space}},  {\em PTEP} {\bf
  2016} (2016), no.~12 12C104, [\href{http://arxiv.org/abs/1606.01857}{{\tt
  arXiv:1606.01857}}].

\bibitem{HarJaf18}
D.~Harlow and D.~Jafferis, {\it {The Factorization Problem in Jackiw-Teitelboim
  Gravity}},  {\em JHEP} {\bf 02} (2020) 177,
  [\href{http://arxiv.org/abs/1804.01081}{{\tt arXiv:1804.01081}}].

\bibitem{Erb21}
H.~Erbin, {\em {String Field Theory: A Modern Introduction}}, vol.~980 of {\em
  Lecture Notes in Physics}.
\newblock 3, 2021.

\bibitem{SenZwi24}
A.~Sen and B.~Zwiebach, {\it {String Field Theory: A Review}},
  \href{http://arxiv.org/abs/2405.19421}{{\tt arXiv:2405.19421}}.

\end{thebibliography}\endgroup
